# Transmission des Ondes Électromagnétiques

Cevdet AKYEL

Slobodan BABIC

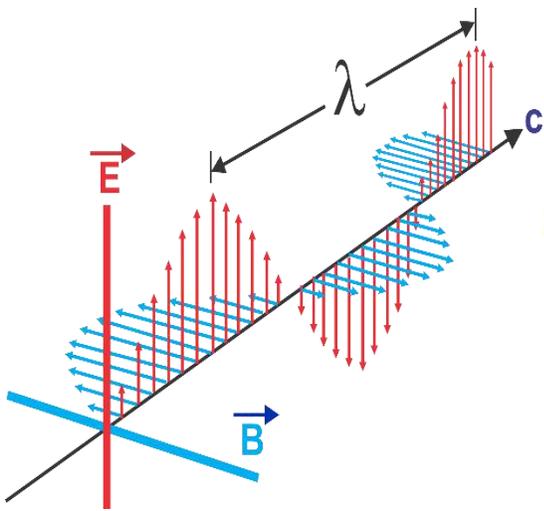

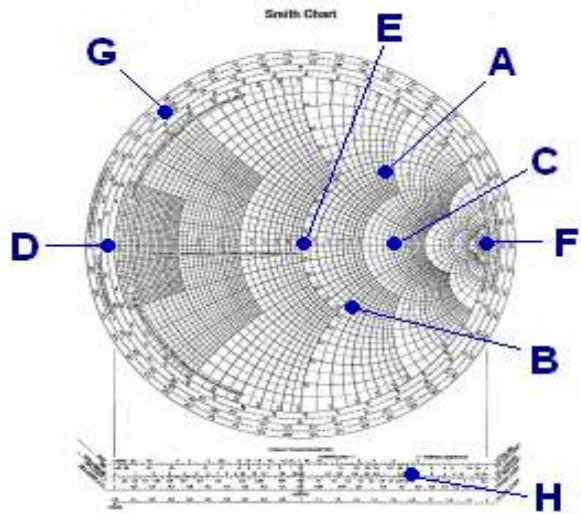

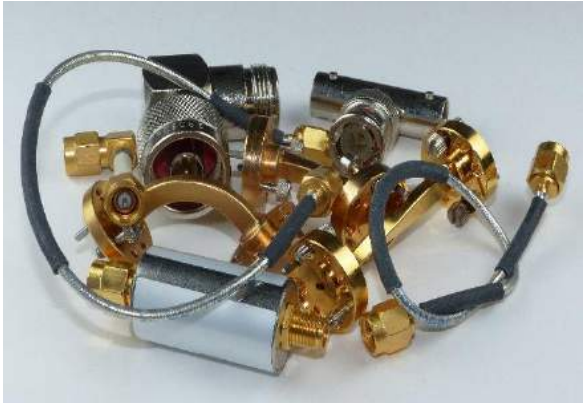

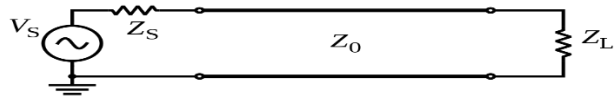

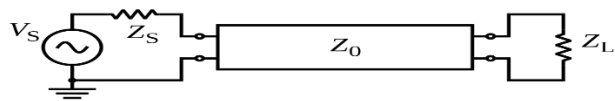

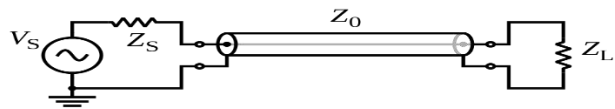

# Transmission des Ondes Électromagnétiques

Cevdet AKYEL

Slobodan BABIC

PRÉFACE

Le recueil présent vise à familiariser les étudiants en Génie Électrique aux applications en exercices et problèmes concernant la transmission des ondes électromagnétiques. Pour faciliter la tâche, une brève introduction théorique avec les formules classiques de l'électromagnétisme est inclue dans la perspective d'afficher leur utilisation pour résoudre ces problèmes. Donc cette partie théorique n'apporte rien de nouveau aux connaissances existantes depuis très longtemps. Sa présence vise à faire le mariage entre la théorie et des applications dans les résolutions des problèmes rencontrés pour les transmissions des ondes électromagnétiques.

Le contenu de ce livre ne couvre pas la source, l'émission ni la réception de ces ondes assurées par les antennes qui est tout un autre domaine en soi. Le domaine des antennes qui exige à lui seul, toute une expertise très étendue, peut amener à la publication de plusieurs livres selon les points de vue considérés dans la technologie de la télécommunication de nos jours

La production des ondes électromagnétiques ne fait pas partie de cet ouvrage non plus.

La transmission des ondes électromagnétiques est essentiellement assurée en espace libre, en câbles coaxiaux et en guides d'ondes de diverses formes. Le matériel qui remplit ces moyens de transmission joue un rôle essentiel dans ce processus de transmission. La permittivité complexe de ces milieux est un acteur principal pour le but visé dans cette opération.

Le livre est conçu de façon à aider le lecteur à résoudre des problèmes à l'aide d'une partie théorique concise pour la compréhension de la solution choisie à travers divers exemples.

Quant à la spécificité du contenu, un étudiant en Génie Électrique projetant de devenir un ingénieur junior dans le domaine peut se poser la question suivante : Sur une ligne bifilaire pourquoi doit-on s'intéresser à la transmission du signal de la source à la charge puisque tout ce qu'on fait on relie la source à charge par ce moyen intermédiaire qui ne doit pas influencer en premier lieu la nature des phénomènes.

Ceci est vrai pour les courtes distances de connexion et surtout à basse fréquence. Lorsque la longueur du fils augmente et la fréquence monte, il y aura le phénomène d'affaiblissement du signal et son changement de phase avec les longueurs d'onde relativement courtes comparé à la longueur de lignes de transmission. Cela va causer un changement d'amplitude et de phase du signal à l'arrivé sur la charge.

D'autre part le matériel qui remplit ces moyens de transmission joue un rôle essentiel dans ce processus de transmission. La permittivité complexe de ces milieux est un acteur principal pour le but visé dans cette opération.
Pour la propagation libre les milieux de propagation jouent le même rôle que les lignes de transmission. En plus les problèmes de réfraction et de réflexions compliquent encore le processus.

Dans les guides d'ondes, la propagation s'effectue d'une façon guidée selon les caractéristiques du milieu de propagation plus les effets des parois qui forment les murs de ces guides. Dans les télécommunications d'aujourd'hui, les fibres optiques jouent aussi un grand rôle dans les transmissions du signal, mais ce sujet est exclu du présent ouvrage.

La préparation de ce recueil se base sur les cours enseignés à l'École Polytechnique de Montréal durant les années 1986-2017.

Les exercices, les devoirs et les problèmes dans les examens de ces cours forment la principale source d'inspiration.

**TABLE DES MATIÈRES**

# Chapitre 1

# 1. Ligne de transmission

## 1.1 DÉFINITION

***Une ligne de transmission est un milieu, qui par son parcours, permet de transmettre un signal d'information de l'énergie d'un point à l'autre sans déformation du signal et avec un minimum de pertes.***

LIGNES TYPIQUES :

- Lignes bifilaires
- Lignes coaxiales
- Lignes à ruban, à micro-ruban
- Guides d'onde
- Fibres optiques
- Etc.

Les lignes à deux ou plusieurs conducteurs peuvent supporter les modes TEM (usage courant).

## 1.2 PARAMÈTRES DISTRIBUÉS

Contrairement aux éléments de circuits électriques formés de composantes discrètes (lumped elements) qui peuvent être placées à différents endroits dans un circuit pour produire les effets $R$, $L$, $C$, $G$, une ligne de transmission peut engendrer les mêmes effets tout au long d'une section donc, par conséquent, créer des pertes et emmagasiner de l'énergie sous forme électrique ou magnétique.

➜ Pertes :
$$\frac{R}{unité\ de\ longueur}$$

$$\frac{G}{unité\ de\ longueur}$$

(Fuites entre les conducteurs)

➜ Énergie emmagasinée : $\dfrac{C,L}{unité\ de\ longueur}$

## 1.3    MODÉLISATION

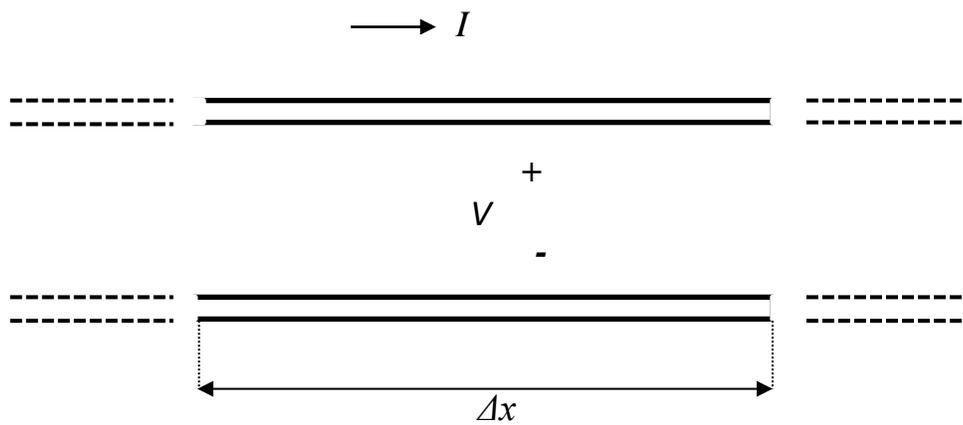

Figure I.1.

On considère une section de ligne (Figure I.1) ayant une longueur

$$\Delta x \ \ll \ \lambda\ (longueur\ d'onde)$$

Le schéma équivalent (Figure I.2) de cette cellule est suivant :

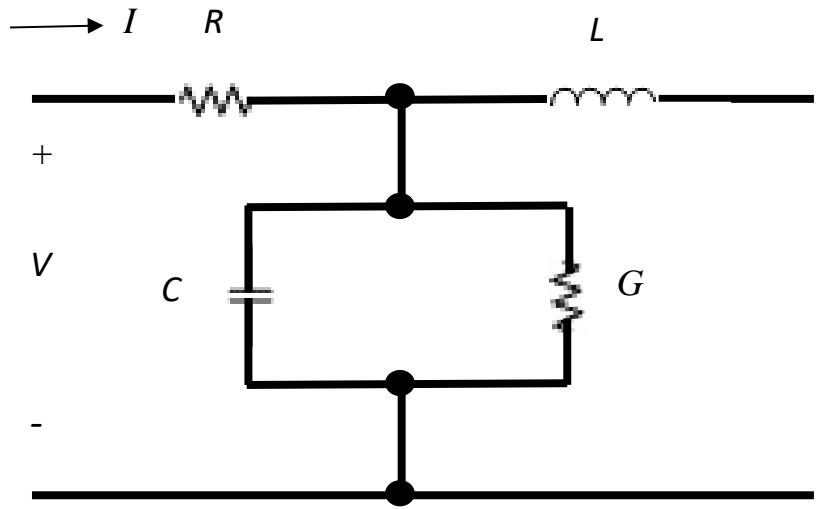

Figure I.2.

## 1.4  VALEURS TYPIQUES

➢ Ancienne ligne télégraphique

-  Distance entre les conducteurs : 30.48 cm
-  Diamètre de fils de fer : 0.26416

$R$  = 10.44        ($\Omega/m$  (chaque conducteur)
$L$  = 3.66        ($milli\ henry/\ m$)
$C$  = 0.00838    ($\mu F/m$)
$G$  ≈ 0

➢ Ligne haute tension

-  Distance entre les conducteurs 990.6 cm
-  Diamètre de chaque conducteur 3.556 cm

$R$  = 0.113        ($\Omega/m$)(chaque conducteur)
$L$  = 0.00212    ($henry/\ m$)
$C$  = 0.0136      ($\mu F/m$)
$G$  ≈ 0

## 1.5    RELATIONS TENSION-COURANT SUR UNE LIGNE

Soit $\qquad\qquad$ $\mathbb{Z} \triangleq R + j\omega L$ $\qquad$ (impédance série)

$\qquad\qquad\qquad$ $Y \triangleq G + j\omega C$ $\qquad$ (admittance shunt)

En régime permanent, la tension et le courant varient en fonction de la position et du temps.

Si l'on considère seulement les phaseurs,

1) $\quad \dfrac{dV}{dx} = -I\mathbb{Z}$

2) $\quad \dfrac{dI}{dx} = -VY$

où x est la variable selon la direction longitudinale de la ligne.

Si l'on procède par une seconde dérivée par rapport à $x$ :

$$\frac{d^2V}{dx^2} = -I\frac{d\mathbb{Z}}{dx} - \mathbb{Z}\frac{dI}{dx} = -I\frac{d\mathbb{Z}}{dx} + \mathbb{Z}VY$$

$$\frac{d^2I}{dx^2} = -V\frac{dY}{dx} - Y\frac{dV}{dx} = -V\frac{dY}{dx} + Y\mathbb{Z}I$$

Pour une ligne de transmission ayant une section uniforme et sans discontinuités,

$$\frac{dZ}{dx} = 0 \qquad ; \qquad \frac{dY}{dx} = 0$$

Les paramètres de la ligne sont indépendants de la position.

Il reste donc :

3) $\quad \dfrac{d^2V}{dx^2} - \mathbb{Z}YV = 0$ $\left.\rule{0pt}{42pt}\right\}$ Équations d'onde pour les lignes de transmission

4) $\quad \dfrac{d^2I}{dx^2} - \mathbb{Z}YI = 0$

Soit une solution de la forme :

$$V = e^{\gamma x}$$

$$\frac{d^2V}{dx^2} = \gamma^2 e^{\gamma x} = \gamma^2 V$$

Si l'on utilise cette dernière expression on obtient:

$$(\gamma^2 - ZY)V = 0$$

$$\gamma^2 - ZY = 0$$

$$\gamma = \pm\sqrt{ZY}$$

où $\gamma$ est le facteur de propagation.

Solution générale :

$$V(x) = V_1 e^{-\sqrt{ZY}x} + V_2 e^{+\sqrt{ZY}x}$$

où $V_1 e^{-\sqrt{ZY}}$ est l'onde incidente et $V_2 e^{+\sqrt{ZY}}$ est l'onde réfléchie.

Le facteur de propagation est :

$$\gamma = \sqrt{ZY} = \sqrt{(R + j\omega L)(G + j\omega C)}$$

$$\gamma = \alpha + j\beta$$

$$\alpha = Re\left\{\sqrt{ZY}\right\} \qquad facteur\ d'atténuation(néper/m)$$

$$\beta = Im\left\{\sqrt{ZY}\right\} \qquad facteur\ de\ phase\left(\frac{rad}{m}\right)$$

Si l'on effectue le développement,

$$(\alpha + j\beta)^2 = (R + j\omega L)(G + j\omega C)$$

$$\alpha^2 - \beta^2 = RG - \omega^2 LC$$

$$2j\alpha\beta = j(\omega LG + \omega RC)$$

En solutionnant pour $\alpha$ et $\beta$ on obtient:

$$\alpha = \sqrt{\frac{1}{2}\left\{RG - \omega^2 LC + \sqrt{(R^2 + \omega^2 L^2)(G^2 + \omega^2 C^2)}\right\}}\ (\frac{Np}{m})$$

$$\beta = \sqrt{\frac{1}{2}\left\{\omega^2 LC - RG + \sqrt{(R^2 + \omega^2 L^2)(G^2 + \omega^2 C^2)}\right\}}\ (\frac{rad}{m})$$

Ces deux paramètres sont fonction de :

- La fréquence
- La température
- L'effet pelliculaire

En régime permanent la tension et le courant instantanés peuvent s'écrire :

$$\mathcal{V}(x,t) = Re\left\{V_1 e^{j\omega t} e^{-(\alpha + j\beta)x} + V_2 e^{j\omega t} e^{(\alpha + j\beta)x}\right\}$$

$$\mathcal{J}(x,t) = Re\left\{I_1 e^{j\omega t} e^{-(\alpha + j\beta)x} + I_2 e^{j\omega t} e^{(\alpha + j\beta)x}\right\}$$

On peut écrire ces deux expressions d'une façon différente, par exemple

$$\mathcal{J}(x,t) = Re\left\{I_1 e^{-\alpha x} e^{j(\omega t + \beta x)} + I_2 e^{\alpha x} e^{j(\omega t + \beta x)}\right\}$$

## 1.6 Relation entre *V* et *I* sur une ligne

$$\frac{dV}{dx} = -I\mathbb{Z}$$

$$V(x) = V_1 e^{-\sqrt{\mathbb{Z}Y}x} + V_2 e^{+\sqrt{\mathbb{Z}Y}x}$$

$$\frac{dV}{dx} = -\sqrt{\mathbb{Z}Y}V_1 e^{-\sqrt{\mathbb{Z}Y}x} + \sqrt{\mathbb{Z}Y}V_2 e^{\sqrt{\mathbb{Z}Y}x}$$

$$I = -\frac{1}{\mathbb{Z}}\frac{dV}{dx}$$

$$I(x) = \frac{\sqrt{\mathbb{Z}Y}}{\mathbb{Z}}V_1 e^{-\sqrt{\mathbb{Z}Y}x} - \frac{\sqrt{\mathbb{Z}Y}}{\mathbb{Z}}V_2 e^{\sqrt{\mathbb{Z}Y}x}$$

$$I(x) = \frac{V_1}{\sqrt{\dfrac{\mathbb{Z}}{Y}}} e^{-\sqrt{ZY}x} - \frac{V_2}{\sqrt{\dfrac{\mathbb{Z}}{Y}}} e^{\sqrt{ZY}x}$$

Si l'on considère seulement l'onde qui se transmet dans la direction (+*x*),

$$\frac{V}{I} = \frac{V_1 e^{-\sqrt{ZY}x}}{\dfrac{V_1}{\sqrt{\dfrac{\mathbb{Z}}{Y}}} e^{-\sqrt{ZY}x}} = \sqrt{\frac{\mathbb{Z}}{Y}} \triangleq \mathbb{Z}_0$$

où $\mathbb{Z}_0$ : impédance caractéristique de la ligne,

$$\mathbb{Z}_0 \triangleq \sqrt{\frac{R + j\omega L}{G + j\omega C}}$$

Étant donné que sur une ligne de transmission on cherche à obtenir des pertes aussi faibles qu'il est économiquement possible de le faire, il arrive souvent le cas suivant :

$$\omega L \gg R$$
$$\omega C \gg G$$

d'où

$$\mathbb{Z}_0 = \sqrt{\frac{L}{C}} \quad \text{(surtout valable à hautes fréquence)}$$

Pour une ligne sans pertes,

$$\gamma = \sqrt{ZY} = \sqrt{(R + j\omega L)(G + j\omega C)}$$

$$\gamma = j\omega\sqrt{LC} \Rightarrow \alpha = 0 \; ; \beta = \omega\sqrt{LC} \; rad/m$$

La vitesse de phase devient alors :

$$v_p = \frac{\omega}{\beta} = \frac{\omega}{\omega\sqrt{LC}} = \frac{1}{\sqrt{LC}}$$

$$\beta = \omega\sqrt{LC}$$

## 1.7    LIGNE DE TRANSMISSION SANS DISTORSION

En pratique, toutes les lignes de transmission ont des pertes et ces pertes varient en fonction de la fréquence. Normalement, les fréquences plus élevées sont plus fortement atténuées causant ainsi une déformation du spectre de fréquences d'où l'expression anglaise « Frequency Distorsion ».

Il est cependant possible (en théorie) de construire une ligne de transmission ayant un facteur d'atténuation indépendant de la fréquence :

$$\alpha = \sqrt{\frac{1}{2}\left\{RG - \omega^2 LC + \sqrt{(R^2 + \omega^2 L^2)(G^2 + \omega^2 C^2)}\right\}}\left(\frac{Np}{m}\right)$$

Si l'on effectue $\frac{d\alpha}{d\omega} = 0$ on obtient la relation:

$$\frac{R}{G} = \frac{L}{C}$$

Le facteur d'atténuation devient alors:

$$\alpha = \sqrt{RG}$$

C'est en même temps un minimum pour le facteur d'atténuation.

Dans telles conditions,

$$\beta = \omega\sqrt{LC} \quad et \quad v_p = \frac{1}{\sqrt{LC}}$$

En pratique les lignes sans distorsion n'existent pas. Il est cependant possible de tendre vers cette condition sur une bande de fréquence donnée.

1)  Compenser à l'aide d'un circuit électronique
2)  Modifier la construction physique de la ligne

Pour satisfaire la relation,

$$\frac{R}{G} = \frac{L}{C}$$

on considère:

$G$ est faible en général, car on veut minimiser les fuites entre les conducteurs formant la ligne.

$$\frac{R}{G} \ \rightarrow \ grand$$

Pour minimiser $R$, il faut utiliser de plus gros conducteurs (très coûteux). Il faut donc que,

$$\frac{L}{C} \ \rightarrow \ grand$$

Si l'on désire un $C$ petit, il faut augmenter la distance entre les conducteurs (cas pratique).

Reste à augmenter $L$ :

1) En enroulant un ruban ferromagnétique autour de chaque conducteur
2) En ajoutant de petites inductances en certain points sur la ligne (Pupinisation – Loading coils de Michel Pupin)

## Exemple I.1.

Une ligne de transmission a un facteur d'atténuation de α = 0.002 (m$^{-1}$). On remarque que les signaux se propagent sans distorsion à une vitesse de $v_p$ = 3x10$^8$ (m/s). L'impédance caractéristique nominale de la ligne est de $Z_0$ =50 ($\Omega$). La ligne peut être considérée comme étant à faibles pertes. Déterminez les paramètres linéiques $R, L, G$ et $G$.

Solution :

Si la ligne est considérée comme étant à faibles pertes et sans distorsion on a :

$$Z_0 = \sqrt{\frac{L}{C}} \qquad , \qquad v_p = \frac{1}{\sqrt{LC}}$$

$$\alpha = \frac{\sqrt{LC}}{2}\left[\frac{R}{L} + \frac{G}{C}\right]$$

$$\frac{L}{C} = \frac{R}{G}$$

Ce system d'équations donne :

$$L = \frac{Z_0}{v_p} = 166.667(\frac{nH}{m}) \ , \ \ C = \frac{1}{Z_0 v_p} = 66.667\left(\frac{pF}{m}\right)$$

$$R = \alpha Z_0 = 0.1(\frac{\Omega}{m}), \qquad G = \frac{R}{Z_0^2} = 40(\frac{\mu S}{m})$$

**Exemple I.2.**

Une ligne de transmission a une impédance caractéristique de 50 (Ω) à très hautes fréquences. À une fréquence relativement basse cette impédance caractéristique devient :

$$Z_0 = \sqrt{\frac{3 + j7500}{4 + jX}}$$

a) Trouver *X*.

b) Quel élément discret représente la valeur da *X* dans la modélisation des lignes de transmission en général ?

Solution :

a) Pour des fréquences assez élevées (pratiquement $\omega \to \infty$ ) l'impédance caractéristique est,

$$Z_0 = \sqrt{\frac{L}{C}}$$

sinon,

$$Z_0 = \sqrt{\frac{Z}{Y}} = \sqrt{\frac{R + j\omega L}{G + j\omega C}} \approx \sqrt{\frac{L}{C}} \text{ pour } R \ll \omega L \text{ et } G \ll \omega C$$

On obtient

$$50^2 = \frac{L}{C}$$

$$Z_0 = \sqrt{\frac{3 + j7500}{4 + jX}} = \sqrt{\frac{\frac{3}{7500} + j}{\frac{4}{7500} + j\frac{X}{7500}}} \approx \sqrt{\frac{j}{j\frac{X}{7500}}} = \sqrt{\frac{7500}{X}}$$

d'ou

$$50^2 = \frac{7500}{X} \rightarrow X = 3 \ (\Omega) = \omega C$$

b) $X$ représente l'effet capacitif de la ligne de transmission.

## 1.8   IMPÉDANCE CARACTÉRISTIQUE DE DIFFÉRENTES LIGNES

Pour toutes les lignes sans pertes avec le mode TEM

$$Z_0 = \sqrt{\frac{L}{C}}$$

$L$ : Inductance par unité de longueur
$C$ : Capacité par unité de longueur
$\varepsilon$ : Constante diélectrique, $\varepsilon = \varepsilon_r.\varepsilon_0$,  $\varepsilon_0 = 10^{-9}/(36\pi)$ (F/m)
$\mu$: Constante magnétique, , $\mu = \mu_r\mu_0$,  $\mu_0 = 4\pi x 10^{-9}$ (H/m)
$\varepsilon_r, \mu_r -$ constantes relatives $de\ \varepsilon$ et $\mu$

Coaxiale

$$C = \frac{2\pi\varepsilon}{\ln\frac{b}{a}} \qquad\qquad \rightarrow \frac{\text{charge totale accumulée}}{\text{différence de tension}}$$

$$L = \frac{\mu}{2\pi} \ln\frac{b}{a} \qquad\qquad \rightarrow \frac{\text{flux magnétique total}}{\text{courant}}$$

$$\mathbb{Z}_0 = \sqrt{\frac{L}{C}} = \left\{ \frac{\mu/_{2\pi} \ln \frac{b}{a}}{2\pi\varepsilon / \ln \frac{b}{a}} \right\}^{1/2}$$

$$\mathbb{Z}_0 = \frac{1}{2\pi} \sqrt{\frac{\mu}{\varepsilon}} \ln \frac{b}{a}$$

$$\mathbb{Z}_0 = \frac{1}{2\pi} \sqrt{\frac{\mu_0}{\varepsilon_0}} \frac{1}{\sqrt{\varepsilon_r}} \ln \frac{b}{a}$$

$$\mathbb{Z}_0 = \frac{120\pi}{2\pi} \frac{1}{\sqrt{\varepsilon_r}} \ln \frac{b}{a} = \frac{60}{\sqrt{\varepsilon_r}} \ln \frac{b}{a}$$

$$\mathbb{Z}_0 = \frac{138}{\sqrt{\varepsilon_r}} \log_{10} \frac{b}{a}$$

Si l'on compare l'expression de *C* et de *L* pour une ligne coaxiale, on remarque qu'une des lignes est réciproque de l'autre par rapport aux spécificités près de ligne.

$$C = \frac{2\pi\varepsilon}{\ln \frac{b}{a}} \qquad\qquad L = \frac{\mu}{2\pi} \ln \frac{b}{a}$$

Il semble que ceci est valable pour toutes les lignes sans pertes. Il est donc suffisant de calculer C et L pour obtenir $\mathbb{Z}_0$.

<u>Lignes bifilaires (Figure I.3)</u>

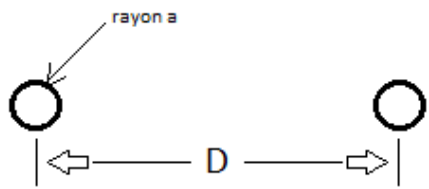

Figure I.3.

$$\mathbb{Z}_0 = \frac{1}{\pi} \sqrt{\frac{\mu}{\varepsilon}} \ln\frac{D}{a}$$

## Elliptique focale (Figure I.4)

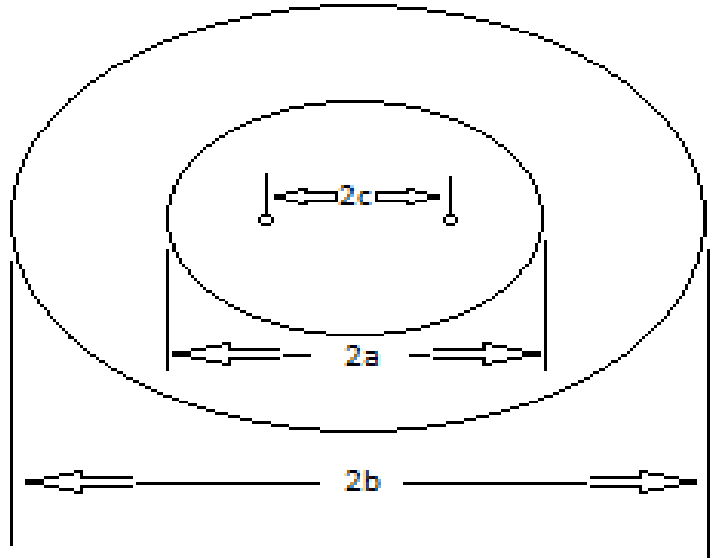

Figure I.4.

$$\mathbb{Z}_0 = \frac{1}{2\pi} \sqrt{\frac{\mu}{\varepsilon}} \ln\left\{\frac{b + \sqrt{b^2 - c^2}}{a + \sqrt{a^2 - c^2}}\right\}$$

## Plaques parallèles (Figures I.5)

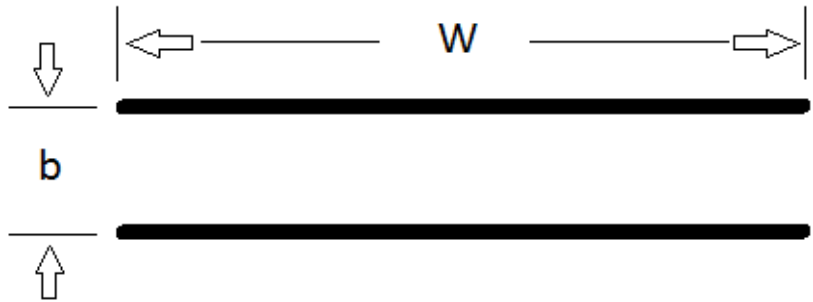

Figure I.5.

$$\mathbb{Z}_0 \approx \sqrt{\frac{\mu}{\varepsilon}} \, \frac{b}{w} \qquad w \gg b$$

## Plaques colinéaires (Figures I.6)

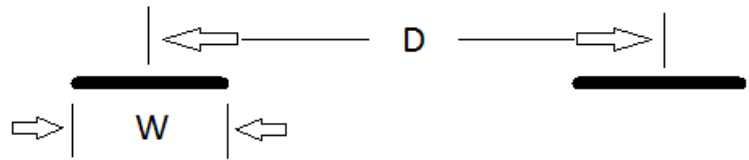

Figure I.6.

$$\mathbb{Z}_0 \approx \sqrt{\frac{\mu}{\varepsilon}} \, \frac{1}{\pi} \ln\left\{\frac{4D}{w}\right\} \qquad D \gg w$$

## Conducteur cylindrique parallèle à un plan conducteur (Figure I.7)

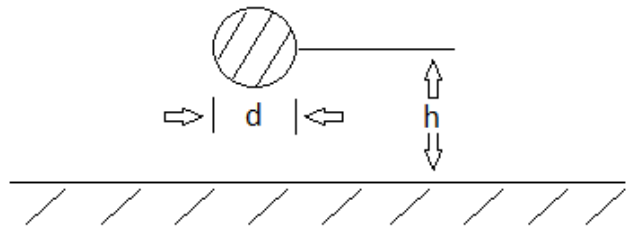

Figure I.7.

$$\mathbb{Z}_0 \approx \frac{1}{2\pi} \sqrt{\frac{\mu}{\varepsilon}} \, \ln\left\{\frac{4h}{d}\right\} \qquad h \gg d$$

## Bifilaire blindée (Figure 1.8)

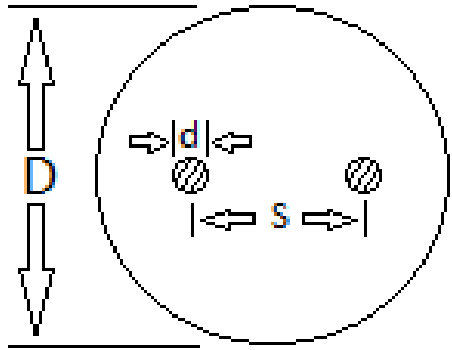

Figure I.8.

$$Z_0 \approx \frac{1}{2\pi} \sqrt{\frac{\mu}{\varepsilon}} \ln\left\{\frac{2S}{d} \frac{D^2 - S^2}{D^2 + S^2}\right\} \qquad D \gg d \qquad S \gg d$$

## Cylindre dans un tunnel (Figure I.9)

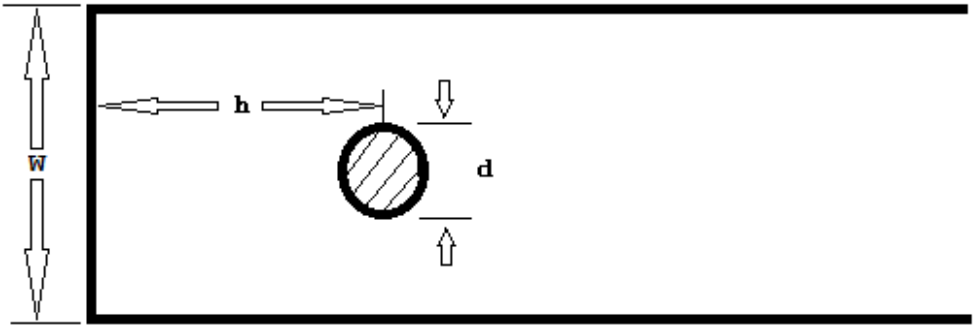

Figure I.9.

$$Z_0 \approx \frac{1}{2\pi} \sqrt{\frac{\mu}{\varepsilon}} \ln\left\{\frac{4W}{\pi d} \tanh(\frac{\pi h}{W})\right\}$$

## Ligne à ruban (Figure I.10)

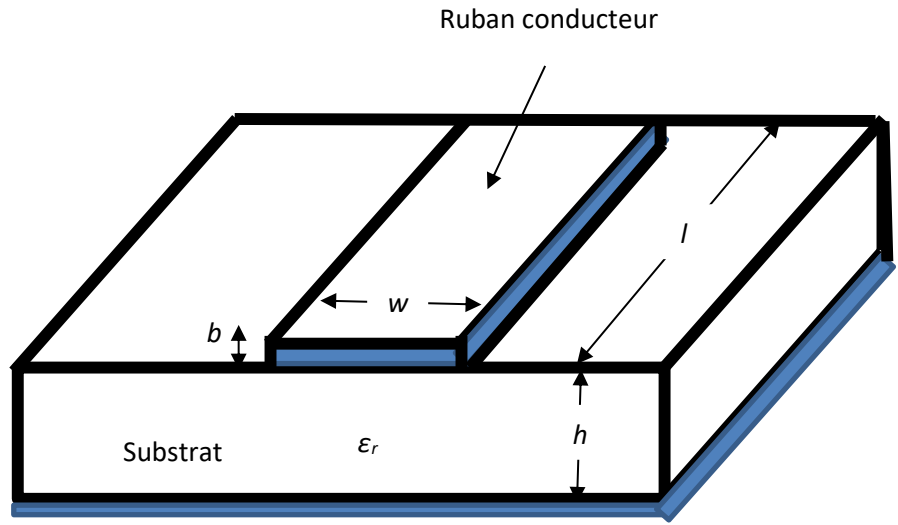

Figure I.10.

$$\mathbb{Z}_0 = \frac{120\pi}{\sqrt{\varepsilon_e}} \ \left\{\frac{w}{h} + 1.393 + 0.667\ln(\frac{w}{h}+1.444)\right\}^{-1} \ \text{si } \frac{w}{h} > 1$$

et

$$\varepsilon_e = \frac{1}{2}(\varepsilon_r + 1) + \frac{1}{2}(\varepsilon_r - 1)(1 + 12\frac{h}{w})^{-0.5}$$

$$\mathbb{Z}_0 = \frac{60}{\sqrt{\varepsilon_e}} \ ln\left\{8\frac{w}{h}+\frac{w}{4h}\right\}^{-1} \ \text{si } \frac{w}{h} < 1$$

et

$$\varepsilon_e = \frac{1}{2}(\varepsilon_r + 1) + \frac{1}{2}(\varepsilon_r - 1)\{\left(1 + 12\frac{h}{w}\right)^{-0.5} + 0.04\left(1 - \frac{w}{h}\right)^2\}$$

Pour tenir compte de l'épaisseur du ruban, il faut remplacer sa largeur réelle *w* par une largeur équivalente $w_e$,

$$w_e = w + \frac{b}{\pi}(1 + ln\frac{2x}{b})$$

avec,

$$x = h \qquad \text{si} \qquad w > \frac{h}{2\pi}$$

$$x = 2\pi w \qquad \text{si} \qquad 2b < w < \frac{h}{2\pi}$$

$\varepsilon_e$ – constante diélectrique effective,

$b$  –  épaisseur du ruban,

$w$  –  largeur du ruban,

$h$  –  hauteur du substrat.

# 2. Réflexion sur les lignes de transmission

On considère la ligne de transmission sur la Figure I.11.

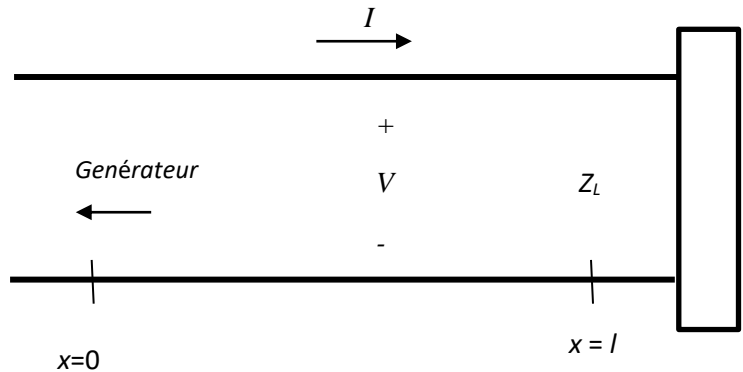

Figure I.11.

La tension et le courant à une position de la ligne sont donnés comme suivi,

$$V = V_1 e^{-j\beta x} + V_2 e^{+j\beta x} \qquad \text{(tension à } x\text{)}$$

$$I = \frac{V_1}{Z_0} e^{-j\beta x} - \frac{V_2}{Z_0} e^{+j\beta x} \qquad \text{(courant à } x\text{)}$$

## 2.1 Définition : coefficient de réflexion

$$\rho = \frac{\text{tension de l'onde réfléchie}}{\text{tension de l'onde incidente}}$$

$$\rho = \frac{V_2 e^{+j\beta x}}{V_1 e^{-j\beta x}} = \frac{V_2}{V_1} e^{2j\beta x}$$

$$\rho = \rho_0 e^{2j\beta x}$$

$$\rho \ : \ \rho_l \quad \text{quelconque distance}$$

$$\rho_0 \ : \ l = 0$$

Nouvelle écriture des équations,

$$V = V_1 e^{-j\beta x}(1 + \rho)$$

$$I = \frac{V_1}{Z_0} e^{-j\beta x}(1 - \rho)$$

## 2.2  Impédance

$$Z_x = \frac{V}{I} = Z_0 \frac{(1 + \rho)}{(1 - \rho)}$$

$$x \to l \qquad Z_x = Z_l$$

$$Z_L = Z_0 \frac{(1 + \rho)}{(1 - \rho)}$$

$$\rho = \rho_0 e^{2j\beta x} \to \rho_l = \rho_0 e^{2j\beta l}$$

$$\rho_0 = \rho_L e^{-2j\beta l}$$

$$\rho = \rho_L e^{-2j\beta(l-x)}$$

$$\rho = \frac{(Z_L - Z_0)}{(Z_L + Z_0)} e^{-2j\beta(l-x)}$$

Remarque :

1) $Z_L = Z_0 \qquad \to \qquad \rho_L = 0$

2) $Z_L = 0 \qquad \to \qquad \rho_L = -1$

3) $Z_L = \infty \qquad \to \qquad \rho_L = +1$

$$Z_x = Z_0 \frac{(1 + \rho)}{(1 - \rho)}$$

Finalement, pour les lignes sans perte, on a :

$$Z_x = Z_0 \frac{\left(1 + \dfrac{(Z_L - Z_0)}{(Z_L + Z_0)} e^{-2j\beta(l-x)}\right)}{\left(1 - \dfrac{(Z_L - Z_0)}{(Z_L + Z_0)} e^{-2j\beta(l-x)}\right)}$$

$$Z_x = Z_0 \frac{\{Z_L + jZ_0 tan\beta(l-x)\}}{\{Z_0 + jZ_L tan\beta(l-x)\}}$$

# 3. Caractéristiques des lignes de transmission

a) Distance de $\lambda/2$ (Figure I.12)

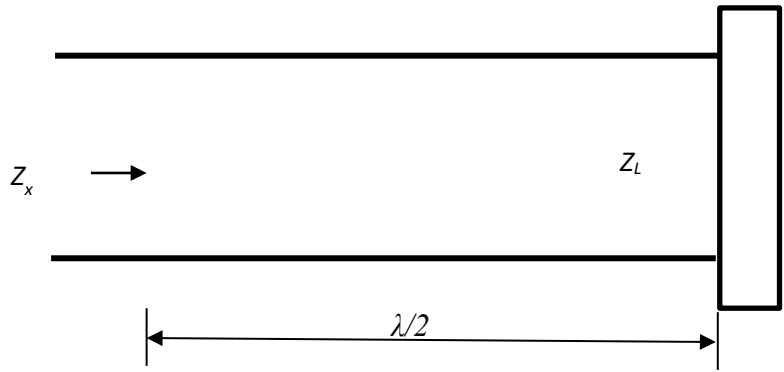

Figure I.12.

$$l - x = \lambda/2 \; ; \; \beta = \frac{2\pi}{\lambda}$$
$$\beta(l-x) = \pi$$

$$Z_x = Z_L$$

Puisque,

$$Z_x = Z_0 \frac{\{Z_L + jZ_0 tan\beta(l-x)\}}{\{Z_0 + jZ_L tan\beta(l-x)\}}$$

b) <u>Court-circuit (Figures I.13 et I.14)</u>

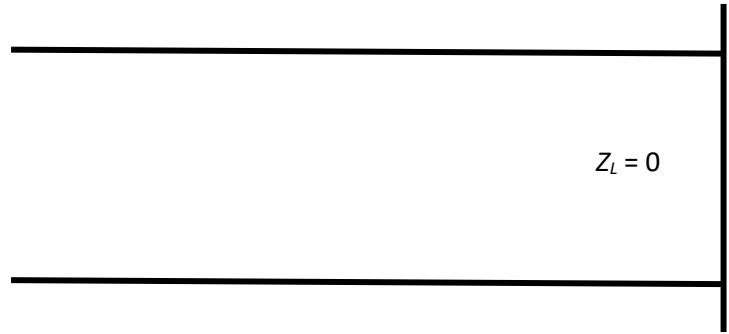

$Z_L = 0$

<div align="center">Figure I.13.</div>

$$Z_x(c.-c.) = Z_0 \frac{\{jZ_0 tan\beta d\}}{\{Z_0\}} = jZ_0 tan\beta d \qquad (toujours\ imaginaire)$$

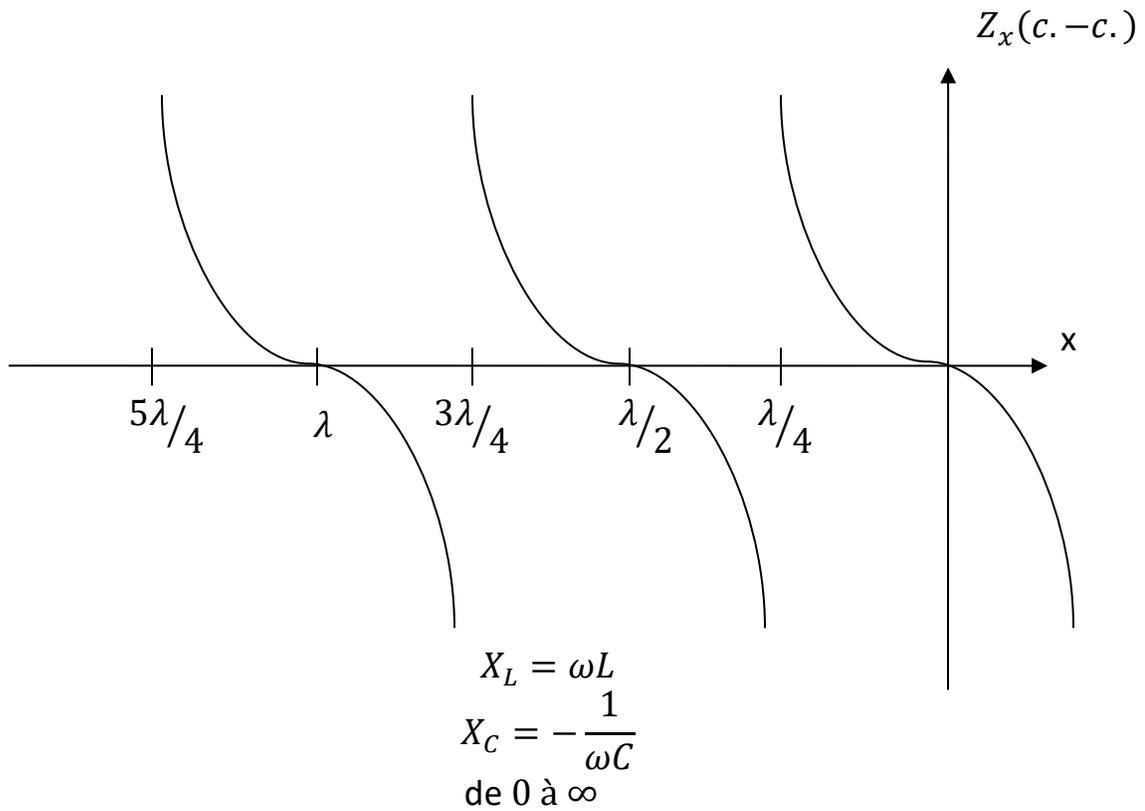

$X_L = \omega L$

$X_C = -\dfrac{1}{\omega C}$

de 0 à $\infty$

<div align="center">Figure I.14.</div>

c)  <u>Circuit ouvert (Figures I.15 et I.16)</u>

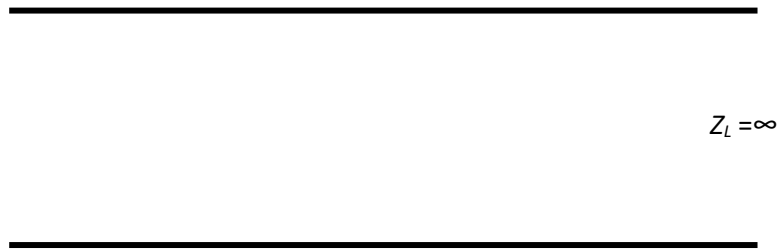

<div align="center">FigureI1.15.</div>

$$Z_x(c.o.) = Z_0 \frac{\{Z_L\}}{\{jZ_L tan\beta d\}} = -jZ_0 cotg\beta d$$

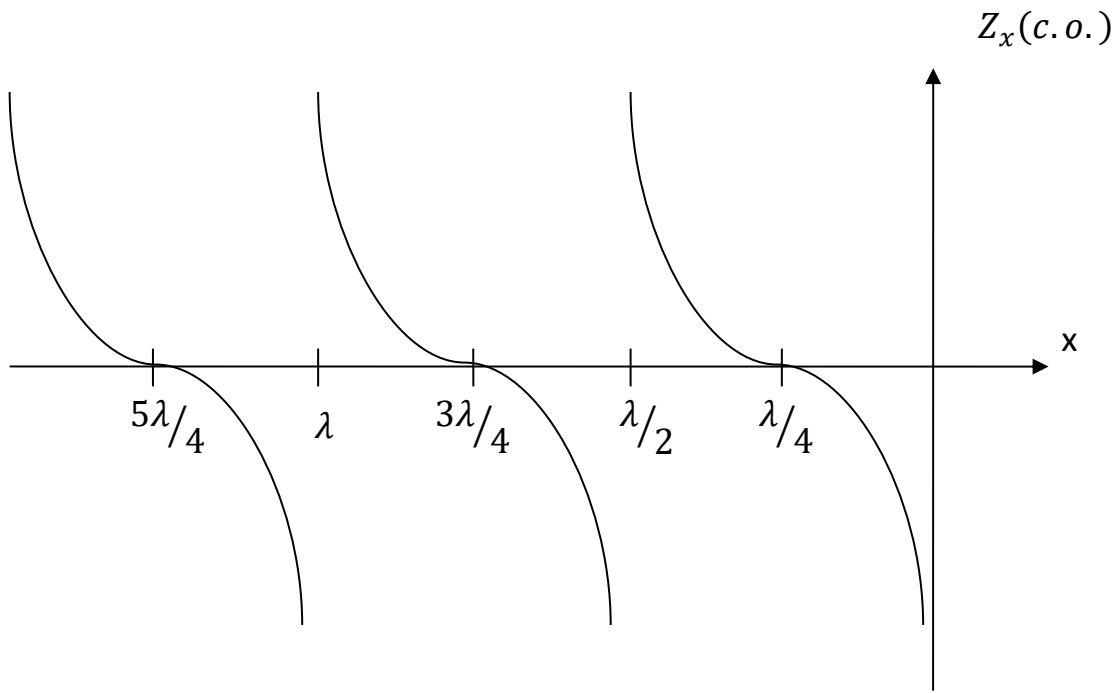

<div align="center">Figure I.16.</div>

Remarque :

1)  $Z_x(c.-c.) \times Z_x(c.o.) = {Z_0}^2$

2)  Admittance

$$Y_x = \frac{1}{Z_x} = \frac{1}{Z_0} \frac{\{Z_L + jZ_0 tan\beta d\}}{\{Z_0 + jZ_L tan\beta d\}} \qquad \leftrightarrow \qquad Y_x = Y_0 \frac{\{Y_L + jY_0 tan\beta d\}}{\{Y_0 + jY_L tan\beta d\}}$$

# 4. Transitoires sur les lignes de transmission

## 4.1 INTRODUCTION

Si une ligne de transmission sans pertes ou sans distorsion, infiniment long ou terminé par son impédance caractéristique le rapport de tout changement de tension au changement du courant, qui en est la conséquence, est égal à l'impédance caractéristique $\mathbb{Z}_0$. Si l'on applique un échelon de tension, le courant qui en résulte est un échelon aussi.

En premier lieu, si la ligne est terminée par une charge quelconque, au début de la ligne, la tension et le courant ne sont pas influencés par la charge qui termine la ligne. Ce n'est qu'une première réflexion qui retourne au point de départ qui va changer la situation avec les conditions frontières appropriées.

Comme exemple, la ligne peut être circuit ouvert ou court-circuit à la fin, une tension en échelon appliquée va causer la même tension et courant initialement vers la charge.

Une étude plus en profondeur nécessiterait les transformées de Laplace. Cependant, si on considère seulement les échelons avec des charges résistives, un traitement dans le domaine du temps est suffisant.

**Exemple I.3.**

Soit un accumulateur idéal avec une tension nominale de $V_0$ et d'impédance interne de $Z_G$. Il alimente une charge $Z_L$ à travers une ligne de transmission sans pertes avec une impédance caractéristique $Z_0$. Un interrupteur $K$ au début de la ligne se ferme à $t = 0$ (Figure I.17). Étudier les tensions et courants aux différentes instances.

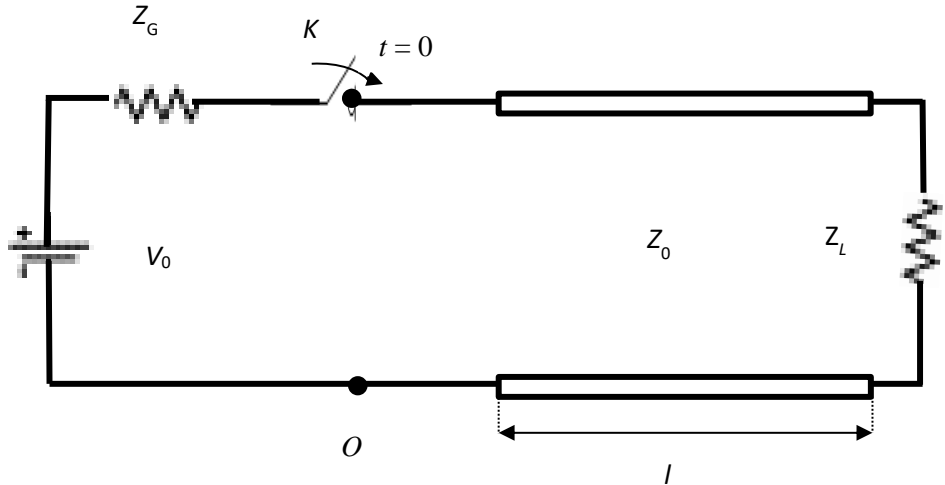

Figure I.17.

Solution:

$$V_{in}(0^+) = (0,0^+) = \frac{V_0 Z_0}{Z_G + Z_0} = V_1 \ (V)$$

$$I_{in}(0^+) = (0,0^+) = \frac{V_0}{Z_G + Z_0} = I_1 \ (A)$$

où $V(0,0^+)$ est la tension à la distance zéro est au temps t = $0^+$. Les ondes d'amplitude $V_1$ et $I_1$ vont se propager vers la fin de la ligne à la vitesse.

$$v = \frac{1}{\sqrt{LC}} \ (\ ^m/_s)$$

Soit $\tau$, le temps de transition des ondes du début à la fin de la ligne et vice versa,

$$\tau = \frac{l}{v} \ (s)$$

Après $\tau$ secondes on aura,

$$V_L(\tau) = V(l,\tau) = \frac{V_0 Z_0}{Z_G + Z_0} \ (1 + \rho_L)$$

$$I_L(\tau) = I(l, \tau) = \frac{V_0}{Z_G + Z_0} \ (1 - \rho_L)$$

où

$$\rho_L = \frac{Z_L - Z_0}{Z_L + Z_0} \quad \text{est le coefficient de réflexion}$$

Du côté générateur on aura,

$$\rho_G = \frac{Z_G - Z_0}{Z_G + Z_0}$$

Les ondes réfléchies à la charge au temps $\tau$ $(ou\ \tau^+)$ sont :

$$V_2 = V_1 \ \rho_L = \frac{V_0 Z_0}{Z_G + Z_0} \rho_L$$

et

$$I_2 = -I_1 \ \rho_L = - \frac{V_0}{Z_G + Z_0} \rho_L$$

Au temps t = 2τ⁺ les ondes vont être réfléchies au début de la ligne.

$$V_3 = V_1 \ \rho_L \rho_G = V_2 \rho_G$$

et

$$I_3 = I_1 \ \rho_L \rho_G = -I_2 \rho_L$$

$$V_1 = \frac{V_0 Z_0}{Z_G + Z_0} \qquad\qquad\qquad I_1 = \frac{V_0}{Z_G + Z_0}$$

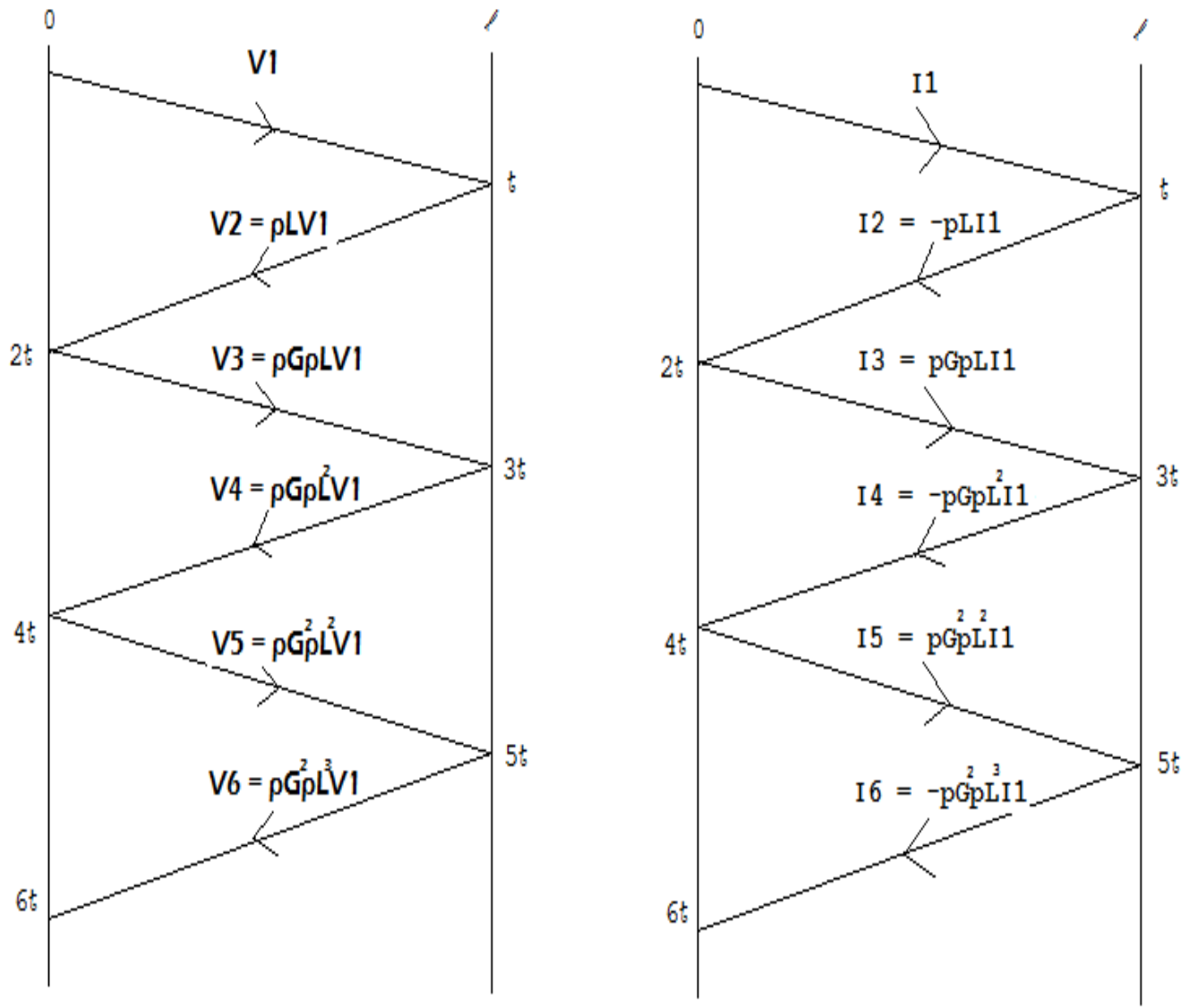

Figure I.18.

Sur la Figure I.18, on donne des tensions et courants aux différentes instances.

**Exemple I.4.**

Dans l'exemple précédant on choisit,

$Z_G$ = 25($\Omega$), $Z_0$ = 50($\Omega$), $Z_L$ = 75($\Omega$)

L'accumulateur a une valeur de 10 ($V$). L'interrupteur se ferme à $t$=0 (sec). Dessiner $V_L$, $I_L$ en fonction du temps et les valeurs pour $t \to \infty$ (état continu; steady-state).

Solution :

$$\rho_L = \frac{75 - 50}{75 + 50} = 0.2$$

$$\rho_G = \frac{25 - 50}{25 + 50} = -0.333$$

$$V_1 = \frac{10(50)}{25 + 50} = 6.67 \ (V)$$

$$I_1 = \frac{10}{25 + 50} = 0.133 \ (A)$$

Ainsi pour les valeurs subséquentes on aura :

$V_1 = 6.67 \ (V)$
$V_2 = 6.67 \ (0.2) = 1.33 \ (V)$
$V_3 = 6.67 \ (0.2) \ (-0.333) = \ -0.445 \ (V)$
$V_4 = 6.67 \ (0.2) \ (-0.333) \ (0.2) = -0.089 \ (V)$
$V_5 = 6.67 \ (0.2) \ (-0.333) \ (0.2) \ (-0.333) = 0.0297 \ (V)$
$V_6 = 6.67 \ (0.2) \ (-0.333) \ (0.2) \ (-0.333) \ (0.2) = 0.0059 \ (V)$

et pour $I$:

$I_1 = 0.133 \ (A)$
$I_2 = 0.133 \ (-0.2) = -0.0367 \ (A)$
$I_3 = 0.133 \ (-0.2) \ (0.333) = \ -0.0089 \ (A)$
$I_4 = 0.133 \ (-0.2) \ (0.333) \ (-0.2) = \ 0.00178 \ (A)$
$I_5 = 0.133 \ (-0.2) \ (0.333) \ (-0.2) \ (0.333) = 0.00059 \ (A)$

$$I_6 = 0.133\,(-0.2)\,(0.333)\,(-0.2)\,(0.333)\,(-0.2) = -0.00018\,(A)$$

La tension aux bornes de la charge peut s'exprimer comme suit (Figure I.19):

$$V_L = 6.67\,u\,(t-\tau) + 1.33\,u\,(t-\tau) - 0.445\,u\,(t-3\tau) -\ 0.089\,u\,(t-3\tau) +\ 0.0297\,u\,(t-5\tau)+...$$

et le courent (Figure I.20):

$$I_L = 0.133\,u\,(t-\tau) - 0.0267\,u\,(t-\tau) - 0.0089\,u\,(t-3\tau) -\ 0.00178\,u\,(t-3\tau) +\ 0.00059\,u\,(t-5\tau)+...$$

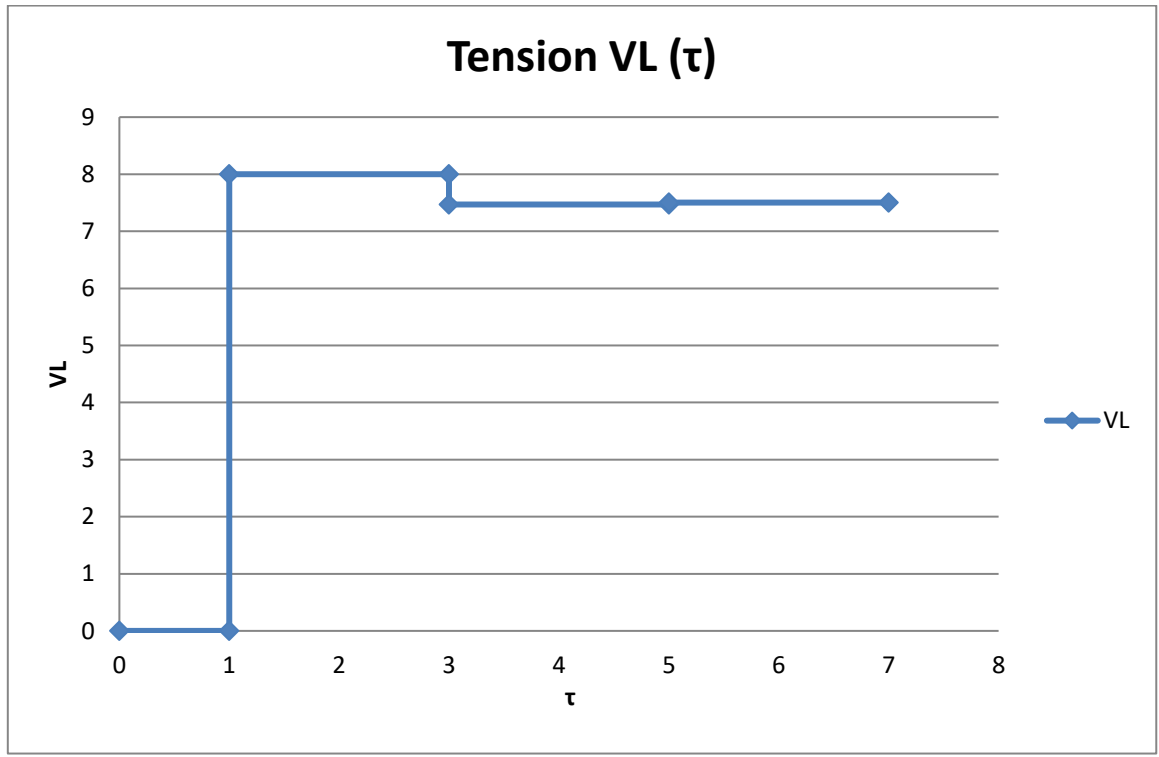

Figure I.19.

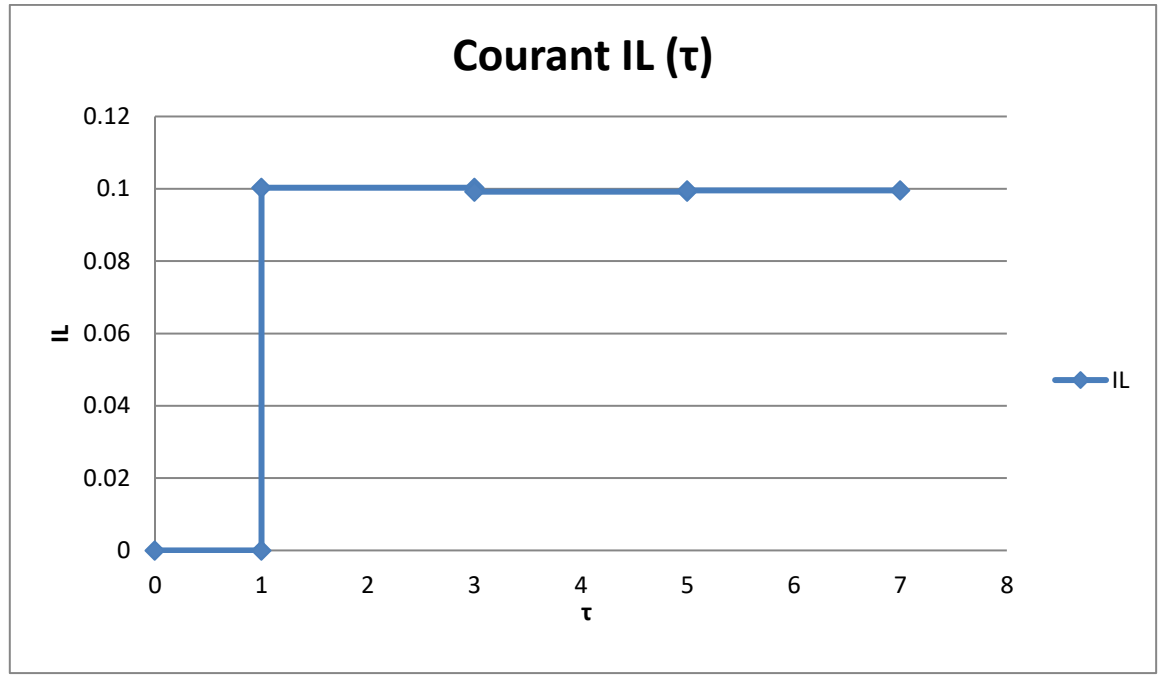

Figure I.20.

En état continu ($t \to \infty$)

$$V_L(t \to \infty) = \frac{10(75)}{25 + 75} = 7.5 \ (V)$$

$$V_L(t \to \infty) = \frac{10}{25 + 75} = 0.1 \ (A)$$

L'influence de la ligne s'enlève. C'est comme s'il y avait une connexion directe entre l'accumulateur et la charge.

Si l'on avait utilisé la transformé de Laplace pour la tension à la charge on aurait:

$$V_L(l, s) = \frac{6.67}{s} (e^{-s\tau} + \rho_L e^{-s\tau} + \rho_L \rho_G e^{-3s\tau} + \rho_L{}^2 \rho_G e^{-3s\tau} + \rho_L{}^2 \rho_G{}^2 e^{-5s\tau} + .)$$

$$= \frac{6{,}67}{s} \left( \frac{e^{-s\tau}(1 + \rho_L)}{1 - \rho_L \rho_G e^{-2s\tau}} \right)$$

$$= \frac{6{,}67}{s}\left(\frac{e^{-s\tau}(1.2)}{1 + 0.0667e^{-2s\tau}}\right)$$

$$V_L(\infty) = \lim_{s \to 0} s\, \mathcal{V}(l,s) = \left(\frac{6{,}67 \times 1{,}2}{1{,}0667}\right) = 7{,}5\ (V)$$

où « $s$ » est la fréquence complexe.

## Exemple I.5.

Reprendre l'exemple précédent. Calculer et dessiner en espace et en temps le comportement de la tension et du courant au milieu de la ligne.

Dans la Figure I.21, le circuit est en état stable. Avant la fermeture de l'interrupteur qui enlève une partie de la charge après $t$ = 0 ($sec$). Dessiner et calculer la tension et le courant à la charge ($V_L$, $I_L$) en fonction du temps dans l'espace et dans le temps.

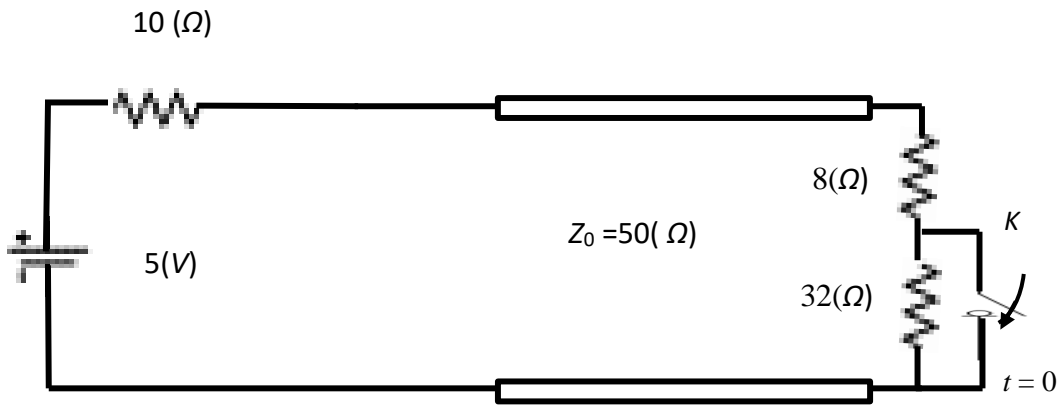

Figure I.21.

## Solution :

À $t$= 0⁻

$$V_L = V^+ + V^- = V^+(1 + \rho_L) = 5 \times \frac{40}{50} = 4\ (V)$$

$$\rho_L = \frac{40 - 50}{40 + 50} = -\frac{1}{9} = -0.1111$$

$$V^+ = \frac{4\,V}{\frac{8}{9}} = 4.5\ (V)$$

$$V^- = -\frac{1}{9} \times (4.5\,V) = -0.5$$

Avant la fermeture de l'interrupteur on peut considérer la tension aux bornes de la charge comme la somme d'une tension incidente de 4.5 ($V$) et d'une tension réfléchie de -0.5 ($V$). Fermer l'interrupteur ne change pas instantanément l'onde incidente mais la tension réfléchie peut changer brusquement pour maintenir les conditions aux frontières.

À $t$=0$^+$

$$V^+ = 4{,}5\ (V)$$

$$\rho_L = \frac{8-50}{8+50} = -0.724$$

$$V^- = -0.724(4.5) = -3.258\ (V)$$

$$\Delta V^- = -3.258 - 0.5 = -2.758\ (V)$$

et

$$V_L(t=0^+) = 4.5 - 3.258 = 1.242 = 4.0(V) - 2.758\ (V)$$

$$= [V_L(t=0^+) - \Delta V^-]\,(V)$$

Le changement dans le courant réfléchi peut être calculé à partir de $\Delta V^-$ et de $Z_0$. Comme le courant et la tension réfléchis ont la polarité contraire :

$$\Delta I_L = \Delta I_L{}^- = -\frac{\Delta V^-}{Z_0} = \frac{2.758}{50} = 0.0552\ (A)$$

Donc,

$$I_L(t=0^+) = I_L(t=0^-) + \Delta I_L = \frac{5}{10+40} + 0.0552 = 0.1552\ (A)$$

D'autre part pour calculer la tension et le courant de la nouvelle charge en fonction du temps pour les moments inférieurs, nous devons tenir compte des réflexions du côté de la source.

$$\rho_G = \frac{10 - 50}{10 + 50} = -0.667 = -\frac{2}{3}$$

$V_L = \left\{ 4.0 + 2.758 \left[ -u(t) + \frac{2}{3} u(t - 2\tau) - 0.724 \left( \frac{2}{3} \right) u(t - 2\tau) + 0.724 \left( \frac{2}{3} \right)^2 u(t - 4\tau) - (0.724)^2 \left( \frac{2}{3} \right)^2 u(t - 4\tau) + \dots \right] \right\}$

$V_L = \{ 4.0 - 2.758 \, u(t) + 0.507 u(t - 2\tau) + 0.245 \, u(t - 4\tau) + \dots \}$

et

$I_L = \left\{ 0.1 + 0.0552 \left[ u(t) + \frac{2}{3} u(t - 2\tau) + 0.724 \left( \frac{2}{3} \right) u(t - 2\tau) + 0.724 \left( \frac{2}{3} \right)^2 u(t - 4\tau) + (0.724)^2 \left( \frac{2}{3} \right)^2 u(t - 4\tau) + \dots \right] \right\}$

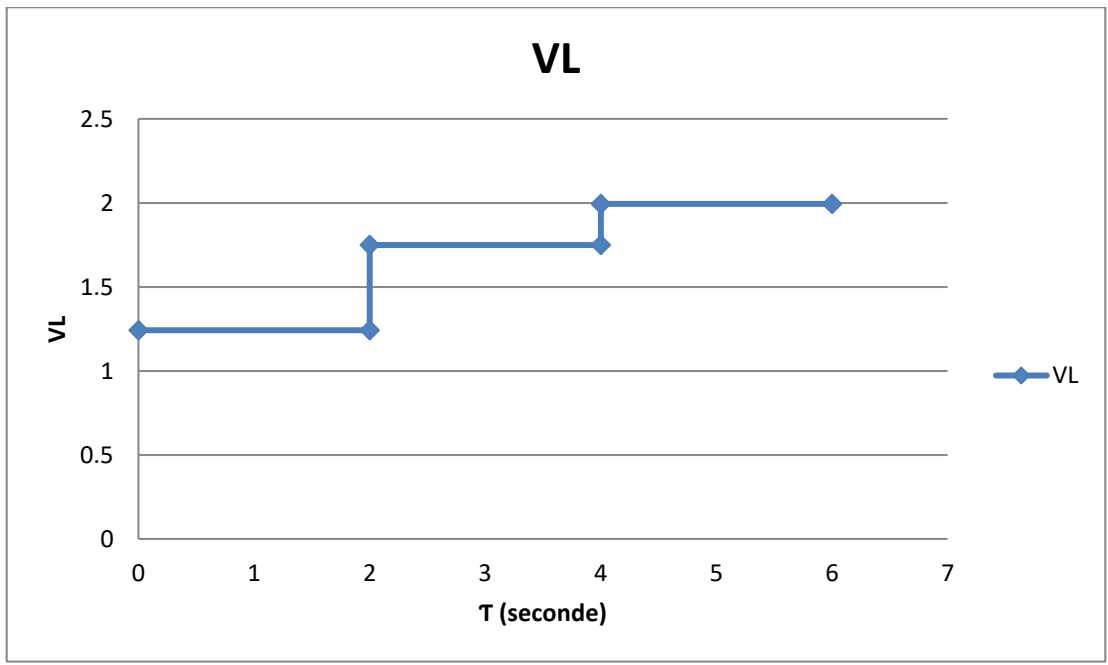

Figure I.22.

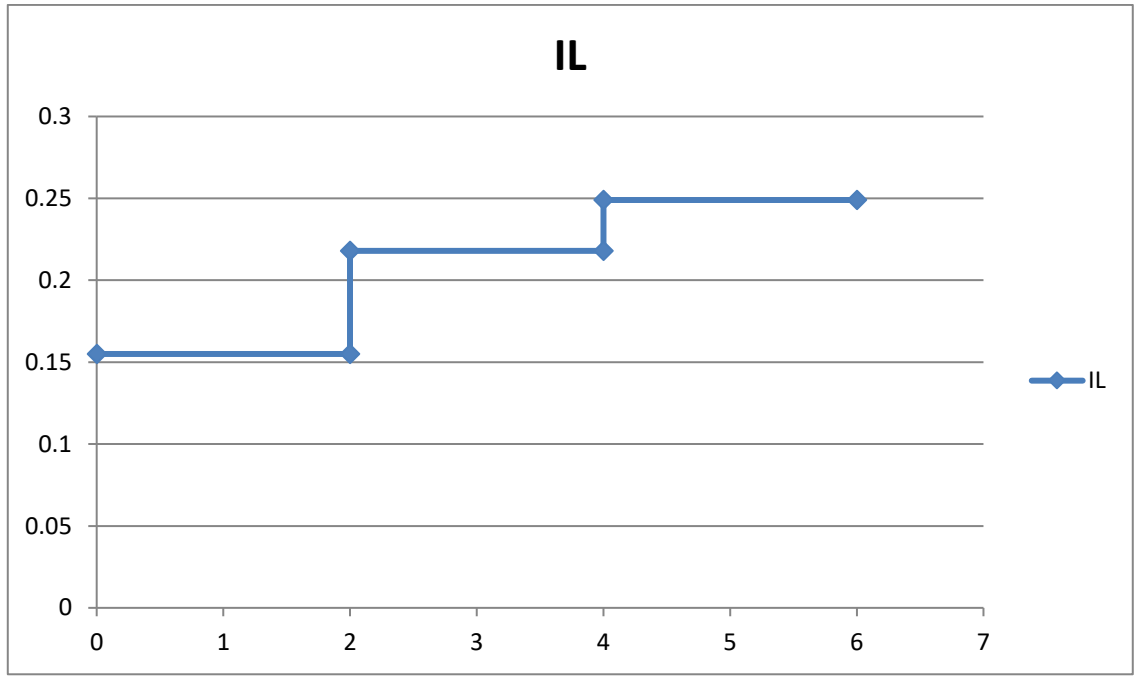

Figure I.23.

La tension et le courant à la charge ($V_L$, $I_L$) en fonction du temps dans l'espace et dans le temps sont dessinées respectivement sur les Figures I.22 et I.23.

$$V_L = (t \to \infty) = ?$$

$$I_L = (t \to \infty) = ?$$

# 5. Variation de la tension le long d'une ligne sans perte

## 5.1    DÉFINITION

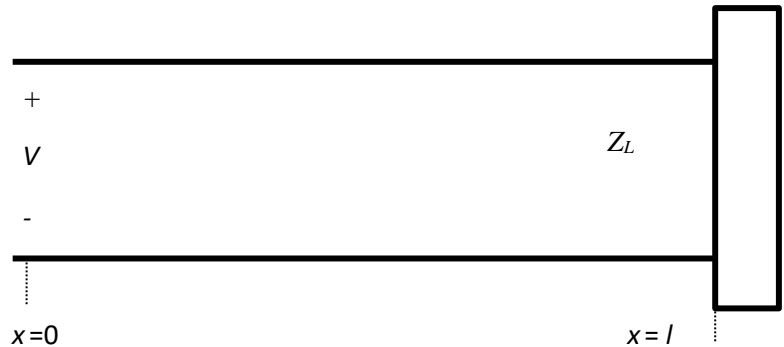

Figure I.24.

Que ça soit le coefficient de réflexion (Figure I.24) $\rho_0 = |\rho_0|e^{j\varphi_0}$ ou $|\rho_0|$ et $\varphi_0$ sont le module et la phase du coefficient de réflexion à $x = 0$.

La tension varie en fonction de la position « $x$ » et du coefficient de réflexion $\rho$ comme,

$$V = V_1(1 + \rho)$$

$$\rho = \rho_0 e^{j2\beta x} = |\rho_0|e^{j\varphi_0}e^{j2\beta x}$$

ou,

$$V = V_1[1 + |\rho_0|e^{j(\varphi_0 + 2\beta x)}]$$

$V_1$ – fonction de générateur.

## 5.2 TERME CARACTÉRISTIQUE

On va considérer le terme caractéristique $1 + |\rho_0| e^{j(\varphi_0 + 2\beta x)}$.

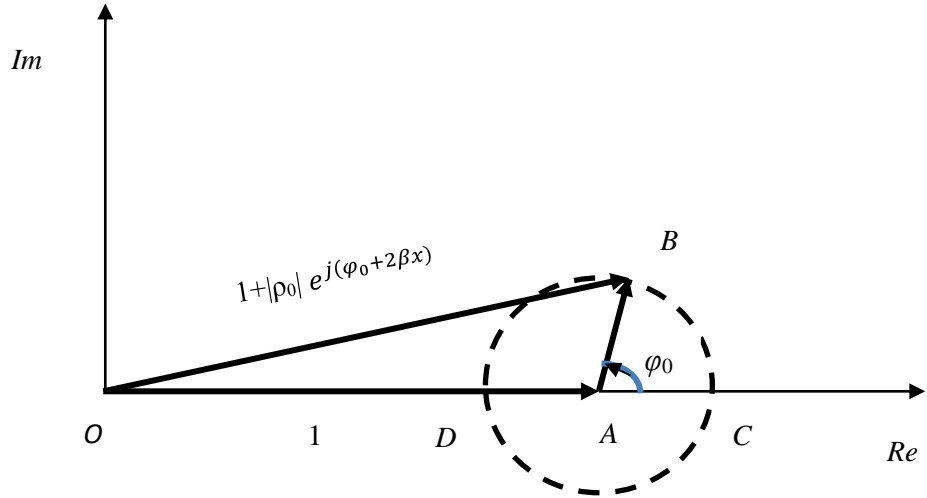

Figure I.25.

On remarque que,

$$|\rho_0| = |\rho| = |\rho_L|$$

Sur la Figure I.25 on note la tension $|V_{max}|$ (point $C$) et la tension $|V_{min}|$ (point $D$).

## Tension maximum

$$|V_{max}| = |V_1||1 + |\rho_0||$$

lorsque,

$$\varphi_0 + 2\beta x = 2\pi n, \quad n = 0,1,2,3,.....$$

À cette position le courant $I$ est minimum.

## Tension minimum

$$|V_{min}| = |V_1||1 - |\rho_0||$$

lorsque,

$$\varphi_0 + 2\beta x = (2n+1)\pi, \quad n = 0,1,2,3,\ldots.$$

Au même endroit, le courant $I$ sera maximum.

### 5.3 DÉFINITION T.O.S.

$$\frac{|V_{max}|}{|V_{min}|} = T.O.S. = s$$

Ce rapport représente le taux d'ondes stationnaires ($V.S.W.R.$ – Voltage Standing Wave Ratio).

La relation entre le taux d'ondes stationnaires $T.O.S.$ et le module du coefficient de réflexion pour une ligne de transmission sans pertes,

$$s = \frac{1 + |\rho_0|}{1 - |\rho_0|} = \frac{1 + |\rho|}{1 - |\rho|}$$

### 5.4 TENSION MAXIMUM (les lieux)

$$\varphi_0 + 2\beta x = 2\pi n, \, n = 0,1,2,3,\ldots$$

$$\rho_L = \frac{Z_L - Z_0}{Z_L - Z_0} = |\rho_L|e^{j\varphi_L}$$

$\rho_L$ – coefficient de réflexion à la charge $Z_L$.

$$\rho_0 = |\rho_0|e^{j\varphi_0} = \rho_L e^{-j2\beta l} = |\rho_L|e^{j(\varphi_L - 2\beta l)} \text{ , } l - \text{longueur de la ligne}$$

d'où,

$$\varphi_0 = \varphi_L - 2\beta l$$

Donc,

$$2\pi n - 2\beta x = \varphi_L - 2\beta l$$

La dernière relation donne,

$$(l - x) = \frac{\varphi_L - 2n\pi}{2\beta}$$

Les lieux de maximum sont,

$$d_{max} = (l - x) = \left[\frac{\varphi_L}{\pi} - 2n\right]\frac{\lambda}{4} \ , n = 0, \pm1, \pm2, \pm3, ....$$

$\varphi_L$ - en radian.

### 5.5 TENSION MINIMUM (les lieux)

$$\varphi_0 + 2\beta x = (2n+1) \ \pi, \ n = 0,1,2, \ 3, .....$$

$$\varphi_0 = \varphi_L - 2\beta l$$

Donc,

$$(2\pi+1)\pi - 2\beta x = \varphi_L - 2\beta l$$

La dernière relation donne,

$$(l - x) = \frac{\varphi_L - (2n+1)\pi}{2\beta}$$

Les lieux de minimum sont,

$$d_{min} = (l - x) = \left[\frac{\varphi_L}{\pi} - (2n + 1)\right]\frac{\lambda}{4} \ , n = 0, \pm1, \pm2, \pm3, ....$$

$\varphi_L$ - en radian.

### Exemple I.6.

Une ligne de transmission sans pertes ayant une impudence caractéristique de $Z_0 = 50 \ (\Omega)$ est terminée par une charge une charge $Z_L = (100+j100) \ (\Omega)$. Trouver les endroits sur la ligne où l'on a des tensions maximums et des tensions minimums.

### Solution :

$$\rho_L = \frac{Z_L - Z_0}{Z_L - Z_0} = \frac{100 + j100 - 50}{100 + j100 + 50} = \frac{50 + j100}{150 + j100} = \frac{1.118e^{j63.4^0}}{1.8e^{j33.7^0}} = 0.62e^{j29.7^0}$$

d'où

$$|\rho_L| = 0.62, \ \varphi_L = 29.7^0 \ \text{ou } 0.518 \ (rad)$$

<u>Tension maximum (les endroits):</u>

$$d_{max} = \left[\frac{\varphi_L}{\pi} - 2n\right]\frac{\lambda}{4} \ , n = 0, \pm 1, \pm 2, \pm 3, \dots \dots$$

Pour *n* = 0; $d_{max\ (1)} = \frac{0.518}{\pi} \cdot \frac{\lambda}{4} = 0.041\lambda$

Pour *n* = 1; $d_{max\ (2)} = \frac{0.518 - 2\pi}{\pi} \cdot \frac{\lambda}{4} = -0.459\lambda$

Pour *n* = -1; $d_{max\ (3)} = \frac{0.518 + 2\pi}{\pi} \cdot \frac{\lambda}{4} = 0.541\lambda$

(Valeurs négatives de *n* convient mieux).

<u>Tension minimum (les endroits):</u>

$$d_{max} = \left[\frac{\varphi_L}{\pi} - (2n + 1)\right]\frac{\lambda}{4} \ , n = 0, \pm 1, \pm 2, \pm 3, \dots \dots$$

Pour *n* = 0; $d_{min\ (1)} = \frac{0.518 - \pi}{\pi} \cdot \frac{\lambda}{4} = -0.208\lambda$

Pour *n* = 1; $d_{min\ (2)} = \frac{0.518 - 3\pi}{\pi} \cdot \frac{\lambda}{4} = -0.709\lambda$

Pour *n* = -1; $d_{min\ (3)} = \frac{0.518 + \pi}{\pi} \cdot \frac{\lambda}{4} = 0.291\lambda$

(Valeurs négatives de *n* convient mieux).

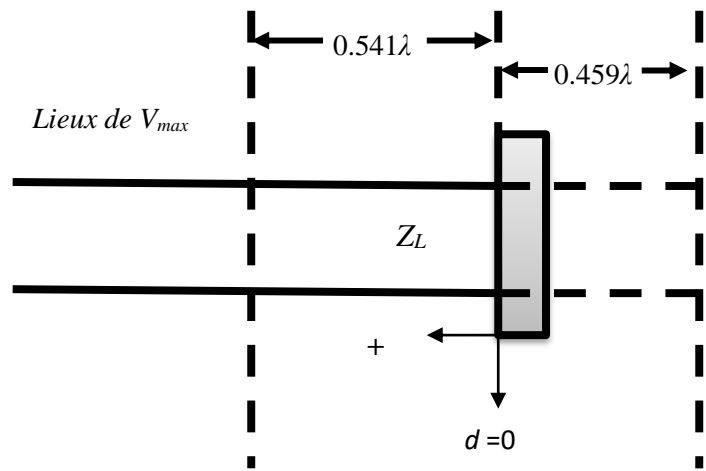

Figure I.26.

Sur la Figure I.26 deux tensions maximum sont montrées.

## 5.6 MESURE DE L'IMPÉNDENCE D'UNE CHARGE

1) Mesure du *T.O.S. = s*

2) Mesure de la distance du premier minimum (ou maximum) à partir de la charge.

$$Z_{max} = \frac{|V_{max}|}{|I_{min}|} = Z_0 \frac{1 + |\rho_0|}{1 - |\rho_0|} = Z_0 s$$

$V_{max}$ et $I_{min}$ sont mesurés au même endroit.

$$Z_{min} = \frac{|V_{min}|}{|I_{max}|} = Z_0 \frac{1 - |\rho_0|}{1 + |\rho_0|} = \frac{1}{s} Z_0$$

$V_{min}$ et $I_{max}$ sont mesurés au même endroit.

3) La formule générale pour l'impédance à n'importe quel point de la ligne est,

$$Z_x = Z_0 \frac{\{Z_L + jZ_0 tan\beta d\}}{\{Z_0 + jZ_L tan\beta d\}}$$

*d* – la distance à partir de la charge.

 À partir de cette formule,

$$Z_{min} = \frac{1}{s} Z_0 = Z_0 \frac{\{Z_L + jZ_0 tan\beta d_{min}\}}{\{Z_0 + jZ_L tan\beta d_{min}\}}$$

ce qui donne,

$$Z_L = Z_0 \frac{\{js \, tan\beta d_{min} - 1\}}{\{j \, tan\beta d_{min} - s\}}$$

### Exemple I.7.

Une ligne de transmission de l'impédance caractéristique $Z_0$ = 50 ($\Omega$) est connectée à une charge inconnue $Z_L$. Les mesures de la tension le long de la ligne démontrent que les valeurs de tension maximum et minimum sont respectivement 1.75 (*V*) et 0.25 (*V*). Aussi, la distance la plus proche de la charge où on détecte la tension maximum est 0.125 $\lambda$.

a) Déterminer le cœfficient de réflexion $\rho_L$, l'impédance inconnue $Z_L$ ainsi que la distance la plus proche de la charge où on détecte la tension minimum,

b) Pour une autre charge inconnue les mêmes minimum et maximum sont détectées, mais la distance la plus proche de la charge où on détecte la tension minimum est 0.125 λ. Déterminer le coefficient de réflexion $\rho_L$ et l'impédance inconnue $Z_L$.

Solution :

a)

$$s = |\frac{V_{max}}{V_{min}}| = \frac{1.75}{0.25} = 7$$

$$|\rho_L| = \frac{s-1}{s+1} = \frac{7-1}{7+1} = 0.75$$

$$d_{max\,(n)} = (\frac{\varphi_L}{\pi} - 2n)\frac{\lambda}{4}$$

$$n = 0 \rightarrow d_{max\,(0)} = \frac{\varphi_L}{\pi}\frac{\lambda}{4} = 0.125\lambda \rightarrow \varphi_L = \frac{\pi}{2}$$

Donc,

$$\rho_L = |\rho_L|e^{j\varphi_L} = 0.75e^{j\frac{\pi}{2}} = j0.75$$

$$Z_L = Z_0\frac{1+\rho_L}{1-\rho_L} = 50\frac{1+j0.75}{1-j0.75} = 50\frac{(1+j0.75)(1+j0.75)}{1+0.76^2} = (14+j48)(\Omega)$$

$$d_{min\,(n)} = [\frac{\varphi_L}{\pi} - (2n+1)]\frac{\lambda}{4}$$

$$n = 0 \rightarrow d_{min\,(0)} = \left(-1 + \frac{\varphi_L}{\pi}\right)\frac{\lambda}{4} = -\frac{3\lambda}{8} = -0.375\lambda$$

$$n = -1 \rightarrow d_{min\,(-1)} = \left(1 + \frac{\varphi_L}{\pi}\right)\frac{\lambda}{4} = \frac{3\lambda}{8} = 0.375\lambda$$

Donc, la distance la plus proche de la charge où on détecte la tension minimum est $d_{min} = 0.375\lambda$.

b)

$$d_{\min} = \frac{\lambda}{8} = 0.125\lambda \;,\; |\rho_L| = 0.75 \;,\; d_{\min(-1)} = \left(1 + \frac{\varphi_L}{\pi}\right)\frac{\lambda}{4} = \frac{\lambda}{8} \rightarrow \varphi_L = -\frac{\pi}{2}$$

$$\rho_L = |\rho_L|e^{j\varphi_L} = 0.75e^{-j\frac{\pi}{2}} = -j0.75$$

$$Z_L = Z_0 \frac{1 + \rho_L}{1 - \rho_L} = 50\frac{1 - j0.75}{1 + j0.75} = 50\frac{(1 - j0.75)(1 - j0.75)}{1 + 0.76^2} = (14 - j48)(\Omega)$$

# 6. Ligne de transmission avec pertes

Dans les lignes de transmission avec pertes le facteur de propagation est complexe,

$$\gamma = \sqrt{ZY} = \sqrt{(R + j\omega L)(G + j\omega C)}$$

L'impédance caractéristique est,

$$Z_0 = \sqrt{\frac{R + j\omega L}{G + j\omega C}}$$

À hautes fréquences,

$$Z_0 \cong \sqrt{\frac{L}{C}}$$

Les variations de la tension et du courent sur la ligne sont respectivement,

$$V(x) = V_1 e^{-\gamma x} + V_2 e^{\gamma x}$$

$$I(x) = \frac{V_1}{Z_0} e^{-\gamma x} - \frac{V_2}{Z_0} e^{\gamma x}$$

$$\rho = \rho_0 e^{j2\gamma x} \ et \ \rho = \rho_L e^{-j2\gamma x}$$

d'où

$$V(x) = V_1 e^{-\gamma x}[1 + \rho]$$

$$I(x) = \frac{V_1}{Z_0} e^{-\gamma x}[1 - \rho]$$

On remarque que,

$$|\rho_0| \neq |\rho| \neq |\rho_L|$$

L'impédance sur la ligne est,

$$Z_x = \frac{V}{I} = Z_0 \frac{1+\rho}{1-\rho}$$

et aussi,

$$Z_x = Z_0 \frac{1 + \rho_L e^{2\gamma(x-l)}}{1 - \rho_L e^{2\gamma(x-l)}}$$

Étant donné que,

$$\rho_L = \frac{Z_L - Z_0}{Z_L - Z_0}$$

on obtient finalement,

$$Z_x = Z_0 \frac{\{Z_L + jZ_0 tanh\beta(l-x)\}}{\{Z_0 + jZ_L tanh\beta(l-x)\}} = Z_0 \frac{\{Z_L + jZ_0 tanh\beta d\}}{\{Z_0 + jZ_L tanh\beta d\}}$$

$l - x = d$  -  distance à partir de la charge.

# 7. Puissance moyenne transportée dans une ligne sans pertes

La puissance moyenne transportée par une ligne de transmission est,

$$P(x) = \frac{1}{2} Re\{V(x)I^*(x)\}$$

On a à une distance $x$ de la ligne,

$$V(x) = V_1 e^{-j\gamma x} + V_2 e^{+j\gamma x} = V_1 e^{-j\gamma x}[1 + \rho(x)]$$

$$I(x) = \frac{V_1}{Z_0} e^{-j\gamma x} - \frac{V_2}{Z_0} e^{+j\gamma x} = \frac{V_1}{Z_0} e^{-j\gamma x}[1 - \rho(x)]$$

où

$$\rho(x) = V_2/V_1 e^{j2\gamma x}$$

Étant donné que la ligne est sans pertes $\gamma = 0$, la puissance moyenne transportée par une ligne de transmission est,

$$P(x) = \frac{|V_1^2|}{2Z_0}[1 - |\rho(x)|^2]$$

La puissance transportée par l'onde incidente est constante et égale à,

$$P_{inc} = \frac{|V_1^2|}{2Z_0}$$

Étant donné que $|\rho(x)| = |\rho_t|$ ne dépend pas de $x$,

$$P = \frac{|V_1^2|}{2Z_0}[1 - |\rho_t|^2] = P_t$$

Donc, dans une ligne sans perte, la puissance dissipée à droite de $x$ ne peut être que dans la ligne, elle est forcément dans la charge.

Si la ligne est chargée par une impédance égale à l'impédance caractéristique ($Z_t = Z_0$) alors $\rho(x) = 0$, la puissance dissipée est égale à la puissance incidente. La puissance incidente est totalement transmise à la charge.

La puissance maximale pouvant être dissipée dans la ligne et la charge, et elle est donc égale à la puissance transportée par l'onde incidente et correspond en $x=0$ à la puissance maximale que peut délivrer le générateur. Au contraire, si la ligne est chargée par une impédance nulle, infinie ou purement complexe, la puissance dissipée dans la charge est nulle puisque $|\rho(x = l)| = |\rho_t = 1|$ ce qui est logique puisque de telles charges ne peuvent consommer d'énergie.

### Exemple I.8.

Un générateur avec $V_G$ = 300 ($V$) et $Z_G$ = 50 ($\Omega$) est connecté à une charge $Z_L$ = 75($\Omega$) à travers une ligne sans perte de 50 ($\Omega$) de longueur $l$ = 0.15 $\lambda$, (Figures I.27 et I.28).

a) Calculer l'impédance d'entrée de la ligne à l'extrémité du générateur, $Z_{in}$,
b) Calculer $I_i$ et $V_i$,
c) Calculer la puissance moyenne délivrée à la ligne, $P_{in}$,
d) Calculer $V_L$ et $I_L$ et la puissance moyenne délivrée à la charge, $P_L$. Comment comparer $P_{in}$ à $P_L$ ?
e) Calculer la puissance moyenne délivrée par le générateur $P_G$ et la puissance moyenne dissipée dans $Z_G$. Est-ce que la loi de conservation est satisfaite ?

### Solution :

a)

$$Z_{in} = Z_0 \frac{Z_L + jZ_0 \tan(\beta l)}{Z_0 + jZ_L \tan(\beta l)} = 50 \frac{75 + j50 \tan(0.3\pi)}{50 + j75 \tan(0.3\pi)} = (41.25 - j16.35)(\Omega)$$

$$Z_{in} = 44.37191 e^{-j21.617^0}$$

b)

$$I_i = \frac{V_G}{Z_G + Z_{in}} = \frac{300}{50 + 41.25 - j16.35} = 3.2361 e^{j10.1584^0} \ (A) \ \text{(Figure 1.28)}$$

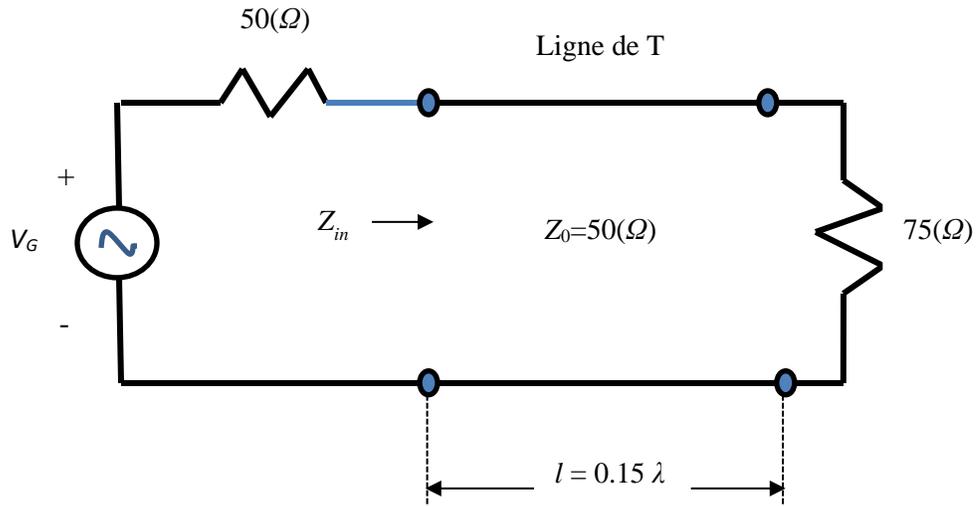

Figure I.27.

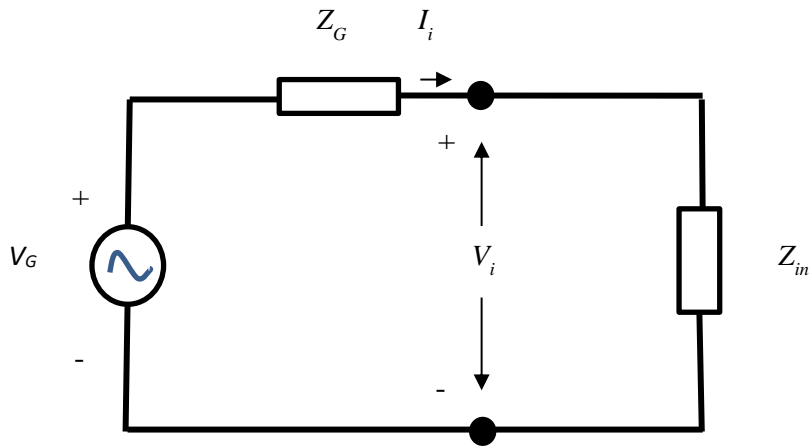

Figure I.28.

$$V_i = I_i Z_{in} = 143.5919 e^{-j11.4586^0} \ (V)$$

c)

$$P_{in} = \frac{1}{2} Re\{V_i I_i^*\} = \frac{1}{2} Re\{143.5919 e^{-j11.4586^0} 3.236 e^{-j10.1584^0}\} = 215.998 (W)$$

d)

$$\rho = \frac{Z_L - Z_0}{Z_L + Z_0} = 0.2$$

$$V_0^+ = V_i \frac{1}{e^{j\beta l} + \rho e^{-j\beta l}} = 149.9999 e^{-j53.9977^0} \ (V)$$

$$V_L = V_0^+(1 + \rho) = 179.9999 e^{-j53.9977^0} \ (V)$$

$$I_L = \frac{V_0^+}{Z_0}(1 - \rho) = 2.399998 e^{-j53.9977^0} \ (V)$$

$$P_L = \frac{1}{2} Re\{V_L I_L^*\} = \frac{1}{2} Re\{180 e^{-j53.9977^0} 2. e^{j53.9977^0}\} = 215.9997(W)$$

Aussi, on peut utiliser la formule suivante

$$P_L = \frac{1}{2} Re\{V_L I_L^*\} = \frac{|V_0^+|^2}{2Z_0}[1 - |\rho|^2] = 215.9997(W)$$

Donc,

$$P_{in} = P_L$$

C'est attendu, parce que la ligne est sans pertes.

e)  La puissance délivrée par le générateur est,

$$P_G = \frac{1}{2} Re\{V_G I_i^*\} = \frac{1}{2} Re\{300 x 3.2361 e^{-j10.1584^0}\} = 477.8056 \ (W)$$

La puissance dissipée dans $Z_G$,

$$P_{Z_G} = \frac{1}{2} Re\{V z_G I_i^*\} = \frac{1}{2} Re\{I_i I_i^* Z_G\} = \frac{1}{2} Re\{|I_i|^2 Z_G\} = \frac{1}{2} 3.24^2 x 50 = 261.81 \ (W)$$

La loi de conservation est satisfaite,

$$P_G = P_{Z_G} + P_{in} \text{ , c'est à dire,}$$

$$477.8056 \approx 261.8086 + 215.9997 = 477.8083$$

# 8. Adaptation d'impédance

On utilise cette technique d'optimiser le transfert d'une puissance électrique entre une source et une charge et d'optimiser la transmission des signaux de télécommunication.

### 8.1 SECTION 'QUART - D'ONDE'

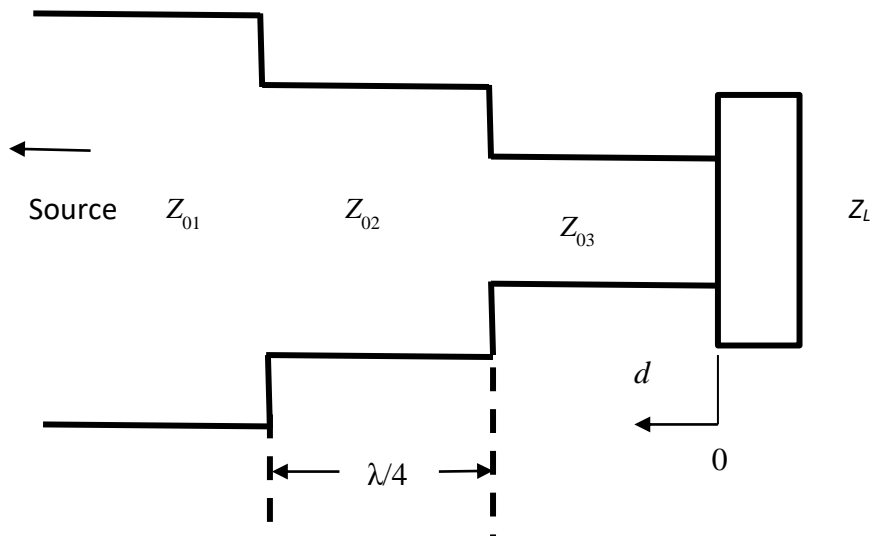

Figure I.29.

Si $Z_{03} = Z_L$ l'impédance $Z_x$ sera à une distance *d* de la charge (Figure I.29),

$$Z_x = Z_0 \frac{\{Z_L + jZ_0 tan\beta d\}}{\{Z_0 + jZ_L tan\beta d\}}$$

On cherche à la distance $\lambda/4$ (Figure I.30) l'impédance $Z_{01}$ avec,

$$Z_{03} = Z_L$$

$$Z_{02} = Z_{02}$$

$$Z_{01} = Z_x$$

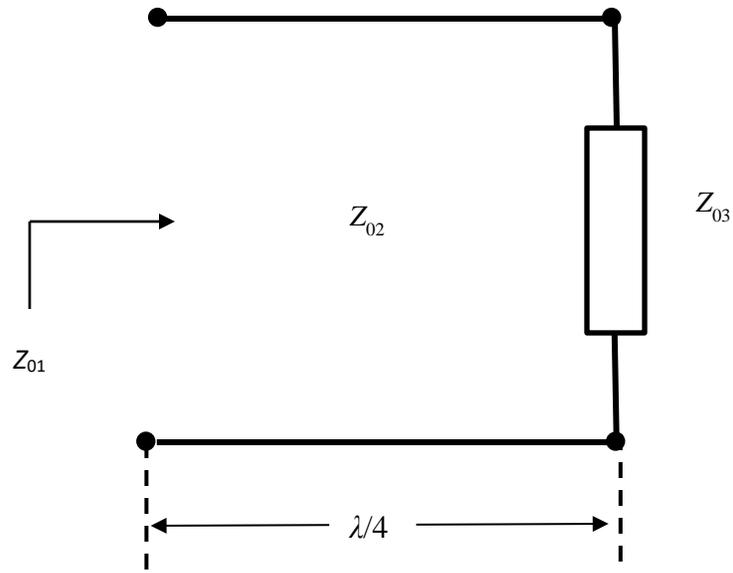

<div align="center">Figure I.30.</div>

Donc,

$$Z_{01} = Z_{02} \frac{\left\{Z_{03} + jZ_{02} tan\frac{\pi}{2}\right\}}{\left\{Z_0 + jZ_L tan\frac{\pi}{2}\right\}} = \frac{Z_{02}^2}{Z_{03}}$$

ce qui donne la condition suivante,

$$Z_{02} = \sqrt{Z_{01}Z_{03}}$$

## Exemple I.9.

Adapter une charge $Z_L$ = (26 − $j$31) ($\Omega$) à 50 ($\Omega$), avec une ligne λ/4, à 5($GHz$).

## Solution :

On ajoute une inductance $L = \frac{|X|}{2\pi f} = 0.987 (nH)$ (en série). On va annuler la partie imaginaire - $j$31 ($\Omega$) et on obtient $Z_{LRe}$ = 26 ($\Omega$). L'impédance de la ligne est,

$$Z_{Ligne} = \sqrt{Z_0 Z_{LRe}} = \sqrt{50 \cdot 26} = 36.06 \ (\Omega)$$

Le circuit qui correspond à cette adaptation est (Figure I.31),

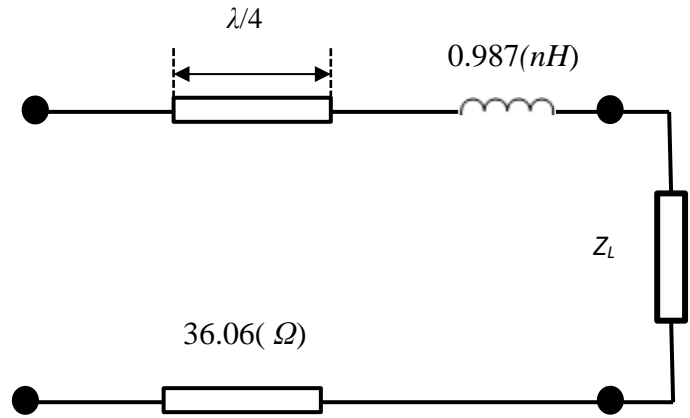

Figure I.31.

## 8.2 ADAPTATION EN SERIE (TRONÇON EN COURT-CIRCUIT)

Un tronçon est un dispositif d'adaptation d'impédance entre une charge et une ligne. On pourrait utiliser un tronçon en court-circuit ou un tronçon ouvert. Le but de cette adaptation est de trouver la longueur du tronçon ainsi que sa position par rapport à la charge. On va faire une adaptation avec un tronçon en court-circuit dont l'impédance caractéristique est égale à $Z_{TRS(c-c)} = kZ_0$ (Figure I.32). Le cas fréquent est $aZ_{TRS(c-c)} = Z_0$. L'impédance caractéristique de la ligne est $Z_0$. Calculer $d_1$ et $d_2$ pour que le générateur voie une charge adaptée (le générateur lui-même est adapté à la ligne).

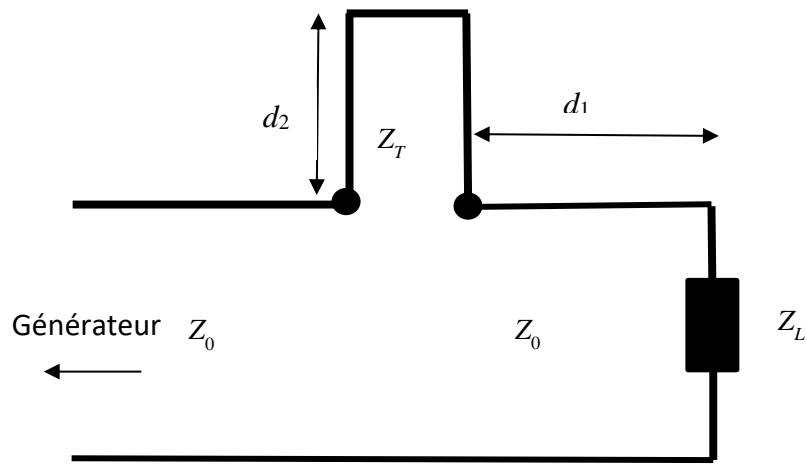

Figure I.32.

Au point de la jonction la condition suivante doit être satisfaite:

$$Z_0 = Z_{TRS(c-c)} + Z_{d_1}$$

ou,

$Z_{TRS(c-c)} = jkZ_0 \tan(\beta d_2)$ - impédance de court-circuit vue au point de la jonction

$$Z_{d_1} = Z_0 \frac{Z_L + jZ_0 \tan(\beta d_1)}{Z_0 + jZ_L \tan(\beta d_1)}$$

$Z_{d_1}$ − impédance de la charge ($Z_L$) vue au point de la jonction

On introduit les changements suivants,

$$a = \tan(\beta d_1)$$

$$b = \tan(\beta d_2)$$

$$Z_L = A + jB$$

La condition à la jonction donne :

$$1 = jkb + \frac{A + j(B + aZ_0)}{(Z_0 - Ba) + jAa}$$

Équation précédente est normalisée par $Z_0$ partout.

Cette relation a deux parties, (partie réelle et partie imaginaire). Il faut égaliser la partie réelle avec 1 et la partie imaginaire avec 0 d'où l'on obtient deux équations,

$$1 = \frac{A(Z_0 - aB) + Aa(B + Z_0)}{(Z_0 - aB)^2 + A^2 a^2}$$

$$0 = bk + \frac{(A + a Z_0)(Z_0 - aB) - aA^2}{(Z_0 - aB)^2 + A^2 a^2}$$

dont les solutions sont,

$$a_{1,2} = \frac{BZ_0 \pm \sqrt{AZ_0[(A-Z_0)^2 + B^2]}}{A^2 + B^2 - AZ_0} = \frac{BZ_0 \pm \sqrt{D}}{A^2 + B^2 - AZ_0}$$

$$D = AZ_0[(A-Z_0)^2 + B^2]$$

$$b_{1,2} = \frac{1}{k} \frac{B\, Z_0 a_{1,2}^2 + a_{1,2}(A^2 + B^2 - Z_0^2) - B\, Z_0}{(Z_0 - a_{1,2}B)^2 + A^2 a_{1,2}^2}$$

À partir de ces solutions on obtient deux couples pour $d_1$ et $d_2$

$$d_1^{(1,2)} = \frac{1}{\beta} \operatorname{atan}(a_{1,2}) = \frac{\lambda}{2\pi} \operatorname{atan}(a_{1,2})$$

$$d_2^{(1,2)} = \frac{1}{\beta} \operatorname{atan}(b_{1,2}) = \frac{\lambda}{2\pi} \operatorname{atan}(b_{1,2})$$

## Cas spéciaux

a)  $A = Z_0$

$$a_1 = \frac{2A}{B}, \ \ a_2 = 0$$

$$b_1 = \frac{B}{kA}, \ \ b_2 = -\frac{B}{kA}$$

et

$$d_1^{(1)} = \frac{\lambda}{2\pi} \operatorname{atan}(a_1), \qquad d_1^{(2)} = 0$$

$$d_2^{(1)} = \frac{\lambda}{2\pi} \operatorname{atan}(b_1), \qquad d_2^{(2)} = \frac{\lambda}{2\pi} \operatorname{atan}(b_2)$$

b)  $A^2 + B^2 - AZ_0 = 0$ et $Z_0 > A$

$$a_1 = \infty, \ \ a_2 = \frac{Z_0 - A}{B + \sqrt{A(Z_0 - A)}}$$

$$b_1 = \frac{Z_0 B}{(A^2 + B^2)k} = \frac{B}{Ak}, \quad b_2 = \frac{1}{k}\frac{B\,Z_0 a_2^2 + a_2(A^2 + B^2 - Z_0^2) - B\,Z_0}{(\,Z_0 - a_2 B)^2 + A^2 a_2^2}$$

$$d_1^{(1)} = \frac{\lambda}{4}, \qquad d_1^{(2)} = \mathrm{atan}(a_2)$$

$$d_2^{(1)} = \frac{\lambda}{2\pi}\mathrm{atan}(b_1), \qquad d_2^{(2)} = \mathrm{atan}(b_2)$$

### Exemple I.10.

Utiliser un tronçon en série pour adapter une ligne de transmission à la charge de l'Impédance $Z_L = (90 + j\,60)\ (\Omega)$. Pour trouver la distance du tronçon à partir de la charge et aussi sa longueur en utilisant un tronçon court-circuit. L'impédance caractéristique de tronçon est égale à celle de la ligne de transmission $Z_0 = 75\ (\Omega)$.

### Solution :

En utilisant les relations précédentes pour l'adaptation d'impédance avec un tronçon en court-circuit on a :

$Z_L = A + jB = (90 + j\,60)\ (\Omega)$; $A = 90\ (\Omega)$, $B = 60\ (\Omega)$, $Z_0 = 75\ (\Omega)$, $k = 1$

$$a_1 = 1.9356, \qquad a_2 = -0.1174$$

$$b_1 = 0.7528, \quad b_2 = -0.7528$$

$$d_1^{(1)} = \frac{1}{\beta}\mathrm{atan}(a_1) = \frac{\lambda}{2\pi}\mathrm{atan}(1.9356) = 0.1741\lambda$$

$$d_1^{(2)} = \frac{1}{\beta}\mathrm{atan}(a_2) = \frac{\lambda}{2\pi}[\pi - \mathrm{atan}(0.1174)] = 0.4814\lambda$$

$$d_2^{(1)} = \frac{1}{\beta}\mathrm{atan}(b_1) = \frac{\lambda}{2\pi}\mathrm{atan}(0.7528) = 0.1027\lambda$$

$$d_2^{(2)} = \frac{1}{\beta}\mathrm{atan}(b_2) = \frac{\lambda}{2\pi}\mathrm{atan}(-0.7528) = \frac{\lambda}{2\pi}[\pi - \mathrm{atan}(0.7528)] = 0.3973\lambda$$

Il y a deux couples de solutions :

*I)*     $d_1^{(1)} = 0.1741\lambda, \ d_2^{(1)} = 0.1027\lambda$

*II)*     $d_1^{(2)} = 0.4814\lambda, \ d_2^{(2)} = 0.3973\lambda$

## Exemple I.11.

Une ligne de transmission sans pertes d'impédance caractéristique $Z_0$ = 50 ($\Omega$) est terminée par une charge $Z_L$ = (50 + $j$40) ($\Omega$). On désire réaliser une adaptation d'impédance utilisant, un tronçon de court-circuit en série ayant la même impédance caractéristique comme la ligne de transmission.

### Solution :

En utilisant les relations précédentes pour l'adaptation d'impédance avec un tronçon an circuit ouvert on a :

$Z_L = A + jB = (50 + j40) \, \Omega$; $A = 50 \ (\Omega) = Z_0$, $B = 40 \ (\Omega)$, $k = 1$

C'est le cas spécial pour lequel,

$$a_1 = \frac{2B}{A} = 2.5, \ a_2 = 0$$

$$b_1 = \frac{B}{A} = 0.8, \ b_2 = -\frac{B}{A} = -0.8$$

$$d_1^{(1)} = \frac{\lambda}{2\pi}\mathrm{atan}(2.5) = 0.1894\lambda, \qquad d_1^{(2)} = 0$$

$$d_2^{(1)} = \frac{\lambda}{2\pi}\mathrm{atan}(0.8) = 0.1074\lambda, \qquad d_2^{(2)} = \frac{\lambda}{2\pi}[\pi - \mathrm{atan}(0.8)\,] = 0.3926\lambda$$

Il y a deux couples de solutions :

*I)*     $d_1^{(1)} = 0.1894\lambda, \ d_2^{(1)} = 0.1074\lambda$

*II)*     $d_1^{(2)} = 0, \ d_2^{(2)} = 0.3926\lambda$

**Exemple I.12.**

Une ligne de transmission sans pertes d'impédance caractéristique $Z_0$ = 50 ($\Omega$) est terminée par une charge $Z_L$ = (50 − $j$50) ($\Omega$). On désire réaliser une adaptation d'impédance utilisant, un tronçon de court-circuit en série ayant la même impédance caractéristique comme la ligne de transmission.

**Solution :**

En utilisant les relations précédentes pour l'adaptation d'impédance avec un tronçon an circuit ouvert on a :

$Z_L = A + jB$ = (50 − $j$50) $\Omega$; $A = 50$ ($\Omega$) $= Z_0$, $B = -50$ ($\Omega$), $k = 1$

C'est le cas spécial pour lequel,

$$a_1 = 0, \ a_2 = -2$$

$$b_1 = 1, \ b_2 = -1$$

$$d_1^{(1)} = 0 \ , \ d_1^{(2)} = \frac{\lambda}{2\pi}[\pi - \operatorname{atan}(2)] = \ 0.3238\lambda,$$

$$d_2^{(1)} = \frac{\lambda}{2\pi}\operatorname{atan}(1) = 0.125\lambda \ , \qquad d_2^{(2)} = \frac{\lambda}{2\pi}[\pi - \operatorname{atan}(1)] \ = \ 0.375\lambda$$

Il y a deux couples de solutions :

*I)* $\qquad\qquad\qquad d_1^{(1)} = \ 0, \ d_2^{(1)} = \ 0.125\lambda$

*II)* $\qquad\qquad\qquad d_1^{(2)} = 0.3238\lambda, \ d_2^{(2)} = 0.375\lambda$

### 8.3 ADAPTATION EN SERIE (TRONÇON EN CIRCUIT OUVERT)

On va faire une adaptation avec un tronçon en circuit ouvert dont l'impédance caractéristique est égale à $Z_{TRS(c-0)} = kZ_0$ (Figure I.33). Le cas fréquent est $Z_{TRS(c-o)} = Z_0$. L'impédance caractéristique de la ligne est égale $Z_0$. Calculer $d_1$ et $d_2$ pour que le générateur voie une charge adaptée (le générateur lui-même est adapté à la ligne).

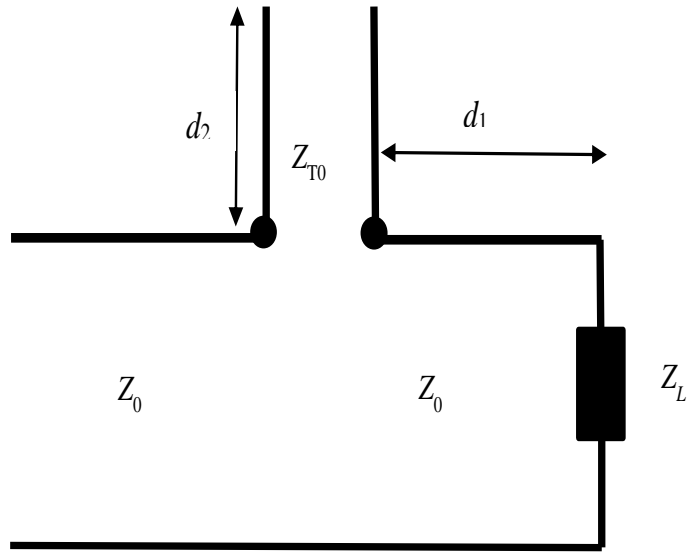

Figure I.33.

Au point de la jonction la condition suivante doit être satisfaite :

$$Z_0 = Z_{TRS(c-o)} + Z_{d_1}$$

ou,

$Z_{TRS(c-o)} = -jkZ_0\text{cotan}(\beta d_2)$ - impédance de circuit ouvert vue au point de la jonction

$$Z_{d_1} = Z_0 \frac{Z_L + jZ_0 \tan(\beta d_1)}{Z_0 + jZ_L \tan(\beta d_1)}$$

$Z_{d_1} -$ Impédance de la charge ($Z_L$) vue au point de la jonction

On introduit les changements suivants,

$$a = \tan(\beta d_1)$$

$$b = \tan(\beta d_2)$$

$$Z_L = A + jB$$

La condition à la jonction donne :

$$1 = -j\frac{k}{b} + \frac{A + j(B + aZ_0)}{(Z_0 - Ba) + jAa}$$

Équation précédente est normalisée par $Z_0$ partout.

Cette relation a deux parties, (partie réelle et partie imaginaire). Il faut égaliser la partie réelle avec 1 et la partie imaginaire avec 0 d'où l'on obtient deux équations,

$$1 = \frac{A(Z_0 - aB) + Aa(B + Z_0)}{(Z_0 - aB)^2 + A^2a^2}$$

$$0 = -\frac{k}{b} + \frac{(A + aZ_0)(Z_0 - aB) - aA^2}{(Z_0 - aB)^2 + A^2a^2}$$

dont les solutions sont,

$$a_{1,2} = \frac{BZ_0 \pm \sqrt{AZ_0[(A - Z_0)^2 + B^2]}}{A^2 + B^2 - AZ_0}$$

$$b_{1,2} = \frac{1}{k}\frac{(Z_0 - a_{1,2}B)^2 + A^2a_{1,2}^2}{-B\,Z_0a_{1,2}^2 - a_{1,2}(A^2 + B^2 - A\,Z_0) + B\,Z_0}$$

À partir de ces solutions on obtient deux couples pour $d_1$ et $d_2$

$$d_1^{(1,2)} = \frac{1}{\beta}\text{atan}\,(a_{1,2}) = \frac{\lambda}{2\pi}\text{atan}\,(a_{1,2})$$

$$d_2^{(1,2)} = \frac{1}{\beta}\text{atan}\,(b_{1,2}) = \frac{\lambda}{2\pi}\text{atan}\,(b_{1,2})$$

Cas spéciaux

a)  $A = Z_0$

$$a_1 = \frac{2A}{B}, \;\; a_2 = 0$$

$$b_1 = -\frac{Ak}{B}, \quad b_2 = \frac{Ak}{B}$$

et

$$d_1^{(1)} = \frac{\lambda}{2\pi} \operatorname{atan}(a_1), \qquad d_1^{(2)} = 0$$

$$d_2^{(1)} = \frac{\lambda}{2\pi} \operatorname{atan}(b_1), \qquad d_2^{(2)} = \frac{\lambda}{2\pi} \operatorname{atan}(b_2)$$

b) $A^2 + B^2 - AZ_0 = 0$ et $Z_0 > A$

$$a_1 = \infty, \quad a_2 = \frac{Z_0 - A}{B + \sqrt{A(Z_0 - A)}}$$

$$b_1 = -\frac{(A^2 + B^2)k}{Z_0 B}, \qquad b_2 = \frac{k[(Z_0 - a_2 B)^2 + A^2 a_2^2]}{-B Z_0 a_2^2 - a_2(A^2 + B^2 - A Z_0) + B Z_0}$$

$$d_1^{(1)} = \frac{\lambda}{4}, \qquad d_1^{(2)} = \operatorname{atan}(a_2)$$

$$d_2^{(1)} = \frac{\lambda}{2\pi} \operatorname{atan}(b_1), \qquad d_2^{(2)} = \operatorname{atan}(b_2)$$

## Exemple I.13.

Une ligne de transmission sans pertes d'impédance caractéristique $Z_0$ = 75 ($\Omega$) est terminée par une charge $Z_L$ = (90 + $j$60) ($\Omega$). On désire réaliser une adaptation d'impédance utilisant, un tronçon de circuit ouvert en série ayant la même impédance caractéristique comme la ligne de transmission.

Solution :

En utilisant les relations précédentes pour l'adaptation d'impédance avec un tronçon en circuit ouvert on a :

$Z_L = A + jB = (90 + j\,60)$ ($\Omega$); $A = 90$ ($\Omega$), $B = 60$ ($\Omega$), $Z_0 = 75$ ($\Omega$), $k$ =1

$$a_1 = 1.9356, \ \ a_2 = -0.1174$$

$$b_{1,2} = -1.3284 \ \text{ et } 1.3284$$

$$d_1^{(1)} = \frac{1}{\beta}\operatorname{atan}(a_1) = \frac{\lambda}{2\pi}\operatorname{atan}(1.9356) = 0.1741\lambda$$

$$d_1^{(2)} = \frac{1}{\beta}\operatorname{atan}(a_2) = \frac{\lambda}{2\pi}[\pi - \operatorname{atan}(0.1174)] = 0.4814\lambda$$

$$d_2^{(1)} = \frac{1}{\beta}\operatorname{atan}(b_1) = \frac{\lambda}{2\pi}[\pi - \operatorname{atan}(1.3284\ )] = 0.3527\lambda$$

$$d_2^{(2)} = \frac{1}{\beta}\operatorname{atan}(ba_2) = \frac{\lambda}{2\pi}\operatorname{atan}(1.3284) = 0.1473\lambda$$

Il y a deux couples de solutions :

*I)* $\qquad\qquad d_1^{(1)} = 0.1741\lambda, \ \ d_2^{(1)} = 0.3527\lambda$

*II)* $\qquad\qquad d_1^{(2)} = 0.4814\lambda, \ \ d_2^{(2)} = 0.1473\lambda$

## Exemple I.14.

Une ligne de transmission sans pertes d'impédance caractéristique $Z_0$ = 50 ($\Omega$) est terminée par une charge $Z_L$ = (50 + $j$40) ($\Omega$). On désire réaliser une adaptation d'impédance utilisant, un tronçon de circuit ouvert en série ayant la même impédance caractéristique comme la ligne de transmission, (Figure 29).

### Solution :

En utilisant les relations précédentes pour l'adaptation d'impédance avec un tronçon an circuit ouvert on a :

$Z_L = A + jB$ = (50 + $j$40) ($\Omega$) ; $\ A = 50$ ($\Omega$) = $Z_0$, $\ B = 40$ ($\Omega$) , $k = 1$

C'est le cas spécial pour lequel,

$$a_1 = \frac{2B}{A} = 2.5, \ \ a_2 = 0$$

$$b_1 = -\frac{A}{B} = -1.25, \ \ b_2 = \frac{A}{B} = 1.25$$

$$d_1^{(1)} = \frac{\lambda}{2\pi}\text{atan}(2.5) = 0.1894\lambda, \qquad d_1^{(2)} = 0$$

$$d_2^{(1)} = \frac{\lambda}{2\pi}\text{atan}(b_1) = \frac{\lambda}{2\pi}[\pi - atan(1.25)] = 0.3574\lambda$$

$$d_2^{(2)} = \frac{\lambda}{2\pi}\text{atan}(b_2) = 0.1426\lambda$$

Il y a deux couples de solutions :

*I)* $\qquad\qquad\qquad d_1^{(1)} = 0.1894\lambda, \ d_2^{(1)} = 0.3574\lambda$

*II)* $\qquad\qquad\qquad d_1^{(2)} = 0, \ d_2^{(2)} = 0.1426\lambda$

## Exemple I.15.

Une ligne de transmission sans pertes d'impédance caractéristique $Z_0$ = 50 ($\Omega$) est terminée par une charge $Z_L$ = (50 $-$ $j$50) ($\Omega$). On désire réaliser une adaptation d'impédance utilisant, un tronçon de circuit ouvert en série ayant la même impédance caractéristique comme la ligne de transmission.

### Solution :

En utilisant les relations précédentes pour l'adaptation d'impédance avec un tronçon an circuit ouvert on a :

$Z_L = A + jB = (50 - j50) \ \Omega; \ A = 50 \ (\Omega) = Z_0, \ B = -50 \ (\Omega), k = 1$

C'est le cas spécial pour lequel,

$$a_1 = 0, \ \ a_2 = -2$$

$$b_1 = -1, \ b_2 = 1$$

$$d_1^{(1)} = 0 \ , \ d_1^{(2)} = \frac{\lambda}{2\pi}[\pi - \text{atan}(2)] \ = \ 0.3238\lambda,$$

$$d_2^{(1)} = \frac{\lambda}{2\pi}[\pi - \text{atan}(1)] = 0.375\lambda \ , \qquad d_2^{(2)} = \frac{\lambda}{2\pi}\text{atan}(1) \ = \ 0.125\lambda$$

Il y a deux couples de solutions :

*I)* $\qquad\qquad\qquad\qquad d_1^{(1)} = \ 0, \ d_2^{(1)} = \ 0.375\lambda$

*II)* $\qquad\qquad\qquad\qquad d_1^{(2)} = 0.3238\lambda, \ d_2^{(2)} = 0.125\lambda$

Remarque : On note que dans le cas d'adaptation en série soit avec un tronçon court-circuité soit avec un tronçon ouvert ses positions on obtient les mêmes distances $d_1^{(1)}$ et $d_1^{(2)}$ à partir de la charge. Seulement les longueurs du tronçon ($d_2^{(1)}$ et $d_2^{(2)}$) dépendent du fait si lui est en court-circuit ou ouvert.

Donc,

a) Si le tronçon est un court-circuit ses positions par rapport à la charge sont $d_1^{(1)}$ et $d_1^{(2)}$ et ces longueurs sont $d_{2(c-c)}^{(1)}$ et $d_{2(c-c)}^{(2)}$.
b) Si le tronçon est un court ouvert ses positions par rapport à la charge sont $d_1^{(1)}$ et $d_1^{(2)}$ et ces longueurs sont $d_{2(c-o)}^{(1)}$ et $d_{2(c-0)}^{(2)}$.

## 8.4 ADAPTATION EN PARALLÈLE (TRONÇON EN COURT-CIRCUIT)

On va faire une adaptation avec un tronçon en court-circuit dont l'impédance caractéristique est égale à $Z_{TRP(c-c)} = kZ_0$ (Figure I.34). Le cas fréquent est $Z_{TRP(c-c)} = Z_0$. L'impédance caractéristique de la ligne est égale $Z_0$. Calculer $d_1$ et $d_2$ pour que le générateur voie une charge adaptée (le générateur lui-même est adapté à la ligne).

Pour avoir l'adaptation à la jonction il faut que la condition suivante soit satisfaite (tronçon est en parallèle avec la charge et on utilise l'approche d'admittance),

$$\frac{1}{Z_0} = \frac{1}{Z_{TRP(c-c)}} + \frac{1}{Z_{d_1}}$$

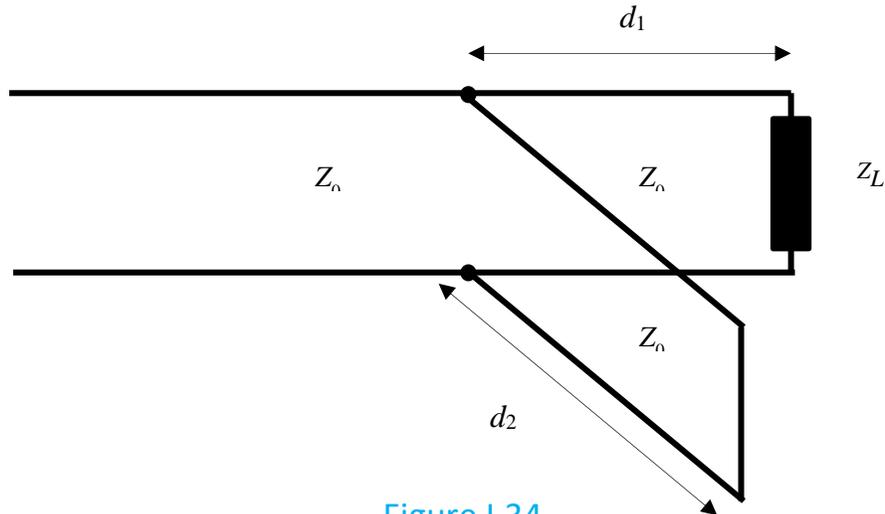

Figure I.34.

avec

$$Z_{TRP(c-c)} = jkZ_0 \tan(\beta d_2)$$

$$Z_{d_1} = Z_d \frac{Z_L + jZ_0 \tan(\beta d_1)}{Z_0 + jZ_L \tan(\beta d_1)}$$

On introduit les changements suivants,

$$a = \tan(\beta d_1)$$

$$b = \tan(\beta d_2)$$

$$Z_L = A + jB$$

La condition de l'adaptation donne le system de deux équations,

$$1 = \frac{A(Z_0 - aB) + Aa(B + Z_0)}{(B + a Z_0)^2 + A^2}$$

$$0 = -\frac{1}{kb} + \frac{aA^2 - (B + a\,Z_0)(Z_0 - aB)}{(Z_0 - aB)^2 + A^2 a^2}$$

dont les solutions sont,

$$a_{1,2} = \frac{-BZ_0 \pm \sqrt{AZ_0[(A - Z_0)^2 + B^2]}}{Z_0^2 - AZ_0}$$

$$b_{1,2} = \frac{1}{k}\frac{(B + a_{1,2}\,Z_0)^2 + A^2}{B\,Z_0 a_{1,2}^2 + a_{1,2}(A^2 + B^2 - Z_0^2) - B\,Z_0}$$

$$d_1^{(1,2)} = \frac{1}{\beta}\operatorname{atan}(a_{1,2}) = \frac{\lambda}{2\pi}\operatorname{atan}(a_{1,2})$$

$$d_2^{(1,2)} = \frac{1}{\beta}\operatorname{atan}(b_{1,2}) = \frac{\lambda}{2\pi}\operatorname{atan}(b_{1,2})$$

Cas spécial

$A = Z_0$

$$a_1 = -\frac{B}{2A}, \;\; a_2 = -\infty$$

$$b_1 = -\frac{A}{Bk}, \;\; b_2 = \frac{A}{Bk}$$

et

$$d_1^{(1)} = \frac{\lambda}{2\pi}\operatorname{atan}(a_1), \qquad d_1^{(2)} = \frac{\lambda}{2\pi}\operatorname{atan}(-\infty) = \frac{\lambda}{2\pi}[\pi - \operatorname{atan}(\infty)] = \frac{\lambda}{4}$$

$$d_2^{(1)} = \frac{\lambda}{2\pi}\operatorname{atan}(b_1), \qquad d_2^{(2)} = \frac{\lambda}{2\pi}\operatorname{atan}(b_2)$$

**Exemple I.16.**

Une ligne de transmission sans pertes est connectée à une antenne de téléphone cellulaire avec l'impédance de charge $Z_L = (60 - j80)(\Omega)$. Trouver la position et la longueur d'un tronçon court-circuit en parallèle pour adapter à 50 $(\Omega)$ ligne.

L'impédance de tronçon est égale à celle de la ligne de transmission.

<span style="color:cyan">Solution :</span>

En utilisant les relations précédentes pour l'adaptation d'impédance avec un tronçon en court-circuit on a :

$Z_L = A + jB = (60 - j\,80)\,(\Omega); A = 60\,(\Omega),\; B = -80\,(\Omega), Z_0 = 50\,(\Omega),\; k = 1$

$$a_1 = -16.8318 \quad a_2 = 0.8318$$

$$b_{1,2} = -0.6794 \quad \text{et} \quad 0.6794$$

$$d_1^{(1)} = \frac{1}{\beta}\,\text{atan}(a_1) = \frac{\lambda}{2\pi}[\pi - \text{atan}(16.8318)] = 0.2594\lambda$$

$$d_1^{(2)} = \frac{1}{\beta}\,\text{atan}(a_2) = \frac{\lambda}{2\pi}\,\text{atan}(0.8318) = 0.1104\lambda$$

$$d_2^{(1)} = \frac{1}{\beta}\,\text{atan}(b_1) = \frac{\lambda}{2\pi}[\pi - \text{atan}\,(0.6794)] = 0.4050\,\lambda$$

$$d_2^{(2)} = \frac{1}{\beta}\,\text{atan}(ba_2) = \frac{\lambda}{2\pi}\,\text{atan}(0.6794) = 0.0950\lambda$$

Il y a deux couples de solutions :

*I)* $\qquad\qquad d_1^{(1)} = 0.2594\lambda,\; d_2^{(1)} = 0.4050\lambda$

*II)* $\qquad\qquad d_1^{(2)} = 0.1104\lambda,\; d_2^{(2)} = 0.0950\lambda$

<span style="color:cyan">**Exemple I.17.**</span>

Une ligne de transmission sans pertes d'impédance caractéristique $Z_0 = 50\,(\Omega)$ est terminée par une charge $Z_l = (50 + j50)\,(\Omega)$. On désire réaliser une adaptation d'impédance utilisant, un tronçon de court-circuit ayant la même impédance caractéristique comme la ligne de transmission.

**Solution :**

En utilisant les relations précédentes pour l'adaptation d'impédance avec un tronçon en court-circuit on a :

$Z_L = A + jB = (50 + j50)\,(\Omega)\,;\;\; A = 50\,(\Omega) = Z_0,\;\; B = 50\,(\Omega)$

C'est le cas spécial pour lequel,

$$a_1 = -0.5,\;\; a_2 = -\infty$$

$$b_1 = -1,\;\; b_2 = 1$$

et

$$d_1^{(1)} = \frac{\lambda}{2\pi}\,\mathrm{atan}(a_1) = \frac{\lambda}{2\pi}[\pi - \mathrm{atan}\,(0.5)] = 0.4262\lambda$$

$$d_1^{(2)} = \frac{\lambda}{2\pi}\,\mathrm{atan}(a_2) = \frac{\lambda}{2\pi}[\pi - \mathrm{atan}\,(\infty)] = 0.25\lambda$$

$$d_2^{(1)} = \frac{\lambda}{2\pi}\,\mathrm{atan}(b_1) = \frac{\lambda}{2\pi}[\pi - \mathrm{atan}(1)] = 0.375\lambda$$

$$d_2^{(2)} = \frac{\lambda}{2\pi}\,\mathrm{atan}(b_1) = 0.125\lambda$$

Il y a deux couples de solutions :

*I)* $\qquad\qquad d_1^{(1)} = 0.4262\lambda,\;\; d_2^{(1)} = 0.375\lambda$

*II)* $\qquad\qquad d_1^{(2)} = 0.25\lambda,\;\; d_2^{(2)} = 0.125\lambda$

**Exemple I.18.**

Une charge $Z_L$ est constituée d'une résistance de $25(\Omega)$ en série avec une inductance de 3.979 $n$H. Un tronçon en parallèle est utilisé pour adapter cette impédance a 50 $(\Omega)$ ligne de transmission a $1GH_z$.

a) Trouver la longueur est la position requises de ce tronçon fait à partir d'une section de la même ligne de 50 $(\Omega)$.

b) Répéter (a) en supposant que le tronçon court-circuite est constitué d'une section de ligne qui a une impédance caractéristique de 75 ($\Omega$).

a) En utilisant les relations précédentes pour l'adaptation d'impédance avec un tronçon en court-circuit on a :

$Z_L = R + j\omega L = R + j2\pi f L = (25 + j25)(\Omega)$ , $A = B = 25\ (\Omega)$

$Z_0 = 50\ (\Omega),\ k = 1$

$$a_1 = 0,\ \ a_2 = -2$$

$$b_1 = -1,\ \ b_2 = 1$$

$$d_1^{(1)} = 0$$

$$d_1^{(2)} = \frac{\lambda}{2\pi}\operatorname{atan}(a_2) = \frac{\lambda}{2\pi}[\pi - \operatorname{atan}(2)] = 0.3238\lambda$$

$$d_2^{(1)} = \frac{\lambda}{2\pi}\operatorname{atan}(b_1) = \frac{\lambda}{2\pi}[\pi - \operatorname{atan}(1)] = 0.375\lambda$$

$$d_2^{(2)} = \frac{\lambda}{2\pi}\operatorname{atan}(b_2) = 0.125\lambda$$

Il y a deux couples de solutions :

*I)* $\qquad\qquad d_1^{(1)} = 0,\ \ d_2^{(1)} = 0.375\lambda$

*II)* $\qquad\qquad d_1^{(2)} = 0.3238\lambda,\ d_2^{(2)} = 0.125\lambda$

b)
$\quad Z_L\ (25 + j25)(\Omega)$ , $A = B = 25\ (\Omega),\ Z_0 = 50\ (\Omega), k = 1.5$

Dans ce cas les impédances de la ligne et du tronçon sont différentes.

$$a_1 = 0,\ \ a_2 = -2$$

$$b_1 = -0.6667, \ b_2 = 0.6667$$

$$d_1^{(1)} = 0$$

$$d_1^{(2)} = \frac{\lambda}{2\pi}\text{atan}(a_2) = \frac{\lambda}{2\pi}[\pi - \text{atan}(2)] = 0.3238\lambda$$

$$d_2^{(1)} = \frac{\lambda}{2\pi}\text{atan}(b_1) = \frac{\lambda}{2\pi}[\pi - \text{atan}(0.6667)] = 0.4064\lambda$$

$$d_2^{(2)} = \frac{\lambda}{2\pi}\text{atan}(b_2) = \frac{\lambda}{2\pi}\text{atan}(0.6667) = 0.0936\lambda$$

Il y a deux couples de solutions :

*I)* $\qquad\qquad\qquad d_1^{(1)} = 0, \ d_2^{(1)} = 0.4064\lambda$

*II)* $\qquad\qquad\qquad d_1^{(2)} = 0.3238\lambda, \ d_2^{(2)} = 0.0936\lambda$

## 8.5 ADAPTATION EN PARALLÈLE (TRONÇON OUVERT)

On va faire une adaptation avec un tronçon en circuit ouvert dont l'impédance caractéristique est égale à celle de la ligne de transmission de $Z_{TRP(c-o)} = Z_0$ (Figure I.35). Le cas fréquent est $Z_{TRP(c-o)} = Z_0$. L'impédance caractéristique de la ligne est égale $Z_0$. Calculer $d_1$ et $d_2$ pour que le générateur voie une charge adaptée (le générateur lui-même est adapté à la ligne).

Pour avoir l'adaptation à la jonction il faut que la condition suivante soit satisfaite (tronçon est en parallèle avec la charge et on utilise l'approche d'admittance),

$$\frac{1}{Z_0} = \frac{1}{Z_{TRP(c-o)}} + \frac{1}{Z_{d_1}}$$

avec

$$Z_{TRP(c-o)} = -jkZ_0 \cot(\beta d_2)$$

$$Z_{d_1} = Z_0 \frac{Z_L + jZ_0 \tan(\beta d_1)}{Z_0 + jZ_L \tan(\beta d_1)}$$

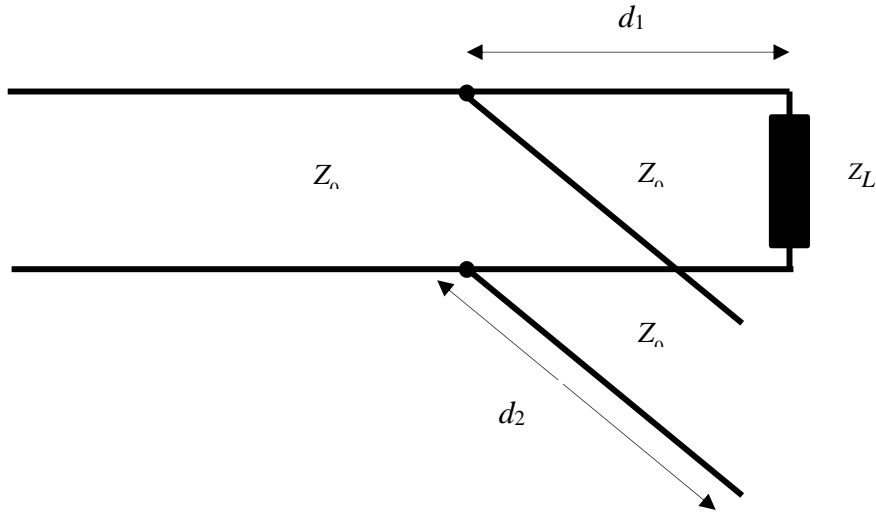

Figure I.35.

On introduit les changements suivants,

$$a = \tan(\beta d_1)$$

$$b = \tan(\beta d_2)$$

$$Z_L = A + jB$$

La condition de l'adaptation donne le system de deux équations,

$$1 = \frac{A(Z_0 - aB) + Aa(B + Z_0)}{(B + a Z_0)^2 + A^2}$$

$$0 = \frac{b}{k} + \frac{aA^2 - (B + a Z_0)(Z_0 - aB)}{(Z_0 - aB)^2 + A^2 a^2}$$

dont les solutions sont,

$$a_{1,2} = \frac{-BZ_0 \pm \sqrt{AZ_0[(A-Z_0)^2 + B^2]}}{Z_0^2 - AZ_0}$$

$$b_{1,2} = -k \frac{B\,Z_0 a_{1,2}^2 + a_{1,2}(A^2 + B^2 - Z_0^2) - B\,Z_0}{(B + a_{1,2}\,Z_0)^2 + A^2}$$

$$d_1^{(1,2)} = \frac{1}{\beta}\operatorname{atan}(a_{1,2}) = \frac{\lambda}{2\pi}\operatorname{atan}(a_{1,2})$$

$$d_2^{(1,2)} = \frac{1}{\beta}\operatorname{atan}(b_{1,2}) = \frac{\lambda}{2\pi}\operatorname{atan}(b_{1,2})$$

### Cas spécial

$A = Z_0$

$$a_1 = -\frac{B}{2A}, \;\; a_2 = -\infty$$

$$b_1 = \frac{B}{A}\,k, \;\; b_2 = -\frac{B}{A}\,k$$

et

$$d_1^{(1)} = \frac{\lambda}{2\pi}\operatorname{atan}(a_1), \qquad d_1^{(2)} = \frac{\lambda}{2\pi}\operatorname{atan}(-\infty) = \frac{\lambda}{2\pi}[\pi - \operatorname{atan}(\infty)] = \frac{\lambda}{4}$$

$$d_2^{(1)} = \frac{\lambda}{2\pi}\operatorname{atan}(b_1), \qquad d_2^{(2)} = \frac{\lambda}{2\pi}\operatorname{atan}(b_2)$$

### Exemple I.19.

Une ligne de transmission sans pertes est connectée à une antenne de téléphone cellulaire avec l'impédance de charge $Z_L = (15 + j10)(\Omega)$. Trouver la position et la longueur d'un tronçon ouvert en parallèle requis pour adapter à 50 ($\Omega$) ligne.

### Solution :

En utilisant les relations précédentes pour l'adaptation avec un tronçon en circuit-ouvert on a :

$Z_L = A + jB = (15 + j\,10)\,\Omega\;;\;\;A = 15\,\Omega,\;B = 10\,\Omega,\,Z_0 = 50\,\Omega$

$$a_1 = 0.2839,\;\;a_2 = -0.8554$$

$$b_{1,2} = 1.3292\;\;\text{et} -1.3292$$

$$d_1^{(1)} = \frac{1}{\beta}\operatorname{atan}(0.2839) = 0.044\lambda$$

$$d_1^{(2)} = \frac{1}{\beta}\operatorname{atan}(a_2) = \frac{1}{\beta}[\pi - \operatorname{atan}(0.8554)] = 0.3874\lambda$$

$$d_2^{(1)} = \frac{1}{\beta}\operatorname{atan}(1.3292) = 0.1473\lambda$$

$$d_2^{(2)} = \frac{1}{\beta}\operatorname{atan}(b_2) = \frac{1}{\beta}[\pi - \operatorname{atan}(1.3292)] = 0.3527\lambda$$

Il y a deux couples de solutions :

I)  $\qquad\qquad d_1^{(1)} = 0.044\lambda,\;\;d_2^{(1)} = 0.1473\lambda$

II)  $\qquad\qquad d_1^{(2)} = 0.3874\lambda,\;d_2^{(2)} = 0.3527\lambda$

### Exemple I.20.

Une ligne de transmission sans pertes d'impédance caractéristique $Z_0$ = 50 ($\Omega$) est terminée par une charge $Z_L$ = (50 + j50) ($\Omega$). On désire réaliser une adaptation d'impédance utilisant, un tronçon de circuit-ouvert en parallèle ayant la même impédance caractéristique comme la ligne de transmission.

### Solution :

En utilisant les relations précédentes pour l'adaptation d'impédance avec un tronçon en court-circuit on a :

$Z_L = A + jB = (50 + j50)\,(\Omega)\;;\;\;A = 50\,(\Omega) = Z_0,\;B = 50\,(\Omega)$

C'est le cas spécial pour lequel,

$$a_1 = -0.5,\;\;a_2 = -\infty$$

$$b_1 = 1, \ b_2 = -1$$

et

$$d_1^{(1)} = \frac{\lambda}{2\pi}[\pi - \text{atan}(\mid a_1 \mid) = 0.4262\lambda$$

$$d_1^{(2)} = \frac{\lambda}{2\pi}[\pi - \text{atan}(\infty)] = 0.25\lambda$$

$$d_2^{(1)} = \frac{\lambda}{2\pi}\text{atan}(b_1) = \frac{\lambda}{2\pi} = 0.1250\lambda$$

$$d_2^{(2)} = \frac{\lambda}{2\pi}[\pi - \text{atan}(b_2)] = 0.3750\lambda$$

Il y a deux couples de solutions :

*I)* $\qquad\qquad d_1^{(1)} = 0.4262\lambda, \ d_2^{(1)} = 0.125\lambda$

*II)* $\qquad\qquad d_1^{(2)} = 0.25\lambda, \ d_2^{(2)} = 0.3750\lambda$

Remarque : On note que dans le cas d'adaptation en parallèle soit avec un tronçon court-circuité soit avec un tronçon ouvert ses positions on obtient les mêmes distances $d_1^{(1)}$ et $d_1^{(2)}$ à partir de la charge. Seulement les longueurs du tronçon ($d_2^{(1)}$ et $d_2^{(2)}$) dépendent du fait si lui est en court-circuit ou ouvert.

Donc,

a) Si le tronçon est un court-circuit ses positions par rapport à la charge sont $d_1^{(1)}$ et $d_1^{(2)}$ et ces longueurs sont $d_{2(c-c)}^{(1)}$ et $d_{2(c-c)}^{(2)}$.

b) Si le tronçon est un court ouvert ses positions par rapport à la charge sont $d_1^{(1)}$ et $d_1^{(2)}$ et ces longueurs sont $d_{2(c-o)}^{(1)}$ et $d_{2(c-0)}^{(2)}$.

Note : Il est recommandé de faire tous les exemples en utilisant l'abaque de Smith (Partie suivante).

# 9. Abaque de Smith

## 9.1    PRINCIPE DE L'ABAGUE DE SMITH

L'abaque de Smith est une représentation graphique de tous les paramètres d'une ligne de transmission. Étant donné qu'avec des ordinateurs puissants il est possible de calculer rapidement et précisément tous les paramètres d'une ligne de transmission cette approche est un outil important pour des circuits *RF* et des circuits micro-ondes.

Soit une ligne sans pertes de longueur « *l* » terminée par la charge $Z_L$ (Figure I.36). L'impédance caractéristique de la ligne est $Z_0$.

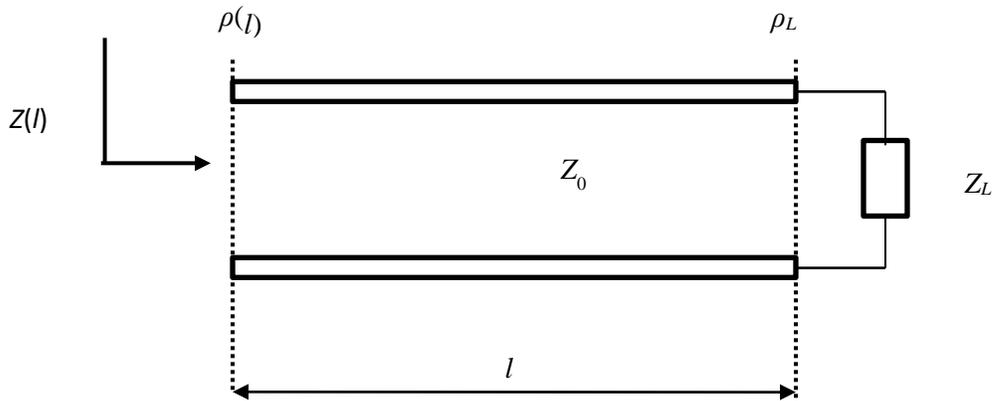

Figure I.36.

L'impédance d'entrée $Z(l)$ est

$$Z(l) = Z_0 \frac{\{Z_L + jZ_0 tan\beta l\}}{\{Z_0 + jZ_L tan\beta l\}} = |Z(l)| \angle \arg(Z(l))$$

C'est une fonction périodique où soit le module $|Z(l)|$ soit l'argument de $Z(l)$ change en fonction de la distance *l*. Cela signifie que pour chaque position sur la ligne il faut calculer soit le module soit l'argument.

D'autre part le coefficient de réflexion peut être écrit sous la forme

$$\rho(l) = \rho_L \angle (-2\beta l) = \rho_L e^{-j2\beta l}$$

$$\rho_L = |\rho_L| \angle (-2\beta l) = |\rho_L| e^{-j2\beta l}$$

Sur la Figure I.37 les positions de $\rho_L$ et $\rho(l)$ sont montrées.

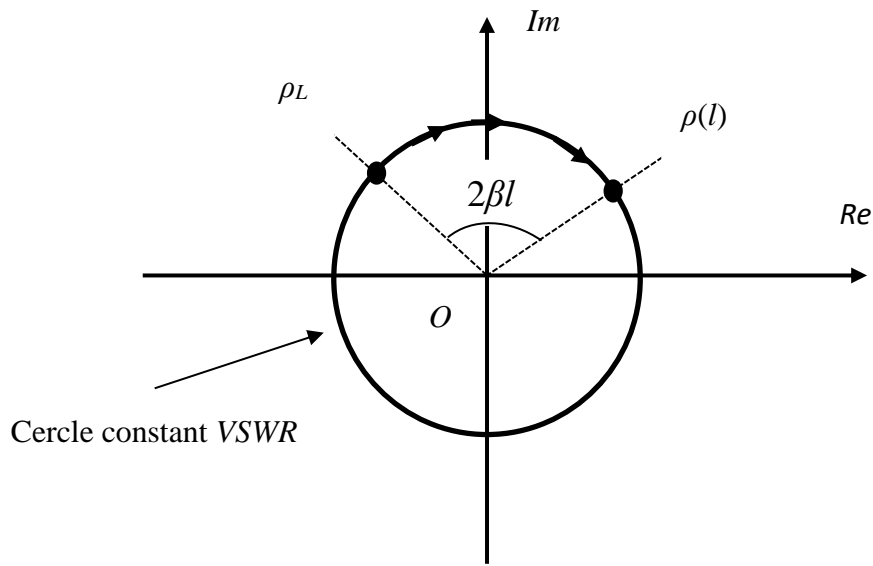

Figure I.37.

On écrit le coefficient de réflexion sous la forme,

$$\rho_L = \frac{Z_L - Z_0}{Z_L + Z_0}$$

On voit que la phase (ou l`argument) du coefficient de réflexion change mais pas le module. Cela signifie que le coefficient de réflexion $\rho_L$ trace dans le plan complexe un cercle chaque fois lorsque la variable *l* augmente par λ/2.

$$2\beta l = 2 \times 2\pi/\lambda \times \lambda/2 = 2\pi$$

Cela signifie qu'un tour de $2\pi$ correspond à $\lambda/2$.

Les cercles $|\rho_L|$ sont nommés **constant $-VSWR$ cercles** (cercles *T.O.S.*). Ils sont seulement la fonction du module $\rho_L$ mais pas de la phase $\varphi(l)$.

Pour trouver la relation graphique entre l`impédance et le coefficient de réflexion à une position de la ligne de transmission le coefficient de réflexion va être exprimé sous la forme complexe:

$$\rho = \rho_r + j\rho_i$$

(On peut omettre l'indice *L*)

Aussi, on va définir l`impédance normalisée,

$$z = \frac{Z_L}{Z_0} = r + jx$$

En tout point d`une ligne de transmission on peut définir un coefficient de réflexion comme le rapport entre la tension de l`onde réfléchit et celle de l`onde incidente en ce point d`abscisse *z* où l`impédance ramenée à *Z*.

Il vient,

$$\rho = \frac{z-1}{z+1}$$

d'où,

$$z = \frac{1+\rho}{1-\rho}$$

ce qui donne

$$r + jx = \frac{1 + \rho_r + j\rho_i}{1 - \rho_r - j\rho_i}$$

On va donc examiner le plan complexe dans lequel est définit $\rho$ représenté par un point *R* d`abscisse *r* et d`ordonné *x*.

Cette identité donne,

$$r + jx = \frac{1 - \rho_r^2 - \rho_i^2 + 2j\rho_i}{(1 - \rho_r)^2 + \rho_i{}^2}$$

On obtient deux relations représentent chacune une famille de cercles dont les expressions générales sont,

$$(\rho_r - \frac{r}{r+1})^2 + \rho_i^2 = \frac{1}{(r+1)^2}, \quad r - \text{cercles (Famille réelle)}$$

$$(\rho_r - 1)^2 + (\rho_i - \frac{1}{x})^2 = \frac{1}{x^2}, \quad x - \text{cercles (Famille imaginaire)}$$

Tous les cercles de la famille réelle ont les centres sur l`axe et un rayon $\frac{r}{r+1} < 1$. De plus ils coupent tous l'axe réel au point 1 et $\frac{r-1}{r+1}$. Les cercles de la famille réelle sont montrés sur la Figure I.38.

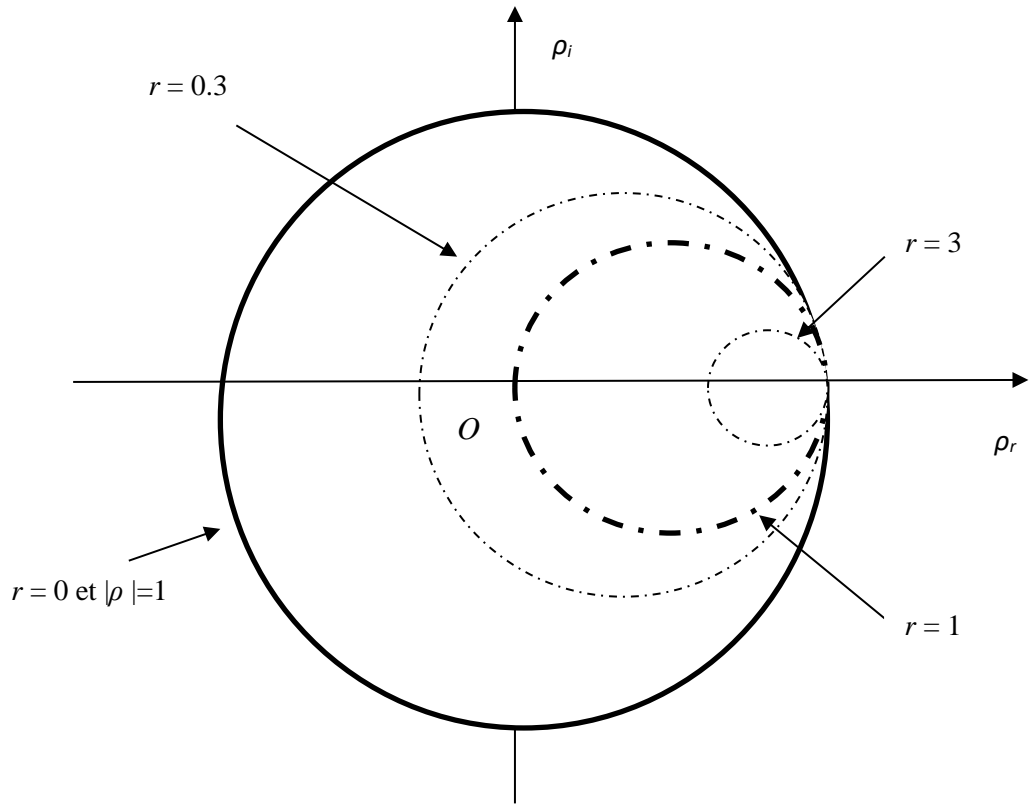

Figure I.38.

En ce qui concerne la partie imaginaire on dispose de deux familles correspondant l'une à $x > 0$, l'autre à $x < 0$, centrées sur l'axe verticale d'abscisse $r = +1$ et passant tous par le point ($r = 1$; $x = 0$) et sont donc tangents à l'axe réel (Figure I.39).

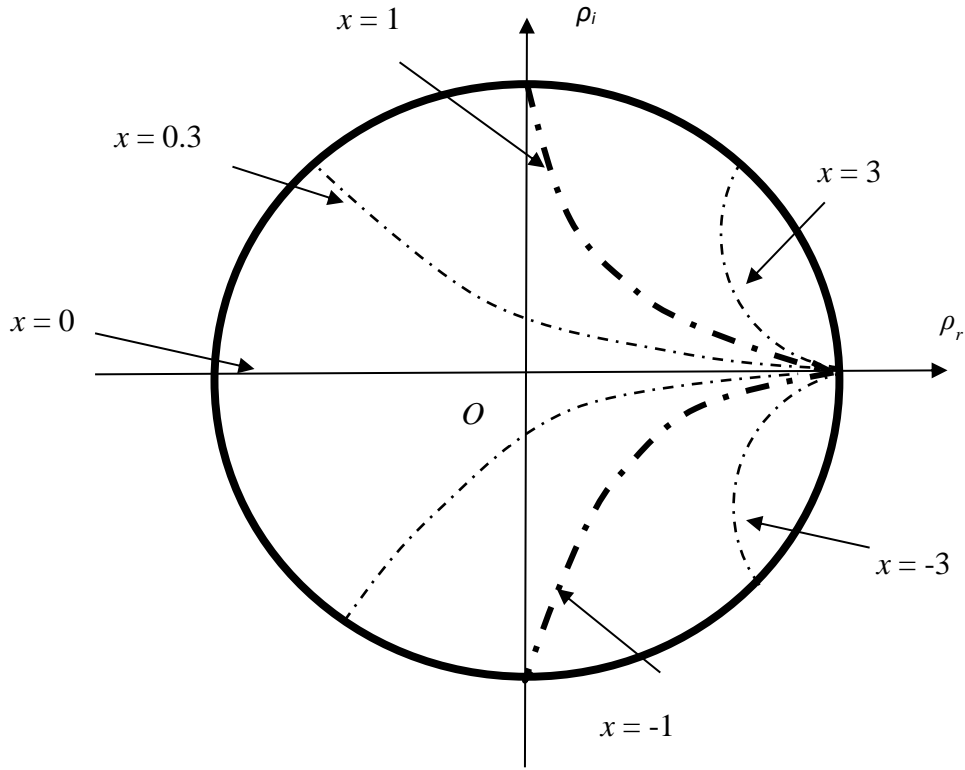

Figure I.39.

Si l'on représente sur le même diagramme ces deux familles de cercles, on constate que la seule partie commune (et donc incluse dans le plus grand cercle possible de rayon 1 de la famille réelle) est utile, d'où, **l'abaque de Smith.**

### 9.2    PROCEDURES À SUIVRE

1) Normaliser l'impédance $Z_L$ en la divisant par $Z_0$.

2) Une impédance inductive va être localisée dans la partie supérieure de l'abaque de Smith ($x$ – cercles).

3) Une impédance capacitive va être localisée dans la partie inférieur de l'abaque de Smith ($x$ – cercles).

4) La même conclusion pour des admittances.

5) On trace une droite entre le centre et le point de l'impédance normalisée.

6) Avec le compas on fait le cercle ($T.O.S$ ou $V.S.W.R$) dont le rayon passe par le centre et le point de l'impédance normalisée. Tous les points sur ce cercle représentent l'impédance normalisée aux points différents de la ligne.

7) Le sens 'horloge', à travers la ligne, représente le mouvement de la charge vers la source, avec un tour de $2\pi$ (rad) ou la moitié de la longueur d'onde $\lambda/2$ ce qui est indiqué aux points externes de l'abaque de Smith.

8) Deux points où le cercle ($T.O.S.$) coupe l'axe réel sont les points de la tension maximum (droit) et de la tension minimum (gauche).

9) Le point symétrique au point de l'impédance normalisée par rapport au centre représente le point de l'admittance.

$$y = \frac{1}{z} = \frac{1}{|z| \angle arg(z)} = |y| \angle [-arg(z)] = |y| \angle [arg(y)]$$

$$\rho = \frac{z-1}{z+1}$$

Sur l'axe réel $x = 0$ est $z = r$. On trouve sur cet axe deux points $z = r_{min} < 1$ et $z = r_{max} > 1$. Le coefficient de réflexion qui correspond à ces points est :

$$|\rho| = \frac{r_{max} - 1}{r_{max} + 1} \rightarrow r_{max} = \frac{1 + |\rho|}{1 - |\rho|} = s = \text{VSWR } (T.O.S)$$

$$s = \text{VSWR } (T.O.S) = \frac{1}{r_{min}}$$

**Exemple I.21.**

La ligne de transmission sans perte de longueur $l = 0.1\lambda$ ayant une impédance caractéristique de $Z_0 = 50\ (\Omega)$ est terminée avec une charge de l'impédance $Z_L = (50 + j25)(\Omega)$, (Figure I.40). En utilisant l'abaque de Smith, il faut trouver :

a)  L'impédance normalisée $z_L$,
b)  L'impédance a la position $l = 0.1\lambda$,
c)  Le taux d'ondes stationnaires (*T.O.S.*),
d)  Le coefficient de réflexion $\rho_L$,
e)  Le coefficient de réflexion $\rho$,
f)  L'admittance normalisée $y_L$.

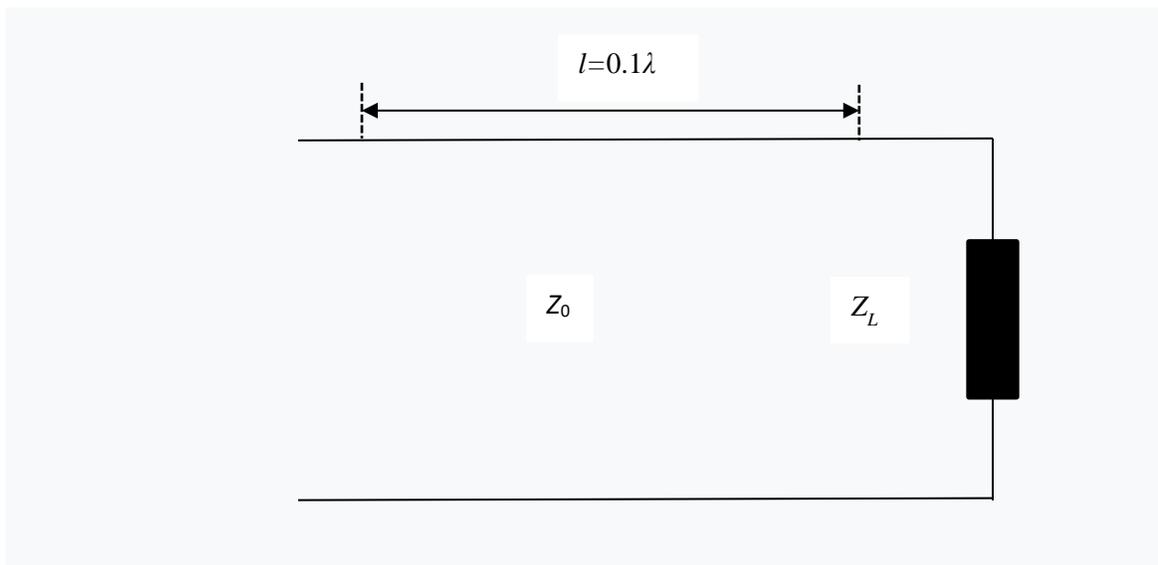

Figure I.40.

**Solution :**

a)

$$z_L = \frac{Z_L}{Z_0} = \frac{5 + j25}{50} = 0.1 + j0.5$$

Sur l'abaque de Smith L'impédance normalisée se trouve au point $A$.

b) Le point $A$ (charge) se trouve sur l'échelle (vers la source) à la position 0.074 $\lambda$. Pour déterminer l'impédance normalisée à la position $l = 0.1\,\lambda$, on ajoute 0.1 $\lambda$ (vers la source) ce qui donne sur l'échelle (vers la source) (0.074 $\lambda$+0.1 $\lambda$=0.174 $\lambda$). On fait une droite entre le centre $O$ et le point 0.174 $\lambda$ (VS). Ensuite, avec le compas on fait la rotation du point $A$ jusqu'à l'intersection avec la droite qui passe par $O$ et 0.174 (VS) ce qui donne le point $B$. On lit au point $B$, $z_B = 0.38 + j1.88$.
L'impédance à ce point est $Z_B = Z_0 z_B = (19 + j94)\Omega$.

c) On trouve $s = T.O.S. = r_{max} = 13$ sur la ligne horizontale à droite du centre $O$. Aussi, on peut utiliser l'échelle $T.O.S.$ Avec le compas on prend la distance entre le centre $O$ et le point $D$ se trouvant à l'intersection de l'axe horizontal et le cercle dont le rayon est ($OA=OB$). Ensuite, on se rend avec le compas sur l'échelle en dessous de l'abaque, $T.O.S.$ et on lit directement $s = T.O.S.(VSWR) = r_{max} = z_{max} = 13$.

d) De l'échelle du coefficient de réflexion ($V$, $I$) en dessous de l'abaque on lit $|\rho_L| = 0.855$. Sur l'échelle du coefficient de l'angle de réflexion sur le périmètre de l'abaque (la position du point $A$) on lit l'angle $\angle(\rho_L) = 126.5^0$. Donc, $\rho_L = 0.855\angle126.5^0$.

e) Étant donné que la ligne de transmission est sans pertes $|\rho| = |\rho_L| = 0.885$. Sur l'échelle du coefficient d'angle de réflexion sur le périmètre de l'abaque (la position du point $B$) on lit l'angle $\angle(\rho) = 55^0$. Donc, $\rho_L = 0.855\angle55^0$.

f) On fait le cercle $|\rho| = OA = OB =$ cte. Dans l'intersection de ce cercle et la droite passant par les points $O$ et $A$ on obtient le point $C$ dans lequel se trouve

Figure I.41.

l'admittance normalisée $y_B$. Donc, l'admittance se trouve diamétralement par rapport au point *A* sur l'abaque de Smith. On lit au point *C* l'admittance normalisée $y_C = 0.4 - j1.9$.

$$Y_C = Y_0 y_C = (8 - j38)\ (mS)$$

Chaque étape peut être suivie sur l'abaque de Smith (Figure I.41).

## Exemple I.22.

Adapter une charge de l'impédance $Z_L = (40 - j40)\ (\Omega)$ à une ligne de l'impédance caractéristique $Z_0 = 10\ (\Omega)$ a l'aide d'un syntoniseur à branche simple (tronçon).

a) À la fréquence $f = 3\ (GH_z)$, trouver deux solutions pour la distance entre le tronçon et la charge ainsi que la longueur du tronçon, (le tronçon en court-circuit et en parallèle) (Figure I.34) et,

b) deux autres solution (le tronçon en circuit ouvert en parallèle) (Figure 1.35). La longueur d'onde est $\lambda = \frac{c}{f} = 0.1\ (m) = 10\ (cm)$.

L'impédance normalisée est :

$$z_L = \frac{Z_L}{Z_0} = \frac{40 - j40}{100} = 0.4 - j0.4$$

Sur l'abaque de Smith l'impédance normalisée se trouve au point *A*.

Ensuite on fait le cercle de rayon $|\rho| = OA =$ cte., qui coupe la droite qui passe par les points *O* et *A* ce qui donne le point *B* dans lequel on lit l'admittance normalisé,

$$y_L = 1.23 + j1.23$$

dont la position sur l'échelle (*LOVS*) est $0.18\ \lambda$.

Le cercle de rayon $|\rho| = OA =$ cte., coupe le cercle $g = 1$ ce qui donne deux points

$C$ et $D$ pour lesquels on lit :

La position du point $C$ sur l'échelle ($LOVS$) est $0.333\ \lambda$ et $y_C = 1 - j1.15$

La position du point $D$ sur l'échelle ($LOVS$) est $0.165\ \lambda$ et $y_C = 1 + j1.15$

Au point $C$, $y_C = 1 - j1.15 \rightarrow, d_1^{(1)}$= (0.333-0.18) $\lambda$ = 0.153$\lambda$ (flèche pointillée rouge)

Au point $D$, $y_D = 1 + j1.15 \rightarrow y_{s2} = -j1.15$, $d_1^{(2)}$= (0.165-0.18+0.5) $\lambda$ =0.485$\lambda$ (flèche pointillée rouge)

a) Tronçon en court-circuit :

Adaptation au point $C$. On ajoute $y_{s1} = j1.15$, $d_2^{(1)}$ =(0.136-0.25+0.25) $\lambda$

=0.386$\lambda$ (flèche pointillée bleu)

Adaptation du point $D$. On ajoute $y_{s2} = -j1.15$, $d_2^{(2)} = $ (0.364-0.5) $\lambda$ =0.114$\lambda$

(flèche pointillée bleu)

Donc, les couples sont :

$d_1^{(1)}$= 0.153$\lambda$=1.53 ($cm$), $d_2^{(1)}$= 0.386$\lambda$=3.86 ($cm$)     et

$d_1^{(2)}$= 0.485$\lambda$=4.85 ($cm$),  $d_2^{(2)}$= 0.114$\lambda$=1.14 ($cm$)

Le calcul analytique donne :

$d_1^{(1)}$= 0.1527$\lambda$=1.527 ($cm$), $d_2^{(1)}$= 0.3854$\lambda$=3.854 ($cm$)     et

$d_1^{(2)}$= 0.4852$\lambda$=4.852 ($cm$),  $d_2^{(2)}$= 0.1146$\lambda$=1.146 ($cm$)

Pour le tronçon en court-circuit on lit ces longueurs à partir du point 0.25 $\lambda$ où $y$= $\infty$.

b) Tronçon ouvert :

Les distances entre le tronçon et la charge restent les mêmes comme dans le calcul précédent.

Adaptation au point $C$. On ajoute $y_{s1} = j1.15$. $d_2^{(1)} = (0.136\text{-}0.0)\,\lambda = 0.136\,\lambda$ (flèche noir foncé)

Adaptation au point $D$. On ajoute $y_{s1} = -j1.15$. $d_2^{(2)} = (0.364\text{-}0.0)\,\lambda = 0.364\,\lambda$ (Flèche noir foncé)

Donc, les couples sont :

$d_1^{(1)}$ = 0.153$\lambda$=1.53 ($cm$), $d_2^{(1)}$ = 0.136$\lambda$=1.36 ($cm$)    et

$d_1^{(2)}$ = 0.485$\lambda$=4.85 ($cm$),  $d_2^{(2)}$ = 0.3646$\lambda$=3.646 ($cm$)

Le calcul analytique donne :

$d_1^{(1)}$ = 0.1527$\lambda$=1.527 ($cm$), $d_2^{(1)}$ = 0.1354$\lambda$=1.354 ($cm$)    et

$d_1^{(2)}$ = 0.4852$\lambda$=4.852 ($cm$),  $d_2^{(2)}$ = 0.3646$\lambda$=3.646 ($cm$)

Pour le tronçon en circuit ouvert on lit ces longueurs à partir du point 0.0 $\lambda$ où $y$= 0.

Les résultats obtenus par le calcul graphique et par le calcul analytique accordent bien. Chaque étape peut être suivie sur l'abaque de Smith (Figure I.42).

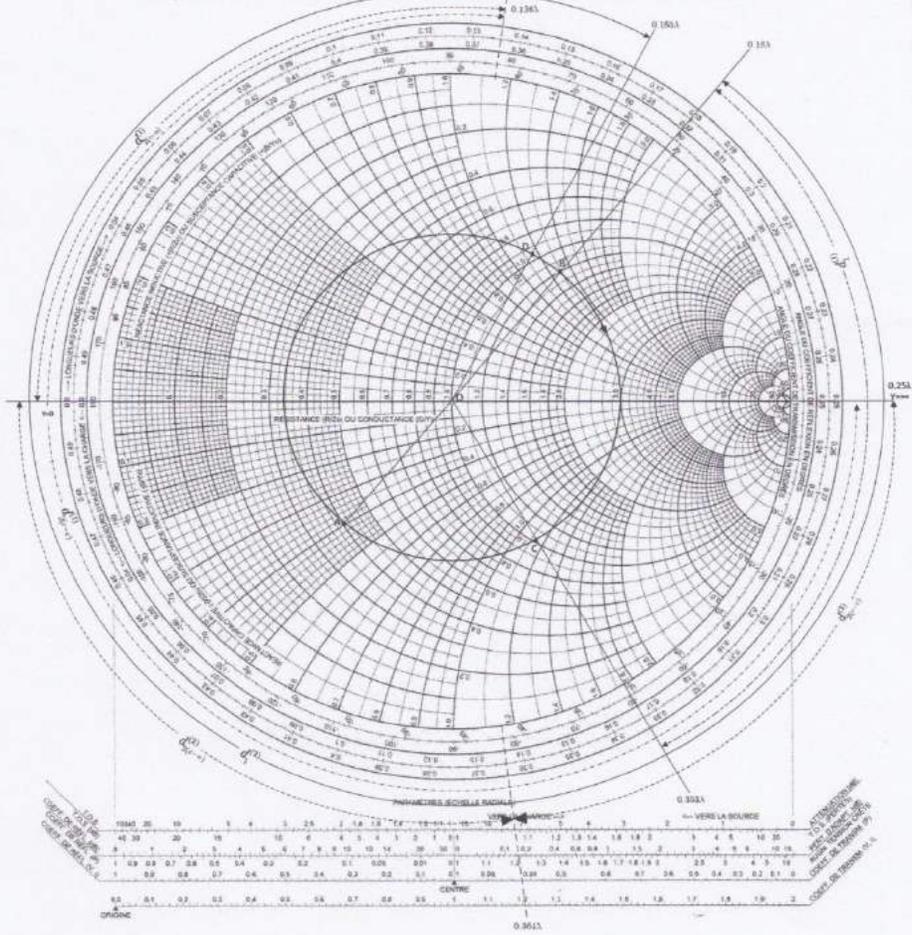

Figure I.42.

**Exemple I.23.**

Adapter une charge de l'impédance $Z_L = (50 + j50)\ (\Omega)$ à une ligne d'impédance caractéristique de $Z_0 = 50\ (\Omega)$ à l'aide d'un syntoniseur à branche simple (tronçon).

a) Trouver deux solutions pour la distance entre le tronçon et la charge ainsi que la longueur du tronçon (le tronçon en court-circuit en série) (Figure 1.32) et,

b) deux autres solutions (le tronçon en circuit ouvert en série) (Figure I.33).

Solution :

L'impédance normalisée est :

$$z_L = \frac{Z_L}{Z_0} = \frac{50 + j50}{50} = 1 + j$$

Sur l'abaque de Smith L'impédance normalisée se trouve au point $A$.

Le cercle de rayon $|\rho| = OA =$ cte., coupe le cercle $r = 1$ ce qui donne deux points $B = A$ et $C$ pour lesquels on lit :

Au point $B$  $z_B = 1 + j1$,  $d_1^{(1)}$= (0.162−0.162) $\lambda$ = 0

Au point $C$  $z_C = 1 − j1$,  $d_1^{(2)}$= (0.337−0.162) $\lambda$ = 0.175 $\lambda$

a) Tronçon en court-circuit :

Adaptation au point $B$. On ajoute $y_{s1} = -j1, d_2^{(1)}$ =(0.375-0.25) $\lambda$=0.125$\lambda$ (flèche pointillée)

Adaptation au point $C$. On ajoute $y_{s2} = j1$, $d_2^{(2)} = (0.125\text{-}0.25\text{+}0.5)$ $\lambda$=0.175$\lambda$

(flèche pointillée)

Donc, les couples sont :

$$d_1^{(1)} = 0, \; d_2^{(1)} = 0.125\lambda \quad \text{et}$$
$$d_1^{(2)} = 0.175\lambda, \; d_2^{(2)} = 0.375\lambda$$

Le calcul analytique donne :

$$d_1^{(1)} = 0.1762\lambda, \; d_2^{(1)} = 0.125\lambda \quad \text{et}$$
$$d_1^{(2)} = 0, \; d_2^{(2)} = 0.375\lambda$$

Pour le tronçon en court-circuit on lit ces longueurs à partir du point 0.25 $\lambda$ où $z$= 0.

b) Tronçon ouvert :

Les distances entre le tronçon et la charge restent les mêmes comme dans le calcul précédent.

Au point $B$, $z_{s1} = -j1$, $d_2^{(1)} = (0.375 - 0.00)\lambda = 0.337\lambda$ (flèche noire)

Au point $C$, $z_{s2} = j1$, $d_2^{(2)} = (0.125 - 0.00)\lambda = 0.162\lambda$ (flèche noire)

Le calcul analytique donne :

$$d_1^{(1)} = 0.1762\lambda, \qquad d_2^{(1)} = 0.3750\lambda \quad \text{et}$$
$$d_1^{(2)} = 0, \; d_2^{(2)} = 0.125\lambda$$

Pour le tronçon en circuit ouvert on lit ces longueurs à partir du point 0.25 λ où $z$= ∞.

Chaque étape peut être suivie sur l'abaque de Smith (Figure 1.43).

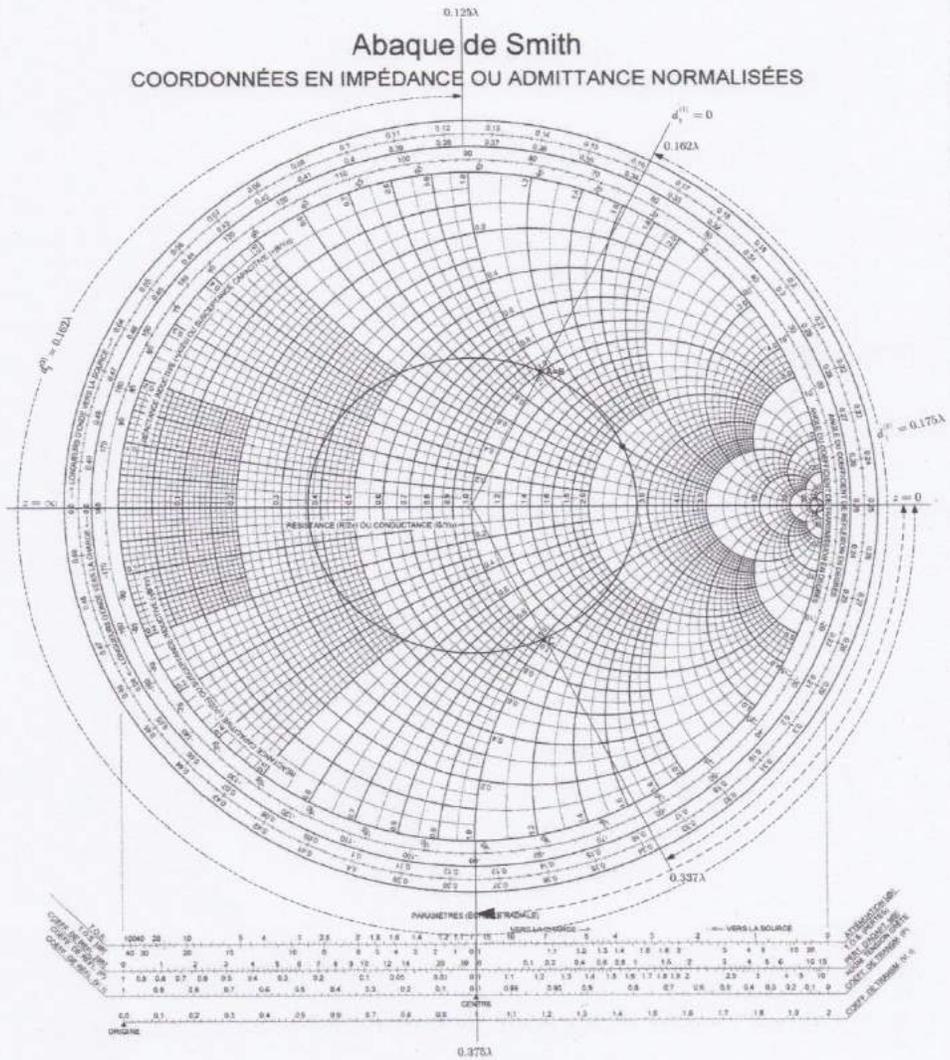

Figure I.43.

## Exemple I.24.

En utilisant l'abaque de Smith déterminer l'impédance d'entrée des deux lignes montrées à la figure I.44.

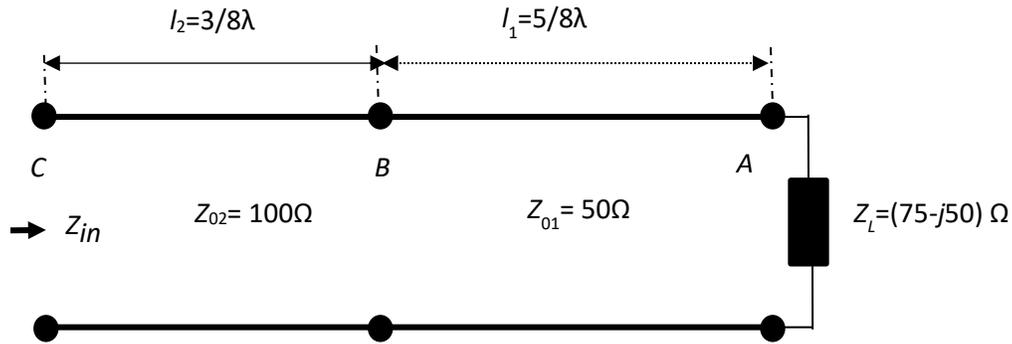

Figure I.44.

Solution:

$$z_{LN} = \frac{Z_L}{Z_{01}} = 1.5 - j1 \quad \text{point } A$$

On se déplace du point $A$, sur le cercle $SWR$, au point $B_{Droit}$ pour la distance $5/8\lambda$. Au point $B$ sur l'abaque de Smith on lit l'impédance normalisée, $z_{N1} = 0.48 - j0.36$.

L'impédance au point $B$ est $Z_1 = Z_{01}z_{N1} = (24 - j18)\Omega$. Pour continuer à travers la deuxième ligne il faut normaliser l'impédance $Z_1$.

$$z_{N2} = \frac{Z_1}{Z_{02}} = 0.24 - j0.18$$

On trace le cercle $SWR$ on se déplace du point $B_{Gauche}$ vers le générateur pour la distance $3/8\lambda$ et on trouve le point $C$. On trouve à ce point l'impédance d'entrée normalisée,

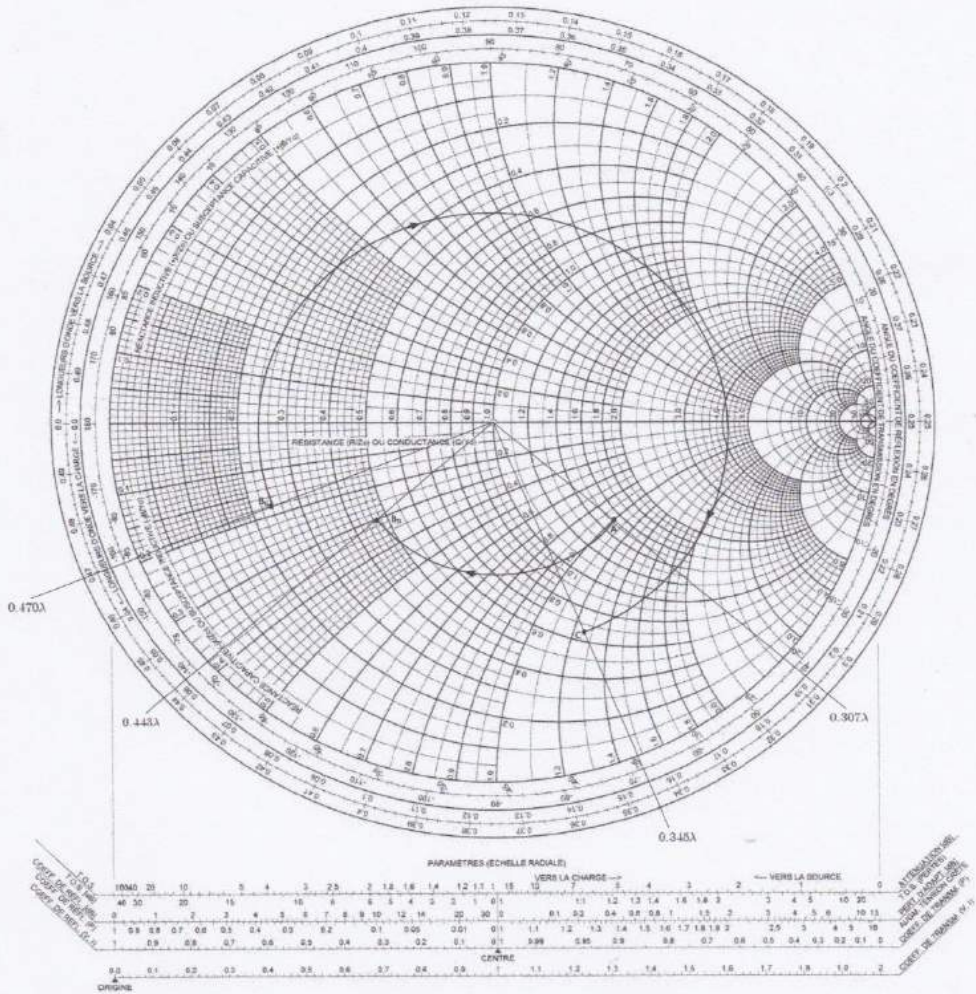

Figure I.45.

$$z_{Nin} = 0.66 - j1.25$$

L'impédance d'entrée est $Z_{in} = Z_{03}z_{Nin} = (66 - j125)\Omega$.

Chaque étape peut être suivie sur l'abaque de Smith (Figure I.45)

## Exemple I.25.

En utilisant l'abaque de Smith déterminer l'impédance d'entrée des de la ligne d'alimentation montrée à la figure I.46. Toutes les lignes sont sans pertes et avec $Z_0 = 50\ \Omega$.

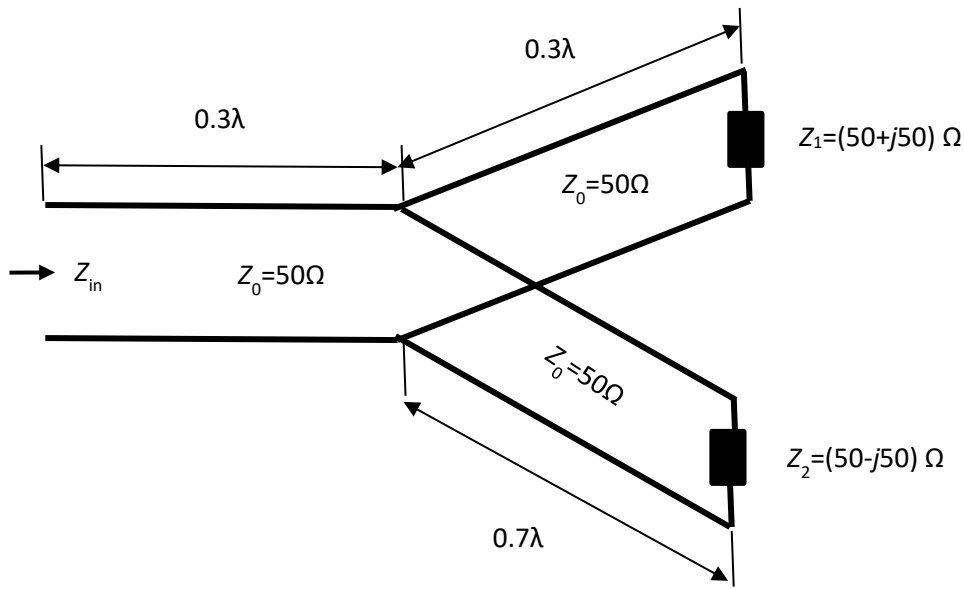

Figure I.46.

Solution:

$$z_1 = \frac{Z_1}{Z_0} = 1 + j1 \quad \text{point } A \text{ sur l'abaque de Smith}$$

$$z_2 = \frac{Z_2}{Z_0} = 1 - j1 \quad \text{point } B \text{ sur l'abaque de Smith}$$

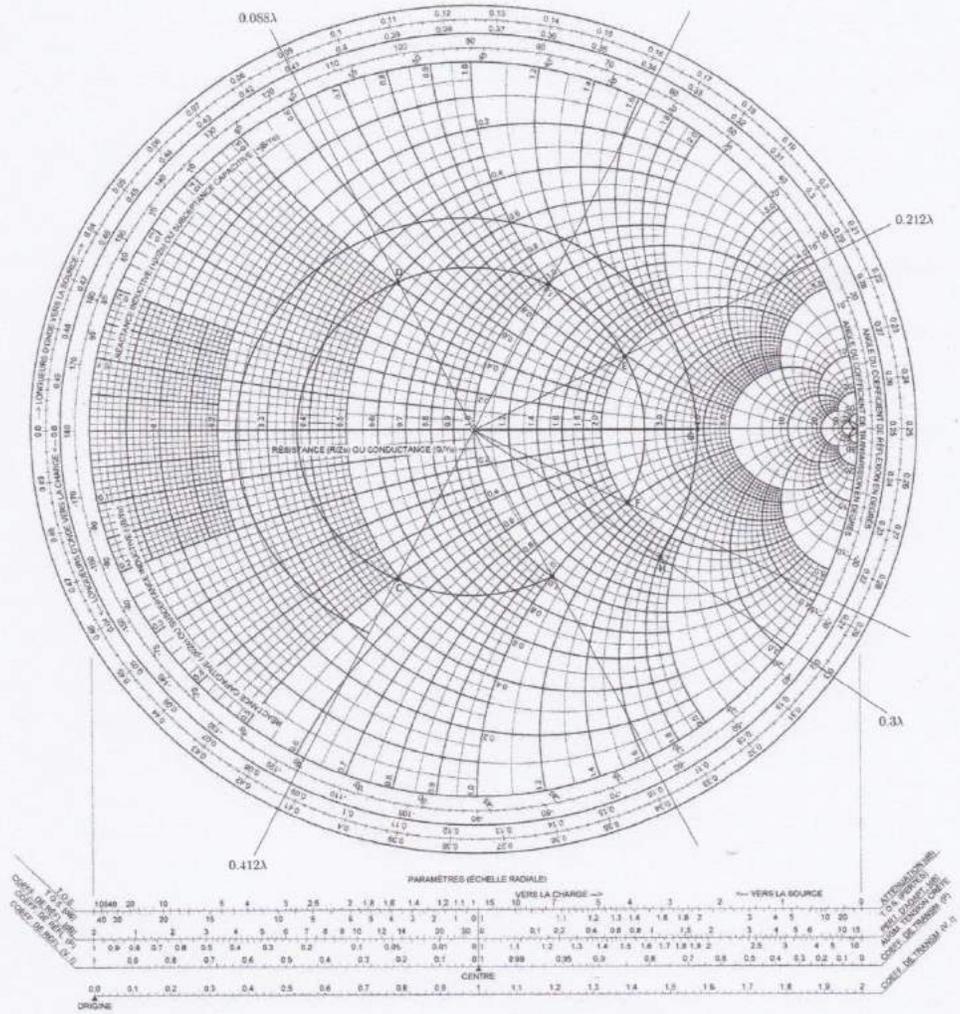

Figure I.47.

À la jonction les lignes sont connectées en parallèle on va travailler avec les admittances. L'admittance $y_1$ est le point *C*, qui est à $0.412\lambda$ sur l'échelle (*WTG-VS*).

L'admittance $y_2$ est le point *D*, qui est à $0.088\lambda$ sur l'échelle (*WTG-VS*). En se déplaçant de $0.3\lambda$ vers la source on obtient l'admittance d'entrée (point *E*) pour la ligne en haut $y_{1h} = 1.97 + j1.02$. Ensuite, on se déplace de $0.7\lambda$ (c'est à dire à $0.2\lambda$) on obtient l'admittance d'entrée (point *F*) pour la ligne en bas $y_{1b} = 1.97 - j1.02$. L'admittance d'entrée à la jonction (Point G) est $y_{Jonction} = y_{1h} + y_{1b} = 3.94 + j0$.

En se déplaçant de $0.3\lambda$ vers la source on obtient l'admittance d'entrée (Point *H*) d'où on trouve l'impédance d'entrée normalisée, $z_{Nin} = 1.65 - j1.79$. L'impédance d'entrée est $Z_{in} = Z_0 z_{Nin} = (82.5 - j89.5)\Omega$. Chaque étape peut être suivie sur l'abaque de Smith (Figure I.47).

# 10. Problèmes

 Pour une fréquence *f*, l'impédance caractéristique sur une ligne de transmission sans pertes est,

$$Z_0 = \sqrt{\frac{3500 + j25000}{1541 + j10}}$$

Trouver l'impédance de cette même ligne pour $f \to \infty$ .

Solution : $[Z_0 = 50 \ (\Omega)]$

Exemple 2. Une ligne de transmission sans distorsion a une impédance caractéristique,

$$Z_0 = \sqrt{\frac{7x10^{-3} + j\omega L}{1.5909x10^{-8} + j10^{-2}}}$$

à une fréquence $f = \frac{1}{\pi}(GHz)$.

a) Calculer la valeur de *L*,
b) Calculer la constante de phase β,
c) Calculer la vitesse de phase $v_p$. Est-ce que cette vitesse est réaliste ?  Pourquoi ?

Solution : [a) *L* = 2.2 (μH/m) ; b) β = 6.633 (rad/m) ; c) $v_p$ = 3.015x10$^8$ $(\frac{m}{s})$. Cela dépasse la vitesse de la lumière]

3) Une ligne de transmission bifilaire est modélisée par les paramètres distribués :

$$R = 1.1x10^{-3}(\frac{\Omega}{m}), L = 0.16(\frac{\mu H}{m}), G = 0.02(\frac{\mu S}{m}), C = 0.17(\frac{nF}{m})$$

Trouver l'impédance caractéristiques, la constante de phase, l'atténuation et la vitesse de propagation à la fréquence de 3 (*kHz*). Quelle la vitesse de propagation sur la ligne sans tenir compte des pertes?

Solution: $[Z_0 = 31.7e^{-j11.3^0}(\Omega), \ \beta = 17.95x10^{-6}(\frac{rad}{m}), \ \alpha = 99.75x10^{-6}\frac{1}{m},$

$$v_p = 1.89x10^8(\frac{m}{s}) , v_{propagation} = 1.917x10^8(\frac{m}{s})]$$

**Exemple 4.** Un générateur d'impulsion de tension $V_G$ = 1 ($V$) et une impulsion de 1 ($\mu s$), est connecté à une ligne de transmission de l'impédance caractéristique $Z_0$ = 50($\Omega$) et la vitesse de propagation est $v_p$ = 3x10$^8$($m/s$). L'impédance interne du générateur d'impulsion est $R_G$ = 75 ($\Omega$), et la résistance qui termine la ligne est $R_L$ = 150 ($\Omega$) (Figure I.48). La longueur de la ligne est $l$ = 900 ($m$). En utilisant le diagramme de réflexion trouver :

a) La tension à l'extrémité d'envoie en fonction du temps pour « $t$ » jusqu'à $t$ = 12 ($\mu s$),
b) La tension à la charge terminant la ligne de transmission en fonction du temps jusqu'à $t$ = 10($\mu s$).

**Solution :** [$\rho_G$ = 0.2, $\rho_L$ = 0.5 ; $v_s$ = 0.4 ($V$) pour 0<$t$<1($\mu s$) ; $v_s$ = 0.24 ($V$) pour 6($\mu s$) <$t$<7($\mu s$); $v_s$ = 0.0024 ($V$) pour 12($\mu s$)<$t$<13($\mu s$)]

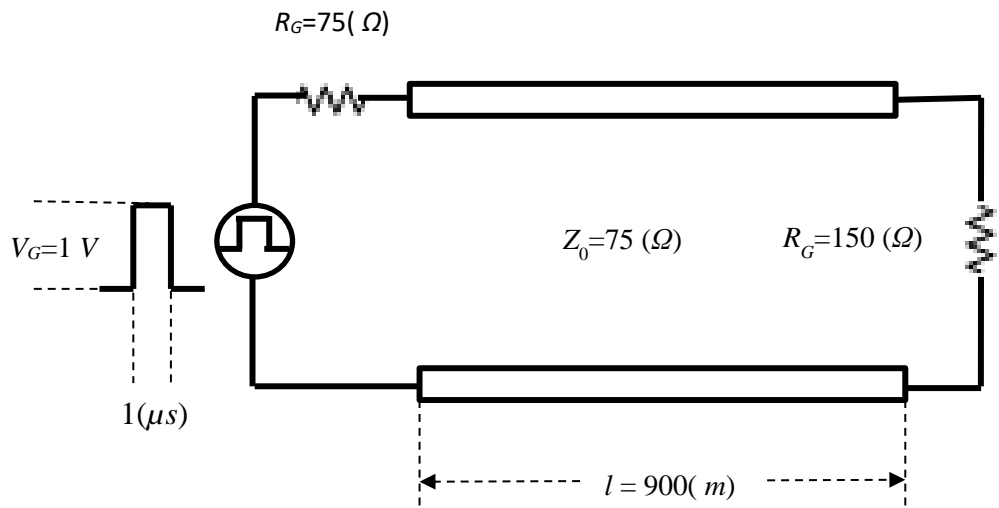

Figure I.48.

**Exemple 5.** Une ligne de transmission sans pertes ayant une impudence caractéristique de 50 ($\Omega$) est terminée par une charge une charge $Z_L$ inconnue. Le taux d'onde stationnaire sur la ligne est $s$ = 3. Les minima successifs de tension sur la ligne sont distants de 20 ($cm$) et le premier minimum se situe à 5 ($cm$) de la charge. Trouver $\rho_L$ (coefficient de réflexion à la charge) et l'impédance $Z_L$.

Solution : [ $\rho_L = 0.5e^{-j\frac{\pi}{2}}$, $Z_L = (30 - j40)$ $(\Omega)$ ]

**Exemple 6.** En mesurant le VSWR le long d'une ligne de transmission fendue, la distance entre les deux successifs minima était de 15 ($cm$) et le VSWR était 3.2. Trouver le module du coefficient de réflexion ainsi que la fréquence d'opération.

Solution : [$|\rho| = 0.52$, $f = 1$ $(GHz)$]

**Exemple 7.** Soit une charge placée au bout de la ligne de transmission. À une certaine distance $d_1$ de la charge le module de l'impédance vue $Z_c(d_1)$ est maximal est vaut 200 ($\Omega$). Cependant, à la distance $d_2$ telle que $d_2 = d_1 + 10$ ($m$), le module de l'impédance vue devient cette fois minimal avec $Z_c(d_2) = 50$ ($\Omega$). (Note : ces deux impédances vues (maximal et minimal) se rencontrent en alternance à intervalle de λ/4).
Trouver :

a) L'impédance caractéristique $Z_0$ de la ligne,
b) Le rapport d'onde stationnaire sur la ligne produit par la charge,
c) Si l'impédance vue à une troisième distance $d_3$ est (100+$j$70.7) ($\Omega$), trouver l'impédance vue à $d_3 + 10$ ($m$).

Solution : [a) Z$_0$ =100 ($\Omega$), b) *SWR* =2, c) Z$_c$($d_3 + 10$ ($m$)) = (66.7- $j$ 47.1) ($\Omega$)]

**Exemple 8.** Une source de tension sinusoïdale, $V_G(t) = 5 \cos(2\pi x 109t)(V)$ et l'impédance interne $Z_G = 50$ ($\Omega$), est connectée à une ligne sans pertes de 50 ($\Omega$). La ligne de longueur $l$ = 5 ($cm$) se termine avec une charge $Z_L = (100 - j100)(\Omega)$. Si la vitesse de propagation est $v = c$, trouver :

a) Le coefficient de réflexion à la charge,
b) L'impédance à l'entrée de la ligne,
c) La tension et le courent à l'entrée de la ligne.

Solution : [a) $\rho_L = 0.62e^{-j27.94^0}$, b) $Z_i = 17.855e^{-j45.44^0}(\Omega) = (12.53 - j12.72)(\Omega)$, c) $V_i = 1.4e^{-j34^0}(V)$, $I_i = 78.4e^{j11.4^0}(mA)$]

**Exemple 9.** Le circuit montré à la Figure I.49, consiste 100 ($\Omega$) ligne de transmission sans pertes qui est terminée par la charge $Z_L = (50 + j100)$ ($\Omega$). Si la valeur crête de la tension de la charge était mesurée $|V_L|$ = 12 *V*, trouver :

a)  Le coefficient de réflexion sur la charge,
b)  La puissance moyenne dissipée dans la charge,
c)  La puissance moyenne incidente sur la ligne,
d)  La puissance moyenne réfléchie par la charge.

**Solution :** [a) $\rho_L = 0.62e^{j82.9^0}$, b) $P_{moyenne} = 0.29$ (*W*), c)$P^i_{moyenne} = 0.47$ (*W*); d)$P^r_{moyenne} = -0.18$ (*W*)]

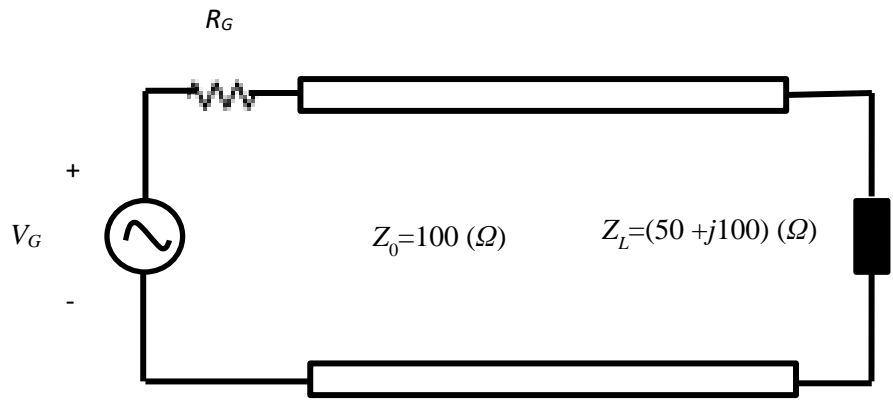

**Figure I.49.**

**Exemple 10.** Une ligne de transmission sans pertes ayant une impédance caractéristique de 50 ($\Omega$) est terminée par une charge une charge $Z_L$ inconnue. Le taux d'onde stationnaire sur la ligne est *s* = 3. Les minima successifs de tension sur la ligne sont distants de 20 cm et le premier minimum se situe à 5 cm de la charge. Trouver :

a) L'impédance $Z_L$ et $\rho_L$ (coefficient de réflexion à la charge),
b)  On veut effectuer une adaptation de la charge à la source en utilisant un tronçon de la ligne de transmission avec les mêmes impédances de $Z_0 = 50(\Omega)$ en montage parallèle sur la ligne principale (Figure I.34). Si ce tronçon se termine par un court-circuit, trouver sa longueur et sa position de jonction de jonction à partir de la charge,

c) Reprendre b) dans le cas où le tronçon de ligne de transmission est terminé par un circuit ouvert (Figure I.35).

Solution [a] $\rho_L = 0.5e^{-j\frac{\pi}{2}}$, $Z_L = (30 - j40)\Omega$, b) $d_1 = 0.0417\lambda$ et $d_2 = 0.1136\lambda$ ; $d_1 = 0.2083\lambda$ et $d_2 = 0.3864\lambda$ ; c) $d_1 = 0.0417\lambda$ et $d_2 = 0.1364\lambda$ ; $d_1 = 0.2083\lambda$ et $d_2 = 0.3636\lambda$ ]

**Exemple 11.** Une ligne de transmission de l'impédance caractéristique $Z_0$ = 50 ($\Omega$) est connectée à une charge inconnue $Z_L$. Les mesures de la tension le long de la ligne démontrent que les valeurs de tension maximum et minimum sont respectivement 1.75 ($V$) et 0.25 ($V$). Aussi, la distance la plus proche de la charge où on détecte la tension maximum est 0.125 $\lambda$.

a) Déterminer le cœfficient de réflexion $\rho_L$, l'impédance inconnue $Z_L$ ainsi que la distance la plus proche de la charge où on détecte la tension minimum,

b) Pour une autre charge inconnue les mêmes minimum et maximum sont détectées, mais la distance la plus proche de la charge où on détecte la tension minimum est 0.125 $\lambda$. Déterminer le coefficient de réflexion $\rho_L$ et l'impédance inconnue $Z_L$

Solution : [a] $\rho_L = j0.75$, $Z_L = (14 + j48)$ ($\Omega$) ; b) $\rho_L = -j0.75$,

$Z_L = (14 - j48)$ ($\Omega$)]

**Exemple 12 :** Une ligne de transmission sans pertes de l'impédance caractéristique $Z_0$ = 50 ($\Omega$) est terminée par une charge $Z_L$ = (25 + $j25$) ($\Omega$). On désire réaliser une adaptation d'impédance utilisant,

    a) Un tronçon de court-circuit (Figure I.32),
    b) Un tronçon de circuit ouvert (Figure I.33),

de l'impédance caractéristique $Z_{T0}$ = 35 ($\Omega$) qui sera placé en série avec la ligne. Déterminer les positions possibles de ce tronçon, calculées à partir de la charge, et également sa longueur pour chacun des cas.

Solution [a] $d_1 = 0.25\lambda$ et $d_2 = 0.1528\lambda$ ; $d_1 = 0.125\lambda$ et $d_2 = 0.3472\lambda$ ; b) $d_1 = 0.25\lambda$ et $d_2 = 0.4028\lambda$ ; $d_1 = 0.125\lambda$ et $d_2 = 0.0972 \lambda$ )]

**Exemple 13.**  25 ($\Omega$) antenne est connecté à 75 ($\Omega$) ligne de transmission sans pertes. Les réflexions vers le générateur peuvent être éliminées en plaçant une impédance $Z$ à la distance $l$ de la charge (Figure I.50).  En utilisant l'abaque de Smith trouver $Z$ et $l$.

Solution : [$l$ = 0.25 $\lambda$, $Z$ = 112.5 ($\Omega$)[

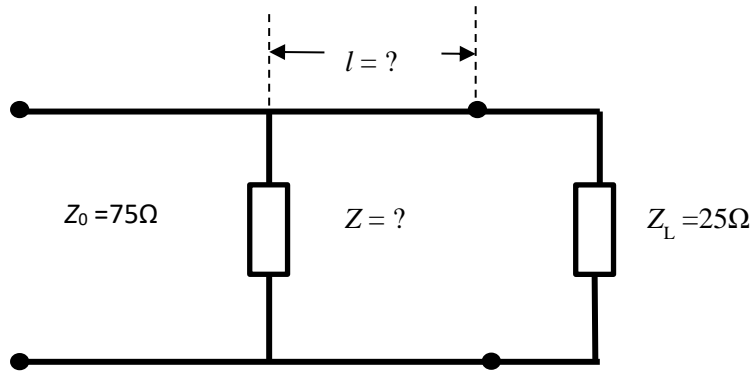

Figure I.50.

**Exemple 14.**  Dans une ligne sans pertes, l'impédance caractéristique $Z_0$ = 100 ($\Omega$) est placée une charge d'impédance : $Z_L$ = (30 + j55) ($\Omega$). La fréquence de travail est $f$ = 1000 ($MH_z$) et la vitesse de phase est $v_p$=2x10$^8$ ($m/s$).

  a) Déterminer la position de $Z_L$ et le coefficient de réflexion $\rho$ (module et argument) sur la charge (abaque de Smith),
  b) Déterminer l'admittance de la charge (abaque de Smith),
  c) Déterminer à l'aide de l'abaque de Smith, la valeur de l'impédance réduite ramenée à 12 (cm) de la charge,
  d) Même question pour $Z_0$ = 70($\Omega$).

Solution :[$a$) $z_L = 0.3 + j0.55$, $\rho_L = |\rho_L|e^{j\varphi_L} = 0.6e^{j119^0}$ ; b) $y_L = 0.78 - j1.4$, $Y_L = (7.8 - j14)(mS)$ ; c) $\lambda = 20\ (cm)$, $z_{L(12\ cm)} = 0.5 + j0.8$, $Z_{L(12\ cm)} = (50 + j80)(\Omega)$]

**Exemple 15.**  Dans la Figure I.34, une adaptation de ligne sans perte de longueur $d_1$ = 0.25 $\lambda$ est réalisée par un tronçon court-circuité de longueur $d_2$ = 0.125 $\lambda$ en parallèle. La ligne et le tronçon sont de même nature avec $Z_0$ = 50 ($\Omega$).

a) Trouver la valeur inconnue de $Z_L$ en utilisant le calcul analytique.
b) Trouver l'autre couple ($d_1$, $d_2$), en utilisant l'abaque de Smith.

Solution [a) $Z_L = 50(1 + j)(\Omega)$ ; b) $d_1 = 0.25\,\lambda$, $d_2 = 0.125\,\lambda$, et $d_1 = 0.425\,\lambda$, $d_2 = 0.375\,\lambda$]

Exemple 16. Deux lignes de transmission sont connectées en tandem avec un connecteur endommagé. Les lignes de transmission et le circuit équivalent du connecteur sont illustrés à la Figure I.51. Le système en cascade est connecté à une charge $Z_L = (75 - j125)$ ($\Omega$) et à un générateur de tension de sortie $V_G = 100\,e^{j0}$ ($V$) avec une résistance interne égale à 180 ($\Omega$). En utilisant l'abaque de Smith, déterminer :

a) l'impédance normalisée de la charge,
b) l'admittance normalisée de la charge,
c) le coefficient de réflexion à la charge,
d) l'impédance normalisée au point $O_2$ juste avant le connecteur,
e) le coefficient de réflexion au point $O_2$ juste avant le connecteur,
f) le coefficient TOS de la ligne 2,
g) l'impédance du circuit équivalent du connecteur, avec l'impédance $Z_{O2}$, ($Z_j$),
h) l'impédance normalisée $z_j$,
i) l'impédance $Z_{O1}$, juste après la ligne 1.

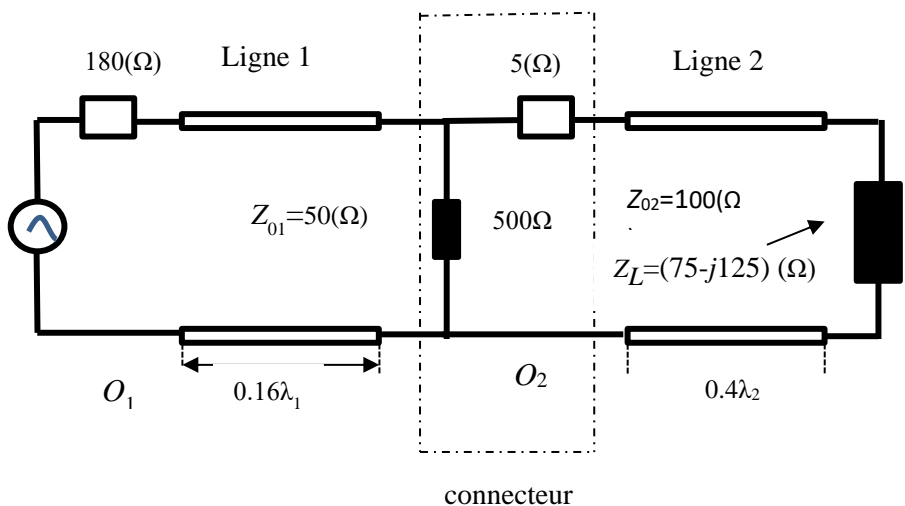

connecteur

Figure I.51.

**Solution :** [a) $z_{Ln} = (0.75 - j1.25)$ ; b) $y_{Ln} = (0.353 + j0.588)$ ; c) $\rho_L(0) = 0.59e^{-j65.8^0}$ ; d) $z_{nO2} = 3.8 + j0.65$, $Z_{O2} = (380 + j65)$ $(\Omega)$ ; e) $\rho(0.4\lambda_2) = \rho_L(0) \, e^{j\theta_{0.4\lambda_2}} = 0.59e^{j5.7^0}$ ; f) $TOS = s = 3.88$ ; g) Pour passer la jonction entre la ligne 1 et la ligne 2, le circuit équivalent du connecteur doit être combiné avec $Z_{O2}$ pour donner la valeur de l'impédance de la ligne de transmission 1, $Z_j = (219 + j20.6)(\Omega)$, h) $z_{nj} = 4.38 + j0.41$ i) $Z_{O1} = (16.5 - j31.5)$ $(\Omega)$]

**Exemple 17.** Une ligne de transmission sans perte avec $Z_0 = 50$ $(\Omega)$, terminée par une charge $Z_L = (72 + j96)$ $(\Omega)$, est adaptée à l'aide d'un transformateur quart d'onde. En utilisant l'abaque de Smith, trouver :

a) L'emplacement du transformateur quart d'onde par rapport à la charge ($d_1$),
b) L'impédance caractéristique de la ligne d'adaptation,
c) Le SWR de la ligne de longueur $d_1$,
d) Le SWR dans la ligne d'adaptation.

**Solution :** [a) $d_1 = 0.053 \, \lambda$; b) $Z'_0 = 106.07$ $(\Omega)$; c) $SWR^{(d_1)} = 4.5$; d) $SWR' = 2.1$]

**Exemple 18.** Une impulsion d'amplitude de 15 ($V$) et d'une durée de $1\mu s$ est appliquée à travers une résistance en série de 25 ($\Omega$) aux bornes d'entrée d'une ligne de transmission sans perte avec 50 ($\Omega$). La ligne mesure 400 $m$ de long et est terminée par un court-circuit (Figure I.52). Déterminer l'évolution temporelle de la tension au point milieu de la ligne pour le temps $t$ entre 0 et $8\mu s$. Le constant diélectrique relatif dans la ligne est 2.25.

**Solution :** [$V(200,t) = 0$ ($V$) pour 0 ($\mu s$) ≤ $t$ < 1 ($\mu s$); $V(200,t) = 10$($V$) pour 1 ($\mu s$) ≤ $t$ <2 ($\mu s$); $V(200,t) = 0$ ($V$) pour 2 ($\mu s$) ≤ $t$ < 3 ($\mu s$); $V(200,t) = $ - 10 ($V$) pour 3 ($\mu s$) ≤ $t$ < 4 ($\mu s$); $V(200,t) = 0$ ($V$) pour 4 ($\mu s$) ≤ $t$ < 5 ($\mu s$); $V(200,t) = 10/3$ ($V$) pour 5 ($\mu s$) ≤ $t$ < 6 ($\mu s$); $V(200,t) = 0$ ($V$) pour 6 ($\mu s$) ≤ $t$ < 7 ($\mu s$); $V(200,t) = -10/3$ ($V$) pour 7 ($\mu s$) ≤ $t$ < 8 ($\mu s$)]

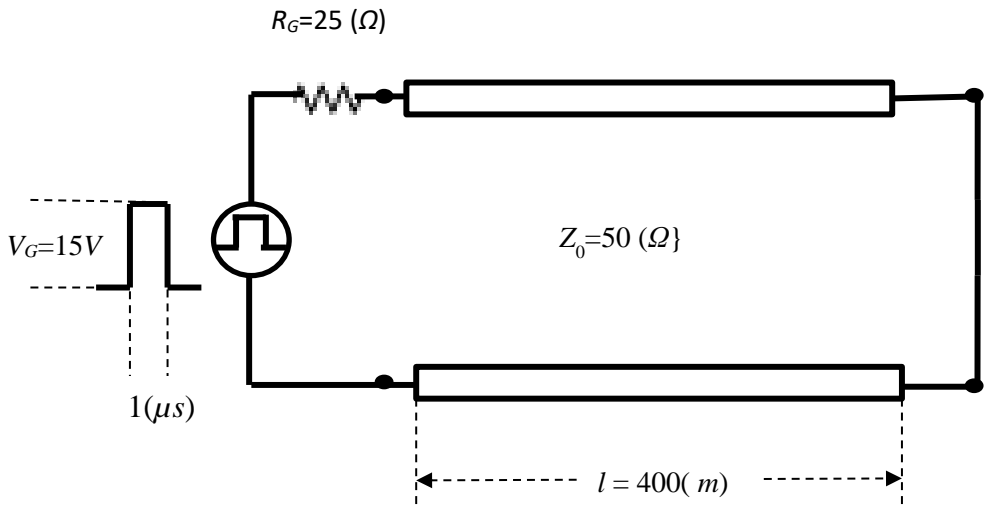

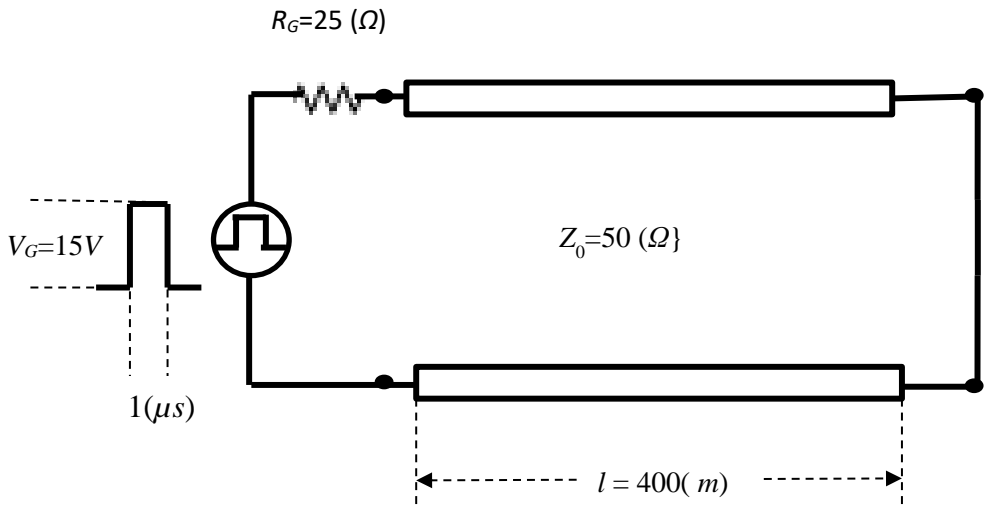

Figure I.52.

**Exemple 19.** Une ligne de transmission sans perte de l'impédance caractéristique de $100(\Omega)$ de longueur $3\lambda/8$ se termine avec une impédance inconnue $Z_L$. Si l'impédance d'entrée est $Z_{in} = -j2.5 \ (\Omega)$, en utilisant l'abaque de Smith, trouver:

a) l'impédance inconnue $Z_L$,
b) Quelle longueur de tronçon ouvert et de court-circuit pourrait-on utiliser pour remplacer $Z_L$.

**Solution**: [a] $Z_L = j95 \ (\Omega)$; $d_{tr_{(c-o)}} = 0.371\lambda$ ; $d_{tr_{(c-c)}} = 0.121\lambda$]

**Exemple 20.** Une ligne de transmission sans perte avec 50 $(\Omega)$ se termine à une charge $Z_L = (50 + j25) \ (\Omega)$. En utilisant l'abaque de Smith, trouver :

a) Le coefficient de réflexion,
b) Le SWR,
c) L'impédance à une distance de 0.35 $\lambda$ de la charge,
d) L'admittance à une distance de 0.35 $\lambda$ de la charge,
e) La longueur la plus courte de la ligne afin d'avoir une impédance d'entrée purement résistive,
f) L'emplacement de la première tension maximale.

Solution: [a] $\rho_L = 0.24 e^{j75^0}$; b) $SWR = 1.65$; c) $Z(0.35\,\lambda) = (30 - j)(\Omega)$; d) $Y(0.35\,\lambda) = (0.034 - j0.0016)(S)$; e) $0.015\,\lambda$; f) $0.015\,\lambda$ ]

**Exemple 21.** Les impédances de circuit ouvert et court-circuit mesurées aux bornes d'entrée d'une ligne de transmission sans perte de longueur $l$ = 1.5 ($m$) (ce qui est inférieur à un quart de longueur d'onde) sont $-j54.6(\Omega)$ et $j103(\Omega)$ respectivement.

a) Trouver $Z_0$, $\beta$ et $\lambda$,
b) Sans changer la fréquence de fonctionnement trouver l'impédance d'entrée d'une ligne court-circuitée deux fois la longueur donnée,
c) Combien de temps la ligne en court-circuit doit-elle être en ordre qu'il apparaisse comme un circuit ouvert à l'entrée ?

Solution:

[a] $Z_0 = \sqrt{Z_{in(c-c)} Z_{in(c-o)}} = 75(\Omega)$, $\beta = \frac{1}{l} \text{atan}\left(\sqrt{-\frac{Z_{in(c-c)}}{Z_{in(c-o)}}}\right) = 0.628 \left(\frac{rad}{m}\right)$;

$\lambda = 10\ (m)$; b) $L = 2l = 3(m)$, $Z_{in(c-c)} = j\,Z_0 \tan(\beta L) = -j232(\Omega)$;

c) $Z_{in(c-c)} = \infty = j\,Z_0 \tan(\beta l) \rightarrow l = (2n+1)\frac{\lambda}{4}$, $n = 0,1,2,\ldots$ ]

**Exemple 22.** Une ligne de transmission a les paramètres suivants $R = 100 \left(\frac{\Omega}{m}\right)$, $L = 80 \left(\frac{nH}{m}\right)$, $G = 1.6 \left(\frac{S}{m}\right)$ et $C = 200 \left(\frac{pF}{m}\right)$. Si une onde progressive de fréquence de 2 ($GH_z$) se propage sur la ligne de transmission,

a) Quelle est la constante de propagation, d'atténuation et phase?
b) Quelle est la constante de propagation, d'atténuation et phase a 1.9 ($GH_z$)?
c) Quelle est la vitesse de phase?
d) Quelle est l'impédance caractéristique de la ligne?
e) Quelle est la vitesse de groupe?

Solution:

[a] À 2 ($GHz$) → $\gamma_1 = (17.94 + j\,51.85) \left(\frac{1}{m}\right)$, $\alpha_1 = 17.94 \left(\frac{Np}{m}\right)$,

$\beta_1 = 51.85 \left(\frac{rad}{m}\right)$; b) À 1.9 ($GHz$) → $\gamma_2 = (17.784 + j\,40.397) \left(\frac{1}{m}\right)$,

$\alpha_2 = 17.784 \left(\frac{Np}{m}\right)$, $\beta_2 = 40.397 \left(\frac{rad}{m}\right)$; c) $\lambda = 0.1212\ (m)$;

d) $v_p = 2.42 x 10^8 \left(\frac{m}{s}\right)$; e) $Z_0 = (17.9 + j4.3)(\Omega)$; f) $v_g = \frac{\Delta\omega}{\Delta\beta} \rightarrow$

$$v_g = \frac{\Delta \omega}{\Delta \beta} = \frac{2\pi(f_2 - f_1)}{\beta_2 - \beta_1} = 2.563x10^8 \left(\frac{m}{s}\right)\ ]$$

**Exemple 23.**
La tension dans une ligne de transmission sans distorsion est donnée par,
$V(x,t) = 60e^{0.0025x} \cos(108t + 2x) + 12e^{-0.0025x} \cos(108t - 2x)\ (V)$ ,où «$x$»
est la distance mesurée par rapport à la charge. Si $Z_L = 200\ (\Omega)$, déterminer :

   a) La constante d'atténuation, la constante de phase et la vitesse de phase.
   b) L'impédance caractéristique $Z_0$ et le courant $I(x,t)$.

**Solution**:
[a] $\alpha = 0.0025 \left(\frac{Np}{m}\right), \beta = 2 \left(\frac{rad}{m}\right), v_p = 0.5x10^8 \left(\frac{m}{s}\right)$; b) $Z_0 = 200(\Omega)$,
   $I(x,t) = 0.3e^{0.0025x} \cos(108t + 2x) - 0.3e^{-0.0025x} \cos(108t - 2x)\ (A)]$

**Exemple 24.** Une ligne de transmission possède une impédance caractéristique de
$50\ (\Omega)$. La ligne est terminée par une résistance de 80 $(\Omega)$ où on mesure une tension
de 5 ($V$). Trouver :

   a) Le coefficient de réflexion à la charge.
   b) Le rapport d'onde stationnaire.
   c) Les valeurs de $V_{max}, V_{min}, I_{max}$ et $I_{min}$.
   d) La valeur de charge vue aux distances $d = \frac{\lambda}{4}, \frac{\lambda}{2}\ et\ \frac{3\lambda}{8}$ de 80 $(\Omega)$.

**Solution**:
[a] $\rho = 0.23$; b) $SWR = 1.6$; c) $V_{max} = 5\ (V), V_{min} = 3.1\ (V), I_{max} = 0.1\ (A)$,
$I_{min} = 0.063\ (A)$; d) $Z\left(\frac{\lambda}{4}\right) = 31.25\ (\Omega), Z\left(\frac{\lambda}{2}\right) = 80\ (\Omega)$,
$Z\left(\frac{3\lambda}{8}\right) = (44.95 + j21.9)\ (\Omega)]$

**Exemple 25.** Une ligne de transmission d'une longueur de 30 ($m$) avec $Z_0 = 50\ (\Omega)$
fonctionne à 2 ($MH_z$) se termine à une charge $Z_L = (60 + j40)\ (\Omega)$. Si $v = 0.6c$
déterminer :
   a) Le coefficient de réflexion,
   b) SWR,
   c) L'impédance d'entrée.

Obtenir les résultats analytiquement et également en utilisant l'abaque de Smith.

<span style="color:cyan">Solution</span>:

[a) $\rho_L = 0.352 e^{j56^0}$; b) $SWR = 2.088$; c) $Z_{in} = (23.97 + j1.35)\ (\Omega)$ ]

# Chapitre II

## 1. ONDES PLANES

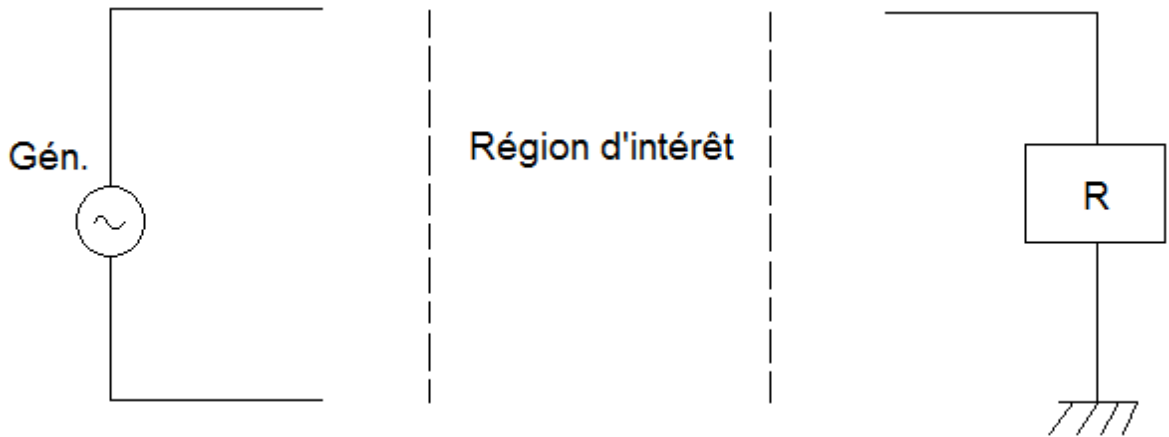

Figure II.1.

On s'intéresse à une zone particulière de l'espace (Figure II.1) entre le générateur et le récepteur. Du côté générateur, une structure d'antenne formée de conducteurs métalliques permet le déplacement des charges électriques en fonction du temps. Une onde électromagnétique est alors émise. On obtient des champs électriques et magnétiques variant dans le temps.

But visé: L'obtention des propriétés de l'onde qui se transmet entre l'émetteur et le récepteur, en espace libre. On va supposer qu'on se trouve à une distance suffisamment éloignée de la source pour ne pas être dérangé des effets du champ proche.

## 1.1 Équation de Maxwell dans ce contexte

$$\nabla \times \bar{H} = \bar{J} + \frac{\partial \bar{D}}{\partial t}$$

$$\nabla \times \bar{E} = -\frac{\partial \bar{B}}{\partial t}$$

$$\nabla \times \bar{D} = \rho$$

$$\nabla \times \bar{B} = 0$$

Pour le milieu homogène, isotrope et linéaire, on a,

$$\nabla \times \bar{H} = \sigma \bar{E} + \frac{\partial \bar{E}}{\partial t}$$

$$\nabla \times \bar{E} = -\mu \frac{\partial \bar{H}}{\partial t}$$

Ce sont des équations couplées en $\bar{H}$ et $\bar{E}$. On essaye d'obtenir une équation qui contient $\bar{E}$ ou seulement $\bar{H}$.

$$\nabla \times \nabla \times \bar{H} = \sigma(\nabla \times \bar{E}) + \varepsilon \frac{\partial(\nabla \times \bar{E})}{\partial t}$$

Puisque

$$\nabla \times \bar{E} = -\mu \frac{\partial \bar{H}}{\partial t}$$

$$\nabla \times \nabla \times \bar{H} = \sigma\left(-\mu \frac{\partial \bar{H}}{\partial t}\right) + \varepsilon \frac{\partial\left(-\mu \frac{\partial \bar{H}}{\partial t}\right)}{\partial t}$$

Pour une identité vectorielle on a :

$$\nabla \times \nabla \times \bar{H} = \nabla(\nabla \cdot \bar{H}) - \nabla^2 \bar{H}$$

Terme nul

Donc,

$$\nabla \times \nabla \times \bar{H} = -\nabla^2 \bar{H}$$

d'où

$$-\nabla^2 \overline{H} = -\sigma\mu \frac{\partial \overline{H}}{\partial t} - \varepsilon\mu \frac{\partial^2 \overline{H}}{\partial t^2}$$

Les équations d'ondes seront alors

$$\nabla^2 \overline{H} - \varepsilon\mu \frac{\partial^2 \overline{H}}{\partial t^2} = \sigma\mu \frac{\partial \overline{H}}{\partial t}$$

$$\nabla^2 \overline{E} - \varepsilon\mu \frac{\partial^2 \overline{E}}{\partial t^2} = \sigma\mu \frac{\partial \overline{E}}{\partial t} - \nabla \left(\frac{\rho}{\varepsilon}\right)$$

On suppose un milieu,

- sans charges d'espace : $\rho = 0$
- non-conducteur : $\sigma = 0$

On aura,

$$\begin{cases} \nabla^2 \overline{H} - \varepsilon\mu \frac{\partial^2 \overline{H}}{\partial t^2} = 0 \\ \nabla^2 \overline{E} - \varepsilon\mu \frac{\partial^2 \overline{E}}{\partial t^2} = 0 \end{cases}$$

À cause de la similarité mathématique, ces deux équations ont le même type de solution. Si l'on s'intéresse seulement à $\overline{E}$ , l'équation vectorielle se met sous forme de trois équations scalaires.

$$\nabla^2 E_x - \varepsilon\mu \frac{\partial^2 E_x}{\partial t^2} = 0$$

$$\nabla^2 E_y - \varepsilon\mu \frac{\partial^2 E_y}{\partial t^2} = 0$$

$$\nabla^2 E_z - \varepsilon\mu \frac{\partial^2 E_z}{\partial t^2} = 0$$

Ces trois expressions du champ électriques ont aussi un type de solution commune.

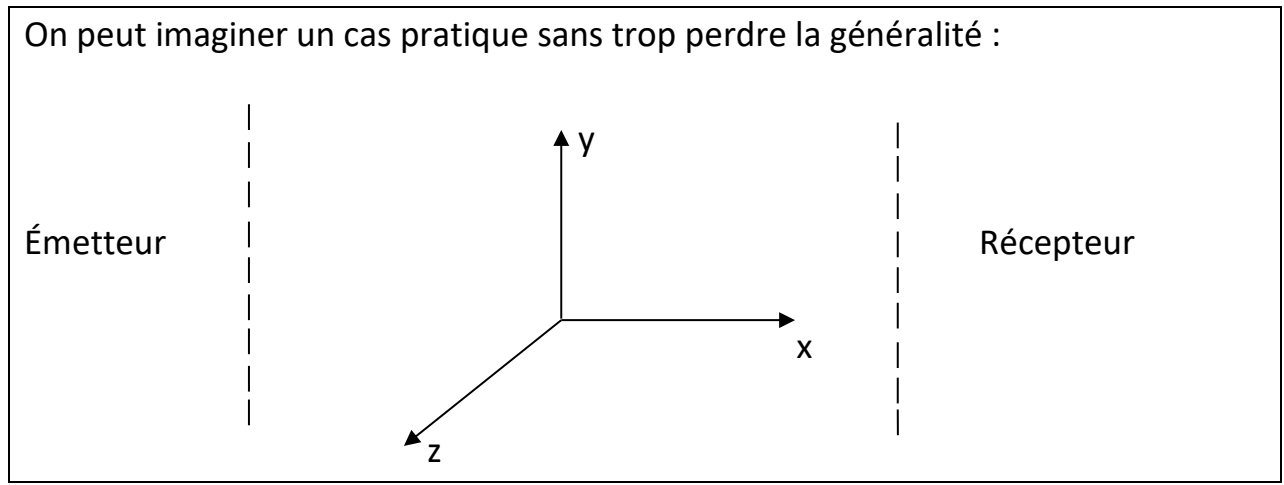

On peut imaginer un cas pratique sans trop perdre la généralité :

Émetteur

Récepteur

Figure II.2.
Soit «*x*», la direction de la propagation de l'onde (Figure II.2) et $E_x = 0$; $E_z = 0$; $E_y \neq 0$.

De plus, on suppose que $E_y$ soit indépendant de « *y* » et de « z » (l'homogénéité dans le plan *yz*).

On va avoir alors seulement une équation pour s'en occuper :

$$\nabla^2 E_y - \mu\varepsilon \frac{\partial^2 E_y}{\partial t^2} = 0$$

En coordonnées cartésiennes, cela donne :

$$\frac{\partial^2 E_y}{\partial x^2} + \underset{0}{\cancel{\frac{\partial^2 E_y}{\partial y^2}}} + \underset{0}{\cancel{\frac{\partial^2 E_y}{\partial z^2}}} = \mu\varepsilon \frac{\partial^2 E_y}{\partial t^2}$$

Finalement il ne reste que :

$$\frac{\partial^2 E_y}{\partial x^2} = \mu\varepsilon \frac{\partial^2 E_y}{\partial t^2}$$

Comme on connait la forme de l'onde lancée par le générateur, on peut supposer une fonction générale du temps et de l'espace avec quand même certaines caractéristiques :

$$E_y(x,t) = E_y{}^+(x - v_p t) + E_y{}^-(x + v_p t)$$

En remplaçant cette forme dans l'équation différentielle on obtient:

$$\frac{\partial^2 E_y}{\partial x^2} = E_y{}^{+''} + E_y{}^{-''} \qquad \text{(par rapport à l'argument)}$$

$$\frac{\partial^2 E_y}{\partial t^2} = v_p^2 E_y{}^{+''} + v_p^2 E_y{}^{-''}$$

d'où

$$E_y{}^{+''} + E_y{}^{-''} - \mu\varepsilon v_p^2 \left( E_y{}^{+''} + E_y{}^{-''} \right) = 0$$

Cette équation « identité » ne peut être satisfaite que si l'on a :

$$v_p = \frac{1}{\sqrt{\mu\varepsilon}} \qquad \text{vitesse de propagation}$$

La vitesse de propagation d'une onde électromagnétique est indépendante de la force du signal émis par le générateur.

Puisque $E_y{}^+(x - v_p t)$ n'est pas connu, on peut imaginer une forme quelconque dans le temps (Figure II.3).

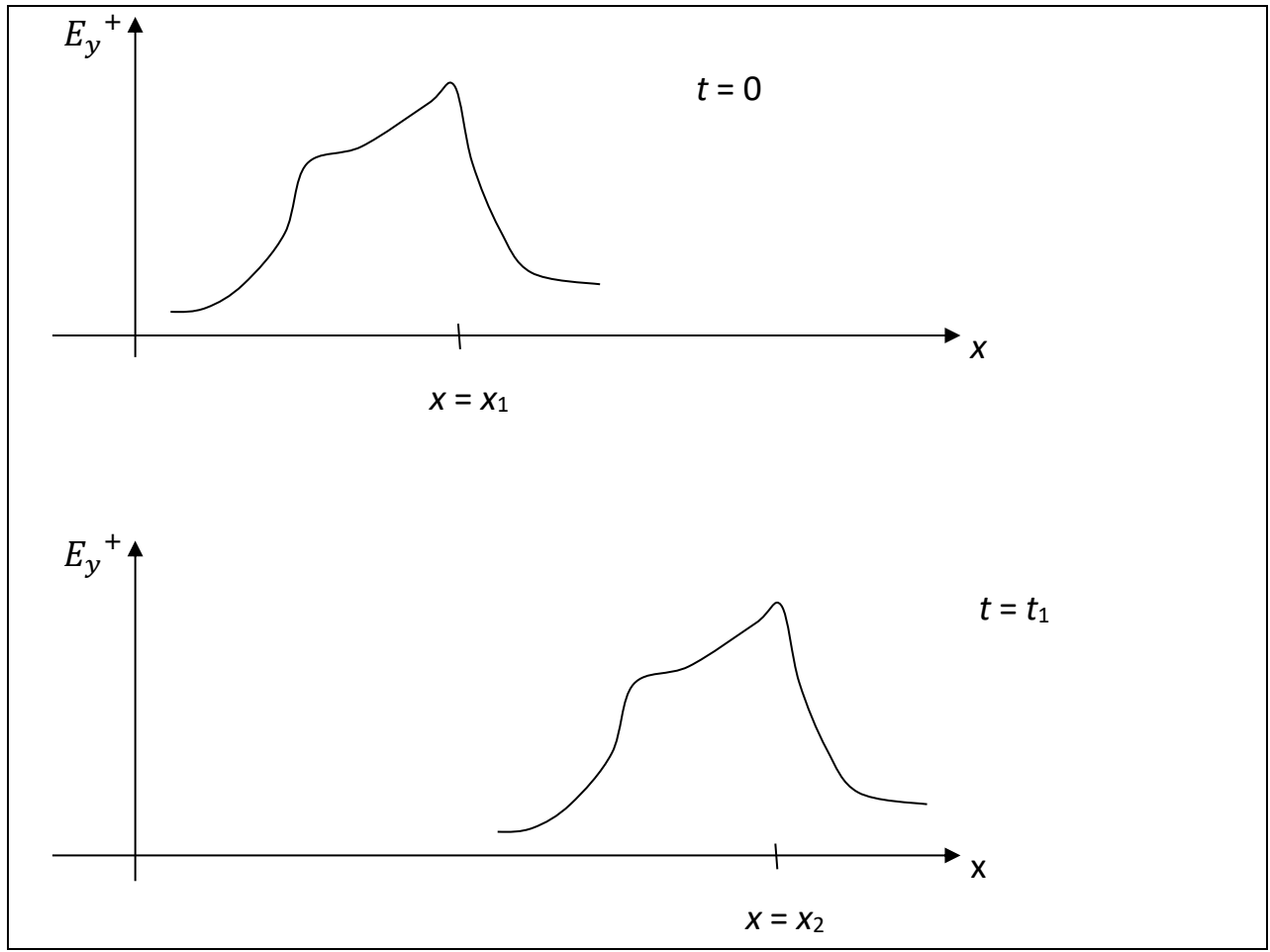

<p style="text-align:center;color:#1a9;">Figure II.3.</p>

Un instant plus tard ($t = t_1$) on retrouve la même forme d'onde à une position en « $x$ » différente de ($x = x_2$).

On peut dire que la fonction $E_y{}^+(x - v_p t)$ représente une onde plane se déplaçant dans la direction positive des « x » à une vitesse $v_p = \frac{1}{\sqrt{\mu\varepsilon}}$.

De la même façon, la fonction $E_y{}^-(x + v_p t)$ représente une onde se déplaçant dans la direction négative de « x » à la même vitesse $v_p$.

Dans un cas pratique les ondes émises par un générateur subissent, en cours de route, des réflexions dues aux obstacles.

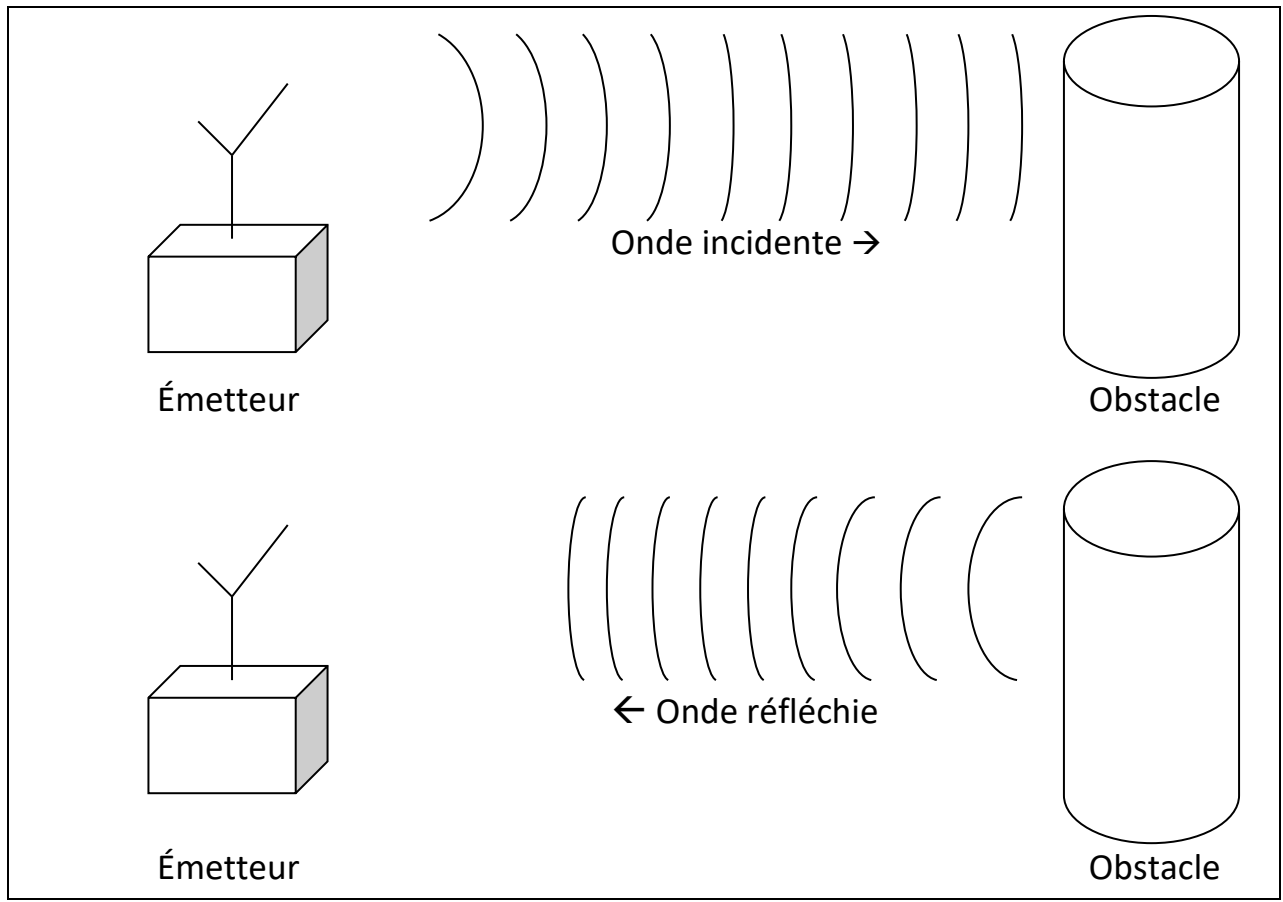

Figure II.4.

En pratique, pour une transmission normale, on essaie d'éviter les obstacles (Figure II.4). Cependant, dans le cas d'un radar, la détection de l'onde réfléchie est capitale.

Dans le cas présent on s'est intéressés à un champ électrique qui est perpendiculaire à la direction de propagation d'où « l'onde transversale » *TE*.

Une analyse identique peut être mise en route pour un champ magnétique $\bar{H}$ avec les mêmes résultats. Nous parlons alors des « ondes transversales » *TM*.

Si l'on considère les deux cas simultanément, on a les ondes de types *TEM*.

## 1.2    Définition de l'onde plane

a) Composantes du champ électromagnétique perpendiculaire à la direction de propagation.
b) Champ électromagnétique uniforme en amplitude et en phase dans un plan perpendiculaire à la direction de la propagation.

## 1.3    Cas plus général pour les ondes planes

Dans le cas précédent $\bar{H}\ et\ \bar{E}$ avaient une seule composante chacune. On peut généraliser les propriétés des ondes planes en considérant les deux composantes de chaque vecteur ($\bar{H}\ et\ \bar{E}$).

Si « $x$ » est la direction de la propagation pour un mode de propagation TEM on a:

1) $E_x = H_x = 0$        Pas de composante dans la direction de propagation

2) $\frac{\partial}{\partial y} = \frac{\partial}{\partial z} = 0$   Champ uniforme dans un plan $\perp$ à la direction de propagation

Donc il reste à considérer les composantes :

$$E_y\ et\ E_z \qquad\qquad H_y\ et\ H_z$$

Relation entre $\bar{H}\ et\ \bar{E}$ :

$$\nabla \times \bar{E} = -\mu \frac{\partial \bar{H}}{\partial t}$$

$$\nabla \times \bar{H} = \varepsilon \frac{\partial \bar{E}}{\partial t}$$

En développant

$$\nabla \times \bar{E} = \begin{vmatrix} \hat{x} & \hat{y} & \hat{z} \\ \frac{\partial}{\partial x} & \frac{\partial}{\partial y} & \frac{\partial}{\partial z} \\ E_x & E_y & E_z \end{vmatrix} \qquad \text{où} \qquad E_x = 0 \text{ et } \frac{\partial}{\partial y} = \frac{\partial}{\partial z} = 0$$

$$\nabla \times \bar{E} = \hat{x}(0) - \hat{y}\left(\frac{\partial E_z}{\partial x}\right) + \hat{z}\left(\frac{\partial E_y}{\partial x}\right)$$

$$\nabla \times \bar{E} = -\hat{y}\left(\frac{\partial E_z}{\partial x}\right) + \hat{z}\left(\frac{\partial E_y}{\partial x}\right) = -\mu \frac{\partial H}{\partial t}$$

Donc,

$$\frac{\partial E_z}{\partial x} = \mu \frac{\partial H_y}{\partial t}$$

$$\frac{\partial E_y}{\partial x} = -\mu \frac{\partial H_z}{\partial t}$$

Par un traitement semblable

$$\nabla \times \bar{H} = \hat{x}(0) - \hat{y}\left(\frac{\partial H_z}{\partial x}\right) + \hat{z}\left(\frac{\partial H_y}{\partial x}\right) = \varepsilon \frac{\partial E}{\partial t}$$

ce qui donne :

$$-\frac{\partial H_z}{\partial x} = \varepsilon \frac{\partial E_y}{\partial t}$$

$$\frac{\partial H_y}{\partial x} = \varepsilon \frac{\partial E_z}{\partial t}$$

On a obtenu ainsi des relations recherchées entre les composantes de $\bar{H}$ et $\bar{E}$, mais elles ne sont pas commodes à utiliser.

On avait déjà suggéré une forme de solution une forme de solution pour,

$$E_y(x,t) = E_y{}^+(x - v_p t) + E_y{}^-(x + v_p t)$$

Une forme semblable peut être envisagée pour $E_z$, $H_y$ et $H_z$.

On considère une des 4 équations :

$$\frac{\partial E_y}{\partial x} = -\mu \frac{\partial H_z}{\partial t} \qquad \text{(par exemple)}$$

On peut simplifier le traitement considérant seulement l'onde émise (on néglige l'onde réfléchie).

On va avoir alors :

$$\frac{\partial E_y{}^+}{\partial x} = -\mu \frac{\partial H_z{}^+}{\partial t}$$

$$E_y{}^{+\prime} = -\mu(-v_p)H_z{}^{+\prime}$$

En intégrant par rapport à « $x$ »

$$E_y{}^+ = \mu v_p H_z{}^+$$

d'où

$$\frac{E_y{}^+}{H_z{}^+} = \frac{\mu}{\sqrt{\mu\varepsilon}} = \sqrt{\frac{\mu}{\varepsilon}} \; ohms$$

$$Z = \sqrt{\frac{\mu}{\varepsilon}}$$

où Z est l'impédance du mileu de propagation (en 1 point)
pour une onde plane

$$Z_0 = \sqrt{\frac{\mu_0}{\varepsilon_0}} = 120\pi = 377\Omega$$

En répétant le même procédé on retrouve :

$$\frac{E_z^+}{H_y^+} = -\sqrt{\frac{\mu}{\varepsilon}}$$

$$\frac{E_z^-}{H_y^-} = \sqrt{\frac{\mu}{\varepsilon}}$$

$$\frac{E_z^-}{H_y^-} = -\sqrt{\frac{\mu}{\varepsilon}}$$

Une forme condensée de ces expressions peut être obtenue en remarquant :

$$\bar{H} = \hat{n} \times \frac{\bar{E}}{Z}$$

$$\bar{E} = Z\bar{H} \times \hat{n}$$

$\hat{n}$ = vecteur unitaire dans la direction de la propagation (Figure II.5)

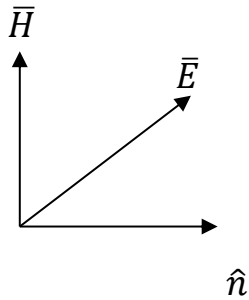

Figure II.5.

**Exemple II.1.** Le champ électrique d'une onde plane se propageant dans l'espace libre est donné par,

$$\bar{E} = \hat{y}\, 120\pi\big(E_0^+ e^{j\beta z} + E_0^- e^{-j\beta z}\big) = \hat{y} 120\pi \left(E_y^+ + E_y^-\right) = 120\pi(\hat{y}\, E_y^+ + \hat{y}\, E_y^-) \left(\frac{V}{m}\right)$$

où $E_0^+$ et $E_0^-$ sont les constants. Trouver le champ magnétique correspondant.

<span style="color:cyan">**Solution :**</span>

Étant donné que,

$$\bar{H} = \hat{n} \times \frac{\bar{E}}{Z} \, , Z = Z_0 = \sqrt{\frac{\mu_0}{\varepsilon_0}} = 120\pi = 377 \, (\Omega), \qquad \bar{E} = 120\pi(\hat{y} \, E_y^+ + \, \hat{y} \, E_y^-)$$

et que pour la direction de propagation dans le sens de l'axe « + z » est $\hat{n} = \hat{z}$ et pour la direction de propagation dans le sens de l'axe « - z » est $\hat{n} = -\hat{z}$.

Cela signifie que,

$$\bar{H}^+ = \hat{z} \times \hat{y} \, 120\pi E_y^+ \frac{1}{Z_0} = -\hat{x} E_0^+ e^{j\beta z} \text{ et } \bar{H}^- = -\hat{z} \times \hat{y} \, 120\pi \, E_y^- \frac{1}{Z_0} = \hat{x} E_0^- e^{-j\beta z} \, .$$

Finalement le champ magnétique correspondant est,

$$\bar{H} = \bar{H}^+ + \bar{H}^- = \hat{x} \left( -E_0^+ e^{j\beta z} + E_0^- e^{-j\beta z} \right) \left(\frac{A}{m}\right)$$

# 2. Notion du vecteur-phaseur

## 2.1 ONDE ÉLECTROMAGNÉTIQUE EN RÉGIME PERMANENT

Dans la plupart des cas pratiques, le générateur relié à l'antenne émettrice génère une onde porteuse sinusoïdale. Soit $\bar{\mathcal{E}}$, la valeur instantanée du champ électrique, fonction du temps et de l'espace.

$$\bar{\mathcal{E}}(x, y, z, t) = \hat{x}\mathcal{E}_x + \hat{y}\mathcal{E}_y + \hat{z}\mathcal{E}_z$$

$$\mathcal{E}_x = \mathcal{E}_{x_0} \cos(\omega t + \varphi x)$$

$$\mathcal{E}_y = \mathcal{E}_{y_0} \cos(\omega t + \varphi y)$$

$$\mathcal{E}_z = \mathcal{E}_{z_0} \cos(\omega t + \varphi z)$$

$\mathcal{E}_x$, $\mathcal{E}_y$ et $\mathcal{E}_z$ sont des valeurs instantanées.

$\mathcal{E}_{x_0}$, $\mathcal{E}_{y_0}$ et $\mathcal{E}_{z_0}$ sont des valeurs crêtes.

Dans la théorie des circuits électriques la solution des problèmes se simplifie lorsque nous adoptons la notion de phaseur ''scalaire'' pour les courants et les tensions.

$$v(t), i(t) \rightarrow \bar{V}, \bar{I} \text{ (nombre complexe)}$$

- Plus riche que la réalité -

En électromagnétisme, on adopte le même principe de calcul à un niveau de calcul un peu plus élevé, leur équivalent sous forme vectorielle.

$$\vec{\mathcal{E}} = Re\{\hat{x}E_{x_0}e^{j(\omega t + \varphi x)} + \hat{y}E_{y_0}e^{j(\omega t + \varphi y)} + \hat{z}E_{z_0}e^{j(\omega t + \varphi z)}\}$$

En isolant le terme $"e^{j\omega t}"$

$$\vec{\mathcal{E}} = Re(\bar{E}e^{j\omega t})$$

Vecteur-phaseur électrique

$$\bar{E} = \hat{x}E_{x_0}e^{j(\varphi x)} + \hat{y}E_{y_0}e^{j(\varphi y)} + \hat{z}E_{z_0}e^{j(\varphi z)}$$

Relation utile,

$$\vec{\mathcal{E}} = Re(\bar{E}e^{j\omega t}) = \frac{1}{2}\{\bar{E}e^{j\omega t} + \bar{E}^*e^{-j\omega t}\}$$

Complexe conjugué

$$\bar{E} = \frac{\begin{cases}\mathcal{E}_{x_0}\cos(\omega t + \varphi x)\hat{x} \\ \mathcal{E}_{y_0}\cos(\omega t + \varphi y)\hat{y} \\ \mathcal{E}_{z_0}\cos(\omega t + \varphi z)\hat{z}\end{cases} + j\begin{cases}\mathcal{E}_{x_0}\sin(\omega t + \varphi x)\hat{x} \\ \mathcal{E}_{y_0}\sin(\omega t + \varphi y)\hat{y} \\ \mathcal{E}_{z_0}\sin(\omega t + \varphi z)\hat{z}\end{cases}}{e^{j\omega t}}$$

## 2.2 Les équations de Maxwell en régime permanent

$$\nabla \times \vec{\mathcal{E}} = -\frac{\partial \vec{\mathcal{B}}}{\partial t}$$

$$\nabla \times \vec{\mathcal{H}} = \vec{\mathcal{J}} + \frac{\partial \vec{\mathcal{D}}}{\partial t}$$

$$\nabla \cdot \vec{\mathcal{B}} = 0$$

$$\nabla \cdot \vec{\mathcal{D}} = \rho$$

Toutes les grandeurs vectorielles sont exprimées en valeurs instantanées

D'autre part on a,

$$\nabla \times (Re\ \bar{A}) = Re(\nabla \times \bar{A})$$

$$\frac{\partial}{\partial t}(Re\ \bar{A}) = Re\left(\frac{\partial \bar{A}}{\partial t}\right)$$

Ceci donne pour les vecteurs-phaseurs des champs EM :

$$\nabla \times Re\big(\bar{E}e^{j\omega t}\big) = -\mu \frac{\partial}{\partial t}\big(Re\ \bar{H}e^{j\omega t}\big)$$

$$Re\big(\bar{E}e^{j\omega t}\big) = \vec{\mathcal{E}}(x,y,z,t) \quad \text{(instantanné)}$$

$$Re\ \bar{H}e^{j\omega t} = \vec{\mathcal{H}}(x,y,z,t) \quad \text{(instantanné)}$$

$$\rightarrow Re\big(\nabla \times \bar{E}e^{j\omega t}\big) = -\mu \frac{\partial}{\partial t}\big(Re\ \bar{H}e^{j\omega t}\big)$$

$$= -\mu\ Re\left(\frac{\partial}{\partial t}\big[\bar{H}e^{j\omega t}\big]\right)$$

$$= -\mu\ Re\big(j\omega \bar{H}e^{j\omega t}\big)$$

Les parties ''$Re$'' et '' $e^{j\omega t}$ ''apparaissent toutes les deux sur les deux côtés de l'équation, on peut les omettre jusqu'à la fin des calculs pour y revenir ensuite. Parfois, les solutions en vecteur-phaseur sont suffisantes puisqu'elles ont toutes l'information nécessaire (amplitudes, phase, fonction des coordonnées de l'espace) sauf dans le temps qui est bien connu (régime permanent = variation sinusoïdale à une fréquence radiale donnée = $\omega$). Donc il suffit de travailler avec les vecteurs-phaseurs à travers les équations de Maxwell jusqu'à l'obtention des résultats finaux désirés.

$$\begin{cases} \nabla \times \bar{E} = -j\omega\,\mu\bar{H} \\ \nabla \times \bar{H} = (\sigma + j\omega\varepsilon)\bar{E} \\ \nabla \cdot \bar{H} = 0 \\ \nabla \cdot \bar{E} = {\rho}/{\varepsilon} \end{cases}$$

Ce sont des équations de Maxwell en régime permanent.

## 2.3    L'équation d'onde en régime permanent

$$\nabla \times \nabla \times \bar{E} = -j\omega\mu \, \nabla \times \bar{H}$$

$$= -j\omega\mu \, (\sigma + j\omega\varepsilon) \, \bar{E}$$

$$\nabla(\nabla \cdot \bar{E}) - \nabla \cdot (\nabla\bar{E}) = -j\omega\mu \, (\sigma + j\omega\varepsilon)\bar{E}$$

$$\nabla\left(\frac{\rho}{\varepsilon}\right) - \nabla^2\bar{E} = -\gamma^2\bar{E}$$

Avec $\gamma \triangleq \sqrt{j\omega\mu \, (\sigma + j\omega\varepsilon)}$

d'où

$$\left.\begin{array}{l} \nabla^2\bar{E} - \gamma^2\bar{E} = 0 \\ \nabla^2\bar{H} - \gamma^2\bar{H} = 0 \end{array}\right\} \text{Équations d'Helmholtz}$$

De la même façon,

$\gamma =$ Facteur de propagation qui indique à toutes fins pratiques la facilité avec laquelle l'onde peut se transmettre dans un milieu donné.

## 2.4 Étude de $\gamma$

Soit une onde plane dans un milieu sans frontières sans sources se propageant dans la direction positive de « $x$ ». Pour cette onde plane on considère :

$$E_x = 0 \; ; \; H_x = 0 \; ; \; \frac{\partial}{\partial y} = 0 \; ; \; \frac{\partial}{\partial z} = 0$$

En imposant ces conditions à l'équation d'Helmholtz on a,

$$\frac{d^2E_y}{dx^2} - \gamma^2E_y = 0 \quad ; \quad \frac{d^2H_y}{dx^2} - \gamma^2H_y = 0$$

$$\frac{d^2E_z}{dx^2} - \gamma^2E_z = 0 \quad ; \quad \frac{d^2H_z}{dx^2} - \gamma^2H_z = 0$$

Une forme générale de solution pour ces équations serait, par exemple :

$$E_y = E_{y_0}{}^+ e^{-\gamma x} + E_{y_0}{}^- e^{+\gamma x}$$

Une composante du vecteur phaseur $\bar{E}$ (Nombre complexe).

Remarque: $E_{y_0}{}^+, E_{y_0}{}^-$ apparaissent comme des constantes d'amplitude vis-à-vis du terme $e^{\pm\gamma x}$ qui symbolise la variation de la phase en ce qui concerne une coordonnée de l'espace : « $x$ ». Mais les composantes $E_{y_0}$ étant complexes peuvent également contenir une information de phase à l'origine ($x$=0).

$$E_{y_0}{}^+ = \left| E_{y_0}{}^+ \right| e^{j(\varphi x_0)} \qquad \text{etc.}$$

Dans le processus inverse,

$$\vec{\mathcal{E}} = Re\left(\bar{E} e^{j\omega t}\right)$$

L'ajout du terme $e^{j\omega t}$ apporte les variations de phase dues à l'écoulement du temps « $t$ » via la fréquence « $\omega$ »: $\omega t$.

Dans l'expression finale de la variation de la phase $\cos(\omega t + \varphi x_0)$, on trouve ces deux termes ensemble. Maintenant, on trouve un troisième type de variation de phase qui est dû à la propagation selon la direction de $x$ : $\varphi_{xp}$ provenant du terme '' $e^{-\gamma x}$ ''. En fait, à la fin on aura $\varphi_x = \varphi_{xo} + \varphi_{xp}$. On omette souvent $\varphi_{xo}$ ($\varphi_{xo} = 0$) et on considère que $\varphi_x$ provient du terme $e^{-\gamma x}$ seulement.

Soit,

$$\gamma^2 = j\omega\mu \left(\sigma + j\omega\varepsilon\right)$$

Pour un milieu non-conducteur ($\sigma = 0$),

$$\gamma^2 = -\omega^2\mu\varepsilon \quad \rightarrow \quad \gamma = j\omega\sqrt{\mu\varepsilon}$$

$\gamma$ est une quantité purement imaginaire dans ce cas.

On l'appelle (partie module)

$$\beta = \omega\sqrt{\mu\varepsilon} \ \left(\frac{rad}{m}\right) \qquad \text{(facteur de phase)}$$

Donc,

$$E_y = E_{y_0}{}^+ e^{-j\beta x} + E_{y_0}{}^- e^{+j\beta x}$$

On peut avoir des formes semblables pour $E_z$, $H_y$ et $H_z$,

## 2.5   Vitesse de phase

On considère une onde qui se transmet dans la direction positive de « $x$ » et qui est polarisée selon « y ».

$$E_y = E_{y_0}{}^+ e^{-j\beta x}$$

Pour figurer une représentation graphique de ce champ orienté selon ''y'' et fonction de « $x$ » et de « $t$ », il vaut mieux considérer les variations instantanées :

$$\mathcal{E}_y = Re\{E_{y_0}{}^+ e^{-j\beta x} e^{j\omega t}\}$$

$$\mathcal{E}_y = E_{y_0}{}^+ \cos(\omega t + \beta y)$$

Remarque : on a considéré $E_{y_0}{}^+$ comme un vrai module.

En premier lieu on fixe « $t$ » ($t = 0$) et on fait varier «x». En deuxième lieu on recommence tout pour un instant plus tard ($t = t_1$)

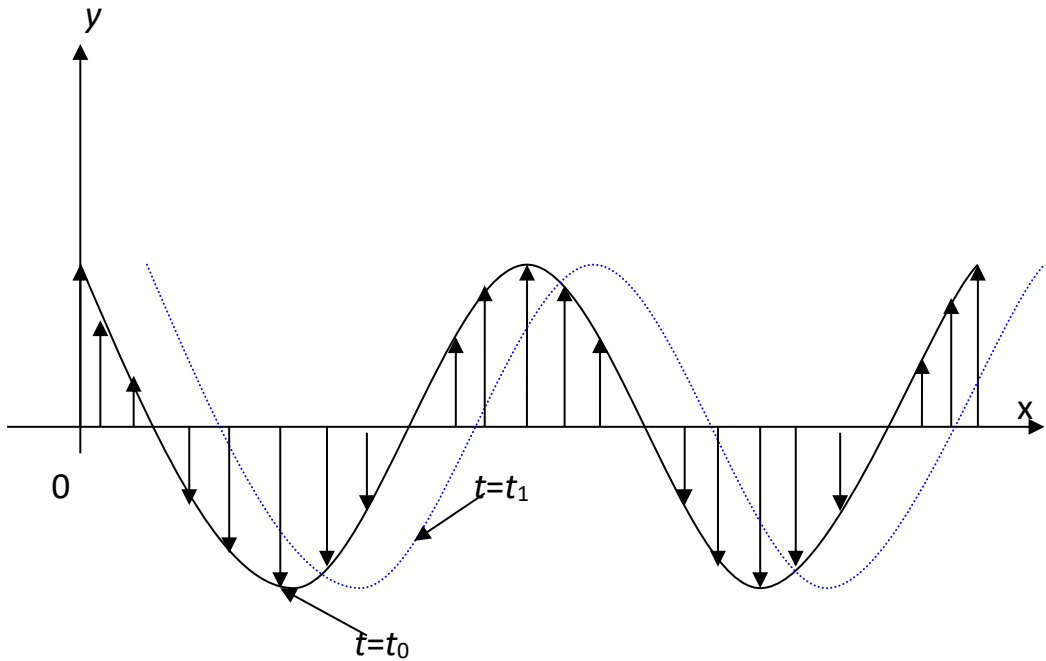

Figure II.6.

Un instant plus tard, on retrouve la même forme mais le tout glisse dans la direction positive de « $x$ » (Figure II.6).

$$v_p = \frac{\omega}{\beta} = \frac{\omega}{\omega\sqrt{\mu\varepsilon}} = \frac{1}{\sqrt{\mu\varepsilon}} \; (\frac{m}{s}) \qquad \text{(vitesse de phase)}$$

Dans l'espace libre (vide),

$$v_p = v_0 = c = 3x10^8 \; (\frac{m}{s}) \qquad \text{(vitesse de phase dans le vide)}$$

## 2.6   Longueur d'onde

À partir des expressions complexes des champs électriques ou magnétiques, on peut voir que lorsque la propagation des ondes est dans la direction « $x$ » le terme $e^{\pm j\beta x}$ change en $\beta x$. La distance « $x$ », que l'onde doit parcourir, pour que la phase change par $2\pi$ (un cycle complet) soit d'un intérêt particulier est appelée la longueur d'onde.

Donc,

$$\beta\lambda = 2\pi \rightarrow \lambda = \frac{2\pi}{\beta} = \frac{2\pi}{\omega\sqrt{\mu\varepsilon}} = \frac{v_h}{f} \; (m)$$

Dans l'espace libre (vide),

$$\lambda_0 = \frac{c}{f} \; (m)$$

**Exemple II.2** : Une onde uniforme se propage dans la direction de « $+z$ » dans le vide. L'amplitude du champ électrique $E_x = 100 \; (\frac{V}{m})$ et la longueur d'onde est 25 (*cm*). Déterminer,

   a) La fréquence et le facteur de phase de cette onde,
   b) Les expressions complexes pour le champ électrique et magnétique,
   c) Les expressions dans le demain temporel pour le champ électrique et magnétique.

**Solution :**

a) Dans l'espace libre (vide) $\lambda_0 = 0.25\ (m)$,

$$\lambda_0 = \frac{c}{f} \rightarrow f = \frac{c}{\lambda_0} = 1.2\ (GH_z)$$

$$\beta = \frac{2\pi}{\lambda_0} = 8\pi\ (\frac{rad}{m})$$

b) L'expression complexe du champ électrique est,

$$\bar{E}(z) = 100 e^{-j8\pi z}\hat{x}\ (\frac{V}{m})$$

L'expression complexe du champ magnétique est,

$$\bar{H}(z) = \hat{n} \times \frac{\bar{E}(z)}{Z} = \hat{z} \times \hat{x}\ \frac{100 e^{-j8\pi z}}{Z_0}$$

d'où,

$$\bar{H} = \frac{100 e^{-j8\pi z}}{377}\hat{y} = 0.265 e^{-j8\pi z}\hat{y}\ (\frac{A}{m})$$

c) $\omega = 2\pi f = 24\pi x 10^8\ (rad)$,

$$E(z,t) = 100 \cos(24\pi x 10^8 t - 8\pi z)\hat{x}\ (\frac{V}{m})$$

$$H(z,t) = 0.265 \cos(24\pi x 10^8 t - 8\pi z)\hat{y}\ (\frac{A}{m})$$

**Exemple II.3** : Une onde plane uniforme se propage dans un milieu sans perte. L'expression du champ magnétique est donnée par :

$$\bar{H} = H_0 \cos(6\pi x 10^6 t + 3\pi y + 0.1\pi)\ \hat{z}\ (\frac{A}{m})$$

Trouver :

a) La fréquence d'opération,

b) La constante de phase,

c) Le vecteur unitaire,

d) Le vecteur unitaire de la direction de propagation du champ magnétique à $t = 0$ et $y = 0$,

e) Le vecteur unitaire de la direction de propagation du champ électrique à $t = 0$ et $y = 0$,

f) La vitesse de propagation,

g) La longueur d'onde.

Solution :

a) $\omega = 2\pi f \rightarrow f = \frac{\omega}{2\pi} = \frac{6\pi x 10^6}{2\pi} = 3 \ (MH_z)$

b) $\beta = 3\pi \ (\frac{rad}{m})$

c) $\hat{n} = -\hat{y}$

d) $\hat{n}_{\bar{H}} = \hat{x}$

e)

$\hat{n} = \hat{n}_{\bar{E}} \times \hat{n}_{\bar{H}} \rightarrow \hat{n} \times \hat{n}_{\bar{H}} = (\hat{n}_{\bar{E}} \times \hat{n}_{\bar{H}}) \times \hat{n}_{\bar{H}} = \hat{n}_{\bar{H}} (\hat{n}_{\bar{H}} \cdot \hat{n}_{\bar{E}}) - \hat{n}_{\bar{E}}(\hat{n}_{\bar{H}} \cdot \hat{n}_{\bar{H}})$

$$\hat{n}_{\bar{E}} = -\hat{n} \times \hat{n}_{\bar{H}} = \hat{y} \times \hat{x} = -\hat{z}$$

f)

$$v_p = \frac{\omega}{\beta} = \frac{6\pi x 10^6}{3\pi} = 2x10^6 \ (\frac{m}{s})$$

g)

$$\lambda = \frac{2\pi}{3\pi} = 0.667 \ (m)$$

### 3. RÉGIME PERMANENT

1)  $\nabla \times \bar{E} = -j\omega \bar{B}$

2)  $\nabla \times \bar{H} = \bar{J} + j\omega \bar{D}$

En effectuant le produit par $\bar{H}^*$ de l'équation 1), on va avoir,

3)  $\bar{H}^* \cdot (\nabla \times \bar{E}) = -j\omega \bar{H}^* \cdot \bar{B}$

On effectue un autre produit scalaire par $\bar{E}$ et la conjugué de l'équation 2):

4)  $\bar{E} \cdot (\nabla \times \bar{H}^*) = \bar{E} \cdot \bar{J}^* + j\omega \bar{E} \cdot \bar{D}^*$

(4)-(3) :

$$\underbrace{\bar{E} \cdot (\nabla \times \bar{H}^*) - \bar{H}^* \cdot (\nabla \times \bar{E})}_{\text{Par identité vectorielle}} = j\omega \bar{H}^* \cdot \bar{B} + \bar{E} \cdot \bar{J}^* - j\omega \bar{E} \cdot \bar{D}^*$$

$$\Downarrow$$

$$-\nabla \cdot (\bar{E} \times \bar{H}^*) = \bar{E} \cdot \bar{J}^* + j\omega(\bar{H}^* \cdot \bar{B} - \bar{E} \cdot \bar{D}^*)$$

On divise par 2 les deux membres et on passe à l'intégrale sur un volume limité par une surface fermée imaginaire.

$$-\int_V \frac{\nabla \cdot (\bar{E} \times \bar{H}^*)}{2} dv = \int_V \frac{\bar{E} \cdot \bar{J}^*}{2} dv + \int_V j\omega \left(\frac{\bar{H}^* \cdot \bar{B}}{2} - \frac{\bar{E} \cdot \bar{D}^*}{2}\right) dv$$

Pour la partie de gauche on a :

$$-\int_V \frac{\nabla \cdot (\bar{E} \times \bar{H}^*)}{2} dv = -\int_S \frac{(\bar{E} \times \bar{H}^*)}{2} \hat{n} ds$$

Ce dernier représente la valeur moyenne de la puissance rayonnée à travers la surface. Elle s'exprime par :

$$P_r = -\frac{1}{2} \int_S (\bar{E} \times \bar{H}^*) \hat{n} ds$$

D'autre part, le terme

$$\int_V \frac{\bar{E} \cdot \bar{J}^*}{2} dv$$

constitue l'énergie communiquée par le champ aux charges du milieu.

Tenir compte du courant de conduction dans le milieu de propagation (s'il y en a) et de courant de surface sur la structure de l'antenne, nous pouvons séparer ces deux aspects :

$$\bar{J} = \bar{J}_c + \bar{J}_s = \sigma \bar{E} + \bar{J}_s$$

Pertes diélectriques et conduction

alimentation à la source

$$\frac{1}{2}(\bar{E} \cdot \bar{J}^*) = \frac{1}{2}\bar{E}(\sigma \bar{E} + \bar{J}_s)$$

$$= \frac{1}{2}\sigma |\bar{E}|^2 - \frac{1}{2}\bar{E} \cdot \bar{J}_s$$

Donc,

$$\int_V \frac{\bar{E} \cdot \bar{J}^*}{2}\, dv = \int_V \frac{1}{2}\, \sigma |\bar{E}|^2\, dv + \int_{V'} \frac{1}{2}\, \bar{E} \cdot \bar{J}^*\, dv'$$

où

$$< P_d > = \int_V \frac{1}{2}\, \sigma |\bar{E}|^2\, dv$$

Ceci représente la puissance moyenne dissipée par effet Joule à l'intérieur de la surface (chauffage hyperfréquence).

$$< P_s > = -Re\left\{ \int_{V'} \frac{1}{2}\, \bar{E} \cdot \bar{J}_s^{\,*}\, dv' \right\}$$

Valeur moyenne de la puissance active
fournie par la source

$$< Q_r > = -Im\left\{ \int_{V'} \frac{1}{2}\, \bar{E} \cdot \bar{J}_s^{\,*}\, dv' \right\}$$

Valeur moyenne de la puissance
réactive fournie par la source

(On considère donc les signes '' - '' pour pouvoir les déplacer à gauche de l'égalité)

- Variation dans l'énergie emmagasinée

Valeur moyenne de l'énergie magnétique emmagasinée
$$< W_m > = \int_V \bar{B} \cdot \bar{H}^*\, dv$$

Valeur moyenne de l'énergie électrique emmagasinée
$$< W_e > = \int_V \bar{E} \cdot \bar{D}^*\, dv$$

$$< P_s > + < Q_r > = < P_r > + < P_d > + \frac{1}{2} j\omega \left\{ < W_m > - < W_e > \right\}$$

### 3.1 Milieu : conducteur – diélectrique (Figure II.7)

$$\gamma^2 = j\omega\mu\sigma - \omega^2\mu\varepsilon$$

$$\varepsilon = \varepsilon' - j\varepsilon'' = \varepsilon_0(\varepsilon_r' - j\varepsilon_r'')$$

$$\mu = \mu' - j\mu'' = \mu_0(\mu_r' - j\mu_r'')$$

$$\gamma = j\omega\sqrt{\mu\varepsilon}\sqrt{1 - j\frac{\sigma}{\varepsilon\omega}}$$

$$\gamma = j\omega\sqrt{\mu}\sqrt{\varepsilon' - j\varepsilon''}\sqrt{1 - j\frac{\sigma}{(\varepsilon' - j\varepsilon'')\omega}}$$

$$\gamma = j\omega\sqrt{\mu}\sqrt{\varepsilon' - j\varepsilon'' - j\frac{\sigma}{\omega}}$$

$$\gamma = j\omega\sqrt{\mu_0(\mu_r' - j\mu_r'')}\sqrt{\varepsilon_0\left[\varepsilon_r' - j\left(\varepsilon_r'' + \frac{\sigma}{\varepsilon_0\omega}\right)\right]}$$

$$\gamma = j\omega\sqrt{\mu_0\varepsilon_0}\sqrt{\mu_r' - j\mu_r''}\sqrt{\varepsilon_r' - j\left(\varepsilon_r'' + \frac{\sigma}{\varepsilon_0\omega}\right)}$$

$$\omega\sqrt{\mu_0\varepsilon_0} = \beta_0 = \frac{2\pi}{\lambda_0}$$

$$\gamma = j\frac{2\pi}{\lambda_0}\sqrt{\mu_r' - j\mu_r''}\sqrt{\varepsilon_r' - j\left(\varepsilon_r'' + \frac{\sigma}{\varepsilon_0\omega}\right)}$$

## Chauffage par hyper fréquence (four micro-ondes)

Fréquences réservées : ISM (Industrial, Scientific & Medical)

- 915 MHz
- 2,45 GHz
- 5,8 GHz
- 22 GHz

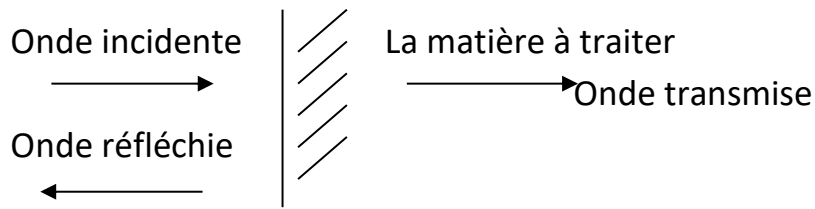

<div align="center">Figure II.7.</div>

Divers matériaux sont utilisés : transparent, absorbant, réfléchissant…

Moment dipolaire qui essaie de suivre le champ électrique à haute fréquence d'où les pertes diélectriques qui causent le chauffage (augmentation de la température).

$$\left.\begin{array}{c} \nabla \times \bar{H} = (\sigma + j\omega\varepsilon)\bar{E} \\ \nabla \times \bar{E} = -j\omega\mu\bar{H} \end{array}\right\} \implies \begin{array}{c} \nabla^2\bar{E} - j\omega\mu(\sigma + j\omega\varepsilon)\bar{E} = 0 \\ \nabla^2\bar{H} - j\omega\mu(\sigma + j\omega\varepsilon)\bar{H} = 0 \end{array}$$

$$\gamma^2 \triangleq j\omega\mu(\sigma + j\omega\varepsilon) = (\alpha + j\beta)^2$$
$$\gamma = \text{facteur de propagation}$$

$$j\omega\mu(\sigma + j\omega\varepsilon) = j\omega\mu\{j\omega(\varepsilon' - j\varepsilon'')\}$$
$$= j\omega\mu\{\sigma' - j\sigma''\}$$

$\varepsilon' = \text{constante diélectrique}$
$\varepsilon'' = \text{perte diélectrique}$ $\Big\}$ $\varepsilon^* = \varepsilon' - j\varepsilon'' = \text{permitivité complexe}$

et

$$\frac{\varepsilon''}{\varepsilon'} = \tan \delta \quad \text{angle de pertes diélectriques}$$

$$\gamma^2 \triangleq j\omega\mu(j\omega\varepsilon' + \omega\varepsilon'') = (\alpha + j\beta)^2$$

$$\gamma^2 = (\alpha + j\beta)^2 = \alpha^2 - \beta^2 + 2j\alpha\beta = j\omega^2\mu\varepsilon'' - \omega^2\mu\varepsilon'$$

| | |
|---|---|
| 1) | $\alpha^2 - \beta^2 = -\omega^2\mu\varepsilon'$ |
| 2) | $2\alpha\beta = \omega^2\mu\varepsilon''$ |

$$\beta = \frac{\omega^2\mu\varepsilon''}{2\alpha}$$

$$\alpha^2 - \left(\frac{\omega^2\mu\varepsilon''}{2\alpha}\right)^2 = -\omega^2\mu\varepsilon'$$

$$4\alpha^4 - \omega^4\mu^2\varepsilon''^2 = -4\alpha^2\omega^2\mu\varepsilon'$$

$$\alpha^4 + \alpha^2\omega^2\mu\varepsilon' - \frac{\omega^4\mu^2\varepsilon''^2}{4} = 0$$

$$\alpha^2 = \frac{-\omega^2\mu\varepsilon' \pm \sqrt{(\omega^2\mu\varepsilon')^2 + \omega^4\mu^2\varepsilon''^2}}{2}$$

$$\alpha^2 = \omega^2\mu\varepsilon' \left\{\frac{-1 + \sqrt{1 + \left(\frac{\varepsilon''}{\varepsilon'}\right)^2}}{2}\right\}$$

$$\alpha = \omega\sqrt{\mu\varepsilon'}\left\{\frac{\sqrt{1+\left(\frac{\varepsilon''}{\varepsilon'}\right)^2}-1}{2}\right\}^{1/2}$$

$$\frac{1}{\sqrt{\mu\varepsilon'}} = \text{vitesse de propagation}$$

$$\frac{\varepsilon''}{\varepsilon'} = \tan\delta$$

$$\omega = 2\pi f$$

$$v_{ph} = \lambda f \;\rightarrow\; \omega\sqrt{\mu\varepsilon'} = \frac{2\pi f}{\vartheta} = \frac{2\pi f}{\lambda f} = \frac{2\pi}{\lambda}$$

$$\alpha = \frac{2\pi}{\lambda}\left\{\frac{\sqrt{1+\tan^2\delta}-1}{2}\right\}^{1/2}$$

Comme $\dfrac{\varepsilon''}{\varepsilon'} \ll 1$ en général

$$\sqrt{1+\tan^2\delta} \approx 1 + \frac{\tan^2\delta}{2} + \cdots$$

le facteur d'atténuation est,

$$\alpha \approx \frac{2\pi}{\lambda}\left\{\frac{1+\frac{\tan^2\delta}{2}-1}{2}\right\}^{\frac{1}{2}}$$

$$\boxed{\alpha \approx \frac{2\pi}{\lambda}\tan\delta \quad (Np/m)}$$

## 3.2 Profondeur de pénétration

$$e^{-\alpha z} = e^{-1}$$

$$z = \frac{1}{\alpha} \quad (m)$$

## 3.3 Puissance moyenne dissipée

$$P_d = -\frac{\partial P_t}{\partial z} = 2\alpha P_t$$

Pour le vecteur de Poynting,

$$P_t = \frac{1}{2} Re(\bar{E} \times \bar{H}^*)$$

si $\bar{E}_0$ et $\bar{H}_0$ sont des valeurs crêtes on obtient,

$$P_t = \frac{E_0^{\;2}}{2Rs}$$

$Rs$ = partie réelle de l'impédance d'une onde plane se propageant en milieu avec perte

$$\frac{\bar{E}_0}{\bar{H}_0} = \eta = \sqrt{\frac{j\omega\mu}{\sigma + j\omega\varepsilon}} = \sqrt{\frac{j\omega\mu}{j\omega(\varepsilon' - j\varepsilon'')}}$$

$$\eta = \sqrt{\frac{\mu}{(\varepsilon' - j\varepsilon'')}} = \sqrt{\frac{\mu}{\varepsilon^*}} = \sqrt{\frac{\mu(\varepsilon' - j\varepsilon'')}{\varepsilon'^2 - \varepsilon''^2}}$$

$$\eta = Rs + jXs \qquad si \; \varepsilon'' \ll \varepsilon'$$

$$Rs \approx \sqrt{\frac{\mu}{\varepsilon'}}$$

Donc,

$$P_t = \frac{1}{2}{E_0}^2 \sqrt{\frac{\varepsilon'}{\mu}}$$

$$P_d = \frac{2\pi}{\lambda}\tan\delta\left(\frac{1}{2}{E_0}^2\sqrt{\frac{\varepsilon'}{\mu}}\right)$$

$$P_d = \frac{\pi}{\lambda}\frac{\varepsilon''}{\varepsilon'}{E_0}^2\sqrt{\frac{\varepsilon'}{\mu}} = \frac{\pi}{\lambda}\frac{\varepsilon''{E_0}^2}{\sqrt{\mu\varepsilon'}}$$

$$\boxed{P_d = \pi f\ \varepsilon''{E_0}^2 \qquad \text{Avec des valeurs efficaces}}$$

$$\boxed{P_d = \pi f\ \varepsilon''{E_{0eff}}^2 \qquad \text{Puissance dissipée en } (W/m^3)}$$

### 3.4 Ondes planes en milieu conducteur

On considère une onde polarisée linéairement selon « $y$ » et se propageant dans la direction de « $x$ ».

$$E_y = E_{yo}^+ e^{-\gamma x}$$

$$\gamma^2 = j\omega\mu\sigma - \omega^2\mu\varepsilon = j\omega(\mu\sigma - \omega\varepsilon)$$

*Si $\sigma \gg \omega\varepsilon$*

On peut considérer le milieu bon conducteur. Cette condition peut aussi être obtenue pour $\omega$ faible.

$$\gamma^2 = j\omega\mu\sigma \ \rightarrow\ \gamma = \pm\sqrt{j\omega\mu\sigma}$$

$$\gamma = \pm\sqrt{\omega\mu\sigma}\ \angle 45^o$$

$$\gamma = \pm\frac{\sqrt{\omega\mu\sigma}}{2}\ (1+j)$$

$$\gamma = \alpha + j\beta$$

$$\alpha \approx \frac{\sqrt{\omega\mu\sigma}}{2} \qquad \text{facteur d'atténuation} \left(\frac{1}{m}\right)$$

$$\beta = \frac{\sqrt{\omega\mu\sigma}}{2} \qquad \text{facteur de phase} \left(\frac{rad}{m}\right)$$

L'expression pour le champ électrique de l'onde plane devient

$$E_y = E_{yo}^+ e^{-\alpha x} e^{-j\beta x}$$

### 3.5 $\alpha$ et $\beta$ pour un milieu quelconque

$$\gamma^2 = (\alpha + j\beta)^2 = j\omega\mu(\sigma + j\omega\varepsilon)$$

$$\alpha^2 - \beta^2 + 2j\alpha\beta = j\omega\mu\sigma - \omega^2\mu\varepsilon$$

On a déjà étudié cette équation en prenant

$$\sigma + j\omega\varepsilon = j\omega(\varepsilon' - j\varepsilon'') = j\omega\varepsilon^*$$

$$\begin{cases} \alpha^2 - \beta^2 = -\omega^2\mu\varepsilon \\ 2\alpha\beta = \omega\mu\sigma \end{cases} \qquad \text{à partir d'une équation complexe}$$

$$4\alpha^2\beta^2 = \omega^2\mu^2\sigma^2$$

$$4(\beta^2 - \omega^2\mu\varepsilon)\beta^2 = \omega^2\mu^2\sigma^2$$

$$4\beta^4 - 4\beta^2\omega^2\mu\varepsilon - \omega^2\mu^2\sigma^2 = 0$$

$$\beta^4 - \beta^2\omega^2\mu\varepsilon - \frac{\omega^2\mu^2\sigma^2}{4} = 0$$

$$\beta^2 = \frac{\omega^2\mu\varepsilon \pm \sqrt{(\omega^2\mu\varepsilon)^2 + \omega^2\mu^2\sigma^2}}{2}$$

$$\beta^2 = \frac{\omega^2\mu\varepsilon + \omega^2\mu\varepsilon\sqrt{1 + \frac{\omega^2\mu^2\sigma^2}{\omega^4\mu^2\varepsilon^2}}}{2}$$

$$\beta^2 = \frac{\omega^2\mu\varepsilon}{2}\left\{1 + \sqrt{1 + \left(\frac{\sigma}{\omega\varepsilon}\right)^2}\right\}$$

$$\beta = \omega\sqrt{\mu\varepsilon}\left\{\frac{1}{2}\left(1 + \sqrt{1 + \left(\frac{\sigma}{\omega\varepsilon}\right)^2}\right)\right\}^{\frac{1}{2}}(\frac{rad}{m})$$

De la même façon,

$$\alpha = \omega\sqrt{\mu\varepsilon}\left\{\frac{1}{2}\left(\sqrt{1 + \left(\frac{\sigma}{\omega\varepsilon}\right)^2} - 1\right)\right\}^{1/2}(\frac{Np}{m})$$

Le terme $\frac{\sigma}{\omega\varepsilon}$ peut être défini comme le rapport entre les courants de conduction et les courants de déplacement.

**Exemple II.4.**

Un ingénieur à conçu une antenne dipôle $(\frac{\lambda}{2})$ dans l'espace libre à 0.6 ($GH_z$).

a) Calculer la longueur de cette antenne,

b) Trouver le module du rapport entre les courants de conduction et les courants de déplacement,

c) Si l'on veut encore utiliser cette antenne dipôle $(\frac{\lambda}{2})$ à la même fréquence pour la communication sous-marine ($\sigma = 4\frac{S}{m}$, $\varepsilon = 8\varepsilon_0$) calculer le change nécessaire dans les dimensions de cette antenne.

**Solution :**

a) $\lambda = \frac{c}{f} = 0.5$ $(m)$. La longueur de l'antenne dans l'espace libre est,

$$l_0 = \frac{\lambda}{2} = 0.25 \ (m)$$

b) Le module du rapport entre les courants de conduction et les courants de déplacement est donné par,

$$\left|\frac{I_{con.}}{I_{dep.}}\right| = \frac{\sigma}{\omega\varepsilon} \ = \ 1.48$$

Cela signifie que l'eau est un conducteur modéré et que ce rapport ne peut pas être négligé dans les expressions de l'atténuation $\alpha$ et du facteur de phase $\beta$.

c) Le facteur de phase doit être calculé par,

$$\beta = \omega\sqrt{\mu\varepsilon}\left\{\frac{1}{2}\left(1 + \sqrt{1 + \left(\frac{\sigma}{\omega\varepsilon}\right)^2}\right)\right\}^{1/2} = 133.02 \ (\frac{rad}{m})$$

ce qui donne,

$$\lambda_{eau} = \frac{2\pi}{\beta} = 0.0472 \ (m)$$

En utilisant cette antenne dipôle ($\frac{\lambda}{2}$) à la même fréquence la longueur requise de l'antenne est,

$$l_{eau} = \frac{\lambda_{eau}}{2} = 0.0236 \ (m)$$

Le rapport de ces deux longueurs est,

$$\frac{l_{eau}}{l_0} = 0.094$$

Cela signifie que $l_{eau}$ est inferieur à 9.4 % de $l_0$.

Note : $\varepsilon_0 = \frac{1}{36\pi} x 10^{-9} \ (\frac{F}{m})$

## 3.6 Profondeur de pénétration

Distance que doit parcourir une onde plane dans un milieu avec pertes ($\sigma \neq 0$) pour que son amplitude devienne $1/e$ de son amplitude originale. Soit $\delta$ cette profondeur de pénétration :

$$e^{-\alpha\delta} = e^{-1} = \frac{1}{e}$$

$$\alpha\delta = 1 \ \rightarrow \ \delta = \frac{1}{\alpha}$$

$$Si \ \sigma \gg \omega\varepsilon \ \rightarrow \ \delta = \sqrt{\frac{2}{\omega\mu\sigma}}$$

$\delta$ très petit pour le bon conducteurs ($\sigma \ élevé$)

## 3.7 Impédance d'un milieu avec pertes pour une onde plane

En régime permanent,

$$\frac{\partial E_y}{\partial x} = -\mu \frac{\partial H_z}{\partial t}$$

$$-\gamma E_y = -\mu(j\omega)H_z$$

$$Z_c = \frac{E_y}{H_z} = \frac{j\omega\mu}{\gamma}$$

$$Z_c = \frac{j\omega\mu}{\sqrt{j\omega\mu\sigma - \omega^2\mu\varepsilon}} = \sqrt{\frac{j\omega\mu}{\sigma + j\omega\varepsilon}}$$

$Si\ \sigma = 0$ $\qquad\qquad Z = \sqrt{\dfrac{\mu}{\varepsilon}}$ milieu sans perte

$Si\ \sigma \gg \omega\varepsilon$ $\qquad\qquad Z_c = \sqrt{\dfrac{j\omega\mu}{\sigma}} = \sqrt{\dfrac{\omega\mu}{\sigma}}\ e^{j\frac{\pi}{4}}$

$$Z_c = \sqrt{\frac{\omega\mu}{\sigma}}\ \angle 45^o \triangleq R + jX = \sqrt{\frac{\omega\mu}{2\sigma}}\ (1 + j)$$

$$R = \sqrt{\frac{\omega\mu}{2\sigma}}$$

$$X = \sqrt{\frac{\omega\mu}{2\sigma}}$$

Pour tous les milieux avec pertes, le champ magnétique est en retard de phase dans le temps d'un angle qui peut être compris entre 0º et 45º.

**Exemple II.5.** Une onde plane de 1 $MH_z$ se propage dans un milieu conducteur ( $\sigma = 4x10^{-2}\ \frac{S}{m}$, $\varepsilon = 2.25\ \varepsilon_0, \mu = \mu_0$, $\varepsilon_0 = \frac{1}{36\pi}x10^{-9}\ \left(\frac{F}{m}\right)$,

$\mu_0 = 4\pi x 10^{-7} \left(\frac{H}{m}\right)$.

a) Trouver le module du rapport entre les courants de conduction et les courants de déplacement,

b) Calculer le coefficient d'atténuation et le facteur de phase,

c) Calculer la profondeur de pénétration.

Solution :

a) Le module du rapport entre les courants de conduction et les courants de déplacement est donné par,

$$\left|\frac{I_{con.}}{I_{dep.}}\right| = \frac{\sigma}{\omega\varepsilon}$$

$$\frac{\sigma}{\omega\varepsilon} = \frac{4x36\pi x10^{-2}}{2\pi x10^6 x2.25x10^{-9}} = 320 \gg 1$$

Cela signifie que ce rapport ne peut pas être négligé dans les expressions de l'atténuation $\alpha$ et du facteur de phase $\beta$.

b)

$$\alpha = \omega\sqrt{\mu\varepsilon}\left\{\frac{1}{2}\left(\sqrt{1+\left(\frac{\sigma}{\omega\varepsilon}\right)^2}-1\right)\right\}^{1/2} = \omega\sqrt{\mu\varepsilon}\sqrt{\frac{\sigma}{\omega\varepsilon}}\frac{1}{\sqrt{2}}$$

$$\alpha = \sqrt{\frac{\omega\sigma\mu}{2}} = \sqrt{\pi f\sigma\mu} = 0.397\left(\frac{Np}{m}\right)$$

$$\beta = \omega\sqrt{\mu\varepsilon}\left\{\frac{1}{2}\left(1+\sqrt{1+\left(\frac{\sigma}{\omega\varepsilon}\right)^2}\right)\right\}^{1/2} = \alpha = 0.397\left(\frac{rad}{m}\right)$$

c)

$$\delta = \frac{1}{\alpha} = 2.52 \ (m)$$

## 3.8 Densité de puissance moyenne à partir du vecteur de Poynting complexe

La puissance transportée par l'onde ou la densité de puissance est une quantité dépendant directement de l'amplitude des champs. Le module du vecteur de Poynting fournit l'information sur la densité se puissance instantanée de l'onde en $(\frac{W}{m^2})$. Le vecteur $\bar{P}$ est dans la direction de son vecteur unitaire.

$$\bar{P} = \bar{E} \times \bar{H} \; (\frac{W}{m^2})$$

En utilisant des phaseurs, on calcule la densité de puissance moyenne à partir du vecteur de Poynting complexe,

$$\bar{P}_{av} = \frac{1}{2} Re\{(\bar{E} \times \bar{H}^*)\} \, (\frac{W}{m^2})$$

Note : La densité de puissance moyenne peut être calculée par

$$\bar{P}_{av} = \frac{1}{T} \int_0^T <\bar{P}> dt$$

où $T$ est la période du vecteur de Poynting sous la forme instantanée.

**Exemple II.6.** Quel est le vecteur de Poynting moyen pour une onde plane se propageant dans l'espace libre avec les champs suivants, lorsque les champs sont complexes,

$$\bar{E}(z,t) = E_0 e^{-j\beta z}\hat{x} \quad \text{et} \; \bar{H}(z,t) = \frac{E_0}{Z_0} e^{-j\beta z}\hat{y}$$

Solution :

$$\bar{P}_{av} = \frac{1}{2} Re\{(\bar{E} \times \bar{H}^*)\} = \frac{1}{2} Re\left\{\left(E_0 e^{-j\beta z}\hat{i} \times \frac{E_0}{Z_0} e^{j\beta z}\hat{j}\right)\right\}$$

$$\bar{P}_{av} = \frac{1}{2}\frac{E_0^2}{Z_0}\hat{k}\ (\frac{W}{m^2})$$

Note : $\hat{x} \times \hat{y} = \hat{z}$

**Exemple II.7.** Le champ électrique associé à une onde plane uniforme se propageant dans l'espace libre (vide) s'écrit :

$$\bar{E} = 0.2\cos(2\pi x10^6 t - 2z)\,\hat{x}\ (\frac{V}{m})$$

Trouver :

  a)  Le champ magnétique correspondant,
  b)  La densité de puissance,
  c)  La densité de puissance moyenne,

Note : Les champs électriques sont donnés dans la forme de temps

Solution :

  a)  Le champ magnétique est,

$$\bar{H} = \hat{n} \times \frac{\bar{E}}{Z} = \hat{n} \times \frac{\bar{E}}{Z_0}$$

parce que le milieu est l'espace libre.

Étant donné que la direction de propagation est celle de « +z », le vecteur unitaire correspondant est $\hat{n} = \hat{k}$ .

$$\bar{H} = \hat{n} \times \frac{\bar{E}}{Z_0} = (\hat{z} \times \hat{x})x0.2\cos(2\pi x10^6 t - 2z)\frac{1}{120\pi} = \hat{y}\frac{0.2}{120\pi}x$$

$$\cos(2\pi x10^6 t - 2z) = 0.531x10^{-3}x\cos(2\pi x10^6 t - 2z)\,\hat{y})(\frac{A}{m})$$

$$\bar{H} = 0.531x10^{-3}\cos(2\pi x10^6 t - 2z)\,\hat{y}(\frac{A}{m})$$

b)

$$\bar{P} = \bar{E} \times \bar{H} = (\hat{x} \times \hat{y})x0.2x0.531x10^{-3}[cos(2\pi x10^6 t - 2z)]^2$$

$$\bar{P} = 0.1062[cos(2\pi x 10^6 t - 2z)]^2 x 10^{-3} \, \hat{z} \, (\frac{W}{m^2})$$

c)

$$\bar{P}_{av} = \frac{1}{T}\int_0^T <\bar{P}> dt = \frac{1}{T}\int_0^T 0.1062[cos(2\pi x 10^6 t - 2z)]^2 x 10^{-3} \, \hat{z} \, dt =$$

$$\bar{P}_{av} = \frac{0.1062}{T}10^{-3} \, \hat{z} \int_0^T [cos(2\pi x 10^6 t - 2z)]^2 \, dt$$

Note : $[cos(2\pi x 10^6 t - 2z)]^2 = \frac{1}{2}[1 + cos(4\pi x 10^6 t - 4z)]$

$$\bar{P}_{av} = \frac{0.1062}{2T}10^{-3} \, \hat{z} \int_0^T [1 + cos(4\pi x 10^6 t - 4z)]dt$$

$$\bar{P}_{av} = \frac{0.1062}{2T}10^{-3} \, \hat{z}[\int_0^T dt + \int_0^T cos(4\pi x 10^6 t - 4z)dt]$$

$$\bar{P}_{av} = \frac{0.1062}{2T}10^{-3} \, \hat{z} \int_0^T dt = 53.1 x 10^{-6} \, \hat{z} \, (\frac{W}{m^2})$$

$$\bar{P}_{av} = 53.1 x 10^{-6} \hat{z} (\frac{W}{m^2})$$

**Exemple II.8.** Le champ électrique complexe d'une onde plane uniforme se déplaçant dans un milieu diélectrique non ferromagnétique non borné est donné par,

$$\bar{E} = \hat{y} \, 10^{-3} e^{-j2\pi z} \, (\frac{V}{m})$$

où « $z$ » est mesuré en mètres. En supposant que la fréquence de fonctionnement est de 100 ($MH_z$). Trouver :

a) La vitesse de phase de l'onde,

b) La constante diélectrique du milieu,

c) Le champ magnétique,

d) La densité de puissance moyenne dans le temps,

e) La densité énergétique totale moyenne dans le temps.

**Solution :**

a) $\beta = 2\pi \left(\frac{rad}{m}\right)$, $\lambda = \frac{2\pi}{\beta} = 1\ (m)$, $\lambda_0 = \frac{c}{f} = 3\ (m)$

b) $\lambda = \frac{\lambda_0}{\sqrt{\varepsilon_r}} \rightarrow \varepsilon_r = [\frac{\lambda_0}{\lambda}]^2 = 3^2 = 9$, $\varepsilon = \varepsilon_r \varepsilon_0 = \frac{1}{4\pi} x 10^{-9} \left(\frac{F}{m}\right)$

c) $\bar{H} = \hat{n} \times \frac{\bar{E}}{Z}$, $\hat{n} = \hat{z}$, $Z = \frac{Z_0}{\sqrt{\varepsilon_r}} = 40\pi\ (\Omega)$,

$$\bar{H} = \frac{1}{40\pi}(\hat{z} \times \hat{y})10^{-3}e^{-j2\pi z} = -\hat{x}7.96x10^{-6}e^{-j2\pi z} \left(\frac{A}{m}\right)$$

$$\bar{E} = \hat{y}\ 10^{-3}\cos(2\pi x10^8 t - 2\pi z)\ \left(\frac{V}{m}\right)$$

$$\bar{H} = -\hat{x}7.96x10^{-6}\cos(2\pi x10^8 t - 2\pi z)\left(\frac{A}{m}\right)$$

$$\bar{P} = \bar{E} \times \bar{H}$$

$$\bar{P} = -\hat{y} \times \hat{x}\ 7.968 10^{-9}\ [\cos(2\pi x10^8 t - 2\pi z)]^2$$

$$\bar{P} = \hat{z}\ 7.968x10^{-9}x0.5[1 + [\cos(4\pi x10^8 t - 4\pi z)]\ \left(\frac{W}{m^2}\right)$$

$$< P_{moyenne} > = 3.984x10^{-9}\left(\frac{W}{m^2}\right)$$

Si l'on utilise,

$$\bar{P} = \frac{1}{2} Re[\bar{E} \times \bar{H}^*] = -\frac{1}{2} Re[\hat{y} 10^{-3}e^{-j2\pi z} \times \hat{x}7.968 10^{-6}e^{j2\pi z}]$$

$$\bar{P} = \hat{z}3.984x10^{-9}(\frac{W}{m^2})$$

où

$$< P_{moyenne} > = 3.984x10^{-9}(\frac{W}{m^2})$$

d) La densité d'énergie d'une électromagnétique onde est,

$$u_{EM} = \frac{1}{2}[\mu H^2 + \varepsilon E^2]$$

$$E^2 = 10^{-6}[\cos(2\pi x10^8 t - 2\pi z)]^2$$

$$H^2 = 7.96^2 x10^{-12}[\cos(2\pi x10^8 t - 2\pi z)]^2$$

$$u_{EM} = \frac{1}{2}[\mu_0 7.96^2 x10^{-12} + 9\varepsilon_0 x10^{-6}][\cos(2\pi x10^8 t - 2\pi z)]^2$$

$$u_{EM} = \frac{1}{4}x10^{-6}\left[7.96^2 x10^{-6}x4\pi x10^{-6} + 9x10^{-9}\frac{1}{36\pi}\right]x[1+$$

$$+\cos(4\pi x10^8 t - 4\pi z)]$$

$$< u_{EM} > = \frac{1}{4}x10^{-6}x10^{-9}[0.743 + 0.08] = 2.0575x10^{-16}(J/m^3)$$

### 3.9 Effet Doppler

L'effet Doppler est essentiellement associé aux radars Doppler qui sont utilisés pour mesurer les vitesses de déplacement de différentes cibles. Similarité avec l'effet acoustique que l'on constate lors de passage d'une voiture de course (qui s'approche et qui s'éloigne d'un observateur). On considère une antenne qui émet un signal et qui est capable de recevoir un écho.

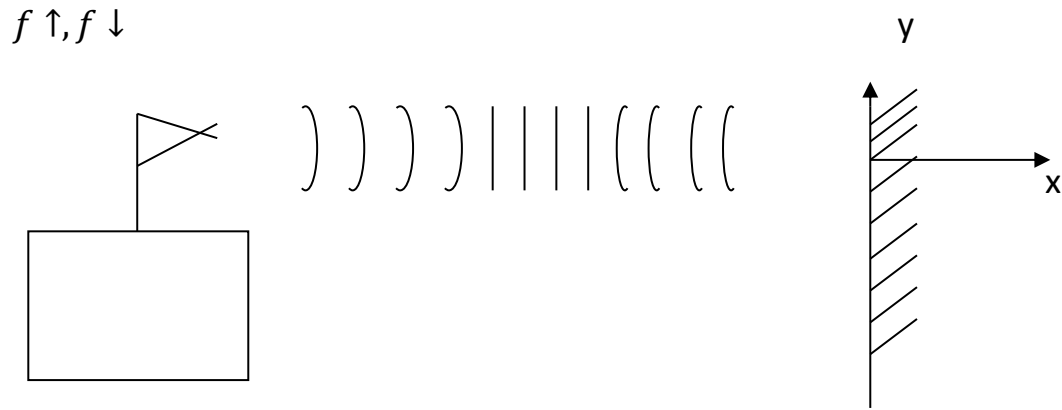

<p style="text-align:center">Figure II.8.</p>

L'obstacle va produire une réflexion d'amplitude plus ou moins grande dépendant des propriétés de la forme et de la grosseur de la cible (Figure II.8). Si la distance entre la cible et l'antenne reste constante, une partie du signal réfléchi sera captée par l'antenne et la fréquence du signal capté sera la même pour celle de l'onde émise.

Dans ce cas on aura :

$$\bar{E}_i = \hat{z} E_i e^{-j\beta x}$$

$$\bar{E}_r = \hat{z} E_r e^{+j\beta x}$$

On suppose que l'obstacle se déplace vers l'antenne à une vitesse « $v_r$ »

$$\bar{E}_r = \hat{z} E_r e^{j\beta(x_o + 2v_r t)}$$

d'où une avance de phase pour l'onde réfléchie.

Le champ électrique instantané sera :

$$\overrightarrow{\mathcal{E}_r}(x,t) = Re\left\{ \bar{E}_r e^{j\omega_o t} e^{j\beta(x_o + 2v_r t)} \right\}$$

$$= |\bar{E}_r| \cos(\omega_o t + \beta x_o + 2\beta v_r t)$$

$$= \bar{E}_r \cos\{(\omega_o + 2\beta v_r)t + \beta x_o\}$$

d'où on peut extraire

$$\omega_o + 2\beta v_r = \omega_1 \qquad \text{une nouvelle fréquence}$$

Cela signifie,

$$\omega_o = \text{fréquence angulaire de l'onde émise}$$

$$\omega_1 = \text{fréquence angulaire de l'onde reçue}$$

$$\Delta f = \frac{\omega_1 - \omega_o}{2\pi} = \frac{\omega_o + 2\beta v_r - \omega_o}{2\pi} = \frac{2\beta v_r}{2\pi} = \frac{2\pi}{\lambda}\frac{1}{\pi}v_r$$

$$\boxed{\Delta f = \frac{2v_r}{\lambda} \times \frac{f_o}{f_o} \quad \rightarrow \quad \Delta f = \frac{2v_r f_o}{c}}$$

$$f_o = \text{fréquence de l'ode émise}$$

$$c = \text{vitesse de propagation dans l'espace libre}$$

$$\Delta f > 0 \text{ pour la cible qui s'approche}$$

$$\Delta f < 0 \text{ pour la cible qui s'éloigne}$$

**Exemple II.9.** Un avion se déplace à une vitesse constante selon une trajectoire rectiligne parallèle à la surface de la Terre (Figure II.9). Un radar Doppler, ayant une fréquence d'opération de 12 ($GH_z$) pointe continuellement ver cet avion. La verticale au point $A$ où se trouve radar coïncide avec la trajectoire de l'avion en point $B$. L'altitude constante de cet avion ($h$) est donnée par la longueur du segment $AB$. Soit un autre point $C$, le point où se trouve l'avion au moment $t$ = 0 (sec). L'angle $\theta$ qui caracterise ce point $C$ qui est en fait l'angle d'élévation du radar vis-à-vis de l'horizon ($\theta$ − c'est l'angle du segment $AC$ avec la surface de la Terre).

a) Calculer la vitesse de cet avion si $\theta = 68.9^0$ à $t$ = 0 (sec) avec un signal écho qu'on reçoit à une fréquence de (12 ($GH_z$) − 6.4 ($kH_z$)),
b) Si à $t$ = 10 (sec), $\theta$ devient $60^0$, calculer ($h$), ainsi que la fréquence reçue à ce moment précis ($t$ = 10 (sec)).

**Solution :**

a)

$$\Delta f = \frac{2v_r}{c} f_0 \;,\; (v_r - \text{vitesse relative})$$

$$f_0 = 12 \; (GH_z), \; v_r = v \cos\theta, \; \Delta f = 6.4 \; (kH_z), \; \theta_1 = 68.9^0$$

$$v_{r1} = \frac{c\Delta f}{2f_0} = 80 \; (\frac{m}{s})$$

$$\text{à } t = 0 \; (sec), \; v_1 = \frac{v_{r1}}{\cos\theta_1} = 800 \; (\frac{km}{s})$$

b)

$$\text{à } t = 10 \; (sec), v - \text{Cte}, \qquad v_{r2} = v_{r1} \cos(\theta_2) = 111.111 \; (\frac{m}{s})$$

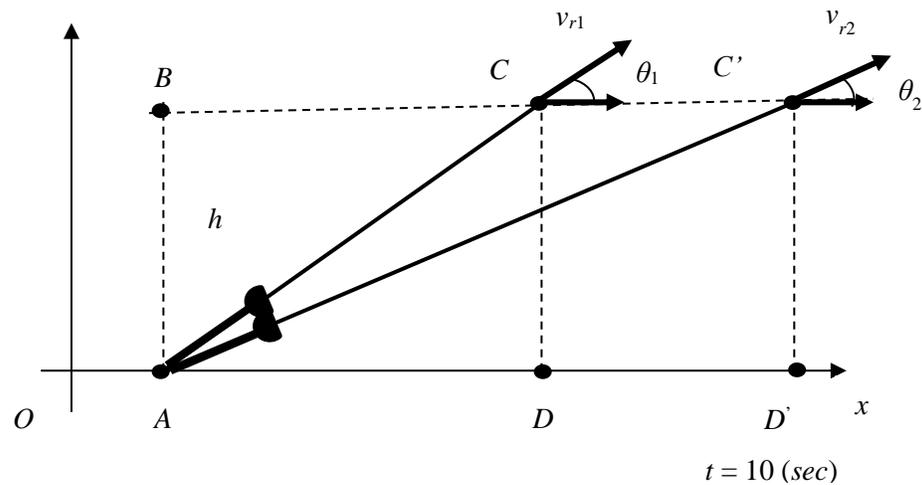

**Figure II.9.**

**Solution :**

$$DD' = v_1 t = 222.222 x 10 = 2222.222 \; (m)$$

$$\text{Pour } \theta_2 = 60^0 \;,\; \tan 60^0 = \frac{h}{AD + DD'} = 1.732$$

$$\frac{h}{AD} = \tan(\theta_1)$$

La combinaison de ces deux dernières équations donne,

$$AD = 4475.583 \ (m)$$

$$h = AD \ tan(\theta_1) = 11600.71 \ (m)$$

$$\Delta f_2 = \frac{2v_{r2}}{c} f_0 = 8.888 \ (kHz)$$

$$f_2 = 12 \ (GH_z) - 8.8888 \ (kH_z) = 11.99999111 (GH_z)$$

**Exemple II.10.** Un observateur immobile sur le bord d'une route très droit entend une fréquence $f_0 = 540 (H_z)$ provenant de la sirène d'une ambulance qui s'approche avec une vitesse $\vartheta_r = 90 \ (\frac{km}{h})$.

a) Calculer la fréquence du son émis $f_e$ par la sirène et perçue par le conducteur du véhicule,

b) Calculer la fréquence $f_0'$ du son perçu par l'observateur quand l'ambulance s'éloignera de lui avec la même vitesse de $90 \ (\frac{km}{h})$. La vitesse du son dans l'air est $v = 340 \ (\frac{m}{s})$.

**Solution :**

a) La relation entre les fréquences émises $f_e$ et perçue $f_0$, est,

$$f_0 = \frac{v}{v - v_r} f_e \ \rightarrow \ f_e = \frac{340 - 25}{340} x 540 = 500.3 \ (H_z)$$

Le son perçu par l'observateur immobile sur le bord de la route a une fréquence $f_0 = 540 \ (H_z)$ plus élevée que celle de son émis $f_e = 500.3 \ (H_z)$ par la sirène. Il est plus aigu.

Note : $v_r = 90 \left(\frac{km}{h}\right) = 25 \left(\frac{m}{s}\right)$.

b) Maintenant, on calcule la fréquence $f_0'$ du son perçu par l'observateur quand l'ambulance s'éloignera de lui avec la même vitesse de $90\ \frac{km}{h}$. La durée de réception d'une seule oscillation sonore est alors plus grande. La fréquence perçue $f_0'$ va diminuer. Donc, cette fréquence se calcule par,

$$f_0' = \frac{v}{v + v_r} f_e = \frac{340}{340 + 25} x 500.3 = 466 (H_z)$$

Le son perçu par l'observateur immobile sur le bord de la route a une fréquence $f_0' = 466\ (H_z)$ plus petite que celle de son émis $f_e = 500.3\ (H_z)$ par la sirène.

## 4. POLARISATION DES ONDES ÉLECTROMAGNÉTIQUES

Définition : La polarisation d'une onde électromagnétique est une indication de l'orientation du champ électrique en fonction du temps.

Soit une onde plane se propageant dans la direction « $+z$ » :

Le champ électrique est,

$$\vec{\mathcal{E}} = \hat{x} E_1 \cos(\omega t - \beta z) + \hat{y} E_2 \cos(\omega t - \beta z + \delta)$$

$E_1$ = valeur crête du champs électrique orienté selon $\hat{x}$

$E_2$ = valeur crête du champs électrique orienté selon $\hat{y}$

$\delta$ = déphasage dans le temps

On examine le champ électrique $\vec{\mathcal{E}}$ en fonction du temps dans le plan perpendiculaire $\perp$ à la direction de propagation (plan $z = 0$), (Figure II.10).

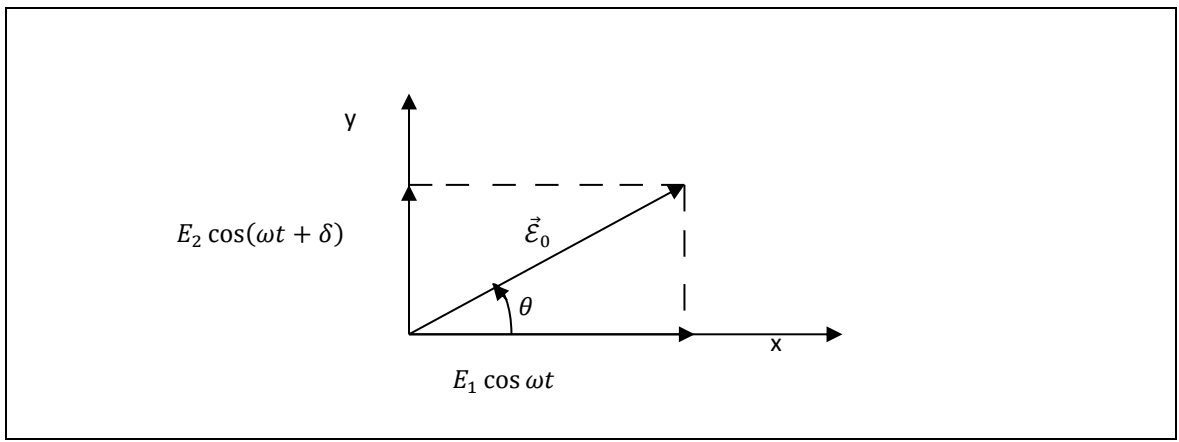

Figure II.10.

La résultante de champ électrique en fonction du temps :

$$\left|\vec{\mathcal{E}_0}\right|^2 = (E_1 \cos \omega t)^2 + (E_2 \cos(\omega t + \delta))^2$$

L'angle entre le champ électrique résultant et l'axe « $x$ » : $\theta$

$$\tan \theta = \frac{E_2 \cos(\omega t + \delta)}{E_1 \cos \omega t}$$

D'une façon générale, la pointe du vecteur champ électrique résultant va dessiner une ellipse dans le plan « *xy* » lorsque « *t* »varie.

Preuve :

$$E_x = E_1 \cos \omega t$$
$$E_y = E_2 \cos(\omega t + \delta)$$

$$\cos(\omega t + \delta) = \cos \omega t \cos \delta - \sin \omega t \, \sin \delta$$
$$\cos \omega t = \frac{E_x}{E_1}$$
$$\cos(\omega t + \delta) = \frac{E_y}{E_2} = \frac{E_x}{E_1} \cos \delta - \sqrt{1 - \left(\frac{E_x}{E_1}\right)^2} \sin \delta$$

En élevant au carré :

$$\left(\frac{E_y}{E_2} - \frac{E_x}{E_1} \cos \delta\right)^2 = \left\{1 - \left(\frac{E_x}{E_1}\right)^2\right\} \sin^2 \delta$$

$$\left(\frac{E_y}{E_2}\right)^2 - 2 \frac{E_x E_y}{E_1 E_2} \cos \delta + \left(\frac{E_x}{E_1}\right)^2 \cos^2 \delta = \sin^2 \delta - \left(\frac{E_x}{E_1}\right)^2 \sin^2 \delta$$

$$\left(\frac{E_y}{E_2}\right)^2 + \left(\frac{E_x}{E_1}\right)^2 (\cos^2 \delta + \sin^2 \delta) - 2 \frac{E_x E_y}{E_1 E_2} \cos \delta - \sin^2 \delta = 0$$

ce qui correspond à l'équation générale d'une ellipse ayant son centre à l'origine du système d'axes.

Pour une meilleur compréhension de la polarisation, on va essayer de regarder de plus près les cas différents.

## 4.1 Cas #1.  Polarisation linéaire

$\delta = 0$ $\qquad$ $E_x$ $et$ $E_y$ sont en phase dans le temps

$$\left.\begin{array}{l} E_x = E_1 \cos \omega t \\ E_y = E_2 \cos \omega t \end{array}\right\} \rightarrow \tan \theta = \frac{E_y}{E_x} = \frac{E_2}{E_1}$$

Le lien de la pointe du vecteur champ électrique résultant est donc une droite faisant un angle $\theta$ avec l'axe « x », (Figure II.11).

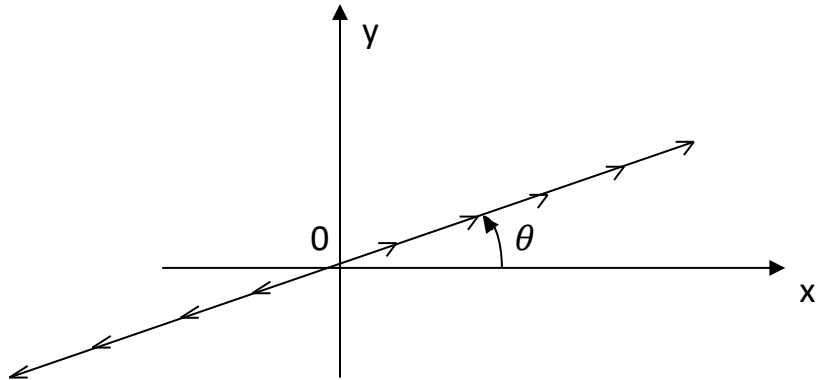

Figure II.11.

## 4.2 Cas #2.  Polarisation circulaire

$\delta = -\frac{\pi}{2}$ $\qquad$ $E_2 = E_1 = E$ (Figure II.12)

$E_y$ est en retard de phase sur $E_x$ de $\frac{\pi}{2}$

$$E_x = E \cos \omega t \,;\, E_y = E \cos\left(\omega t - \frac{\pi}{2}\right) = E \sin \omega t$$

$$\tan \theta = \frac{E \sin \omega t}{E \cos \omega t} = \tan \omega t$$

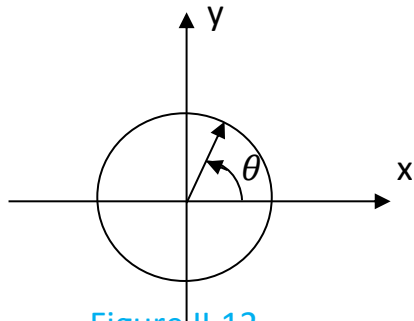

Figure II.12.

Le champ électrique résultant a toujours la même amplitude. Ce vecteur tourne à une vitesse angulaire de $\omega$ rad/sec où $\omega$ est la fréquence angulaire de l'emetteur.

Pour un observateur placé en arrière de l'emetteur et qui voit l'onde s'éloigner de lui le vecteur champ électrique tourne dans le sens horaire (dextrogyre).

Si on fait $\delta = \frac{\pi}{2}$ en maintenant $E_1 = E_2$, on obtient encore une polarisation circulaire, mais cette fois-ci, la rotation se fait dans l'autre sens (lévogyre).

### 4.3 Cas #3.   Polarisation elliptique

Soit $E_1 \neq E_2$   ,   $E_1 > E_2$ ;  $\delta = -\frac{\pi}{2}$ (Figure II.13)

$$E_x = E_1 \cos \omega t$$
$$E_y = E_2 \sin \omega t$$

$$\tan \theta = \frac{E_2}{E_1} \tan \omega t$$

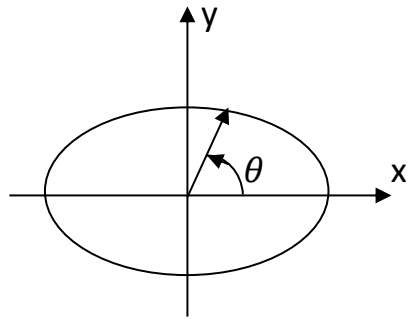

Figure II.13.

Le vecteur champ électrique résultant fait une rotation complète à chaque cycle d'oscillation du générateur, mais la vitesse angulaire de rotation n'est pas constante.

## 4.4 Cas #4. Polarisation elliptique (cas général)

Soit $E_1 \neq E_2$ $\qquad \delta =$ quelconque différent de $\pm \frac{\pi}{2}$ (Figure II.14)

$$\left(\frac{E_x}{E_1}\right)^2 - 2\frac{E_x E_y}{E_1 E_2}\cos\delta + \left(\frac{E_y}{E_2}\right)^2 = \sin\delta$$

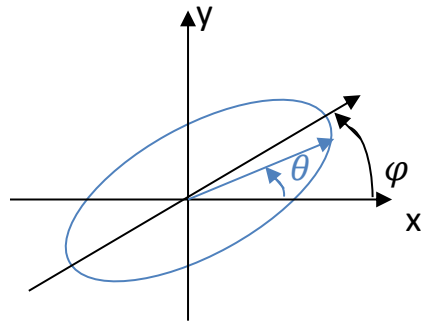

Figure II.14.

L'axe majeur fera un angle $\varphi$ par rapport à l'axe « x ».

$$\tan 2\varphi = 2\frac{E_1 E_2 \cos\delta}{E_1{}^2 - E_2{}^2}$$

Discussion :

a) Si on fixe « $t$ » et laisse varier « $z$ »
b) Si on laisse varier « $\underline{t}$ » et « $z$ » en même temps

Cas pratiques :

Poste émetteur :
- Bande AM (550-1600 KHz)
- Bande FM (88-108 MHz)
- Bande TV
- Radio Mobile
- Satellites
- Tel. Cellulaires

## Exemple II.11.

Déterminer la polarisation d'une onde plane uniforme dont les composantes du champ électrique sont $E_x = 4\cos(\omega t - \beta z)\left(\frac{V}{m}\right)$,

$E_y = 8\cos(\omega t - \beta z + \varphi)\left(\frac{V}{m}\right)$, $E_z = 0$.

**Solution :**

Si $\varphi = 2n\pi, n = 0,1,2,\ldots$, $\frac{E_y}{E_x} = 2$ = Cte. (Figure II.15)

Si $\varphi = (2n+1)\pi, n = 0,1,2,\ldots$, $\frac{E_y}{E_x} = -2$ = Cte. (Figure II.16)

Donc, c'est la polarisation linéaire.

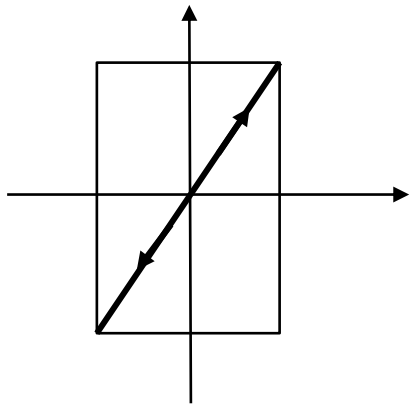

Figure II.15.

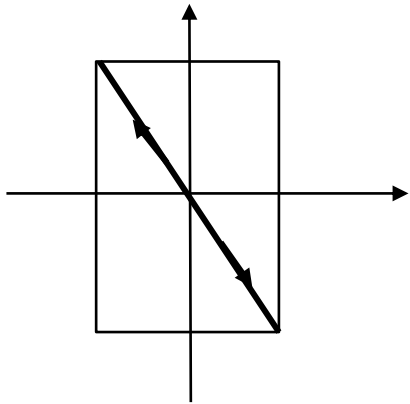

Figure II.16.

**Exemple II.12.**

Quelle est la polarisation d'une onde plane uniforme dont les composantes du champ électrique sont $E_x = 4\cos(\omega t - \beta z)\left(\frac{V}{m}\right)$, $E_y = 4\cos(\omega t - \beta z + \varphi)\left(\frac{V}{m}\right)$, $E_z = 0$.

**Solution :**

On voit que, $|E_{x0}| = |E_{y0}| = 4$

$$\text{Si } \varphi = (4n + 1)\frac{\pi}{2}, n = 0,1,2,\ldots, \text{ (Figure II.17)}$$

C'est la polarisation circulaire à gauche,

$$\text{Si } \varphi = (2n + 3)\frac{\pi}{2}, n = 0,1,2,\ldots, \text{ (Figure II.18)}$$

C'est la polarisation circulaire à droite.

On parle de la polarisation circulaire droite si le champ électrique tourne dans le sens horaire pour un observateur qui reçoit l'onde. Dans le cas contraire elle est circulaire gauche.

Dans les deux cas,

$$E_x^2 + E_y^2 = 16.$$

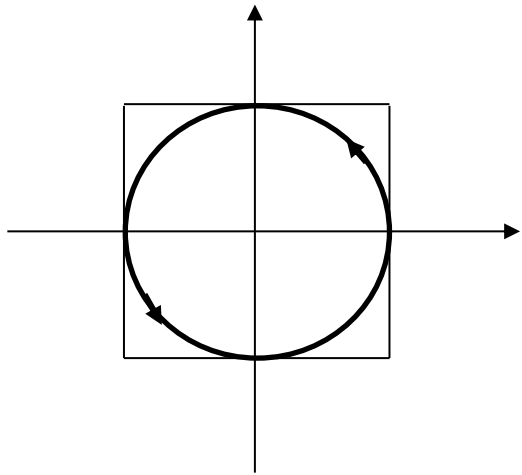

Figure II.17.

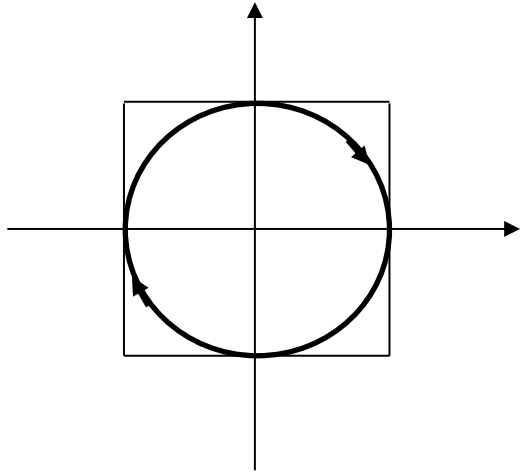

Figure II.18.

## Exemple II.13.

Quelle est la polarisation d'une onde plane uniforme dont les composantes du champ électrique sont $E_x = 4\cos(\omega t - \beta z)\ (\frac{V}{m})$, $E_y = 2\cos(\omega t - \beta z + \varphi)\ (\frac{V}{m})$, $E_z = 0$.

**Solution :**

On voit que, $|E_{x0}| \neq |E_{y0}|$

$$\text{Si } \varphi = (4n + 1)\frac{\pi}{2}, n = 0,1,2, …,\ \text{(Figure II.19)}$$

C'est la polarisation elliptique à gauche,

$$\text{Si } \varphi = (2n + 3)\frac{\pi}{2}, n = 0,1,2, …,\ \text{(Figure II.20)}$$

C'est la polarisation elliptique à droite.

Dans les deux cas,

$$\frac{E_x^2}{16} + \frac{E_y^2}{4} = 1.$$

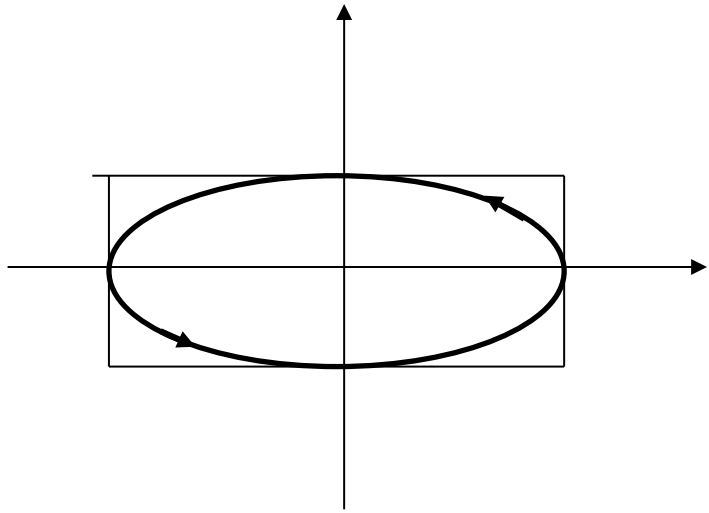

Figure II.19.

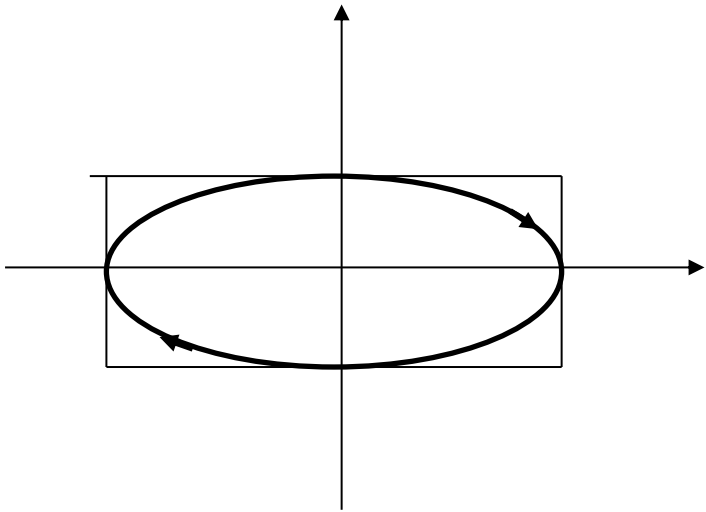

Figure II.20.

## Exemple II.14.

Dans une région dans l'espace libre sans source le champ magnétique complexe d'un champ harmonique temporel est donné par,

$$\bar{H} = \left[ \hat{x} \left( 1 + j \right) + \hat{z} \sqrt{2} \, e^{j\frac{\pi}{4}} \right] e^{-j\beta_0 y} \frac{E_0}{Z_0} \; \left( \frac{A}{m} \right)$$

où $E_0$ est une constante et $Z_0$ l'impédances de l'espace libre. Trouver :

a) Le champ électrique correspondant,
b) La polarisation de l'onde,
c) Le sens de rotation.

**Solution :**

a)
$$\bar{E} = Z_0 \, (\bar{H} \times \hat{n}) \ , \ \hat{n} = \hat{y}$$

$$\bar{E} = E_0[\hat{x}\,(1+j) + \hat{z}\,\sqrt{2}\,e^{j\frac{\pi}{4}}\,]e^{-j\beta_0 y} \times \hat{y}$$

$$\bar{E} = E_0 \left[\hat{z}\,(1+j) - \hat{x}\,\sqrt{2}\,e^{j\frac{\pi}{4}}\,\right] e^{-j\beta_0 y} \, (\frac{V}{m})$$

b)
$$\bar{E} = e^{-j\beta_0 y}[E_x + E_z]$$

où

$$E_x = -\sqrt{2}\,e^{j\frac{\pi}{4}} \ , \ E_z = (1+j) = \frac{\sqrt{2}}{2}\,e^{j\frac{\pi}{4}} \ , \ E_y = 0$$

$$\frac{E_z}{E_x} = -\frac{1}{2} = \text{Cte.}$$

C'est la polarisation linéaire.

c)
$$\frac{E_z}{E_x} = \tan(\alpha_{Lin}) = -\frac{1}{2} \ \rightarrow \ \alpha_{Lin} = 180^0 - \text{atan}(0.5) = 180^0 - 26.57^0$$

$$\alpha_{Lin} = 153.43^0$$

$$\hat{n}_{Lin} = \cos(\alpha_{Lin})\,\hat{x} + \sin(\alpha_{Lin})\hat{z}$$

$$\hat{n}_{Lin} = -0.894\,\hat{x} + 0.447\hat{z}$$

Donc, l'onde est polarisée linéairement dans la direction du vecteur unitaire $\hat{n}_{Lin}$.

## 5. VECTEUR DE POYNTING

Pour les ondes de polarisation :

- Linéaire $\quad\quad <P> = \frac{1}{2} Re\ (\bar{E} \times \bar{H}^*) = \frac{1}{2} \frac{E^2}{Z}$

- Elliptique $\quad\quad <P> = \frac{1}{2} \frac{E_1^2}{Z} + \frac{1}{2} \frac{E_2^2}{Z}$

- Circulaire $\quad\quad <P> = \frac{E^2}{Z}$ $\quad\quad\quad$ puisque $E_1 = E_2 = E$

## 6. PROPAGATION D'UNE ONDE PLANE DANS UNE DIRECTION ARBITRAIRE DE L'ESPACE

On a développé jusqu'ici des expressions mathématiques pour représenter des ondes planes dont la direction de propagation à l'un des axes du système de coordonnées cartésiens. Par exemple, si la direction de propagation est selon l'axe « x », le champ électrique instantané peut s'écrire :

$$\bar{\mathcal{E}} = Re\ (E_0 e^{-j\beta x} e^{j\omega t})$$

$$\bar{\mathcal{E}} = E_0 \cos(\omega t - \beta x)$$

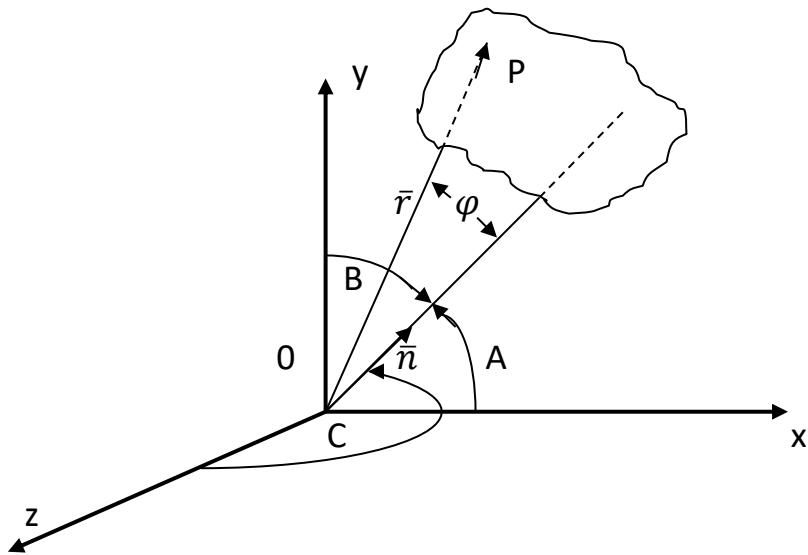

Figure II.21.

La même forme d'expression si l'onde se transmet selon « $y$ » ou « $z$ ».

Qu'arrive-t-il si la direction de propagation de l'onde plane fait un angle arbitraire avec l'axe du système de coordonnées (Figure II.21)?

Soit        $\hat{n}$ = vecteur unitaire orienté, dans la direction de propagation
           A = Angle entre la direction de propagation et l'axe « x »
           B = Angle entre la direction de propagation et l'axe « y »
           C = Angle entre la direction de propagation et l'axe « z »

Puisque on considère une onde plane, les composantes du champ électromagnétique sont uniformes dans un plan perpendiculaire à la direction de propagation.

           $\hat{r}$ = vecteur reliant l'origine du système d'axes à un point P, dans le plan perpendiculaire à la direction de propagation.

Sous forme mathématique, le fait de dire que les composantes du champ sont uniformes dans le plan perpendiculaire à la direction de propagation correspond à dire :

$$\hat{r} \cdot \hat{n} = \text{Cte.}$$

$\hat{r} = x\hat{x} + y\hat{y} + z\hat{z}$

$\hat{n} \cdot r = n\, r \cos\varphi$

$\varphi$ : Angle entre les deux vecteurs $\hat{r}\ et\ \hat{n}$

Le produit scalaire peut ainsi être obtenu en faisant la somme des produits des composantes des deux vecteurs:

$$\hat{n} = \hat{x}\cos(A) + \hat{y}\cos(B) + \hat{z}\cos(C)$$

Cos ($A$), cos ($B$) et cos ($C$) sont des cosinus du vecteur unitaire $\hat{n}$.

$$\cos^2(A) + \cos^2(B) + \cos^2(C) = 1$$

d'où

$$\hat{r} \cdot \hat{n} = x\cos(A) + y\cos(B) + z\cos(C)$$

D'une façon générale, les phaseurs représentant le champ électrique d'une onde plane peut s'écrire :

$$\bar{E} = E_0 e^{-j\beta\hat{n}\cdot\hat{r}}$$

En développant :

$$\bar{E} = E_0 e^{-j\beta[x\cos(A) + y\cos(B) + z\cos(C)]}$$

## Exemples II.15.

Une onde électromagnétique de polarisation circulaire se propage dans un diélectrique sans perte dont la perméabilité relative est $\mu_{r1} = 1$. La fréquence de l'onde est 500 ($MH_z$). Le vecteur-phaseur concernant le champ magnétique a l'expression suivante :

$$\bar{H} = (K\hat{x} + \frac{4}{j3\pi}\hat{y} + M\hat{z})e^{-j\beta(\hat{n}\cdot\vec{r})}$$

où   $\hat{n} = \frac{3}{5}\hat{y} - \frac{4}{5}\hat{z}$ .  La longueur d'onde dans ce diélectrique est 20 ($cm$).

a) Calculer β et $\varepsilon_r$ du diélectrique,
b) Déterminer $M$ en se servant de la relation: $\overline{H} \perp \hat{n}$,
c) À partir de l'information sur la polarisation, déterminer $K$,
d) Calculer le vecteur-phaseur $\bar{E}$,
e) Donner les valeurs réelles des composants $E_x, E_y$ et $E_z$ du champ électrique instantané en fonction des coordonnées d'espace et du temps.

**Solution:**

a) À partir de l'expression donnée, $\lambda = \frac{2\pi}{\beta} \rightarrow \beta = \frac{2\pi}{\lambda}$ on obtient, $\beta = 31.42 \left(\frac{rad}{m}\right)$, $\beta = \omega\sqrt{\varepsilon\mu} = \frac{2\pi f}{c}\sqrt{\varepsilon_r}$, $\sqrt{\varepsilon_r} = \frac{\beta c}{2\pi f} \rightarrow \varepsilon_r = \left(\frac{\beta c}{2\pi f}\right)^2 = 9$

b) La condition $\bar{H} \perp \hat{n}$ donne $\bar{H} \cdot \hat{n} = 0$, c'est-à-dire,

$$(K\hat{x} + \frac{4}{j3\pi}\hat{y} + M\hat{z})(\frac{3}{5}\hat{y} - \frac{4}{5}\hat{z}) = 0$$

$$M = -j\frac{1}{\pi} \ \left(\frac{A}{m}\right)$$

c)

$$Z = \sqrt{\frac{\mu}{\varepsilon}} = \sqrt{\frac{\mu_0}{9\varepsilon_0}} = \frac{Z_0}{3} = 40\pi \ (\Omega) = 125.664 \ (\Omega)$$

$$\bar{E} = Z(\bar{H} \times \hat{n})$$

$$\bar{E} = 125.664 \ [(K\hat{x} - j\frac{4}{3\pi}\hat{y} - \frac{1}{\pi}\hat{z})e^{-j31.42\left(\frac{3}{5}y - \frac{4}{5}z\right)} \times (\frac{3}{5}\hat{y} - \frac{4}{5}\hat{z})]$$

$$\bar{E} = 25.133 \left[j\frac{25}{3\pi}\hat{x} + 4K\hat{y} + 3K\hat{z}\right]e^{-j2\pi(3y-4z)}\left(\frac{V}{m}\right)$$

De la polarisation circulaire on obtint,

$$16K^2 + 9K^2 = \frac{25^2}{(3\pi)^2} \rightarrow K = \pm\frac{5}{3\pi}\left(\frac{A}{m}\right)$$

Donc,

$$\bar{E} = 25.133 \left[j\frac{25}{3\pi}\hat{x} \pm \frac{20}{3\pi}\hat{y} \pm \frac{15}{3\pi}\hat{z}\right]e^{-j2\pi(3y-4z)}\left(\frac{V}{m}\right)$$

$$\bar{E} = 13.333[j5\hat{x} \pm 4\hat{y} \pm 3\hat{z}]e^{-j2\pi(3y-4z)}\left(\frac{V}{m}\right)$$

e) Il y a deux possibilités:

1)
$$E_x = 66.665 \cos\left(10\pi x10^8 t - 6\pi y + 8\pi z + \frac{\pi}{2}\right)\left(\frac{V}{m}\right)$$

$$E_y = 53.332 \cos(10\pi x10^8 t - 6\pi y + 8\pi z)\left(\frac{V}{m}\right)$$

$$E_z = 39.999 \cos(10\pi x10^8 t - 6\pi y + 8\pi z)\left(\frac{V}{m}\right)$$

2)
$$E_x = 66.665 \cos\left(10\pi x10^8 t - 6\pi y + 8\pi z + \frac{\pi}{2}\right)\left(\frac{V}{m}\right)$$

$$E_y = 53.332 \cos(10\pi x10^8 t - 6\pi y + 8\pi z - \pi)\left(\frac{V}{m}\right)$$

$$E_z = 39.999 \cos(10\pi x10^8 t - 6\pi y + 8\pi z - \pi)\left(\frac{V}{m}\right)$$

## Exemples II.16.

Une onde électromagnétique de fréquence 600 ($MH_z$) se propage dans un diélectrique qui a des pertes. Le vecteur-phaseur pour le champ électrique a l'expression suivante :

$$\bar{E} = E_0 \hat{y}\, e^{-j\frac{2\pi}{\lambda_0}[2+2j]x}\left(\frac{V}{m}\right).$$

a) Quelle est la direction de propagation?
b) Calculer la longueur d'onde $\lambda_0$ de cette onde en espace libre (air),
c) Calculer α et β (facteur d'atténuation et facteur de phase) pour cette onde dans le diélectrique en question,
d) Calculer la longueur d'onde λ de cette onde dans le diélectrique,
e) Calculer l'atténuation en d$B$ de cette onde pour une distance de 20 cm à l'intérieur de ce diélectrique.

## Solution :

a) $\hat{n} \cdot \hat{r} = x \rightarrow \left(\frac{n_x}{|\hat{n}|}, \frac{n_y}{|\hat{n}|}, \frac{n_z}{|\hat{n}|}\right)(\hat{x}, \hat{y}, \hat{z}) = \hat{x}$

$$n_x = \frac{n_x}{|\hat{n}|} = 1,$$
$$n_y = n_z = 0$$

$$|\hat{n}| = \sqrt{n_x^2 + n_y^2 + n_z^2}$$

$$\hat{n} = \hat{x}$$

b) Dans l'espace libre $\beta_0 = \omega\sqrt{\mu_0 \varepsilon_0} = \frac{2\pi f}{c} = \frac{2\pi x 6 x 10^8}{3 x 10^8} = 4\pi \left(\frac{rad}{m}\right)$

$$\lambda_0 = \frac{2\pi}{\beta_0} = 0.5 \ m$$

Aussi,

$$\lambda_0 = \frac{c}{f} = 0.5 \ m$$

c) Étant donné que,

$$\bar{E} = E_0 \hat{y} \ e^{-j \gamma x} \ \left(\frac{V}{m}\right)$$

où,

$$\gamma = \alpha + j\beta = \frac{2\pi}{\lambda_0} \ [2 + 2j] = (8\pi + j8\pi)(1/m)$$

ce qui donne,

$$\alpha = 8\pi \left(\frac{Np}{m}\right)$$

$$\beta = 8\pi \left(\frac{rad}{m}\right)$$

d)

$$\beta = \frac{2\pi}{\lambda}$$

d'où,

$$\lambda = \frac{2\pi}{\beta} = \frac{2\pi}{8\pi} = 0.25 \ (m)$$

e)

$$Att(dB) = 20log_{10}\left(e^{-\lambda d}\right) = 20log_{10}\left(e^{-8\pi x 0.2}\right) = 43.66 \ (dB)$$

# 7. LONGUEUR D'ONDE ET VITESSE DE PHASE

Jusqu'à maintenant on a toujours traité le cas de longueur d'onde et de vitesse de phase dans la direction où l'onde se propage. Que deviennent ces deux paramètres lorsque l'onde se transmet dans une direction arbitraire (Figure II.22)?

$$v_p = \frac{\omega}{\beta}$$

$$\lambda = \frac{2\pi}{\beta}$$

$$\lambda_x = \frac{\lambda}{\cos(A)}$$

$$\lambda_y = \frac{\lambda}{\cos(B)}$$

$$\lambda_z = \frac{\lambda}{\cos(C)}$$

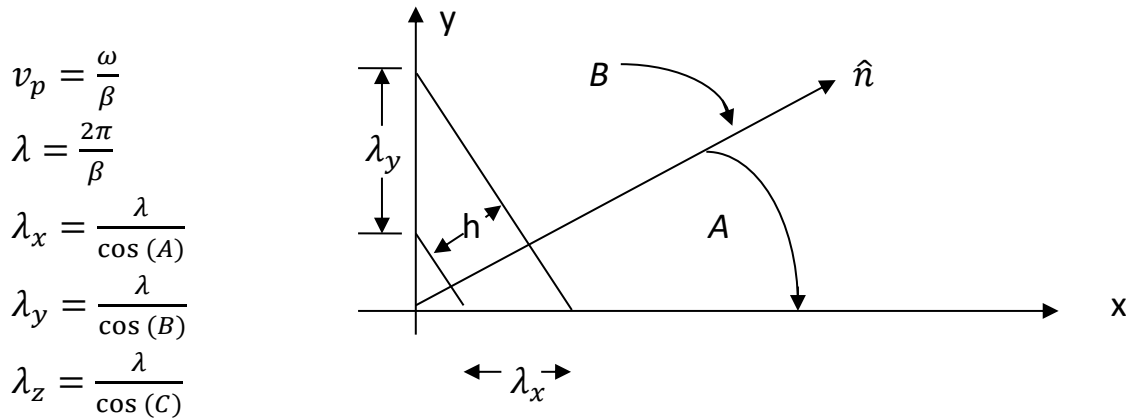

Longueur d'onde plus grande que lorsque mesurée dans une direction de propagation.

Figure II.22.

## 7.1 Vitesse de phase

Soit $v_p$, la vitesse de phase mesurée dans la direction de propagation,

$$v_p(x) = \frac{\omega}{\beta_x}$$

où

$$\beta_x = \frac{2\pi}{\lambda_x} = \frac{2\pi}{\lambda}\cos(A)$$

ce qui donne,

$$v_p(x) = \frac{\omega x \lambda}{2\pi \cos(A)} = \frac{2\pi f}{2\pi} \frac{\lambda}{\cos(A)}$$

$$v_p(x) = \frac{v_p}{\cos(A)}; \qquad v_p(y) = \frac{v_p}{\cos(B)}; \qquad v_p(z) = \frac{v_p}{\cos(C)}$$

## Exemple II.17.

Le champ électrique d'une onde électromagnétique est donné par l'expression suivante :

$$\bar{E} = (2.5\hat{x} + 4.33\hat{y})e^{-j0.2094(0.75\,x\,-0.433y+0.5z)}(\frac{V}{m})$$

Trouver les longueurs d'onde et les vitesses de phase dans la direction de propagation. La fréquence d'onde est de 10 $MH_z$.

Solution :

$$\lambda = \frac{2\pi}{\beta} = \frac{2\pi}{0.2094} = 30\ (m)$$

$$v_p = \frac{\omega}{\beta} = \frac{2\pi f}{\beta} = \frac{2\pi x 10^7}{0.2094} = c = 3x10^8 (\frac{m}{s})$$

$$\cos(A) = 0.75\ , \cos(B) = 0.433\ , \cos(C) = 0.5$$

$$\beta_x = \frac{2\pi}{\lambda}\cos(A) = \frac{2\pi}{30}0.75 = 0.1571\ (\frac{rad}{m})$$

$$\beta_y = \frac{2\pi}{\lambda}\cos(B) = \frac{2\pi}{30}0.433 = 0.0907\ (\frac{rad}{m})$$

$$\beta_z = \frac{2\pi}{\lambda}\cos(C) = \frac{2\pi}{30}0.5 = 0.1047\ (\frac{rad}{m})$$

$$\lambda_x = \frac{\lambda}{\cos(A)} = 40\ (m)$$

$$\lambda_y = \frac{\lambda}{\cos(B)} = 69.28\ (m)$$

$$\lambda_z = \frac{\lambda}{\cos(C)} = 60 \ (m)$$

$$v_{px} = \frac{v_p}{\cos(A)} = 4x10^8 \left(\frac{m}{s}\right)$$

$$v_{py} = \frac{v_p}{\cos(B)} = 6.938x10^8 \left(\frac{m}{s}\right)$$

$$v_{pz} = \frac{v_p}{\cos(C)} = 6x10^8 \left(\frac{m}{s}\right)$$

## 8. RÉFLEXION ET TRANSMISSION D'UNE ONDE PLANE À INCIDENCE NORMALE SUR UNE SURFACE DE SÉPARATION ENTRE DEUX MILIEUX

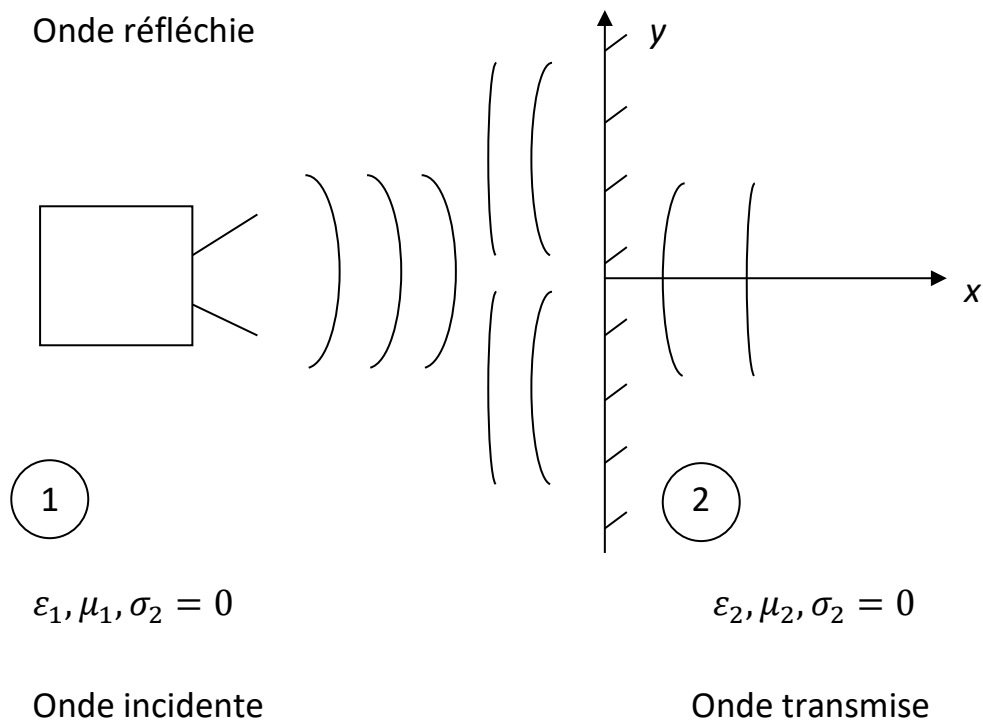

Figure II.23.

Soit une onde polarisée selon « $y$ » et se propageant dans la direction positive de « $x$ » dans un milieu sans pertes (Figure II.23),

L'onde incidente :

$$\bar{E}_i = \hat{y} E_i e^{-j\beta_1 x}$$

$$\bar{H}_i = \hat{z} \frac{E_i}{Z_i} e^{-j\beta_1 x}$$

$$\text{avec } Z_i = \sqrt{\frac{\mu_1}{\varepsilon_1}} \qquad \text{et} \qquad \bar{H}_i = \hat{n} \times \frac{\bar{E}_i}{Z_i}$$

L'onde réfléchie :

$$\bar{E}_r = \hat{y} E_r e^{+j\beta_1 x}$$

$$\bar{H}_r = -\hat{z} \bar{H}_r e^{+j\beta_1 x} = -\hat{z} \frac{E_r}{Z_1} e^{+j\beta_1 x}$$

L'onde transmise :

$$\bar{E}_t = \hat{y} E_t e^{-j\beta_2 x}$$

$$\bar{H}_t = -\hat{z} \frac{E_t}{Z_2} e^{-j\beta_2 x}$$

À la surface de séparation, lorsque $x = 0$ les composantes tangentielles de $\bar{E}$ et de $\bar{H}$ doivent être continues.

$$E_i + E_r = E_t$$

$$H_i - H_r = H_t$$

$$\frac{E_i}{Z_1} - \frac{E_r}{Z_1} = \frac{E_t}{Z_2}$$

On obtient,

$$\frac{E_r}{E_i} \quad \text{et} \quad \frac{E_t}{E_i}$$

$$\frac{E_i}{Z_1} - \frac{E_r}{Z_1} = \frac{E_i + E_r}{Z_2} \quad \Rightarrow \quad E_i\left(\frac{1}{z_1} - \frac{1}{z_2}\right) = E_r\left(\frac{1}{z_2} + \frac{1}{z_1}\right)$$

$$\Rightarrow E_i(Z_2 - Z_1) = E_r(Z_2 + Z_1)$$

d'où

$$\frac{E_r}{E_i} = \rho = \frac{Z_2 - Z_1}{Z_2 + Z_1} \qquad \text{coefficient de réflexion}$$

De la même façon :

$$\frac{E_t}{E_i} = T = \frac{2Z_2}{Z_2 + Z_1} \qquad \text{coefficient de transmission}$$

Valeur moyenne de l'intensité de puissance

$$<\bar{P}_i> = \frac{1}{2}Re\left(\bar{E}_i \times \bar{H}_i^{\,*}\right) = \frac{E_i^{\,2}}{2Z_1}\hat{x}$$

$$<\bar{P}_r> = \frac{1}{2}Re\left(\bar{E}_r \times \bar{H}_r^{\,*}\right) = -\frac{E_r^{\,2}}{2Z_1}\hat{x}$$

$$<\bar{P}_t> = \frac{1}{2}Re\left(\bar{E}_t \times \bar{H}_t^{\,*}\right) = \frac{E_t^{\,2}}{2Z_2}\hat{x}$$

$$\left|\frac{\bar{P}_r}{\bar{P}_i}\right| = \frac{\left(\frac{E_r^{\,2}}{2Z_1}\right)}{\left(\frac{E_i^{\,2}}{2Z_1}\right)} = \frac{E_r^{\,2}}{E_i^{\,2}} = \left(\frac{Z_2 - Z_1}{Z_2 + Z_1}\right)^2 = |\rho|^2$$

$$\frac{P_t}{P_i} = \frac{E_t^{\,2}}{Z_2}\frac{Z_1}{E_i^{\,2}} = \frac{Z_1}{Z_2}\left(\frac{E_t}{E_i}\right)^2$$

**ExempleII.18.**

L'intensité du champ électrique d'une onde électromagnétique de polarisation circulaire à gauche de fréquence de 200 ($MH_z$) est 10 ($V/m$). Cette onde se propage dans l'air et frappe normalement un milieu diélectrique (Figure II.24) avec $\varepsilon_r = 4$ situé dans la région $z \geq 0$. La composante $x$ du champ électrique incident- phaseur a un maximum à $z = 0$ lorsque $t = 0$.

a) Trouver le champ électrique phaseur de l'onde incidente,
b) Trouver les coefficients de réflexion et transmission,
c) Trouver le champ électrique phaseur des ondes réfléchie et transmise, et le champ total dans la région $z \leq 0$,
d) Trouver la fraction de la puissance moyenne incidente qui est réfléchie par l'interface et celle qui est transmise au milieu diélectrique.

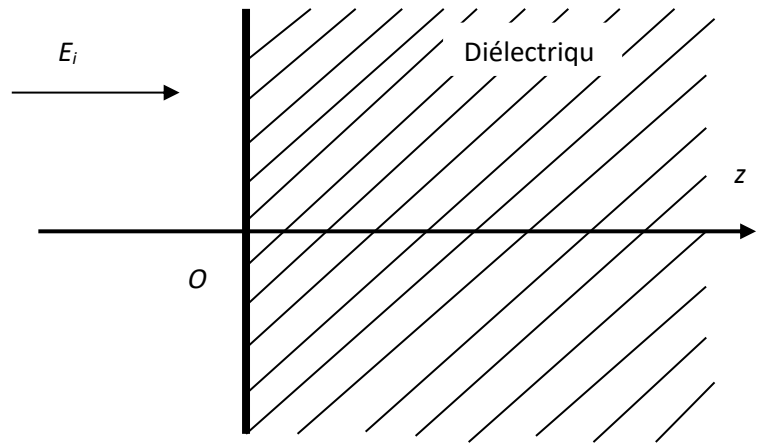

Figure II.24.

Solution :

a) Dans l'air,

$$\beta_0 = \omega\sqrt{\mu_0\varepsilon_0} = \frac{2\pi f}{c} = \frac{2\pi x 2x10^8}{3x10^8} = \frac{4\pi}{3} \left(\frac{rad}{m}\right)$$

L'onde incidente est de la polarisation circulaire à gauche. Ça signifie,

$$E_x = E_0 \cos(\omega t - \beta_0 z), \quad E_y = E_0 \cos\left(\omega t - \beta_0 z + \frac{\pi}{2}\right)$$

En forme phaseur on peut écrire,

$$E_x = E_0 e^{-j\beta_0 z}, \quad E_y = E_0 e^{-j\beta_0 z} e^{j\frac{\pi}{2}}$$

Finalement,

$$\bar{E}_i(z) = (E_x \hat{x} + E_y \hat{y}) = E_0 e^{-j\beta_0 z}(\hat{x} + j\hat{y})$$

ou

$$\bar{E}_i(z) = 10 e^{-j\frac{4\pi}{3}z}(\hat{x} + j\hat{y}) \left(\frac{V}{m}\right)$$

b)

$$\frac{E_r}{E_i} = \rho = \frac{Z_2 - Z_1}{Z_2 + Z_1}$$

$$\frac{E_t}{E_i} = T = \frac{2Z_2}{Z_2 + Z_1}$$

$$Z_1 = Z_0 = 120\pi \ (\Omega)$$

$$Z_2 = \sqrt{\frac{\mu_0}{4\varepsilon_0}} = \frac{Z_0}{2} = 60\pi \ (\Omega)$$

$$\rho = \frac{Z_2 - Z_1}{Z_2 + Z_1} = \frac{60\pi - 120\pi}{60\pi + 120\pi} = -\frac{1}{3}$$

$$T = \frac{2Z_2}{Z_2 + Z_1} = \frac{120\pi}{60\pi + 120\pi} = \frac{2}{3}$$

c)

$$\frac{E_r}{E_i} = \rho \ \rightarrow \ \frac{\bar{E}_r}{\bar{E}_i} = \rho \ \rightarrow \ \bar{E}_r = \rho \bar{E}_i$$

$$\bar{E}_r = -\frac{10}{3} e^{j\frac{4\pi}{3}z}(\hat{x} + j\hat{y}) \left(\frac{V}{m}\right)$$

$$\frac{E_t}{E_i} = T \ \rightarrow \ \frac{\bar{E}_t}{\bar{E}_i} = T \ \rightarrow \ \bar{E}_t = T\bar{E}_i$$

$$\bar{E}_t = \frac{20}{3} e^{-j\frac{4\pi}{3}z} (\hat{x} + j\hat{y}) \left(\frac{V}{m}\right)$$

e) Pour la polarisation circulaire,

$$< P_{moy,i} > = \frac{E_{0,i}^2}{Z_1} = \frac{E_{0,i}^2}{Z_0}$$

$$< P_{moy,r} > = \frac{E_{0,r}^2}{Z_1} = \frac{E_{0,r}^2}{Z_0}$$

$$\frac{< P_{moy,r} >}{< P_{moy,i} >} = \frac{E_{0,r}^2}{E_{0,i}^2} = \frac{\frac{100}{9}}{100} = \frac{1}{9}$$

$$< P_{moy,t} > = \frac{E_{0,t}^2}{Z_2} = \frac{2E_{0,t}^2}{Z_0}$$

$$\frac{< P_{moy,t} >}{< P_{moy,i} >} = \frac{2E_{0,t}^2}{Z_0} \frac{Z_0}{E_{0,i}^2} = \frac{2E_{0,t}^2}{E_{0,i}^2} = \frac{8}{9}$$

**Exemple II.19.**

À 200 ($MH_z$), un milieu donné est caractérisé par $\sigma = 0, \mu_r = 15(1 - j3)$ et $\varepsilon_r = 50(1 - j)$ dans la région $z \geq 0$. Trouver :

a) $Z/Z_0$, $\lambda/\lambda_0$ et $v/v_0$, où $Z$ est l'impédance intrinsèque et $v$ la vitesse de phase (l'indice ''0'' correspond aux valeurs dans l'air).

b) La profondeur de pénétration $\delta$.

c) L'affaiblissement en $dB$ sur une distance de 5 ($mm$).

d) Le coefficient de réflexion d'une onde se propageant dans l'air qui empiète normalement sur ce milieu.

**Solution :**

a) $Z = \sqrt{\frac{\mu}{\varepsilon}} = \sqrt{\frac{\mu_0 \mu_r}{\varepsilon_0 \varepsilon_r}} = \sqrt{\frac{\mu_0}{\varepsilon_0}} \sqrt{\frac{\mu_r}{\varepsilon_r}} = Z_0 \sqrt{\frac{15(1-j3)}{50(1-j)}} = Z_0 \sqrt{\frac{3(1-j3)(1+j)}{20}}$

$Z = Z_0 \sqrt{\frac{3(4-j2)}{20}} = Z_0 \sqrt{0.3\sqrt{5} \, e^{-jatan(\frac{1}{2})}} = Z_0 \sqrt{0.3\sqrt{5} \, e^{-j26.56^0}}$

$$\frac{Z}{Z_0} = 0.82e^{-j13.28^0} = 0.798 - j0.188$$

Dans un milieu diélectrique complexe où $\sigma = 0$ la constante de propagation est donnée par,

$$\gamma = j\frac{2\pi}{\lambda_0}\sqrt{\mu_r{}' - j\mu_r{}''}\sqrt{\varepsilon_r{}' - j\left(\varepsilon_r{}'' + \frac{\sigma}{\varepsilon_0\omega}\right)}$$

$$\gamma = j\frac{2\pi}{\lambda_0}\sqrt{\mu_r{}' - j\mu_r{}''}\sqrt{\varepsilon_r{}' - j\varepsilon_r{}''}$$

$$\gamma = j\frac{2\pi}{\lambda_0}\sqrt{(\mu_r{}'^2 + \mu_r{}''^2)(\varepsilon_r{}'^2 + \varepsilon_r{}''^2)}\, e^{-j\frac{\text{atan}\,(\mu_r{}''/\mu_r{}')}{2}} e^{-j\frac{\text{atan}\,(\varepsilon_r{}''/\varepsilon_r{}')}{2}}$$

$$\gamma = j\frac{2\pi}{\lambda_0}\sqrt{(\mu_r{}'^2 + \mu_r{}''^2)(\varepsilon_r{}'^2 + \varepsilon_r{}''^2)}\, [\cos\left(\frac{\text{atan}\left(\frac{\mu_r{}''}{\mu_r{}'}\right) + \text{atan}\left(\frac{\varepsilon_r{}''}{\varepsilon_r{}'}\right)}{2}\right) -$$

$$\sin\left(\frac{\text{atan}\left(\frac{\mu_r{}''}{\mu_r{}'}\right) + \text{atan}\left(\frac{\varepsilon_r{}''}{\varepsilon_r{}'}\right)}{2}\right)]$$

$$\alpha = \frac{2\pi}{\lambda_0}\sqrt{(\mu_r{}'^2 + \mu_r{}''^2)(\varepsilon_r{}'^2 + \varepsilon_r{}''^2)}\, \sin\left(\frac{\text{atan}\left(\frac{\mu_r{}''}{\mu_r{}'}\right) + \text{atan}\left(\frac{\varepsilon_r{}''}{\varepsilon_r{}'}\right)}{2}\right)$$

$$\beta = \frac{2\pi}{\lambda_0}\sqrt{(\mu_r{}'^2 + \mu_r{}''^2)(\varepsilon_r{}'^2 + \varepsilon_r{}''^2)}\, \cos\left(\frac{\text{atan}\left(\frac{\mu_r{}''}{\mu_r{}'}\right) + \text{atan}\left(\frac{\varepsilon_r{}''}{\varepsilon_r{}'}\right)}{2}\right)$$

$$\beta = \beta_0\sqrt{(\mu_r{}'^2 + \mu_r{}''^2)(\varepsilon_r{}'^2 + \varepsilon_r{}''^2)}\, \cos\left(\frac{\text{atan}\left(\frac{\mu_r{}''}{\mu_r{}'}\right) + \text{atan}\left(\frac{\varepsilon_r{}''}{\varepsilon_r{}'}\right)}{2}\right)$$

$$\frac{\beta}{\beta_0} = \sqrt{(\mu_r{}'^2 + \mu_r{}''^2)(\varepsilon_r{}'^2 + \varepsilon_r{}''^2)}\, \cos\left(\frac{\text{atan}\left(\frac{\mu_r{}''}{\mu_r{}'}\right) + \text{atan}\left(\frac{\varepsilon_r{}''}{\varepsilon_r{}'}\right)}{2}\right)$$

$$\frac{\lambda}{\lambda_0} = \frac{\beta_0}{\beta} = \frac{1}{\sqrt{(\mu_r'^2 + \mu_r''^2)(\varepsilon_r'^2 + \varepsilon_r''^2)}\cos\left(\dfrac{\text{atan}\left(\dfrac{\mu_r''}{\mu_r'}\right) + \text{atan}\left(\dfrac{\varepsilon_r''}{\varepsilon_r'}\right)}{2}\right)}$$

$$\frac{\lambda}{\lambda_0} = \frac{1}{\sqrt{15\sqrt{10}}\sqrt{50\sqrt{2}}\cos\left(\dfrac{\text{atan}(3) + \text{atan}(1)}{2}\right)}$$

$$\frac{\lambda}{\lambda_0} = 0.0173\frac{1}{\cos(58.28^0)} = 32.9x10^{-3}$$

$$\lambda = \frac{v_p}{f}, \quad \lambda_0 = \frac{v_o}{f} \rightarrow \frac{\lambda}{\lambda_0} = \frac{v_p}{v_o}$$

$$\frac{v_p}{v_o} = 32.9x10^{-3}$$

b)

$$\alpha = \frac{2\pi}{\lambda_0}\sqrt{(\mu_r'^2 + \mu_r''^2)(\varepsilon_r'^2 + \varepsilon_r''^2)}\sin\left(\dfrac{\text{atan}\left(\dfrac{\mu_r''}{\mu_r'}\right) + \text{atan}\left(\dfrac{\varepsilon_r''}{\varepsilon_r'}\right)}{2}\right)$$

$$\lambda_o = \frac{c}{f} = \frac{3x10^8}{2x10^8} = 1.5 \ (m)$$

$$\alpha = \frac{2\pi}{1.5}\sqrt{15\sqrt{10}}\sqrt{50\sqrt{2}}\sin\left(\frac{\text{atan}(3)+\text{atan}(1)}{2}\right) = 242.59\sin(58.28^0)$$

$$\alpha = 206.36(\frac{Np}{m})$$

$$\delta = \frac{1}{\alpha} = 4.85 \ (mm)$$

c)

Affaiblissent $(dB)$ = $20\log_{10}(e^{-\alpha d})$ = $20\log 10\ (e^{-\alpha d}) = -8.96\ (dB)$

d)

$$\rho = \frac{Z_2 - Z_1}{Z_2 + Z_1} = \frac{Z - Z_0}{Z + Z_0} = \frac{0.798 - j0.188 - 1}{0.798 - j0.188 + 1} = \frac{-0.202 - j0.188}{1.798 - j0.188}$$

$$\rho = -\sqrt{\frac{0.202^2 + 0.188^2}{1.798^2 + 0.188^2}} \quad \frac{e^{j\frac{\text{atan}(\frac{0.188}{0.202})}{2}}}{e^{-j\frac{\text{atan}(\frac{0.188}{1.798})}{2}}}$$

$$\rho = -\frac{0.276}{1.808}e^{j42.94^0}e^{j5.97^0} = -0.153e^{j48.91^0}$$

$$\rho = -0.153(0.657 + j0.754) = -0.101 - j0.115$$

$$\rho = 0.153e^{j(180^0 + 48.71^0)} = 0.15e^{j277.42^0}$$

## Exemple II.20.

Un four à micro-ondes fonctionnant à 2.45 ($GH_z$) est utilisé pour décongeler un aliment surgelé ayant la permittivité complexe $\varepsilon_c = (4 - j)\varepsilon_0 \left(\frac{F}{m}\right)$. Déterminer l'amplitude du champ électrique à une profondeur de 1 ($cm$) et l'exprimer en $dB$ et en pourcentage de sa valeur à la surface. Répéter si $\varepsilon_c = (45 - j15)\varepsilon_0 \left(\frac{F}{m}\right)$.

Solution :

$$\beta_0 = \omega\sqrt{\varepsilon_0\mu_0} = \frac{2\pi f}{c} = 51.31 \left(\frac{rad}{m}\right)$$

$$\beta_c = \omega\sqrt{\varepsilon_c\mu_c} = \frac{2\pi f}{c}\sqrt{\varepsilon_{cr}} = 51.31\sqrt{4 - j} = (103.41 - j12.73)\left(\frac{1}{m}\right)$$

$$\beta_c = \beta - j\alpha = 103.41 - j12.73, \beta = 103.41\left(\frac{rad}{m}\right)$$

$$\alpha = 12.73 \left(\frac{Np}{m}\right)$$

$$\delta = \frac{1}{\alpha} = 7.86 \ (cm)$$

L'atténuation à 1 ($cm$) sera :

$$Att = 8.686x \ \frac{z}{\delta} = 8.686x \ \frac{1}{7.86} = 1.1 \ (dB) \ \rightarrow 10^{-\frac{Att}{20}} = 0.88 = 88 \ \%$$

Le champ à la profondeur de 1 ($cm$) et de 88% de sa valeur sur la surface.

Si $\varepsilon_c = (45 - j15)\varepsilon_0 \left(\frac{F}{m}\right)$,

$$\beta_c = \omega\sqrt{\varepsilon_c \mu_c} = \frac{2\pi f}{c}\sqrt{\varepsilon_{cr}} = 51.31\sqrt{45 - j15} = (348.84 - j56.61)(\frac{1}{m})$$

$$\beta = 348.84(\frac{rad}{m})$$

$$\alpha = 56.61 \left(\frac{Np}{m}\right)$$

$$\delta = \frac{1}{\alpha} = 1.77 \ (cm)$$

L'atténuation à 1 ($cm$) sera :

$$Att = 8.686x \ \frac{z}{\delta} = 8.686x \ \frac{1}{1.77} = 4.9 \ (dB) \ \rightarrow 10^{-\frac{Att}{20}} = 0.57 = 57 \ \%$$

Le champ à la profondeur de 1 ($cm$) est de 57% de sa valeur sur la surface.

### Exemple II.21.

Une feuille de matériau ferromagnétique et conducteur est suspendue dans l'air. Une onde électromagnétique plane la traverse à incidence normale. On considère les dimensions latérales (perpendiculaires à la direction de propagation) comme étant infinies. Les propriétés caractéristiques du matériau en question sont :

$$\sigma = 100 \left(\frac{S}{m}\right), \qquad \mu_r = 25, \ \ \varepsilon_r = 1$$

L'épaisseur de la feuille est 1 ($mm$). Trouver l'atténuation de la puissance de l'onde qui la traverse et ayant une fréquence de 80 ($MH_z$). (Tenir compte uniquement des transmissions à l'entrée et à la sortie plus la perte de puissance durant le trajet à travers la feuille).

Solution :

Air :  $Z_1 = Z_0 = 120\pi(\Omega) = 377 \ (\Omega), \ \varepsilon_0 = \frac{10^{-9}}{36\pi} \left(\frac{F}{m}\right)$

Feuille : $Z_2 = \sqrt{\frac{j\omega\mu}{\sigma + j\omega\varepsilon}}, \ \ \omega\varepsilon = 2\pi f \varepsilon_0 \varepsilon_r = 2\pi x8x10^7 \frac{10^{-9}}{36\pi} = 0.004$

$$\omega\varepsilon \ll \sigma$$

Cela implique,

$$Z_2 \cong \sqrt{\frac{j\omega\mu}{\sigma}} = \sqrt{\frac{\omega\mu}{\sigma}} \ e^{j\frac{\pi}{4}}$$

$$\mu_0 = 4\pi x10^{-7}(\frac{H}{m})$$

$$Z_2 = 12.566 e^{j\frac{\pi}{4}} = (8.886 + j8.886)(\Omega)$$

À l'entrée de la feuille : $T_1 = \frac{2Z_2}{Z_2 + Z_1}$

$$T_1 = \frac{2}{1 + \frac{Z_1}{Z_2}} = \frac{2}{1 + \frac{120\pi}{4\pi e^{j\frac{\pi}{4}}}} = \frac{2}{1 + 30 e^{-j\frac{\pi}{4}}}$$

$$T_1 = 0.065 e^{j43.681^0}$$

À la sortie de la feuille : $T_2 = \frac{2Z_0}{Z_2 + Z_0}$

$$T_2 = \frac{2}{1 + \dfrac{Z_2}{Z_0}} = \frac{2}{1 + \dfrac{4\pi e^{j\frac{\pi}{4}}}{120\pi}} = \frac{2}{1 + \dfrac{1}{30}\, e^{-j\frac{\pi}{4}}}$$

$$T_2 = 1.953 e^{-j1.319^0}$$

$$\alpha = \frac{1}{\delta} = \sqrt{\pi f \mu \sigma} = 888.575 \left(\frac{Np}{m}\right)$$

$$e^{-\alpha d} = 0.4112 \ (Np)$$

Le rapport entre l'amplitude de l'onde sortante par rapport à l'onde incidente est :

$$\frac{|E_s|}{|E_i|} = |T_1||T_2|e^{-\alpha d} = 0.065 x 1.953 x 0.4112 = 0.052$$

Pour la puissance on a,

$$\text{Att}(dB) = 10\log\left[\frac{|E_i|}{|E_s|}\right]^2 = 20\log\left[\frac{1}{0.052}\right] = 25.65 \ (dB)$$

## 8.1. Représentation graphique du champ électrique dans la région 1 (champ total).

$$\bar{\mathcal{E}}_T = \bar{\mathcal{E}}_i + \bar{\mathcal{E}}_r = Re\left(E_i e^{j(\omega t - \beta_1 x)}\right) + Re\left(E_r e^{j(\omega t + \beta_1 x)}\right)$$

$$= E_i \cos(\omega t - \beta_1 x) + E_r \cos(\omega t + \beta_1 x)$$

$$= E_i(\cos \omega t \cos \beta_1 x + \sin \omega t \sin \beta_1 x) +$$

$$+E_r \left(\cos \omega t \cos \beta_1 x - \sin \omega t \sin \beta_1 x\right)$$

$$\bar{\mathcal{E}}_T = (E_i + E_r) \cos \omega t \cos \beta_1 x + (E_i - E_r) \sin \omega t \sin \beta_1 x$$

## Exemple :

Milieu ① parfait diélectrique

Milieu ② parfait conducteur

$$\frac{E_r}{E_i} = \frac{Z_2 - Z_1}{Z_2 + Z_1} \ ; \ Z_1 = \sqrt{\frac{\mu_0}{\varepsilon_r}} \ ; \ Z_2 = \sqrt{\frac{j\omega\mu}{\sigma + j\omega\varepsilon}} ; \ Z_2 = 0 \ \Rightarrow \frac{E_r}{E_i} = -1$$

À la surface de séparation, le champ électrique réfléchi est déphasé de 180° par rapport au champ électrique incident (Figure II.24).

$$\mathcal{E}_T = 2E_i \sin \omega t \sin \beta x$$

Posons « $t$ » de façon que $\sin \omega t = 1$

$$\mathcal{E}_T = 2E_i \sin \beta x$$

$$x = 0 , \qquad\qquad \sin \beta_1 x = 0$$
$$x = -\frac{\lambda}{4} , \qquad\qquad \sin \beta_1 x = -1$$
$$x = -\frac{\lambda}{4} , \qquad\qquad \sin \beta_1 x = 0$$

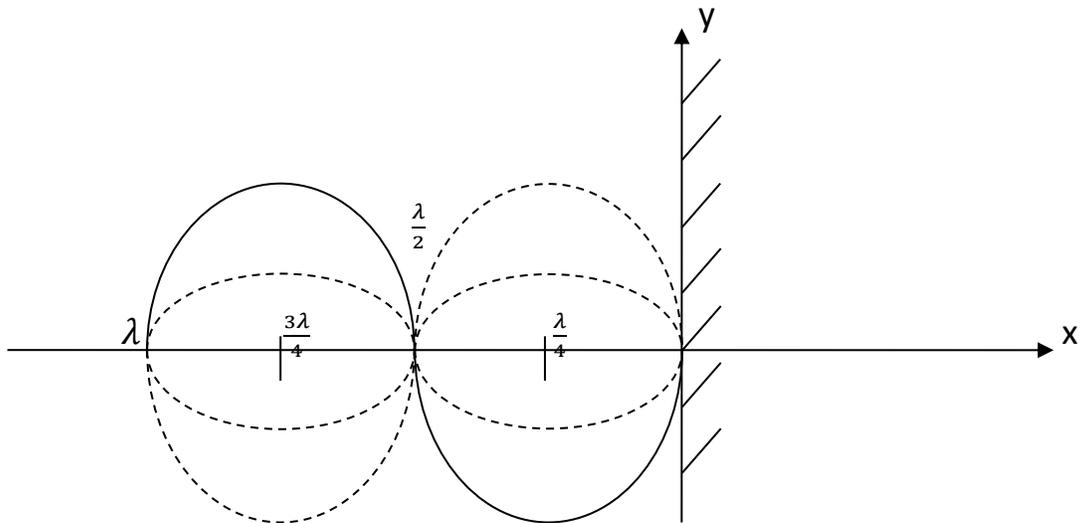

Figure II.24.

Si l'on laisse varier « $t$ », on remarque :

1) Le champ électrique total parallèle à la surface parfaitement conductrice est nul près de cette dernière.
2) Ce nul se répète à chaque $\lambda/2$ à partir du plan conducteur.

3) Le champ électrique est maximum à une distance de $\lambda/4$ du plan conducteur et son amplitude est $2E_i$

- comparaison avec une corde vibrante
- mesure avec une antenne et un oscilloscope

Note : 1) La profondeur de pénétration du champ est nulle dans le conducteur parfait.

   2) Dans un conducteur parfait, le champ électrique et le champ magnétique sont nuls.

   3) Dans un conducteur parfait il n'a pas de puissance dissipée.

   4) Sur la surface conductrice le champ magnétique n'est pas nul,

$$\bar{H}_{Total}(0,t) = \frac{2E_0}{Z_0} e^{j\omega t} \hat{z} = \bar{H}_{Tang}(0,t)$$

   5) Le champ magnétique est donc tangent à la surface, perpendiculaire au courant surfacique et son intensité est proportionnelle à celle du courant surfacique.

   6) Dans le conducteur parfait il n'a pas de courants volumiques et de charges.

   7) Le seul courant qui existe, c'est le courant surfacique qui est exprimé en (*A/m*),

$$\bar{J}_s = \hat{n} \times \bar{H}_{Tang}$$

Si le milieu ② est un diélectrique différent du milieu ① ou un conducteur quelconque.

$$-1 < \frac{E_r}{E_i} < +1$$

Et la seule façon de faire une représentation graphique est d'utiliser l'expression de $\bar{\mathcal{E}}_T$ au complet :

$$\bar{\mathcal{E}}_T = (E_i + E_r) \cos \omega t \cos \beta_1 x + (E_i - E_r) \sin \omega t \sin \beta_1 x$$

Si on considère seulement l'enveloppe de la courbe obtenue pour plusieurs valeurs de x et de t (Figure II.25) :

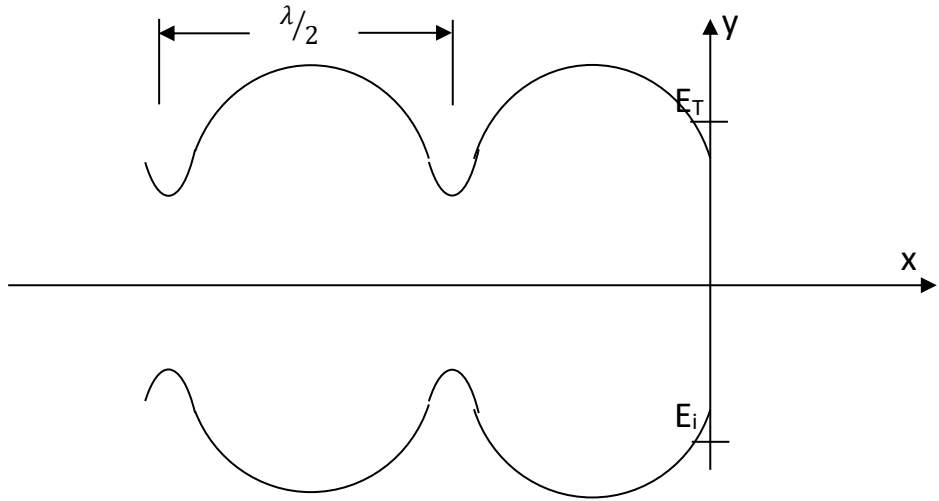

Figure II.25.

Facile à mesurer

$$T.O.S.\,(\text{Taux d'Ondes Stationnaires}),\ \text{ou}$$

$$V.S.W.R.\,(\text{Voltage Standing Wave Ratio}) = \frac{|E_{max}|}{|E_{min}|}$$

Est-il possible de déterminer les propriétés du milieu ② à partir des mesures dans le milieu ① ?

a) T.O.S.
b) Position du maxima ou du minima par rapport à la surface de séparation.

### Exemple II.22.

Trouver le courant sur la surface conductrice infiniment longue qui est frappée par une onde électromagnétique dont l'amplitude du champ électrique est $E_i = E_0 = 10\ \left(\frac{V}{m}\right)$,(Figure II.24).

Solution :

$$\bar{E}_i = \hat{y} E_i e^{-j\beta_1 x} e^{j\omega t}$$

$$\bar{H}_i = \hat{z} \frac{E_i}{Z_i} e^{-j\beta_1 x} e^{j\omega t}$$

$$\bar{E}_r = \hat{y} E_r e^{+j\beta_1 x} e^{j\omega t}$$

$$\bar{H}_r = -\hat{z} \frac{E_r}{Z_1} e^{+j\beta_1 x} e^{j\omega t}$$

Étant donné que le deuxième milieu est un conducteur parfait $\rho = -1$ et $E_r = -E_i$. Le premier milieu est l'espace libre, $Z_i = 377 \ (\Omega)$.

$$\bar{E}_i = \hat{y} E_0 e^{-j\beta_1 x} e^{j\omega t}$$

$$\bar{H}_i = \hat{z} \frac{E_0}{Z_0} e^{-j\beta_1 x} e^{j\omega t}$$

$$\bar{E}_r = -\hat{y} E_0 e^{+j\beta_1 x} e^{j\omega t}$$

$$\bar{H}_r = \hat{z} \frac{E_0}{Z_0} e^{+j\beta_1 x} e^{j\omega t}$$

Il n'a pas de composantes normales sur la surface conductrice ($x$ = 0) et la seule condition est $\bar{E}_i + \bar{E}_r = 0$ soit $\bar{E}_r = -\bar{E}_i$ .

Le champ magnétique n'est pas nul sur la surface et il est tangent à la surface. Il est sur la surface conductrice ($x$ = 0), $\bar{H}_r = \bar{H}_i$.

Donc, sur la surface conductrice,

$$\bar{H}_T(0, t) = \frac{2E_0}{Z_0} e^{j\omega t} \hat{z} = \bar{H}_{tang}(0, t)$$

Le courant surfacique ($x$ =0) est donc,

$$\bar{J}_{s(x=0)} = \hat{n} \times \bar{H}_{tan}(0, t)$$

où $\hat{n} = -\hat{x}$ (normale sur le plan $x$ =0).

$$\bar{J}_{s(x=0)} = -\hat{x} \times \hat{z} \frac{2E_0}{Z_0} e^{j\omega t} = \frac{2E_0}{Z_0} e^{j\omega t} \hat{y} = 53.1x10^{-3} \, e^{j\omega t} \left(\frac{A}{m}\right)$$

$$\left|\bar{J}_{s(x=0)}\right| = 53.1x10^{-3} \, \left(\frac{A}{m}\right)$$

# 9. INCIDENCE OBLIQUE

## 9.1 Cas #1. Champ électrique ⊥ au plan d'incidence (Figure 2. 26)

Champ électrique perpendiculaire au plan d'incidence (parallèle à la surface), (Figure 2.26).

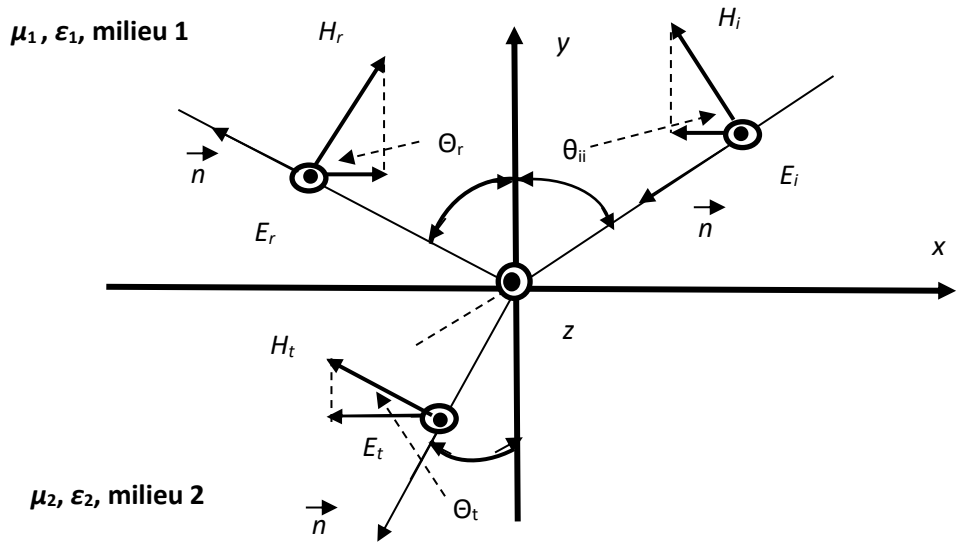

Figure 2.26.

**Expressions pour l'onde incidente, l'onde réfléchie et l'onde transmise.**

D'une façon générale, une onde plane qui se transmet dans une direction arbitraire de l'espace:

$$\bar{E} = E_0 e^{-j\beta[xcos(A)+ycos(B)+zcos(C)]}$$

$$C = \frac{\pi}{2} \quad , \quad \cos(C) = 0 \quad , \quad \hat{n} \, dans \, le \, plan \, x,y$$

Onde incidente

$$\bar{E}_i = \hat{z}E_i e^{-j\beta_1(xcosA+ycosB)}$$

$$A = \frac{\pi}{2} + \theta_i$$

$$B = \pi - \theta_i$$

$$\overline{E}_i = \hat{z} E_i e^{j\beta_1 (x sin\theta_i + y cos\theta_i)}$$

$$\overline{H}_i = (-\hat{x} cos\theta_i + \hat{y} sin\theta_i) H_i e^{j\beta_1 (x sin\theta_i + y cos\theta_i)}$$

**Onde réfléchie**

$$\overline{E}_r = \hat{z} E_r e^{j\beta_1 (x sin\theta_r - y cos\theta_r)}$$

$$A = \frac{\pi}{2} + \theta_r$$

$$B = \theta_r$$

$$\overline{H}_r = (\hat{x} cos\theta_r + \hat{y} sin\theta_r) H_r e^{j\beta_1 (x sin\theta_r - y cos\theta_r)}$$

**Onde transmise**

$$\overline{E}_t = \hat{z} E_t e^{j\beta_2 (x sin\theta_t + y cos\theta_t)}$$

$$A = \frac{\pi}{2} + \theta_t$$

$$B = \pi - \theta_t$$

$$\overline{H}_t = (-\hat{x} cos\theta_t + \hat{y} sin\theta_t) H_t e^{j\beta_2 (x sin\theta_t + y cos\theta_t)}$$

**\* Conditions frontières**

Les composantes tangentielles du champ électrique et du champ magnétique doivent être continues à la surface de séparation (*y*=0)

$$E_i e^{j\beta_1 x sin\theta_i} + E_r e^{j\beta_1 x sin\theta_r} = E_t e^{j\beta_2 x sin\theta_t}$$

Continuité de phase

$$j\beta_1 x sin\theta_i = j\beta_1 x sin\theta_r$$

$$\boxed{\theta_i = \theta_r}$$

$$j\beta_1 x sin\theta_i = j\beta_2 x sin\theta_t$$

$$\boxed{\frac{sin\theta_i}{sin\theta_t} = \frac{\beta_2}{\beta_1} \qquad \textit{Loi de Snell}}$$

Si les milieux de propagation ont des pertes,

$$\frac{sin\theta_i}{sin\theta_t} = \frac{\delta_2}{\delta_1}$$

Pour deux parfaits diélectriques:

$$\frac{sin\theta_i}{sin\theta_t} = \frac{\omega\sqrt{\mu_0\varepsilon_2}}{\omega\sqrt{\mu_0\varepsilon_1}} = \sqrt{\frac{\varepsilon_2}{\varepsilon_1}}$$

Si l'on fait ensuite $x = 0$ et $y = 0$ et que l'on considère seulement l'amplitude des champs:

$$E_i + E_r = E_t$$

$$H_i(-\hat{x}cos\theta_i) + H_r(\hat{x}cos\theta_i) = H_t(-\hat{x}cos\theta_t)$$

$$-\frac{E_i}{Z_1}cos\theta_i + \frac{E_r}{Z_1}cos\theta_i = -\frac{E_t}{Z_2}cos\theta_t$$

On calcule :

$$\frac{E_r}{E_i} = \rho_\perp \quad et \quad \frac{E_t}{E_i} = T_\perp$$

$$-\frac{E_i}{Z_1}cos\theta_i + \frac{E_r}{Z_1}cos\theta_i = -\frac{(E_r + E_i)}{Z_2}cos\theta_t$$

$$E_i\left(\frac{-cos\theta_i}{Z_1} + \frac{cos\theta_t}{Z_2}\right) = E_r\left(\frac{-cos\theta_t}{Z_2} - \frac{cos\theta_i}{Z_1}\right)$$

$$\frac{E_r}{E_i} = \frac{\left(\frac{-cos\theta_i}{Z_1} + \frac{cos\theta_t}{Z_2}\right)}{\left(\frac{-cos\theta_t}{Z_2} - \frac{cos\theta_i}{Z_1}\right)}$$

$$\boxed{\frac{E_r}{E_i} = \rho_\perp = \frac{Z_2cos\theta_i - Z_1cos\theta_t}{Z_1cos\theta_t + Z_2cos\theta_i}}$$

Expressions valables seulement si le champ électrique est $\perp$ au plan d'incidence.

De la même façon,

$$T_\perp = \frac{E_t}{E_i} = \frac{2Z_2 cos\theta_i}{Z_1 cos\theta_t + Z_2 cos\theta_i}$$

Remarque:

$$1 + \rho_\perp = T_\perp$$

## 9.2 Cas intéressant (réflexion totale)

Si l'on a une surface de séparation entre deux parfaits diélectriques et que le milieu ① est plus dense que le milieu ②

$$( \varepsilon_1 > \varepsilon_2 ),$$

Il peut arriver que pour certaines valeurs de $\theta_i$, $\rho = 1$ et que toute l'énergie soit réfléchie. Quelles sont ces valeurs de $\theta_i$ ?

$$\rho_\perp = \frac{\sqrt{\frac{\mu_0}{\varepsilon_2}} cos\theta_i - \sqrt{\frac{\mu_0}{\varepsilon_1}} cos\theta_t}{\sqrt{\frac{\mu_0}{\varepsilon_1}} cos\theta_t + \sqrt{\frac{\mu_0}{\varepsilon_2}} cos\theta_i}$$

$$\rho_\perp = \frac{\frac{cos\theta_i}{\sqrt{\varepsilon_2}} - \frac{cos\theta_t}{\sqrt{\varepsilon_1}}}{\frac{cos\theta_t}{\sqrt{\varepsilon_1}} + \frac{cos\theta_i}{\sqrt{\varepsilon_2}}}$$

En utilisant la loi de Snell

$$\frac{sin\theta_i}{sin\theta_t} = \sqrt{\frac{\varepsilon_2}{\varepsilon_1}} \quad \rightarrow \quad cos\theta_t = \sqrt{1 - \frac{\varepsilon_1}{\varepsilon_2}\sin^2\theta_i}$$

$$\rho_\perp = \frac{\frac{cos\theta_i}{\sqrt{\varepsilon_2}} - \frac{\sqrt{1 - \frac{\varepsilon_1}{\varepsilon_2}\sin^2\theta_i}}{\sqrt{\varepsilon_1}}}{\frac{\sqrt{1 - \frac{\varepsilon_1}{\varepsilon_2}\sin^2\theta_i}}{\sqrt{\varepsilon_1}} + \frac{cos\theta_i}{\sqrt{\varepsilon_2}}}$$

$$\rho_\perp = \frac{\sqrt{\varepsilon_1}\cos\theta_i - \sqrt{\varepsilon_2}\sqrt{1 - \frac{\varepsilon_1}{\varepsilon_2}\sin^2\theta_i}}{\sqrt{\varepsilon_1}\cos\theta_i + \sqrt{\varepsilon_2}\sqrt{1 - \frac{\varepsilon_1}{\varepsilon_2}\sin^2\theta_i}}$$

$$\rho_\perp = \frac{\cos\theta_i - \sqrt{\frac{\varepsilon_2}{\varepsilon_1} - \sin^2\theta_i}}{\cos\theta_i + \sqrt{\frac{\varepsilon_2}{\varepsilon_1} - \sin^2\theta_i}}$$

---

Lorsque

$$\sin^2\theta_i = \frac{\varepsilon_2}{\varepsilon_1} \rightarrow \rho_\perp = 1$$

Réflexion totale

$$\theta_{i_C} = \sin^{-1}\sqrt{\frac{\varepsilon_2}{\varepsilon_1}} \; - \; \text{angle critique}$$

---

Si

$$\theta_i > \theta_{i_C}$$

Est-ce que l'on obtient aussi une réflexion totale?

$$\cos\theta_t = \sqrt{1 - \frac{\varepsilon_1}{\varepsilon_2}\sin^2\theta_i}$$

Si

$$\theta_i = \theta_{i_C} \; \rightarrow \; \cos\theta_t = 0 \; \rightarrow \; \theta_t = \frac{\pi}{2}$$

Si

$$\theta_i > \theta_{i_C} \; \rightarrow \; \cos\theta_t = \text{quantité imaginaire}$$

(Onde transmise n'existe pas)

Que devient l'onde transmise si la réflexion est totale?

$$T_\perp = \frac{2Z_2 cos\theta_i}{Z_2 cos\theta_i} = 2 \quad \rightarrow \quad E_t = 2E_i$$

Est-ce que la condition frontière est satisfaite?

$$① \qquad ②$$

$$E_i + E_r = E_t$$

$$2E_i = 2E_i$$

Valeur moyenne de la puissance transportée dans le milieu 2 lorsque $\theta_i = \theta_{i_C}$,

$$<P> = \frac{1}{2} Re \, (\bar{E}_t \times \bar{H}_t^{\,*})$$

Ce produit est nul partout sauf si on se trouve infiniment près de la surface de séparation, d'où la puissance transportée dans le milieu 2 est nulle. Ce phénomène se répète même si $E_i$ est ∥ au plan d'incidence. Il est connu sous le nom d'onde de surface (dans le milieu 1). Ceci permet ainsi d'expliquer en partie la possibilité de transmettre une onde lumineuse dans une fibre optique sur de grandes distances sans trop de pertes.

**Exemple II.23.**

Soit une interface air-silicium. Le silicium est assumé sans perte avec une constante diélectrique relative de 11.7. Trouver l'angle auquel la réflexion sera totale si une onde plane à la polarisation parallèle arrive à incidence,

  a) de l'air vers le silicium,
  b) du silicium vers l'air

**Solution :**

  a)  Air-Silicium : $\varepsilon_1 = \varepsilon_0$ , $\varepsilon_2 = 11.7\varepsilon_0$

$$\theta_{i_C} = \sin^{-1} \sqrt{\frac{\varepsilon_2}{\varepsilon_1}} = arcsin\sqrt{11.7}$$

C'est impossible parce que $-1 \leq \sin(x) \leq 1$.

  b)  Silicium − Air : $\varepsilon_1 = 11.7\varepsilon_0$ , $\varepsilon_2 = \varepsilon_0$

$$\theta_{i_C} = \sin^{-1}\sqrt{\frac{\varepsilon_2}{\varepsilon_1}} = arcsin\sqrt{\frac{1}{11.7}} = 17^0$$

### 9.3 Cas #2   Champ électrique ∥ au plan d'incidence (Figure 2.27)

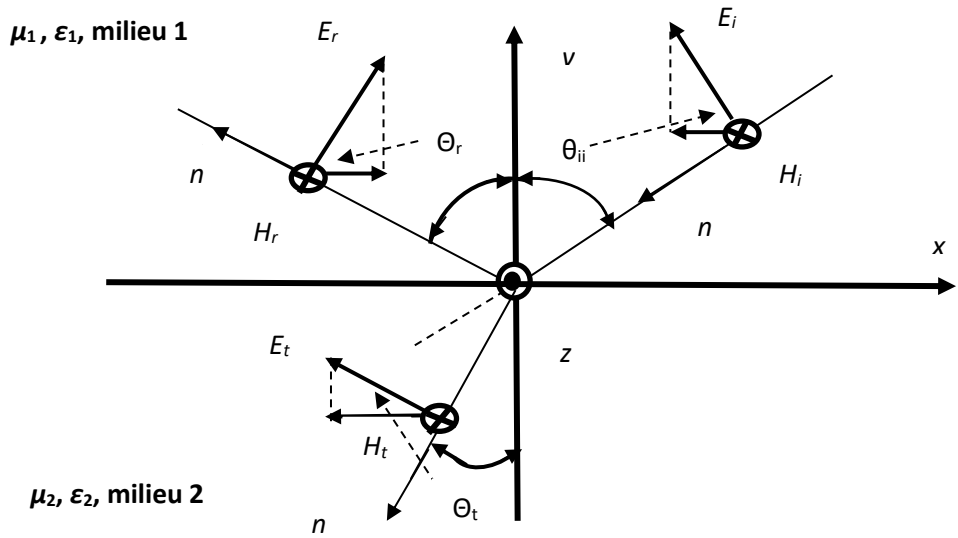

Figures 2.27.

<u>Onde incidente</u>

$$\overline{H}_i = -\hat{z}\frac{E_i}{Z_1}e^{j\beta_1(xsin\theta_i + ycos\theta_i)}$$

$$A = \frac{\pi}{2} + \theta_i$$

$$B = \pi - \theta_i$$

$$\overline{E}_i = (-\hat{x}cos\theta_i + \hat{y}sin\theta_i)E_i e^{j\beta_1(xsin\theta_i + ycos\theta_i)}$$

<u>Onde réfléchie</u>

$$\overline{H}_r = +\hat{y}\frac{E_r}{Z_1}e^{j\beta_1(xsin\theta_i - ycos\theta_i)}$$

$$A = \frac{\pi}{2} + \theta_i$$

$$B = \theta_i$$

$$\bar{E}_r = (-\hat{x}cos\theta_i - \hat{y}sin\theta_i)E_r e^{j\beta_1(xsin\theta_i - ycos\theta_i)}$$

**Onde transmise**

$$\bar{H}_t = -\hat{z}\frac{E_t}{Z_1}e^{j\beta_2(xsin\theta_t + ycos\theta_t)}$$

$$A = \frac{\pi}{2} + \theta_t$$

$$B = \pi - \theta_t$$

$$\bar{E}_t = (-\hat{x}cos\theta_t + \hat{y}sin\theta_t)E_t e^{j\beta_2(xsin\theta_t + ycos\theta_t)}$$

$$\theta_i = \theta_r \quad , \quad Loi\ de\ Snell\ toujours\ valable$$

On cherche,

$$\rho_\parallel = \frac{E_r}{E_i} \quad \text{et} \quad T_\parallel = \frac{E_t}{E_i}$$

Lorsque $x$ = 0 et $y$ = 0 à la surface de séparation

$$-\frac{E_i}{Z_1} + \frac{E_r}{Z_1} = -\frac{E_t}{Z_2}$$

$$-E_i cos\theta_i - E_r cos\theta_i = -E_t cos\theta_t$$

$$\frac{(-E_i + E_r)}{Z_1}Z_2 = -E_t$$

$$-E_i cos\theta_i - E_r cos\theta_i = \frac{(-E_i + E_r)}{Z_1}Z_2 cos\theta_t$$

$$E_r(-cos\theta_i - \frac{Z_2}{Z_1}cos\theta_t) = E_i(-\frac{Z_2}{Z_1}cos\theta_t + cos\theta_i)$$

$$E_r(-Z_1 cos\theta_i - Z_2 cos\theta_t) = E_i(-Z_2 cos\theta_t + Z_1 cos\theta_i)$$

$$\boxed{\frac{E_r}{E_i} = \rho_\parallel = \frac{Z_2 cos\theta_t - Z_1 cos\theta_i}{Z_1 cos\theta_i + Z_2 cos\theta_t}}$$

$$\frac{E_r}{E_i} = T_\parallel = \frac{2Z_2 cos\theta_i}{Z_1 cos\theta_i + Z_2 cos\theta_t}$$

Ces deux expressions sont valables pour le choix des orientations de $\bar{E}$ et de $\bar{H}$ posée au départ.

## 9.4 Angle de Brewster

Valeur de $\theta_i$ qui rend $\rho_\parallel$ = 0

(Réflexion nulle, transmission totale)

Pour obtenir $\rho_\parallel$ = 0

$$Z_1 cos\theta_i = Z_2 cos\theta_t$$

$$Z_1{}^2 cos^2\theta_i = Z_2{}^2 cos^2\theta_t$$

$$Z_1{}^2(1 - sin^2\theta_i) = Z_2{}^2(1 - sin^2\theta_t)$$

Pour une surface de séparation entre 2 parfaits diélectriques

$$\frac{\sin\theta_i}{\sin\theta_t} = \sqrt{\frac{\varepsilon_2}{\varepsilon_1}}$$

$$\sin\theta_t = \sqrt{\frac{\varepsilon_1}{\varepsilon_2}}\sin\theta_i$$

$$\sin^2\theta_t = \frac{\varepsilon_1}{\varepsilon_2}\sin^2\theta_i$$

$$\frac{\mu_0}{\varepsilon_1}(1 - sin^2\theta_i) = \frac{\mu_0}{\varepsilon_2}\left(1 - \frac{\varepsilon_1}{\varepsilon_2}sin^2\theta_i\right)$$

$$sin^2\theta_i\left(-\frac{1}{\varepsilon_1} + \frac{\varepsilon_1}{\varepsilon_2{}^2}\right) = \frac{1}{\varepsilon_2} - \frac{1}{\varepsilon_1}$$

$$sin^2\theta_i\left(\frac{-\varepsilon_2^2+\varepsilon_1^2}{\varepsilon_1\varepsilon_2^2}\right)=\frac{\varepsilon_1-\varepsilon_2}{\varepsilon_1\varepsilon_2}$$

$$\sin^2\theta_i=\frac{(\varepsilon_1-\varepsilon_2)\varepsilon_2}{\varepsilon_1^2-\varepsilon_2^2}$$

$$\sin^2\theta_{i_B}=\frac{\varepsilon_2}{\varepsilon_1+\varepsilon_2}$$

$$\frac{1}{\sin^2\theta_{i_B}}=1+\frac{\varepsilon_1}{\varepsilon_2}$$

$$\frac{\cos^2\theta_{i_B}}{\sin^2\theta_{i_B}}=\frac{\varepsilon_1}{\varepsilon_2}$$

$$\tan\theta_{i_B}=\sqrt{\frac{\varepsilon_2}{\varepsilon_1}}$$

$$\theta_{i_B}=\text{angle de Brewster}$$

À discuter :

a) Onde de polarisation circulaire ou elliptique incidentes sur une surface de séparation

b) Milieu ② ayant une certaine conductivité

c) Milieux non réflecteurs (absorbants)

d) Dômes pour antennes

e) Avions invisibles aux radars

f) Transmission à travers une plaque mince

g) Transmission à travers plusieurs plaques minces

## Exemple II.24.

Soit une interface air-silicium. Le silicium est assumé sans perte avec une constante diélectrique relative de 11.7. Trouver l'angle auquel la transmission sera totale si une onde plane à la polarisation parallèle arrive à incidence,

    a) de l'air vers le silicium,
    b) du silicium vers l'air

**Solution :**

    a) Air-Silicium : $\varepsilon_1 = \varepsilon_0$ , $\varepsilon_2 = 11.7\varepsilon_0$

$$\tan \theta_{i_{BII}} = \sqrt{\frac{\varepsilon_2}{\varepsilon_1}}$$

$$\theta_{i_{BII}} = \text{angle de Brewster}$$

$$\tan \theta_{i_{BII}} = \sqrt{\frac{\varepsilon_2}{\varepsilon_1}} \rightarrow \theta_{i_{BII}} = atan\sqrt{\frac{\varepsilon_2}{\varepsilon_1}} = atan\sqrt{11.7} \rightarrow \theta_{i_{BII}} = 73.7^0$$

    b) Silicium − Air : $\varepsilon_1 = 11.7\varepsilon_0$ , $\varepsilon_2 = \varepsilon_0$

$$\tan \theta_{i_{BII}} = \sqrt{\frac{\varepsilon_2}{\varepsilon_1}} \rightarrow \theta_{i_{BII}} = atan\sqrt{\frac{\varepsilon_2}{\varepsilon_1}} = atan\sqrt{\frac{1}{11.7}} \rightarrow \theta_{i_{BII}} = 16.3^0$$

On voit que ces angles sont complémentaires.

**Exemple II.25.** Une onde électromagnétique plane dont le vecteur-phaseur électrique est

$$\bar{E} = (-j4\hat{x} + j3\hat{y} + 5\hat{z})e^{j(12x+16y)} \left(\frac{V}{m}\right)$$

se propage dans un milieu diélectrique ($\varepsilon = 2\varepsilon_0$) qui se trouve dans le demi-espace ($y \geq 0$) du système cartésien ($x$, $y$, $z$). L'autre milieu de l'espace ($y < 0$) est de l'air. On considère $O$, l'origine des coordonnées comme point d'impact et l'axe $Oy$ comme normale au point d'incidence.

Dans le diélectrique ($\varepsilon = 2\varepsilon_0$) :

a) Quelle est la direction (le vecteur $\hat{n}$) de la propagation?
b) Quelle est la fréquence du signal?
c) Quel est l'angle d'incidence?
d) Quelle est la polarisation du signal?
e) Calculer le vecteur-phaseur du champ magnétique,
f) Calculer la puissance moyenne transportée par unité de surface dont la normale est dans le sens de la propagation.

Dans l'air ($\varepsilon = \varepsilon_0$) :

g) Quel est l'angle de transmission?
h) Quelle est l'expression du vecteur-phaseur champ électrique?
i) Quelle est la polarisation d'onde?
j) Calculer la puissance moyenne transportée par unité de surface dont la normale est dans le sens de la propagation.

**Solution :**

a) Pour trouver le vecteur $\hat{n}$ on utilise la forme canonique concernant le champ électrique :

$$\bar{E} = (-j4\hat{x} + j3\hat{y} + 5\hat{z})e^{-j\beta(n_x x + n_y y + n_z z)} \left(\frac{V}{m}\right)$$

On a,

$$-\beta n_x = 12, \quad -\beta n_y = 16, \quad -\beta n_z = 0$$

avec

$$n_x^2 + n_y^2 + n_z^2 = 1$$

Ce système d'équations donne,

$$\beta^2(n_x^2 + n_y^2) = 400 \rightarrow \beta = 20 \left(\frac{rad}{m}\right) = \beta_1$$

Finalement,

$$n_x = -0.6, \ n_y = -0.8, \ n_z = 0$$

ou

$$\hat{n} = -0.6\,\hat{x} - 0.8\,\hat{y}$$

b)

$$\beta = \frac{2\pi}{\lambda} = \frac{2\pi f}{v_p}, v_p = \frac{c}{\sqrt{2}} \rightarrow f = \frac{c\beta}{2\pi\sqrt{2}}$$

$$f = \frac{c\beta}{2\pi\sqrt{2}} = \frac{2x10^8 x 20}{2\pi\sqrt{2}} = 675.2 \ (MH_z)$$

c) Dans l'expression de la phase pour l'onde incidente oblique on a

$e^{j\beta_1(xsin\theta_i + ycos\theta_i)}$.

C'est aussi bien pour les cas || et $\perp$.

Donc,

$$sin\theta_i = 0.6, \ \ cos\theta_i = 0.8 \ \rightarrow \theta_i = \ 36.87^0$$

d) $\bar{E}_1 = -j4\hat{x} + j3\hat{y}, \ \bar{E}_2 = 5\hat{z}, \rightarrow \bar{E}_1 \cdot \bar{E}_2 = 0 \rightarrow \ \bar{E}_1 \perp \bar{E}_2$

$$|\bar{E}_1| = \sqrt{4^2 + 3^2} = 5, \qquad |\bar{E}_2| = 5, \rightarrow \ |\bar{E}_1| = |\bar{E}_2|$$

$$\angle\bar{E}_1 = e^{j\frac{3\pi}{2}}, \qquad \angle\bar{E}_2 = 0, \quad \bar{E}_1 \neq \bar{E}_2 e^{j\frac{3\pi}{2}}$$

Donc, c'est la polarisation elliptique.

e)

$$\bar{H} = \hat{n} \times \frac{\bar{E}}{Z}, \ \ Z = \frac{1}{\sqrt{2}} Z_0 = \frac{377}{\sqrt{2}} \rightarrow Z = 266.6 \ (\Omega)$$

$$\bar{H} = \frac{1}{266.6} \begin{vmatrix} \hat{x} & \hat{y} & \hat{z} \\ -0.6 & -0.8 & 0 \\ -j4 & j3 & 0 \end{vmatrix} e^{j(12x+16y)}$$

$$\bar{H} = \frac{1}{266.6}[-4\hat{x} + 3\hat{y} - j5\hat{z}]e^{j(12x+16y)} \ \ \left(\frac{A}{m}\right)$$

f)

$$< P(\varepsilon = 2\varepsilon_0) > = \frac{1}{2} \frac{E_1^{\ 2}}{Z} + \frac{1}{2} \frac{E_2^{\ 2}}{Z}$$

$$< P(\varepsilon = 2\varepsilon_0) > = \frac{1}{2x266.6}[25 + 25] = 93.9x10^{-3}\left(\frac{W}{m^2}\right)$$

g)

$$\frac{sin\theta_i}{sin\theta_i} = \frac{\beta_2}{\beta_1} = \frac{\omega\sqrt{\mu_0\varepsilon_0}}{\omega\sqrt{\mu_0\varepsilon_0}\sqrt{\varepsilon_r}} = \frac{1}{\sqrt{\varepsilon_r}}$$

$$sin\theta_t = \sqrt{\varepsilon_r}sin\theta_i = \sqrt{2}x0.6 = 0.8485$$

$$\theta_t = 58.052^0$$

h)

$$Z_1 = \frac{120\pi}{\sqrt{2}}, \quad Z_1 = 120\pi$$

$$T_\perp = \frac{2Z_2cos\theta_i}{Z_1cos\theta_t + Z_2cos\theta_i} = \frac{2}{\frac{Z_1cos\theta_t}{Z_2cos\theta_i} + 1}$$

$$T_\parallel = \frac{2Z_2cos\theta_i}{Z_1cos\theta_i + Z_2cos\theta_t} = \frac{2}{\frac{Z_1}{Z_2} + \frac{cos\theta_t}{cos\theta_i}}$$

$$\frac{Z_1}{Z_2} = \frac{1}{\sqrt{2}}, \quad cos\theta_t = 0.529, \quad cos\theta_i = 0.6$$

$$T_\perp = 1.363, \quad T_\parallel = 1.462$$

$$\frac{\beta_2}{\beta_1} = \frac{1}{\sqrt{2}} \rightarrow \beta_2 = \frac{\beta_1}{\sqrt{2}} = 14.14\left(\frac{rad}{m}\right)$$

$$sin\theta_t = 0.8485, \quad cos\theta_t = 0.5292$$

$$\bar{E}_{t||} = T_\parallel \left(-cos\theta_t\hat{x} + sin\theta_t\hat{y}\right)\bar{E}_{i||}e^{j\beta_2(xsin\theta_t + ycos\theta_t)}$$

$$\bar{E}_{t||} = 7.31j \left(-0.5292\hat{x} + 0.8485\hat{y}\right)e^{j14.14(0.85x + 0.53y)}\left(\frac{V}{m}\right)$$

$$\bar{E}_{t\perp} = T_\perp \hat{z}\bar{E}_{i\perp}e^{j\beta_2(xsin\theta_t + ycos\theta_t)}$$

$$\bar{E}_{t\perp} = 6.815 \, \hat{z} \, e^{j14.14(0.85x+0.53y)}(\frac{V}{m})$$

$$\bar{E}_{Total} = \bar{E}_{t||} + \bar{E}_{t\perp}$$

$$\bar{E}_T = (-0.5292\hat{x} + 0.8485\hat{y} + 6.815\,\hat{z})e^{j14.14(0.85x+0.53y)}(\frac{V}{m})$$

i)

$$|\bar{E}_{t\perp}| = 6.815, \qquad |\bar{E}_{t||}| = \sqrt{3.876^2 + 6.203^2} = 7.309$$

La différence en angle est $\frac{\pi}{2}$ , mais les modules sont différents. Donc, c'est une polarisation elliptique à droite.

j)

$$< P(\varepsilon_0) > = \frac{1}{2}\frac{|E_{t||}|^2}{Z_0} + \frac{1}{2}\frac{|E_{t\perp}|^2}{Z_0}$$

$$< P(\varepsilon = \varepsilon_0) > = \frac{1}{2}\frac{6.815^2}{120\pi} + \frac{1}{2}\frac{7.309^2}{120\pi} = 132.5x10^{-3}(\frac{W}{m^2})$$

**Exemple II.26.**

Une onde plane de fréquence 500 ($MH_z$) frappe un plan de séparation ($xOz$) entre deux diélectriques sans pertes comme indiqué à la Figure 26. On a $\varepsilon'_2 = (\frac{3}{4})\varepsilon'_1$ et $\mu'_{1r} = \mu'_{2r} = 1$. Le premier diélectrique est de l'air. Le champ magnétique transmis dans le deuxième milieu et le champ magnétique réfléchi dans le premier milieu est obtenu comme suit :

$$\bar{H}_t = \left[-\frac{\sqrt{3}}{2}\hat{x} + \frac{1}{2}\hat{y} - \hat{z}\right]H_2 e^{j\frac{2\pi}{\lambda_2}(\frac{x}{2}+y\frac{\sqrt{3}}{2})}(\frac{A}{m})$$

et le champ magnétique réfléchi dans le premier milieu est obtenu comme suit :

$$\bar{H}_r = \left[\frac{1}{2}\hat{x} + \frac{\sqrt{3}}{2}\hat{y}\right] H_1 e^{j\frac{2\pi}{\lambda_1}(\sqrt{3}\frac{x}{2} - y\frac{1}{2})} \left(\frac{A}{m}\right)$$

a) Calculer les longueurs d'onde $\lambda_1$ et $\lambda_2$ dans les deux milieux,

b) Trouver $\theta_t$ et $\theta_i$. Que peut-on dire de $\theta_i$ ?

c) Si $H_1 = 10\left(\frac{A}{m}\right)$ et $H_2 = 30\left(\frac{A}{m}\right)$ calculer sous forme phaseur $\bar{E}_t$ et $\bar{E}_r$,

d) Calculer $E_i$ et donner la polarisation de l'onde électromagnétique.

**Solution :**

a)

$$\gamma = j\omega\sqrt{\mu_r' - j\mu_r''}\sqrt{\varepsilon_r' - j\left(\varepsilon_r'' + \frac{\sigma}{\varepsilon_0\omega}\right)}$$

$$\gamma = j\omega\sqrt{\varepsilon_0\mu_0}\sqrt{\varepsilon_r'\mu_r'}$$

parce que,

$$\mu'_{1r} = \mu'_{2r} = 1, \ \mu''_{1r} = \mu''_{2r} = 0,$$

$$\varepsilon'_{2r} = 3\,\varepsilon'_{1r}\ , \ \varepsilon''_{1r} = \varepsilon''_{2r} = 0, \sigma = 0$$

$$\gamma_1 = j\frac{\omega}{c}\sqrt{\varepsilon'_{1r}\mu'_{1r}} = j\frac{2\pi x 5x10^8}{3x10^8}\sqrt{1x1} = j10.472\left(\frac{1}{m}\right) = j\beta_1$$

$$\beta_1 = 10.472\left(\frac{rad}{m}\right), \quad \lambda_1 = \frac{2\pi}{\beta_1} = 60\,(cm)$$

$$\beta_2 = \sqrt{3}\beta_1$$

$$\beta_2 = 18.14\left(\frac{rad}{m}\right), \ \lambda_2 = \frac{2\pi}{\beta_2} = 34.64\,(cm)$$

b) Que ce soit pour $\bar{H}_{r\perp}$ ou $\bar{H}_{r\|}$ on a comme la phase,

$$e^{j\beta(x\sin\theta_i - y\cos\theta_i)} \text{ avec } \theta_i = \theta_r$$

Donc,

$$sin\theta_i = sin\theta_r = \frac{\sqrt{3}}{2}$$

$$cos\theta_i = cos\theta_r = \frac{1}{2}$$

ce qui donne,

$$\theta_i = \theta_r = 60^0$$

Pour l'onde transmise, la phase de $\bar{H}_{t\perp}$ ou $\bar{H}_{t||}$ est,

$$e^{j\beta(x\,sin\theta_t + y cos\theta_t)}$$

Donc,

$$sin\theta_t = \frac{1}{2}$$

$$cos\theta_t = \frac{\sqrt{3}}{2}$$

ce qui donne,

$$\theta_t = 30^0$$

On voit que ces deux angles sont complémentaires ($\theta_i + \theta_t = 90^0$)

$$sin\theta_i = \frac{\sqrt{3}}{2} \rightarrow (sin\theta_i)^2 = \frac{3}{4} = \frac{\varepsilon_2}{\varepsilon_2 + \varepsilon_1}$$

L'angle $\theta_i$ est l'angle de Brewster.

c) $Z_1 = Z_0 = 120\pi\ (\Omega),\ Z_2 = \frac{Z_0}{\sqrt{\varepsilon'_{2r}}} = \frac{120\pi}{\sqrt{3}}\ (\Omega)$

$$E_r = Z_1 H_r = Z_1 H_1 = 1200\pi = 3769.911\ (\frac{V}{m})$$

$$E_t = Z_2 H_t = Z_2 H_2 = 1200\pi\sqrt{3} = 6529.678\ (\frac{V}{m})$$

$$\bar{E}_t = Z_2 H_2(-cos\theta_t \hat{x} + sin\theta_t \hat{y} + \hat{z})e^{j\beta_2(sin\theta_t x + cos\theta_t y)}\ (V/m)$$

$$\bar{E}_t = 1200\pi\sqrt{3}\left(-\frac{\sqrt{3}}{2}\hat{x} + \frac{1}{2}\hat{y} + \hat{z}\right)e^{j\frac{10\pi}{\sqrt{3}}\left(\frac{1}{2}x + \frac{\sqrt{3}}{2}y\right)} \ (V/m)$$

$$\bar{E}_r = Z_1 H_1(-cos\theta_i\hat{x} - sin\theta_i\hat{y} + \hat{z})e^{j\beta_2(sin\theta_i x - cos\theta_i y)} \ (V/m)$$

$$\bar{E}_r = 1200\pi(-\frac{1}{2}\hat{x} - \frac{\sqrt{3}}{2}\hat{y} + \hat{z})e^{j\frac{10\pi}{3}(\frac{\sqrt{3}}{2}x - \frac{1}{2}y)} \ (V/m)$$

d)

$$\bar{E}_i = E_i(-cos\theta_i\hat{x} + sin\theta_i\hat{y} + \hat{z})e^{j\beta_1(sin\theta_i x + cos\theta_i y)} \ (V/m)$$

où

$$E_i = \frac{E_r}{\rho_{||}} + \frac{E_r}{\rho_\perp}$$

$$\rho_\perp = \frac{Z_2 cos\theta_i - Z_1 cos\theta_t}{Z_1 cos\theta_t + Z_2 cos\theta_i} = \frac{\frac{Z_2 cos\theta_i}{Z_1 cos\theta_t} - 1}{\frac{Z_2 cos\theta_i}{Z_1 cos\theta_t} + 1}$$

$$\frac{Z_2}{Z_1} = \frac{1}{\sqrt{3}} \ , \qquad \frac{cos\theta_i}{cos\theta_t} = \frac{1}{\sqrt{3}}$$

$$\rho_\perp = \frac{\frac{1}{3} - 1}{\frac{1}{3} + 1} = -\frac{1}{2}$$

$$\rho_{||} = \frac{Z_2 cos\theta_t - Z_1 cos\theta_i}{Z_1 cos\theta_i + Z_2 cos\theta_t} = \frac{\frac{Z_2 cos\theta_t}{Z_1 cos\theta_i} - 1}{\frac{Z_2 cos\theta_t}{Z_1 cos\theta_i} + 1} = \frac{1 - 1}{1 + 1} = 0$$

Ce coefficient $\rho_{||} = 0$ à cause de l'angle de Brewster.

Donc,

$$\bar{E}_{i\perp} = -2\bar{E}_{r\perp} = -2\bar{E}_r$$

$$\bar{E}_{i\perp} = -2400\pi(-\frac{1}{2}\hat{x} - \frac{\sqrt{3}}{2}\hat{y} + \hat{z})e^{j\,5\pi(\frac{1}{\sqrt{3}}x - \frac{1}{3}y)}\ (V/m)$$

Pour $\bar{E}_{i||}$ on doit faire appel à $\bar{E}_{t||}$. La partie || de $\bar{E}_{t||}$ est,

$$\bar{E}_{t||} = 1200\pi(-\frac{3}{2}\hat{x} + \frac{\sqrt{3}}{2}\hat{y})e^{j\,5\pi(\frac{1}{\sqrt{3}}x + y)}\ (V/m)$$

Pour ce cas,

$$T_{||} = \frac{2Z_2 cos\theta_i}{Z_1 cos\theta_i + Z_2 cos\theta_t} = \frac{2\frac{120\pi}{\sqrt{3}}\frac{1}{2}}{\frac{120\pi}{2} + \frac{120\pi}{\sqrt{3}}\frac{\sqrt{3}}{2}} = \frac{1}{\sqrt{3}}$$

$$\bar{E}_{i||} = \frac{1}{T_{||}}\bar{E}_{t||} = 1200\pi\sqrt{3}(-\frac{3}{2}\hat{x} + \frac{\sqrt{3}}{2}\hat{y})e^{j\,5\pi(\frac{1}{\sqrt{3}}x + y)}\ (V/m)$$

Donc, pour le champ total de $\bar{E}_i$ on a,

$$\bar{E}_i = 1200\pi\sqrt{3}(-\frac{3}{2}\hat{x} + \frac{\sqrt{3}}{2}\hat{y} - 2\hat{z})e^{j\,5\pi(\frac{1}{\sqrt{3}}x + y)}\ (V/m)$$

Maintenant, on va faire une autre vérification. On peut remonter à $\bar{E}_{i\perp}$ à partir de la partie transmise. Pour cela on a besoin de $T_\perp$.

$$T_\perp = \frac{2Z_2 cos\theta_i}{Z_1 cos\theta_t + Z_2 cos\theta_i} = \frac{2\frac{120\pi}{\sqrt{3}}\frac{1}{2}}{\frac{120\pi}{\sqrt{3}}\frac{1}{2} + \frac{120\pi}{2}\sqrt{3}} = \frac{1}{2}$$

$$\bar{E}_{i\perp} = \frac{\bar{E}_{t\perp}}{T_\perp} = 2400\pi\hat{z}e^{j\,5\pi(\frac{1}{\sqrt{3}}x - \frac{1}{3}y)}\ (V/m)$$

On se pose la question: Est-ce que $H_2$ et $H_1$ peuvent être choisis indépendamment? Il faut vérifier les conditions aux frontières à $x = 0, y = 0, z = 0$. Pour $\bar{E}_{i\perp}$ et $\bar{E}_{t\perp}$.

$$\bar{E}_{i\perp}(0,0,0) = Z_0 H_1 x(-2)\hat{z} \qquad (1)$$

$$\bar{E}_{t\perp}(0,0,0) = 2\,Z_2 H_2 \hat{z} \qquad\qquad (2)$$

On doit avoir $(1) = (2)$ ce qui donne,

$$Z_0 H_1 x(-2)\hat{z} = 2\,Z_2 H_2 \hat{z}$$

$$-H_1 = \frac{H_2}{\sqrt{3}}$$

Donc on aurait dû choisir,

$$H_2 = 10\sqrt{3}\,\left(\frac{A}{m}\right)$$

On a posé $H_2 = 30\,\left(\frac{A}{m}\right)$ d'où cette différence.

# 10. Transmission à travers une plaque mince

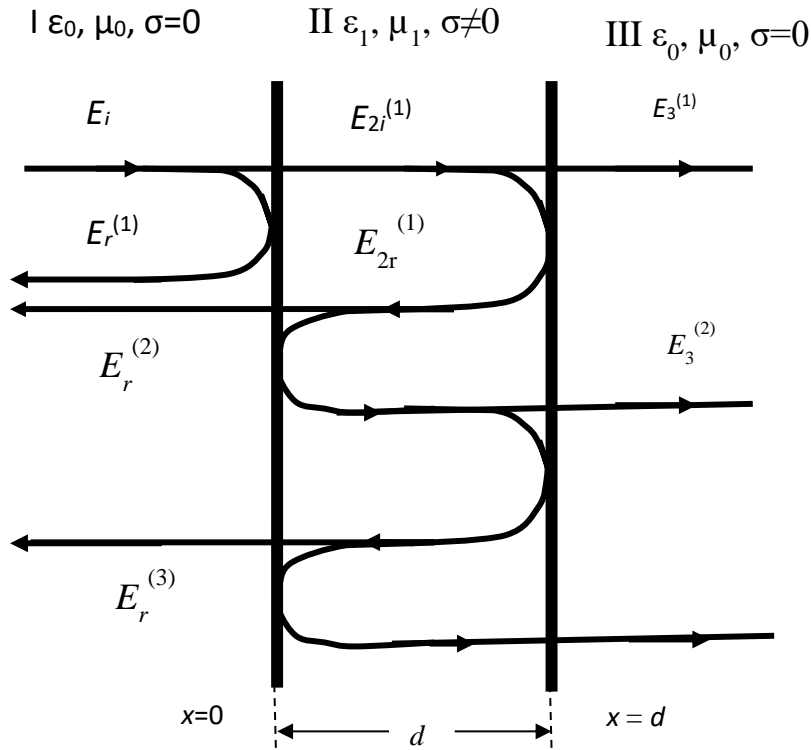

On considère une plaque mince (Figure II.28) de caractéristiques ($\varepsilon_0$, $\mu_0$, $\sigma \neq 0$) se trouvant dans l'espace libre (air). Si une onde électromagnétique frape perpendiculairement la surface entre les régions I et II, il faut analyser les transmissions à travers la plaque.

$$E_r^{(1)} = \rho_{12} E_i = \frac{Z_2 - Z_1}{Z_2 + Z_1} E_i \rightarrow E_r^{(1)} = \rho_{12} E_i$$

$$\rho_{12} = \frac{Z_2 - Z_1}{Z_2 + Z_1}$$

$$E_{2i}^{(1)} = \frac{2Z_2}{Z_2 + Z_1} E_i = T_{12} E_i \ \rightarrow \ E_{2i}^{(1)} = T_{12} E_i$$

$$T_{12} = \frac{2Z_2}{Z_2 + Z_1}$$

Lorsque l'onde $E_{2i}^{(1)}$ arrive à la surface $x = d$ son amplitude tombe à,

$$E_{2i}^{(1)}(x = d) = E_{2i}^{(1)} e^{-\gamma_2 d} E_i = T_{12} e^{-\gamma_2 d} E_i$$

L'onde qui passe au milieu III sera,

$$E_3^{(1)} = E_{2i}^{(1)}(x = d) T_{23}$$

et l'onde réfléchie de la surface II/III sera,

$$E_{2r}^{(1)} = T_{12} \rho_{23} e^{-\gamma_2 d} E_i$$

Une partie de cette onde $E_{2r}^{(1)}$ passe au milieu I et l'autre va être réfléchie à la surface I/II, etc.

Donc,

À $x = 0$,

$$E_r^{(2)}(x = 0) = E_{2r}^{(1)} e^{-\gamma_2 d} T_{21} \ \text{ou}$$

$$E_r^{(2)} = T_{12} T_{21} E_{2r}^{(1)} e^{-\gamma_2 d} E_i \ \text{ou}$$

$$E_r^{(2)} = T_{12} e^{-\gamma_2 d} \rho_{23} e^{-\gamma_2 d} T_{21} E_i$$

Donc, l'onde réfléchie dans le milieu I peut être écrite étant comme la somme,

$$E_r = \rho_{12} E_i + T_{12} e^{-\gamma_2 d} \rho_{23} e^{-\gamma_2 d} T_{21} E_i +$$

$$+ T_{12} e^{-\gamma_2 d} \rho_{23} e^{-\gamma_2 d} \rho_{21} e^{-\gamma_2 d} \rho_{23} e^{-\gamma_2 d} T_{21} E_i + \cdots +$$

$$E_r = \rho_{12} E_i + E_i T_{12} T_{22} \rho_{23} e^{-\gamma_2 d} [1 + \rho_{21} \rho_{23} e^{-2\gamma_2 d}] + (\rho_{21} \rho_{23} e^{-2\gamma_2 d})^2 +$$

$$+(\rho_{21}\rho_{23}e^{-2\gamma_2 d})^3 + \cdots]E_i$$

$$\frac{E_r}{E_i} = \rho_{12}E_i + T_{12}\,T_{21}\rho_{23}e^{-2\gamma_2 d}[1 + \delta + \delta^2 + \delta^3 + \cdots]$$

où

$$\delta = \rho_{21}\rho_{23}e^{-2\gamma_2 d}, \qquad |\delta| < 1$$

$$\frac{E_r}{E_i} = \rho_{12}E_i + E_i T_{12}\,T_{22}\rho_{23}e^{-2\gamma_2 d}\frac{1}{1-\delta}$$

$$\frac{E_r}{E_i} = \rho_{12} + T_{12}\,T_{22}\rho_{23}e^{-2\gamma_2 d}\frac{1}{1-\rho_{21}\rho_{23}e^{-2\gamma_2 d}}$$

$$\rho_{12} = \frac{Z_2 - Z_1}{Z_2 + Z_1}\,, \qquad \rho_{21} = \frac{Z_1 - Z_2}{Z_1 + Z_2} \to \rho_{12} = -\rho_{21}$$

$$T_{12} = \frac{2Z_2}{Z_2 + Z_1}\,, \qquad T_{21} = \frac{2Z_1}{Z_1 + Z_2}\,, \quad \rho_{23} = \frac{Z_3 - Z_2}{Z_3 + Z_2}$$

$$Z_1 = Z_3 \;\to\; \rho_{23} = \frac{Z_1 - Z_2}{Z_1 + Z_2} = -\rho_{12}$$

Donc,

$$\frac{E_r}{E_i} = \rho_{12} - \frac{T_{12}\,T_{22}\rho_{12}e^{-2\gamma_2 d}}{1 - \rho_{21}{}^2 e^{-2\gamma_2 d}}$$

$$\gamma_2 = \alpha_2 + j\beta_2$$

$$\frac{E_r}{E_i} = \frac{Z_2 - Z_1}{Z_2 + Z_1} - \frac{\dfrac{4Z_2 Z_1(Z_2 - Z_1)}{(Z_2 + Z_1)^3}e^{-2\gamma_2 d}}{1 - (\dfrac{Z_2 - Z_1}{Z_2 + Z_1})^2 e^{-2\gamma_2 d}}$$

$$\frac{E_r}{E_i} = \frac{Z_2 - Z_1}{Z_2 + Z_1}[1 - \frac{4Z_2 Z_1 e^{-2\gamma_2 d}}{(Z_2 + Z_1)^2 - (Z_2 - Z_1)^2 e^{-2\gamma_2 d}}]$$

$$\frac{E_r}{E_i} = \frac{Z_2 - Z_1}{Z_2 + Z_1} \frac{(Z_2 + Z_1)^2 - (Z_2 + Z_1)^2 e^{-2\gamma_2 d}}{(Z_2 + Z_1)^2 - (Z_2 - Z_1)^2 e^{-2\gamma_2 d}}$$

$$\frac{E_r}{E_i} = \frac{(Z_2 - Z_1)(Z_2 + Z_1)^2}{Z_2 + Z_1} \frac{1 - e^{-2\gamma_2 d}}{(Z_2 + Z_1)^2 - (Z_2 - Z_1)^2 e^{-2\gamma_2 d}}$$

On multiplie partout par $e^{\gamma_2 d}$ ce qui donne,

$$\frac{E_r}{E_i} = (Z_2 - Z_1)(Z_2 + Z_1) \frac{[1 - e^{-2\gamma_2 d}]e^{\gamma_2 d}}{[(Z_2 + Z_1)^2 - (Z_2 - Z_1)^2 e^{-2\gamma_2 d}]e^{\gamma_2 d}}$$

$$\frac{E_r}{E_i} = (Z_2^2 - Z_1^2) \frac{e^{\gamma_2 d} - e^{-\gamma_2 d}}{[(Z_2 + Z_1)^2 e^{\gamma_2 d} - (Z_2 - Z_1)^2 e^{-\gamma_2 d}]}$$

$$\sinh(\gamma_2 d) = \frac{e^{\gamma_2 d} - e^{-\gamma_2 d}}{2} \ , \qquad \cosh(\gamma_2 d) = \frac{e^{\gamma_2 d} + e^{-\gamma_2 d}}{2}$$

$$\frac{E_r}{E_i} = (Z_2^2 - Z_1^2) \frac{2\sinh(\gamma_2 d)}{(Z_1^2 + Z_2^2)(e^{\gamma_2 d} - e^{-\gamma_2 d}) + 2Z_2 Z_1 (e^{\gamma_2 d} + e^{-\gamma_2 d})}$$

Finalement,

$$\frac{E_r}{E_i} = (Z_2^2 - Z_1^2) \frac{\sinh(\gamma_2 d)}{(Z_1^2 + Z_2^2)\sinh(\gamma_2 d) + 2Z_2 Z_1 \ \cosh(\gamma_2 d)}$$

Le rapport $\frac{E_r}{E_i}$ peut être écrit sous la forme suivante,

$$\frac{E_r}{E_i} = \frac{Z_e - Z_1}{Z_e + Z_1}$$

où $Z_e$ représente une impédance ''caractéristique'' équivalente des milieux II et III par rapport au milieu I.

Le calcul de $Z_e$.

$$\frac{Z_e - Z_1}{Z_e + Z_1} = (Z_2^2 - Z_1^2)\frac{\sinh(\gamma_2 d)}{(Z_1^2 + Z_2^2)\sinh(\gamma_2 d) + 2Z_2 Z_1 \ \cosh(\gamma_2 d)}$$

$(Z_e - Z_1)[(Z_1^2 + Z_2^2)\sinh(\gamma_2 d) + 2Z_2 Z_1 \ \cosh(\gamma_2 d)] =$

$= (Z_2^2 - Z_1^2)\sinh(\gamma_2 d)(Z_e + Z_1)$

$Z_e[2Z_1^2 \sinh(\gamma_2 d) + 2Z_2 Z_1 \ \cosh(\gamma_2 d)] =$

$= Z_1[2Z_2^2 \sinh(\gamma_2 d) + 2Z_2 Z_1 \ \cosh(\gamma_2 d)]$

$$Z_e = Z_2 \frac{Z_2 \sinh(\gamma_2 d) + Z_1 \ \cosh(\gamma_2 d)}{Z_1 \sinh(\gamma_2 d) + Z_2 \ \cosh(\gamma_2 d)}$$

On divise partout par $\sinh(\gamma_2 d)$ ce qui donne,

$$Z_e = Z_2 \frac{Z_2 + Z_1 \ \tanh(\gamma_2 d)}{Z_1 + Z_2 \ \tanh(\gamma_2 d)}$$

On a déjà vu cette expression (ligne de transmission).

Questions à réfléchir:

1) Si $\gamma_2 = \alpha_2 + j\beta_2 = j\beta_2$, $(\alpha_2 = 0)$

$Z_e = ?$

Réponse: ($Z_e = Z_2 \frac{Z_2 + jZ_1 \ \tan(\beta_2 d)}{Z_1 + jZ_2 \ \tan(\beta_2 d)}$)

2) Si $Z_3 \neq Z_1$, quelle sera l'impédance $Z_e$?

Réponse: ($Z_e = Z_2 \frac{Z_2 + Z_3 \ \tanh(\beta_2 d)}{Z_3 + Z_2 \ \tanh(\beta_2 d)}$)

3) Si $d \to \infty$, quelle sera l'impédance $Z_e$ et quel sera le rapport $\frac{E_r}{E_i}$ ?

Réponse: ($Z_e = Z_2$, $\frac{E_r}{E_i} = \frac{Z_2 - Z_1}{Z_2 + Z_1}$)

**Exemple II.27.**

Une plaque d'épaisseur « $d$ » avec des dimensions infinies dans les deux autres directions est suspendue dans l'air comme indiqué à la Figure 28. Une onde électromagnétique de fréquence $f$ = 1 *GHz* frappe cette plaque à l'incidence normale. La plaque est formée d'un diélectrique avec pertes dont la permittivité complexe peut s'exprimer comme :

$$\varepsilon^* = \varepsilon_0 \left[ \varepsilon_r' - j \left( \varepsilon_r'' + \frac{\sigma}{\omega \varepsilon_0} \right) \right]$$

avec $\mu_r' = 10$ et $\mu_r'' = 0.02$ .

a) Si σ = 0 (la conductivité à $f$ = 1 *GHz*), calculer $\varepsilon^* = \varepsilon_0 [\varepsilon_r' - \varepsilon_r'']$ pour qu'il n'y ait pas de réflexion à la première interface air-diélectrique.

b) Si $\sigma \neq 0$ et $\varepsilon_r'' = 0.01$ calculer σ pour le même effet qu'en (a).

c) Si $d$ = 10 cm, calculer l'atténuation causée par cette plaque sur le chemin de l'onde électromagnétique dans les conditions de (a) ou de (b).

d) Pour une atténuation de 1 *dB* calculer « $d$ ».

Note : $\varepsilon_0 = \frac{10^{-9}}{36\pi} \left( \frac{F}{m} \right)$, $\mu_0 = 4\pi x 10^{-7} \left( \frac{H}{m} \right)$

**Solution :**

a)

$$\frac{E_r}{E_i} = \frac{Z_e - Z_1}{Z_e + Z_1} = 0 \ \rightarrow \ Z_e - Z_1$$

$$Z_e = Z_2 \frac{Z_2 + Z_1 \ \tanh(\gamma_2 d)}{Z_2 + Z_1 \ \tanh(\gamma_2 d)} \ \rightarrow \ Z_1 = Z_2 \frac{Z_2 + Z_1 \ \tanh(\gamma_2 d)}{Z_2 + Z_1 \ \tanh(\gamma_2 d)}$$

$$Z_1 = Z_0 = 120 \ \pi \ (\Omega)$$

$$Z_2 = \sqrt{\frac{\mu^*}{\varepsilon^*}} = Z_0 \sqrt{\frac{\mu_r^*}{\varepsilon_r^*}} = 120\,\pi \sqrt{\frac{\mu_r' - j\,\mu_r''}{\varepsilon_r^*}} = 120\,\pi \sqrt{\frac{10 - j\,0.02}{\varepsilon_r^*}}$$

$$Z_2 = 120\,\pi x A$$

où

$$A = \sqrt{\frac{10 - j\,0.02}{\varepsilon_r^*}} = \frac{10 e^{-j0.11^0}}{\sqrt{\varepsilon_r^*}} \;\rightarrow\; A^2 = \frac{10 - j\,0.02}{\varepsilon_r^*}$$

$$\frac{Z_2}{Z_1} = \frac{Z_2}{Z_0} = A$$

On divise partout par $Z_1$ on obtient:

$$Z_1 = Z_2 \frac{Z_2 + Z_1\,\tanh(\gamma_2 d)}{Z_2 + Z_1\,\tanh(\gamma_2 d)}$$

$$1 = A \frac{A +\,\tanh(\gamma_2 d)}{A +\,\tanh(\gamma_2 d)}$$

$$\gamma_2 = \alpha_2 + j\beta_2$$

$$A +\,\tanh(\gamma_2 d) = A^2 + A \tanh(\gamma_2 d)$$

$$A^2 - [1 - \tanh(\gamma_2 d)]A - \tanh(\gamma_2 d) = 0$$

$$A_{1,2} = \frac{1 - \tanh(\gamma_2 d) \pm (1 + \tanh(\gamma_2 d))}{2}$$

$$A_1 = 1$$

$$A_2 = -\tanh(\gamma_2 d)$$

I)

$$A_1 = 1 \;\rightarrow\; A^2 = 1 = \frac{10 - j\,0.02}{\varepsilon_r^*} \;\rightarrow\; \varepsilon_r^* = 10 - j\,0.02$$

$$\varepsilon^* = \varepsilon_0 \varepsilon_r^* = \varepsilon_0 (10 - j\,0.02)$$

II) Ce cas n'est pas possible parce que $A_2 = - \tanh(\gamma_2 d)$ et la racine d'un numéro ne peut pas être un nombre négatif.

b)

$$\varepsilon_r'' + \frac{\sigma}{\omega \varepsilon_0} = 0.01 + 18\,\sigma$$

$$Z_2 = 120\,\pi \sqrt{\frac{10 - j\,0.02}{\varepsilon_r' - j(0.01 + 18\sigma)}} = 120\,\pi x B$$

où

$$B = \sqrt{\frac{10 - j\,0.02}{\varepsilon_r' - j(0.01 + 18\sigma)}}$$

En utilisant la même condition qu'en (b),

$$1 = B\,\frac{B + \tanh(\gamma_2 d)}{B + \tanh(\gamma_2 d)}$$

où,

$$B^2 - [1 - \tanh(\gamma_2 d)]B - \tanh(\gamma_2 d) = 0$$

Ce qui donne,

$$B_1 = 1$$

$$B_2 = - \tanh(\gamma_2 d)$$

III)

$$B_1 = 1 = \sqrt{\frac{10 - j\,0.02}{\varepsilon_r' - j(0.01 + 18\sigma)}}$$

$$\varepsilon_r' - j(0.01 + 18\sigma) = 10 - j\,0.02$$

$$\varepsilon_r' = 10$$

$$0.01 + 18\sigma = 0.02$$

$$\sigma = 5.556x10^{-4} \left(\frac{S}{m}\right)$$

$$\varepsilon^* = \varepsilon_0 \varepsilon_r^* = \varepsilon_0(10 - j\,0.0201)$$

c)

$$\gamma_2 = \alpha_2 + j\beta_2$$

$$\gamma_2 = j\frac{2\pi}{\lambda_0}\sqrt{\mu_r' - j\mu_r''}\sqrt{\varepsilon_r' - j\left(\varepsilon_r'' + \frac{\sigma}{\varepsilon_0\omega}\right)}$$

a) $\sigma = 0$

$$\gamma_2 = j\frac{2\pi}{\lambda_0}\sqrt{\mu_r' - j\mu_r''}\sqrt{\varepsilon_r' - j\varepsilon_r''}$$

$$\gamma_2 = j\frac{2\pi}{\lambda_0}\sqrt{10 - j0.02}\sqrt{10 - j0.02} = j\frac{2\pi}{\lambda_0}(10 - j0.02)$$

$$\frac{2\pi}{\lambda_0} = \frac{2\pi f}{c} = \frac{2\pi x10^9}{3x10^8} = 20.94 \left(\frac{1}{m}\right)$$

$$\gamma_2 = \alpha_2 + j\beta_2 = 0.4188 + j209.4 \left(\frac{1}{m}\right)$$

$$\alpha_2 = 0.4188 \left(\frac{Np}{m}\right)$$

$$\beta_2 = 209.4 \left(\frac{rad}{m}\right)$$

$d = 0.1\ (m)$,

$$Att(dB) = 20\,log_{10}(e^{-\alpha_2 d}) = -0.364\ (dB)$$

b) $\sigma \neq 0$

$$\gamma_2 = j \frac{2\pi}{\lambda_0} \sqrt{\mu_r{}' - j\mu_r{}''} \sqrt{\varepsilon_r{}' - j\left(\varepsilon_r{}'' + \frac{\sigma}{\varepsilon_0 \omega}\right)}$$

$$\gamma_2 = 20.94j\sqrt{10 - j0.02} \ \sqrt{10 - j0.02} = j \frac{2\pi}{\lambda_0}(10 - j0.02)$$

$$\gamma_2 = \alpha_2 + j\beta_2 = 0.4188 + j209.4 \ (\frac{1}{m})$$

$$\alpha_2 = 0.4188 \ (\frac{Np}{m})$$

$$\beta_2 = \ 209.4 \ (\frac{rad}{m})$$

$d$ = 0.1 ($m$),

$$Att(dB) = 20 \ log_{10}(e^{-\alpha_2 d}) = -0.364 \ (dB)$$

Donc, on obtient le même résultat.

d)

$$1 \ dB \ \leftrightarrow \ \frac{1}{8.686} \ Np = 0.115 \ Np = \alpha_2 d$$

$$d = \frac{0.115}{\alpha_2} = 27.6 \ (cm)$$

## Exemple II.28.

Les trois régions montrées sur la figure II.29 sont des diélectriques parfaits. Une onde électromagnétique plane se propageant dans le milieu 1 et frappant normalement l'interface située à $z = 0$. Trouver $\varepsilon_{r2} \ et \ «\ d\ »$ tel qu'il n'y ait pas d'onde réfléchie, ($\varepsilon_{r1} = 4, \varepsilon_{r3} = 2.25$).

Solution :

Les diélectriques sont parfaits ce qui donne,

$$Z_1 = \ Z_0 \frac{1}{\sqrt{\varepsilon_{r1}}} \ , \quad Z_2 = \ Z_0 \frac{1}{\sqrt{\varepsilon_{r2}}} \ , \qquad Z_3 = \ Z_0 \frac{1}{\sqrt{\varepsilon_{r3}}} \ , \quad \gamma_2 = j\beta_2$$

$$\gamma_2 = \alpha_2 + j\beta_2 = j\beta_2$$

En utilisant les résultats obtenus pour la transmission à travers une plaque mince (Figure II.28) et appliquant la condition qu'il n'a pas d'onde réfléchie on obtient (ce sont les calculs assez compliquées) :

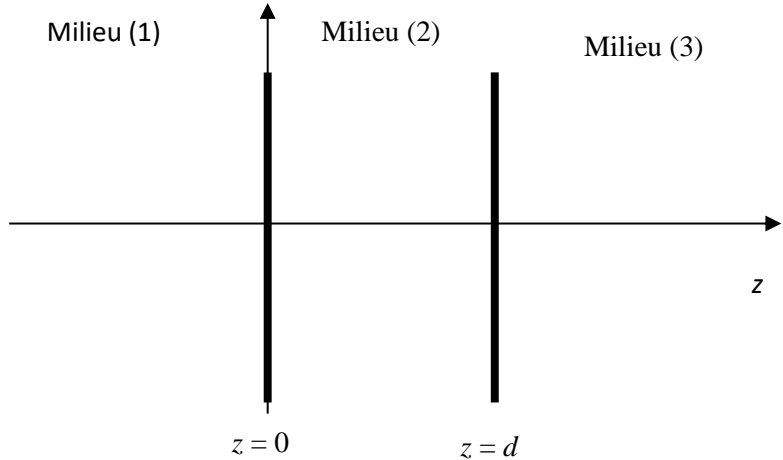

Milieu (1)    Milieu (2)    Milieu (3)

$z$

$z = 0$    $z = d$

Figure II.29.

$$Z_3 Z_2 + Z_1 Z_3 - Z_1 Z_2 - Z_2^2 = -(Z_2^2 + Z_3 Z_2 - Z_1 Z_2 - Z_1 Z_3)e^{-j2\beta_2 d}$$

Toutes impédances sont réelles. Ça signifie que la dernière identité peut être consistante si

$$e^{-j2\beta_2 d} = \pm 1$$

Dont la solution est $d = n\frac{\lambda_2}{4}$ , $n = 0,1,2,...,$

Si $n = 2k$ , $k = 1,2,3,...,$  on obtient,

$$Z_3 Z_2 + Z_1 Z_2 - Z_1 Z_3 - Z_2^2 = -Z_2^2 - Z_3 Z_2 + Z_1 Z_2 + Z_1 Z_3$$

d'où

$$Z_1 = Z_3 \text{ , pour chaque } Z_2$$

C'est une solution triviale. Dans le problème (Figure 28) on utilise $Z_1 = Z_3 = Z_0$.

Si $n = (2k + 1)\pi$ , $k = 1,2,3, …,$ on obtient,

$$Z_3 Z_2 + Z_1 Z_2 - Z_1 Z_3 - Z_2^2 = Z_2^2 + Z_3 Z_2 - Z_1 Z_2 - Z_1 Z_3$$

ou,

$$Z_2^2 = Z_1 Z_3 \ \rightarrow \ Z_2 = \sqrt{Z_1 Z_3}$$

On a déjà vu cette expression dans le traitement de lignes de transmission (Adaptation: $\frac{\lambda}{4}$)

Donc,

$$Z_2 = \sqrt{Z_1 Z_3}$$

$$Z_0 \frac{1}{\sqrt{\varepsilon_{r2}}} = Z_0 \sqrt{\frac{1}{\sqrt{\varepsilon_{r1}}} \frac{1}{\sqrt{\varepsilon_{r3}}}}$$

ce qui donne,

$$\varepsilon_{r2} = \sqrt{\varepsilon_{r1} \, \varepsilon_{r3}}$$

Finalement :

$$\varepsilon_{r2} = \sqrt{4x2.25} = 3 , \qquad d = (2k + 1) \frac{\lambda_2}{4} , k = 0,1,2, …,$$

Exemple II.29.

On considère une plaque de verre plate de 5 ($mm$) d'épaisseur avec $\varepsilon_r = 2.56$.
   a) Si un faisceau de lumière verte ($\lambda_0 = 0.52$ ($\mu m$)) est normalement incident sur un des côtés de la plaque quelle pourcentage de la puissance est réfléchie par le verre ?
   b) Pour éliminer les reflets il est souhaitable d'ajouter une fine couche de matériau de revêtement antireflet de chaque côté du verre. Si l'on est libre de spécifier l'épaisseur du matériau antireflet ainsi que sa permittivité relative, quelles seraient ces spécifications?

Solution :

a) L'impédance d'entrée sur le côté gauche de l'interface air-verre est,

$$Z_{en} = Z_v \frac{Z_0 + jZ_v \ \tan(\beta_2 d)}{Z_v + jZ_0 \ \tan(\beta_2 d)}$$

$$Z_0 = 120\pi \ (\Omega), \qquad Z_v = \frac{Z_0}{\sqrt{\varepsilon_r}} = \frac{120\pi}{1.6} \ (\Omega)$$

$$\beta_2 d = \frac{2\pi}{\lambda_2} \ d = \frac{2\pi}{\lambda_0} \sqrt{\varepsilon_r} d = 30769.23 \ \pi$$

Ici, on doit soustraire le maximum possible de multiples de $2\pi$.

$$\beta_2 d = 1.23 \ \pi + 15384\text{x}2\pi = 1.23 \ \pi$$

$$\beta_2 d = 1.23 \ \pi \ (rad)$$

$$Z_{en} = Z_v \frac{Z_0 + jZ_v \ \tan(\beta_2 d)}{Z_v + jZ_0 \ \tan(\beta_2 d)} = (224.2 - j108.4)(\Omega)$$

Le coefficient de réflexion sur la surface air-verre est,

$$\rho = \frac{Z_{en} - Z_0}{Z_{en} + Z_0} = 0.3067 e^{-j154.8^0}$$

Le pourcentage de la puissance réfléchie (%) = $|\rho|^2 \ x100 = 9.4\%$ .

b) Pour éviter les reflets, on peut ajouter un transformateur quart d'onde de chaque côté du verre (Voir Exemple 12). Cela nécessite que,

$$d = (2k + 1)\frac{\lambda_2}{4} = \frac{\lambda_2}{4} + k \frac{\lambda_2}{2} \ , k = 0,1,2, \dots,$$

où $\lambda_2$ est la longueur d'onde dans ce matériau, c'est-à-dire $\lambda_2 = \lambda_0 \frac{1}{\sqrt{\varepsilon_r}}$.

La constante $\varepsilon_r$ est la permittivité du matériau de revêtement. Aussi, il faut trouver $Z_c$ du matériau de revêtement qui se calcule selon,

$$Z_c^2 = Z_0 Z_v$$

ou,

$$Z_c^2 \frac{1}{\varepsilon_{rc}} = Z_c^2 \frac{1}{\sqrt{\varepsilon_r}}$$

d'où

$$\varepsilon_{rc} = \sqrt{\varepsilon_r} = 1.6$$

$$\lambda_2 = \lambda_0 \frac{1}{\sqrt{\varepsilon_r}} = 0.411 \ (\mu m)$$

$$d = (2k+1)\frac{\lambda_2}{4} = (2k+1)x0.1028 \ (\mu m) \, , k = 0,1,2,\dots,$$

## Exemple II.30.

Un prisme triangulaire dont la section est un triangle isocèle est utilisé dans un instrument optique (Figure II.30). Aux fréquences optiques, on considère $\varepsilon_r = 4$ pour le verre. Calculer le pourcentage de la puissance lumineuse réfléchie par ce prisme.

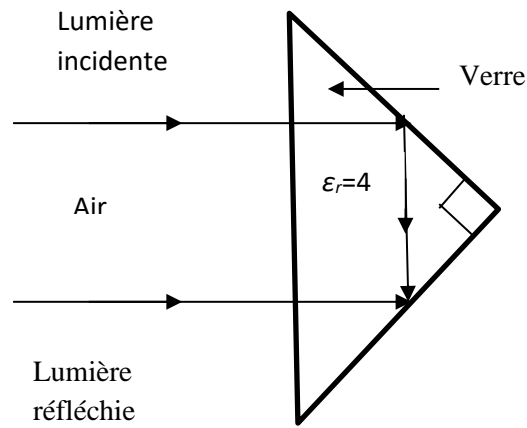

Figure II.30.

Solution:

Sur l'hypoténuse on aura:

$$K = \frac{<P_t>}{<P_i>} = \frac{1}{2}[\bar{E}_t \times \bar{H}_t^*]x2\frac{1}{[\bar{E}_i \times \bar{H}_i^*]} = \frac{Z_0}{Z_v}T_{0-v}^2$$

$$T_{0-v} = \frac{2Z_v}{Z_v + Z_0}$$

$$K = 4\frac{Z_vZ_0}{(Z_v + Z_0)^2}$$

$$S = \frac{<P_0>}{<P_{tv}>} = \frac{Z_v}{Z_0}T_{v-0}^2$$

$$T_{v-0} = \frac{2Z_0}{Z_v + Z_0}$$

$$K = 4\frac{Z_vZ_0}{(Z_v + Z_0)^2}$$

La réflexion totale sur les autres surfaces,

$$\theta_i = 45^0, \qquad \theta_c = \arcsin\left(\frac{1}{2}\right) = 30^0$$

$$\theta_i > \theta_c \ \rightarrow \frac{Z_v}{Z_0} = \frac{1}{\sqrt{\varepsilon_r}} = \frac{1}{2} = L$$

Donc,

$$\frac{<P_0>}{<P_i>} = KS = 16\frac{Z_v^2Z_0^2}{(Z_v + Z_0)^4} = 16\frac{L^2}{(1 + L)^4} = 0.79 = 79\ \%$$

# 11. Problèmes

**Exemple 1.** Une onde électromagnétique plane se propageant dans l'air le long de + z (z < 0) frappe normalement à z = 0 sur un conducteur avec $\sigma$ = 61.7 (*MS/m*), $\mu_r = 1$. L'onde incidente à une fréquence de *f* = 1.5 (*MHz*) et le champ électrique à l'interface z = 0 est donné par,

$$\bar{E}\,(0,t) = 10 \cos\left(\omega t + \frac{\pi}{2}\right)\hat{y}\,(\frac{V}{m})$$

a) Déterminer les champs électrique et magnétique dans l'air et dans le conducteur,

b) Montrer que le coefficient de réflexion est approximativement donné par $\rho \approx (Z_2 - 1)(1 - Z_2) \approx -1 + 2Z_2 = \sqrt{2\frac{\omega\varepsilon_0}{\sigma}}\,(1 + j) - 1$ , où $Z_2 = \sqrt{j\frac{\omega\varepsilon_0}{\sigma}}$,

c) Trouver la fraction de la puissance incidente qui est perdue pour le conducteur après la réflexion,

c) Trouver la profondeur de pénétration dans le matériau conducteur.

**Solution :**

[a)
$$Z_1 = Z_0 = 120\pi\,(\Omega), \quad Z_2 = \sqrt{j\frac{\omega\mu}{\sigma+j\omega\varepsilon}}$$

$$\frac{\sigma}{\omega\varepsilon} = \frac{\sigma}{\omega\varepsilon_0} = \frac{61.7x10^6x36\pi}{2\pi x1.5x10^6x10^{-9}} = 740.4x10^9 \gg 1$$

$$Z_2 = \sqrt{j\frac{\omega\mu}{\sigma+j\omega\varepsilon}} \cong \sqrt{j\frac{\omega\mu}{\sigma}} = Z_0\sqrt{j\frac{\omega\varepsilon_0}{\sigma}} = 120\pi e^{j\frac{\pi}{4}}\sqrt{\frac{2\pi x1.5x10^6x10^{-9}}{36\pi x61.7x10^6}}$$

$$Z_2 = 120\pi e^{j\frac{\pi}{4}}x1.6x10^{-6} = 0.603x10^{-3}e^{j\frac{\pi}{4}}(\Omega)$$

$$\rho = \frac{Z_2 - Z_1}{Z_2 + Z_1} = \frac{\frac{Z_2}{Z_1} - 1}{\frac{Z_2}{Z_1} + 1}$$

$$\frac{Z_2}{Z_1} = 1.6e^{j\frac{\pi}{4}}x10^{-6} \ll 1$$

$$\rho = -1$$

Air :

$$\beta_0 = \omega\sqrt{\mu_0\varepsilon_0} = \frac{2\pi x1.5x10^6}{3x10^8} = 0.0314 = \frac{\pi}{100} \left(\frac{rad}{m}\right)$$

$$\lambda_0 = \frac{2\pi}{\beta_0} = 200 \ (m)$$

$$\bar{E}_{air}\ (z,t) = 10e^{j\frac{\pi}{100}z}\ e^{j\frac{\pi}{2}}\hat{y}\left(\frac{V}{m}\right)$$

$$\bar{H}_{air}\ (z,t) = \frac{1}{Z_1}[\hat{n}\ \times\ \bar{E}_{air}\ (z,t)] = \frac{1}{Z_0}[\hat{z}\ \times\hat{y}]10e^{j\frac{\pi}{100}z}\ e^{j\frac{\pi}{2}}$$

$$\bar{H}_{air}\ (z,t) = -\frac{10}{120\pi}e^{j\frac{\pi}{100}z}\ e^{j\frac{\pi}{2}}\hat{x}\ \left(\frac{A}{m}\right)$$

parce que $\hat{n} = \ \hat{z}$

$$\bar{H}_{air}\ (z,t) = 26.5\ e^{j\frac{\pi}{100}z}\ e^{j\frac{3\pi}{2}}\hat{x}\ (10^{-3}\frac{A}{m})$$

$$T = \frac{2Z_2}{Z_2 + Z_1} = \frac{2\frac{Z_2}{Z_1}}{\frac{Z_2}{Z_1} + 1} \cong 0$$

Donc,

$$\bar{E}_{conducteur}\ (z,t) = \ \bar{H}_{conducteur}\ (z,t) = 0$$

$$\bar{E}_{air(r)}\ (0,t) = -\ \bar{E}_{air(i)}\ (0,t) = 10\ e^{j\frac{3\pi}{2}}\hat{y}\left(\frac{V}{m}\right)$$

b)

$$Z_2 = Z_0\sqrt{j\frac{\omega\varepsilon_0}{\sigma}}\ \ \rightarrow\frac{Z_2}{Z_1} = \frac{Z_2}{Z_0} = \sqrt{j\frac{\omega\varepsilon_0}{\sigma}} = e^{j\frac{\pi}{4}}\sqrt{\frac{\omega\varepsilon_0}{\sigma}} = q$$

$$|q| < 1$$

$$\rho = \frac{Z_2 - Z_1}{Z_2 + Z_1} = \frac{q-1}{q+1} = (q-1)\frac{1}{1+q}$$

$$\frac{1}{1+q} = \sum_0^\infty (-q)^n = 1 - q + q^2 - \cdots$$

$$|q| < 1 \quad \rightarrow \frac{1}{1+q} \approx 1 - q$$

$$\rho = (q-1)(1-q) = -1 + 2q - q^2 \approx -1 + 2q$$

$$\rho = (q-1)(1-q) \approx -1 + 2q = -1 + e^{j\frac{\pi}{4}}\sqrt{\frac{\omega \varepsilon_0}{\sigma}} = -1 + (\frac{\sqrt{2}}{2} + j\frac{\sqrt{2}}{2})\sqrt{\frac{\omega \varepsilon_0}{\sigma}}$$

$$\rho = \sqrt{\frac{2\omega \varepsilon_0}{\sigma}} \ (1+j) - 1$$

c)

$$< P_{moy,i} > = \frac{1}{2} Re[\bar{E}_{air(i)}(z,t) \times \bar{H}_{air(i)}^*]$$

$$< P_{moy,i} > = \frac{1}{2} Re\left[(\hat{y} \times \hat{x})x10\ e^{j\frac{\pi}{2}} x26.5xe^{j\frac{3\pi}{2}}\right]x10^{-3}$$

$$< P_{moy,i} > = -\frac{1}{2} Re\left[(\hat{y} \times \hat{x})x10\ e^{j\frac{\pi}{2}} x26.5xe^{j\frac{3\pi}{2}}\right]x10^{-3}$$

$$< P_{moy,i} > = \frac{1}{2} Re\left[(\hat{y} \times \hat{x})x10\ e^{j\frac{\pi}{2}} x26.5xe^{j\frac{3\pi}{2}}e^{j\pi}\right]x10^{-3}$$

$$< P_{moy,i} > = -0.1325\ (\frac{W}{m^2})$$

d)

$$\delta = 0$$

**Exemple 2.** Une onde plane uniforme se propage dans un diélectrique parfait de constante $\varepsilon_r = 2.25$ non-magnétique. Le champ électrique de l'onde s'exprime comme'

$$\bar{E}(t) = 10(4\hat{x} + 5\hat{y} - 3\hat{z})\cos[2\pi x10^7 t - 0.02\pi(3x + 4z)]x10^{-3}\left(\frac{V}{m}\right).$$ Trouver,

a) Le vecteur unitaire de propagation.
b) La longueur d'onde suivant la direction de propagation.
c) Les longueurs d'onde apparentes suivant les trois axes.
d) Les vitesses de phase suivant les trois axes.

**Solution :**

a) Le vecteur-phaseur $\bar{E}(t) = 10(4\hat{x} + 5\hat{y} - 3\hat{z})e^{-j0.02\pi\,(3x+4z)}x10^{-3}\left(\frac{V}{m}\right)$

$$\hat{n} = \frac{3}{5}\hat{x} + \frac{4}{5}\hat{z}$$

b)

$$\beta_x = 0.02\pi x3 = 0.06\pi\left(\frac{rad}{m}\right), \beta_y = 0, \beta_z = 0.02\pi x4 = 0.08\pi(\frac{rad}{m})$$

$$\beta = \sqrt{\beta_x^2 + \beta_y^2 + \beta_z^2} = 0.1\pi(\frac{rad}{m})$$

$$\lambda = \frac{2\pi}{\beta} = \frac{2\pi}{0.1\pi} = 20\ (m)$$

c)

$$\lambda_x = \frac{2\pi}{\beta_x} = \frac{2\pi}{0.06\pi} = 33.33\ (m)$$

$$\lambda_y = \frac{2\pi}{\beta_y} = \infty$$

$$\lambda_z = \frac{2\pi}{\beta_z} = \frac{2\pi}{0.08\pi} = 25\ (m)$$

d)

$$v_x = \frac{\omega}{\beta_x} = \frac{2\pi x 10^7}{0.06\pi} = 3.333 x 10^8 (m/s)$$

$$v_y = \frac{2\pi f}{\beta_y} = \infty$$

$$v_z = \frac{\omega}{\beta_z} = \frac{2\pi x 10^7}{0.08\pi} = 2.5 \; x 10^8 (m/s)$$

**Exemple 3.** Une onde plane se propageant dans un milieu non-magnétique avec une constante diélectrique $\varepsilon_r = 2.25$ est caractérisée par,

$$\bar{E}_i(x,y) = 10(-0.8\hat{x} + 0.6\hat{y})e^{-j4\pi(3x+4y)} \; (\frac{V}{m})$$

Cette onde est incidente obliquement à l'interface dans la région $x > 0$, rempli d'un non-magnétique à indice de réfraction 2.

a) Quelle est la fréquence de cette onde?
b) Quelle est la direction de propagation des ondes incidentes, réfléchies et transmises?
c) Obtenir les phaseurs de champ électrique des ondes réfléchies et transmises.

Solution :

a) De $e^{-j4\pi(3x+4y)} = e^{-j(12\pi x + 16\pi y)} \rightarrow \beta_{x1} = 12\pi, \beta_{y1} = 16\pi, \beta_{z1} = 0$

$$\beta = \beta_1 = \sqrt{\beta_{x1}^2 + \beta_{y1}^2 + \beta_{z1}^2} = 20\pi(\frac{rad}{m})$$

$$\lambda = \frac{2\pi}{\beta} = \frac{2\pi}{20\pi} = 0.1 \; (m)$$

$$\lambda = \frac{v}{f} = \frac{c}{f\sqrt{\varepsilon_{r1}}} = \frac{3x10^8}{f\sqrt{2.25}} = \frac{3x10^8}{fx1.5} = \frac{2x10^8}{f} = 0.1(m)$$

$$v_1 = v = \frac{c}{\sqrt{\varepsilon_{r1}}} = \frac{3x10^8}{\sqrt{4}} = 2x10^8 \left(\frac{m}{s}\right)$$

$$f = 2GH_z$$

b) De $e^{-j4\pi(3x+4y)} = e^{-j\frac{4\pi}{5}\left(\frac{3}{5}x+\frac{4}{5}y\right)} \rightarrow n_{x(i)} = \frac{3}{5}, n_{y(i)} = \frac{4}{5}, n_{z(i)} = 0$

$$\hat{n}_{(i)} = \frac{3}{5}\hat{x} + \frac{4}{5}\hat{y} = 0.6\hat{x} + 0.8\hat{y}$$

$$\hat{n}_{(i)} = sin\theta_i \hat{x} + cos\theta_i \hat{y}$$

$$\theta_r = \theta_i$$

$$\hat{n}_{(r)} = -sin\theta_r \hat{x} + cos\theta_r \hat{y} = -0.6\hat{x} + 0.8\hat{y}$$

Loi de Snell − Descartes :

$$\frac{sin\theta_i}{sin\theta_t} = \frac{\beta_2}{\beta_1} = \sqrt{\frac{\varepsilon_{r2}}{\varepsilon_{r1}}}$$

Indice de réfraction $n = \frac{c}{v_2} = \frac{c}{c}\sqrt{\varepsilon_{r2}} \rightarrow n = \sqrt{\varepsilon_{r2}} \rightarrow \varepsilon_{r2} = 4$

$$v_2 = \frac{c}{\sqrt{\varepsilon_{r2}}} = 1.5x10^8 \left(\frac{m}{s}\right)$$

$$\lambda_2 = \frac{v_2}{f} = \frac{c}{2f} = \frac{1.5x10^8}{2x10^9} = 0.075 \ (m)$$

$$\lambda_2 = \frac{2\pi}{\beta_2} \rightarrow \beta_2 = \frac{2\pi}{\lambda_2} = \frac{80\pi}{3} \left(\frac{rad}{m}\right)$$

$$\frac{sin\theta_i}{sin\theta_t} = \sqrt{\frac{\varepsilon_{r2}}{\varepsilon_{r1}}} = \frac{2}{1.5} = \frac{4}{3}$$

$$sin\theta_t = \frac{4}{3}sin\theta_i = \frac{4}{3}x0.6 = 0.8$$

$$\hat{n}_{(t)} = sin\theta_t\hat{x} + cos\theta_t\hat{y} = 0.8\hat{x} + 0.6\hat{y}$$

c)    $sin\theta_t = 0.8, \quad cos\theta_t = 0.6$

$$Z_1 = Z_0 \frac{1}{\sqrt{\varepsilon_{r1}}} = \frac{120\pi}{\sqrt{2.25}} = \frac{120\pi}{1.5} = 120\pi\frac{4}{3} \ (\Omega)$$

$$Z_2 = Z_0 \frac{1}{\sqrt{\varepsilon_{r2}}} = \frac{120\pi}{\sqrt{4}} = 120\pi\frac{1}{2} \ (\Omega)$$

$$\rho_\parallel = \frac{Z_2 cos\theta_t - Z_1 cos\theta_i}{Z_1 cos\theta_i + Z_2 cos\theta_t} = 0$$

$$Z_2 cos\theta_t - Z_1 cos\theta_i = Z_1 cos\theta_t \left[\frac{Z_2}{Z_1} - \frac{cos\theta_i}{cos\theta_t}\right] = \left[\frac{15}{20}x - \frac{0.6}{0.8}\right] = 0$$

$$\frac{E_r}{E_i} = T_\parallel = \frac{2Z_2 cos\theta_i}{Z_1 cos\theta_i + Z_2 cos\theta_t} = \frac{2\frac{Z_2}{Z_1}}{1 + \frac{cos\theta_t}{cos\theta_i}\frac{Z_2}{Z_1}}$$

$$T_\parallel = \frac{3}{4}, \quad \rho_\parallel = 0$$

$$\bar{E}_r(x, y) = 0$$

$$\bar{E}_t(x, y) = 7.5(-0.6\hat{x} + 0.8\hat{y})e^{-j\frac{80\pi}{3}(0.8x + 0.6y)} \ \left(\frac{V}{m}\right)$$

$$\bar{E}_t = (-\hat{x}cos\theta_t + \hat{y}sin\theta_t)E_t e^{j\beta_2(xsin\theta_t + ycos\theta_t)}$$

$$T_\parallel = \frac{3}{4}, \quad \rho_\parallel = 0$$

**Exemple 4.**

La permittivité de l'eau aux fréquences optiques est $1.75 \ \varepsilon_0$. Une source lumineuse isotropique se trouve à une distance « $d$ » de la surface à l'intérieur de l'eau d'un bassin très large. Un observateur à l'extérieur du bassin constate la nuit une tache lumineuse circulaire de rayon 5 ($m$) à la surface de l'eau (Figure II.31). Déterminer « $d$ ».

Solution :

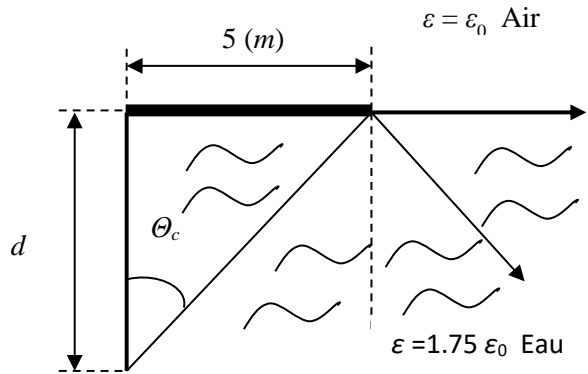

Figure II.31.

[θ ≥ θ$_c$, Réflexion totale, θ$_c$ = 49.107⁰, $d$ = 4.33 $(m)$]

**Exemple 5.** Une onde plane se propage dans l'air suivant la direction $\hat{n}$ . Dans un système de coordonnées cartésiennes, on observe les longueurs d'onde $\lambda_x = 40 \ (cm)$ , $\lambda_y = 50 \ (cm)$ Si la fréquence est de 1 $GHz$, calculer $\lambda_z$ et déterminer $\hat{n}$.

Solution :
($\lambda_z = 69.27 \ (cm)$, $\hat{n} = 0.75 \ \hat{x} + 0.5 \ \hat{y} \pm 0.433 \ \hat{z}$)

**Exemple 6.** Une onde électromagnétique plane à la fréquence de 300 MHz, arrive sur une surface de séparation (air – diélectrique parfait) avec un angle d'incidence $\theta_i = 63.435⁰$ . Le premier milieu de propagation étant l'air, l'onde rencontre le deuxième milieu au plan *xOz*, considéré comme plan d'interface dans un système cartésien *Oxyz*. Le point d'impact *O* est l'origine des coordonnées et la normale à la surface de séparation est l'axe *Oy*. Les trois composantes du champ électrique incident (sous forme de vecteur-phaseur) s'expriment par les relations suivantes :

$$E_z = \ E_0 e^{j\frac{2\pi}{\lambda}(x\sqrt{0.8}+y\sqrt{0.2})} \ (\frac{V}{m})$$

$$E_x = \ -\sqrt{0.2}E_z e^{j\frac{\pi}{2}} \ (\frac{V}{m})$$

$$E_y = \sqrt{0.8}E_z e^{j\frac{\pi}{2}}\left(\frac{V}{m}\right)$$

D'autre part $\varepsilon_{r2}$ du diélectrique (deuxième milieu) est 4.

a) Calculer l'expression du vecteur-phaseur concernant le champ électrique du signal réfléchi au point :

$$x = \frac{\sqrt{5}}{2}\lambda \,, y = \sqrt{5}\lambda \,,$$

si $E_0 = 100 \left(\frac{V}{m}\right)$,

b) Calculer l'expression générale du vecteur-phaseur relatif au champ électrique total transmis dans le deuxième milieu,

c) Commenter sur la polarisation du signal incident, réfléchi et transmis,

d) Calculer la densité de puissance moyenne transmise dans le diélectrique.

Solution :

[ $\bar{E}_x + \bar{E}_y = \bar{E}_{||}$ , $\bar{E}_z = \bar{E}_\perp$, $a)$ $\operatorname{atan}\left(\sqrt{\frac{\varepsilon_{r2}}{\varepsilon_{r1}}}\right) = \operatorname{atan}(2) = 63.435^0, \theta_i \to$

angle de Brewster $\to \rho_{||} = 0$, $\sin\theta_t = 0.45$, $\cos\theta_t = 0.89, \rho_\perp = -0.6$,

$\bar{E}_r = -0.6E_0 e^{j\frac{2\pi}{\lambda}(x\sin\theta_r - y\cos\theta_r)}\hat{z}\left(\frac{V}{m}\right), si \ x = \frac{\sqrt{5}}{2}\lambda \,, y = \sqrt{5}\lambda \to$

$\bar{E}_r = -60 \ \hat{z} \left(\frac{V}{m}\right); b) \ T_{||} = 0.5$, $\rho_\perp = 0.4$, $\beta_2 = \omega\sqrt{\mu_0\varepsilon_0\varepsilon_{r2}} = \frac{4\pi}{\lambda_2}$ ;

$(\perp) \ \bar{E}_{z t\perp} = \hat{z} \frac{2}{5}E_0 e^{j\frac{2\pi}{\lambda_2}(\sqrt{0.2}x + \sqrt{0.8}y)}\left(\frac{V}{m}\right),$

$(||) \ \bar{E}_{t_{||}} = \left(-\sqrt{0.8}\,\hat{x} + \sqrt{0.2}\,\hat{y}\right)0.5E_0 e^{j\frac{2\pi}{\lambda_2}(\sqrt{0.2}x + \sqrt{0.8}y)}e^{j\frac{\pi}{2}}\left(\frac{V}{m}\right),$

$$\beta_1 = \omega\sqrt{\mu_0 \varepsilon_0} = 2\pi \;, \beta_2 = \beta_1\sqrt{\varepsilon_{r2}} = 4\pi = \frac{2\pi}{\lambda_2} \;\rightarrow\; \lambda_2 = 0.5\;(m),$$

$$\bar{E}_{t_{Total}} = \bar{E}_{t_{||}} + \bar{E}_{z_{t\perp}} = 100(-j\sqrt{0.2}\,\hat{x} + j\sqrt{0.05}\,\hat{y} + 0.4\hat{z}\,)e^{j\,4\pi(\sqrt{0.2}x+\sqrt{0.8}y)}\left(\frac{V}{m}\right),$$

c) Onde incidente : Polarisation circulaire, Onde réfléchie : Polarisation linéaire,

Onde transmise : Polarisation elliptique à gauche;

$$< P > = \frac{1}{2Z_2}\left|\bar{E}_{t_{||}}^2\right| + \frac{1}{2Z_2}\left|\bar{E}_{t\perp}^2\right| = 10.876 \left(\frac{W}{m^2}\right) ]$$

## Exemple 7.

Une plaque d'un matériau spécial a une épaisseur de 5cm et des dimensions infinies dans les autres sens. Elle est suspendue dans l'air et reçoit une onde électromagnétique plane de 500 MHz à incidence normale sur sa surface. La perméabilité et la permittivité complexes relatives de cette plaque sont identiques et valent :

$$\mu_r = \mu_r = 3 - j0.5$$

a) Calculer le facteur d'atténuation $\alpha$ dans cette plaque pour la fréquence donnée.

b) Quelle est l'atténuation totale de l'onde (en *dB*) transmise à l'autre côté de la plaque, dans l'air en comparaison avec l'onde incidente? (On considère les puissances).

c) Quelle est l'atténuation totale (en *dB*) pour l'onde réfléchie à la première interface air-matériau?

## Solution:

[ a) $\alpha = 5.236 \left(\frac{Np}{m}\right)$; b) $Att = 5.924\;(dB)$, Il n'ya pas la reflexion nulle part;

c) $\frac{E_r}{E_i} = 0, \qquad Att = \infty$ ]

## Exemple 8.

Déterminer le type de polarisation et l'amplitude maximale du champ pour le phaseurs suivants du champ électrique d'une onde plane uniforme :

a) $\bar{E}_1 = 10\,(\hat{x} + e^{j\frac{5\pi}{2}}\hat{z})e^{-j4\pi y}\,(\frac{V}{m})$

b) $\bar{E}_2 = \bar{E}_1 + 10\,(\hat{x} + e^{j\frac{7\pi}{2}}\hat{z})e^{-j4\pi y}\,(\frac{V}{m})$

c) $\bar{E}_3 = 10\,(\hat{x} - 5j\hat{y})e^{-j50z}\,(\frac{V}{m})$

d) $\bar{E}_4 = 10\,(e^{j\frac{\pi}{4}}\hat{x} - e^{-j\frac{\pi}{4}}\hat{y})e^{j10z}\,(\frac{V}{m})$

Solution :

[a) Circulaire droite, $E_{max}$ = 10 (V/m); b) linéaire en $x$, $E_{max}$ = 20 (V/m);
c) Elliptique droite, $E_{max}$ = 50 (V/m); d) Circulaire gauche, $E_{max}$ = 10 (V/m)]

**Exemple 9.** Un milieu exerce des pertes de conductivité ($\sigma = 1.112 x 10^{-2}\,(\frac{S}{m})$) et des pertes diélectriques ($tg\delta = \frac{2}{5}$). La permittivité diélectrique relative du milieu est $\varepsilon'_r = 5$ et la perméabilité magnétique complexe relative $\mu_r^* = 5 - j4$. La fréquence de l'onde se propageant dans ce milieu est de 100 $MH_z$. Trouver :

a) L'impédance caractéristique du milieu,
b) La constante de propagation,
c) Le facteur d'atténuation,
d) Le facteur de phase
e) La profondeur de pénétration.

Solution :

( a) $\varepsilon''_r = 2$, $\varepsilon_r^* = 5 - j4$, $Z = Z_0 = 377\,(\Omega)$; b) $\gamma = (8.38 + j10.472)\left(\frac{1}{m}\right)$

c) $\alpha = 8.38\,\left(\frac{Np}{m}\right)$;  d) $\beta = 10.472\,\left(\frac{rad}{m}\right)$;  $\delta = 0.119\,(m)$)

**Exemple 9.** Une onde de polarisation circulaire à gauche frappe une interface entre deux milieux différents sous un angle de $45^0$. Si ces milieux sont de l'air (milieux 1) et un conducteur parfait (milieu 2) il faut déterminer :

a) La polarisation de l'onde réfléchie.
b) Pour l'interface air-polystyrène ($\varepsilon_{rp} = 2.25$) la polarisation de l'onde réfléchie et de l'onde transmise.

Solution : [a)Polarisation circulaire à droite; b) Onde réfléchie : Polarisation elliptique à droite, Onde transmise : Polarisation elliptique à gauche]

**Exemple 10.** Une onde électromagnétique plane, de polarisation linéaire, sinusoïdale, de fréquence 1 MHz, se propageant dans l'air, frappe la surface plane d'un diélectrique sous un angle d'incidence de $50^0$. Le champ électrique incident est normal au plan d'incidence et son amplitude est:

$$|\overline{E}_i| = 1 \left(\frac{V}{m}\right)$$

Le milieu diélectrique a les propriétés suivantes:

$$\sigma = 0, \qquad \mu_r = 1, \; Z = 24\,\pi\,(\Omega)$$

a) Trouver $\varepsilon_{r2}$ du diélectrique (le second milieu),

b) Trouver $\beta_2$, le facteur de phase pour le diélectrique,

c) Trouver l'angle de transmission ($\theta_t$),

d) Calculer le coefficient de transmission,

e) Trouver les amplitudes des champs électriques réfléchi et transmis,

Le system de coordonnées cartésiennes est choisi de telle façon que :

i) Le champ électrique incident est dans la direction $O_z$,

ii) Le point d'impact est le point $O$, origine des coordonnées,

iii) Le plan de séparation entre deux milieux est le plan *xOz*,

iv) Le plan d'incidence est le plan *xOy* avec la normale Oy au plan de séparation.

Après avoir dessiné le schéma de cette incidence oblique, donner

f) Les expressions des vecteurs unitaires incident, réfléchi et transmis ($\hat{n}_i$ , $\hat{n}_r$ , $\hat{n}_t$ ) en fonction de (*x, y, z*),

g) Donner l'expression du vecteur phaseur $\overline{E}_i$ selon votre schéma,

h) Donner l'expression du vecteur phaseur du champ magnétique transmis $\overline{H}_t$,

i) Trouver le rapport entre les puissances transmise et incidente ($\frac{P_t}{P_i}$).

**Solution :**

[a) $\varepsilon_{r2} = 25$; b) $\beta_2 = 0.105 \left(\frac{rad}{m}\right)$; c) $\theta_t = 8.813^0$; d) $T_\perp = 0.23, \rho_\perp = -0.77$;

e) $|\bar{E}_t| = |\bar{E}_i||T_\perp| = 0.23 \left(\frac{V}{m}\right)$, $|\bar{E}_r| = |\bar{E}_i||\rho_{\perp}| = 0.77 \left(\frac{V}{m}\right)$;

f) $\hat{n}_i = -0.76604\hat{x} - 0.64279\,\hat{y}$, $\hat{n}_r = -0.76604\hat{x} + 0.64279\,\hat{y}$,

$\hat{n}_t = -0.15321\hat{x} - 0.98819\,\hat{y}$; g) $\bar{E}_i = e^{-j0.021(-0.77x - 0.64y)}\hat{z} \left(\frac{V}{m}\right)$,

$\beta_1 = 0.021 \left(\frac{rad}{m}\right)$ h) $|\bar{H}_t| \frac{|=|\bar{E}_i|T_\perp|}{Z_2} = 0.00305 \left(\frac{A}{m}\right)$,

$\bar{H}_i = (-0.00301\hat{x} + 0.00047\hat{y})e^{j(0.10609x + 0.10376y)} \left(\frac{A}{m}\right)$; i) $\frac{P_t}{P_i} = 0.265$)

**Exemple 11.** Trouver la valeur crête d'un courant surfacique induit lorsqu'une onde plane est incidente à un angle sur une grande feuille parfaitement conductrice. La surface de la feuille est située à $z = 0$ et

$$\bar{E}_i = 10 \cos\left(10^{10}t - \beta\frac{x}{\sqrt{2}} - \beta\frac{z}{\sqrt{2}}\right)\hat{x}\left(\frac{V}{m}\right)$$

**Solution :**

[Le vecteur de propagation est donné par,

$$\bar{\beta} = \frac{\beta}{\sqrt{2}}\hat{x} + \frac{\beta}{\sqrt{2}}\hat{z} = \beta\left(sin\frac{\pi}{4}\hat{x} + cos\frac{\pi}{4}\hat{z}\right)$$

Donc,

$$\hat{n}_i = sin\frac{\pi}{4}\hat{x} + cos\frac{\pi}{4}\hat{z}$$

et

$$\theta_i = \frac{\pi}{4}$$

Le champ électrique est le long de direction « $y$ » est perpendiculaire à la plane d'incidence.

Sur la surface conductrice à $z = 0$ (Figure II.32), la condition suivante doit être valable,

$$\bar{E}_y(x,z) = \bar{E}_{y(i)} + \bar{E}_{y(r)} = 0 \text{ à } z = 0$$

$$\bar{E}_y(x,0) = \bar{E}_{m(i)}e^{-j\beta x sin\theta_i} + \bar{E}_{m(r)}e^{-j\beta x sin\theta_r} = 0$$

$$\bar{H}_z(x,0) = \frac{1}{Z_0}\bar{E}_{m(i)}\,sin\theta_i e^{-j\beta xsin\theta_i} + \frac{1}{Z_0}\bar{E}_{m(r)}\,sin\theta_r e^{-j\beta xsin\theta_r} = 0$$

$$\theta_r = \theta_i = \frac{\pi}{4}$$

$$\bar{E}_{m(r)} = -\bar{E}_{m(i)}$$

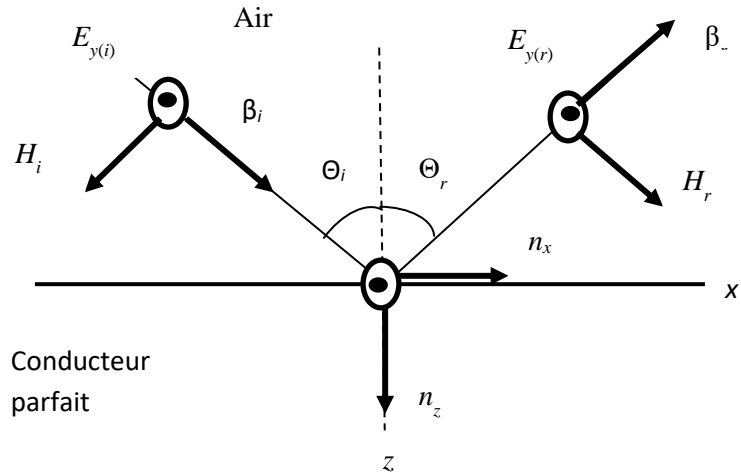

Figure II.32.

$$\bar{E}_y(x,y) = \bar{E}_{m(i)}e^{-j\beta xsin\theta_i}[e^{-j\beta zsin\theta_i} - e^{j\beta zsin\theta_i}]$$

$$\bar{E}_y(x,y) = -2j\bar{E}_{m(i)}e^{-j\beta xsin\theta_i}[\sin{(\beta zcos\theta_i)}]$$

$$\hat{n}_{\beta(i,r)} = sin\frac{\pi}{4}\hat{x} + cos\frac{\pi}{4}\hat{z}$$

$$\bar{H} = \bar{H}_i + \bar{H}_r = \frac{1}{Z_0}[\hat{n}_{\beta(i)} \times \hat{y}]\bar{E}_{m(i)}e^{-j\beta(i)z} - \frac{1}{Z_0}[\hat{n}_{\beta(r)} \times \hat{y}]\bar{E}_{m(i)}e^{-j\beta(r)z}$$

$$\bar{H}_x(x,z) = -2\frac{\bar{E}_{m(i)}}{Z_0}cos\theta_i\,\sin{(\beta zcos\theta_i)}\,e^{-j\beta xsin\theta_i}$$

$$\bar{H}_z(x,z) = -2\frac{\bar{E}_{m(i)}}{Z_0}sin\theta_i\,\sin{(\beta zcos\theta_i)}\,e^{-j\beta xsin\theta_i}$$

$$\bar{J}_{z=0} = \hat{n} \times \bar{H} = -\hat{z} \times \bar{H} = \hat{x}\frac{2\bar{E}_{m(i)}}{Z_0}\frac{\bar{E}_{m(i)}}{Z_0}\cos\theta_i \ e^{-j\beta x sin\theta_i}$$

$$\bar{J}_{z=0} = 3.75x10^{-2} \ (\frac{A}{m})$$

**Exemple 12.** Une onde plane uniforme ayant le champ électrique donne par

$$\bar{E}_i = 10 \cos\big(10^9 t - 10\pi \ (x + \sqrt{3}z)\big)(\frac{\sqrt{3}}{2}\hat{x} \ -\frac{1}{2}\hat{z}) \ (\frac{V}{m})$$

est incidente à l'interface entre l'espace libre et un diélectrique de permittivité $\varepsilon = 1.5 \ \varepsilon_0$ comme montré à la Figure II.33. Trouver :

a) L'angle $\theta_t$,
b) La direction de propagation $\hat{\beta}_i$ et $\hat{n}_i$,
c) Le champ électrique – phaseur,
d) Les coefficients $\rho_{||}$ et $T_{||}$,
e) L'expression du champ réfléchie – phaseur et sinusoïdale,
f) La direction de propagation $\hat{\beta}_t$ et $\hat{n}_t$,
g) L'expression du champ transmis– phaseur et sinusoïdale.

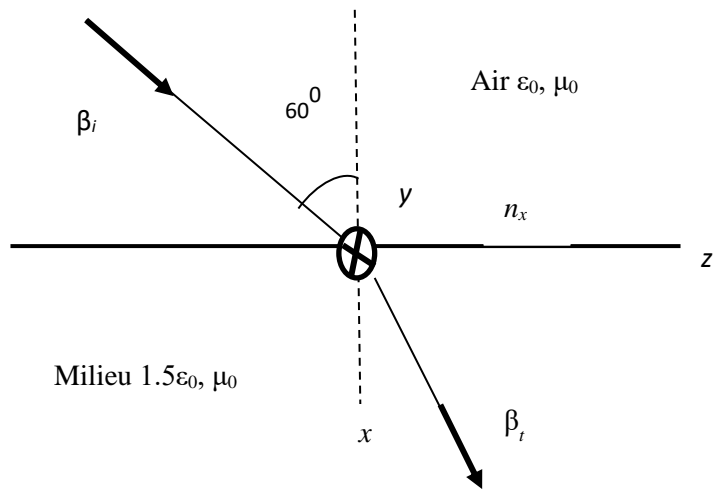

**Figure II.33.**

**Solution :**

[ a) $\theta_t = \frac{\pi}{4}$ ; b) $\hat{\beta}_i = 10\pi\hat{x} + 10\pi\sqrt{3}\ \hat{z}, \beta_i = 20\pi\left(\frac{rad}{m}\right), \hat{n}_i = \frac{1}{2}\hat{x} + \frac{\sqrt{3}}{2}\ \hat{z}$ ;

c) $\bar{E}_i = 10\left(\frac{\sqrt{3}}{2}\hat{x} - \frac{1}{2}\hat{z}\right)e^{-j10\pi\ (x+\sqrt{3}z)}\ \left(\frac{V}{m}\right)$ ; d), $\rho_{||} = 0.0781$ , $T_{||} = 0.758$,

e) $\bar{E}_r = 0.781\left(\frac{\sqrt{3}}{2}\hat{x} + \frac{1}{2}\hat{z}\right)e^{-j10\pi(-x+\sqrt{3}z)}\ \left(\frac{V}{m}\right)$,

$\bar{E}_r = 0.781\left(\frac{\sqrt{3}}{2}\hat{x} + \frac{1}{2}\hat{z}\right)cos\left(10^9 t - 10\pi\left(-x + \sqrt{3}z\right)\right)\left(\frac{V}{m}\right)$ ;

f) $\hat{n}_t = \frac{\sqrt{2}}{2}\hat{x} + \frac{\sqrt{2}}{2}\ \hat{z}, \beta_t = \beta_i\sqrt{1.5} = 12.24\pi\left(\frac{rad}{m}\right),\ \hat{\beta}_t = 8.66\pi\hat{x} + $

$+8.66\pi\ \hat{z}$ ;  g) $\bar{E}_t = (\hat{x} + \hat{z})e^{-j8.66\pi(x+z)}\ \left(\frac{V}{m}\right)$,

$\bar{E}_t = 5.36(\hat{x} - \hat{z})cos\ \left(10^9 t - 8.66\pi\left(-x + \sqrt{3}z\right)\right)\ \left(\frac{V}{m}\right)$ ]

**Exemple 13.** Une onde électromagnétique qui se propage dans l'eau de mer dont les caractéristiques sont $\sigma = 4\left(\frac{S}{m}\right)$, $\mu = \mu_0$ et  $\varepsilon = 80\varepsilon_0$. La propagation est dans la direction des $z$ négatifs. L'expression du champ électrique est,

$$|\bar{E}(z = 0, t)| = \cos(2\pi x 10^5 t)\ \left(\frac{V}{m}\right)$$

Trouver :

- a) Le coefficient d'atténuation.
- b) La constante de phase.
- c) L'impédance caractéristique du milieu.
- d) Le champ électrique dans le milieu.
- e) Le champ magnétique dans le milieu.
- f) La densité de puissance instantanée.
- g) La densité de puissance moyenne.

Solution :

[a) $\alpha = 1.257\left(\frac{Np}{m}\right)$ ; b) $\beta = 1.257\left(\frac{rad}{m}\right)$ ; c) $Z = 0.1x\pi(1 + j) = 0.444e^{j\frac{\pi}{4}}\ (\Omega)$;

d) $\bar{E}(z,t) = \cos(2\pi x 10^5 t + 1.257z)\, e^{1.257z}\, \hat{x}\; \left(\frac{V}{m}\right)$ ;

e) $\bar{H}(z,t) = -2.25 \cos(2\pi x 10^5 t + 1.257z) \cos\left(2\pi x 10^5 t + 1.257z - \frac{\pi}{4}\right) x$

$e^{1.257z} \hat{y}\left(\frac{A}{m}\right)$ ;  f) $\bar{P} = \bar{E}(z,t) \times \bar{H}(z,t) = [-1.125 e^{2.514z} \cos\left(\frac{\pi}{4}\right)\hat{z} - 1.125 x$

$e^{2.514z} \cos\left(4\pi x 10^5 t + 2.514z - \frac{\pi}{4}\right)\hat{z}]\; \left(\frac{W}{m^2}\right), cosa x cosb = \frac{1}{2}\,[\cos(a+b) +$

$\cos(a-b)]$ ;  g) $< P_{moy} > = -1.125 e^{2.514z} \cos\left(\frac{\pi}{4}\right) = -0.8 e^{2.514z}\;(W/m^2)]$

**Exemple 14.** Décrire l'état de polarisation de chacune de ces ondes représentées par les champs électriques suivants ($\bar{E}_0$ , $\bar{E}_1$ , $\bar{E}_2$, $\bar{E}$):

a)  $E_{x0} = E_0 \cos(\theta) \cos(\omega t - \beta z), E_{y0} = E_0 \sin(\theta) \cos(\omega t - \beta z), E_{z0} = 0$
b)  $E_{x1} = E_1 \cos(\omega t - \beta z), E_{y1} = E_1 \cos(\omega t - \beta z), E_{z1} = 0$
c)  $E_{x2} = E_2 \cos(\omega t - \beta z), E_{y2} = -E_2 \sin(\omega t - \beta z), E_{z2} = 0$
d)  Quel est l'état de polarisation de l'onde dont le champ électrique est $\bar{E} = \bar{E}_1 + \bar{E}_2$ en supposant $|\bar{E}_1| = |\bar{E}_2|$ .

Solution :

[a) Polarisation linéaire suivant le vecteur unitaire $\hat{n} = \cos(\theta)\,\hat{x} + \sin(\theta)\,\hat{y}$ ;
b) Polarisation circulaire à gauche; c) Polarisation circulaire à droite;
   e) $\bar{E} = 2E_1 \cos(\omega t - \beta z)\,\hat{x}$ , polarisation linéaire suivant $\hat{x}$ ]

**Exemple 15.** Un train roulant à 25 $\left(\frac{m}{s}\right)$ sonne à une fréquence de 800 $(H_z)$  à l'approche d'un tunnel dans une falaise.

a)  Quelle est la fréquence observée par une personne se tenant près l'entrée du tunnel ?
b)  Le son du cor se reflète da la falaise vers le conducteur de train. Qu'entend le conducteur du train ?
   La vitesse du son dans l'air est de 345 $\left(\frac{m}{s}\right)$.

Solution :

a) $f_s = 800\ H_z$ , $v_s = 25\ \left(\frac{m}{s}\right)$ , $v_0 = 0$ , $v = 345\ \left(\frac{m}{s}\right), f_0 = ?$

Le train s'approche d'observateur, $f_{01} \uparrow$ .

$$f_{01} = f_s \frac{v + v_0}{v - v_s}\quad - \text{ Effet Doppler}$$

$$f_{01} = 800 \frac{345}{345 - 25} = 863\ (H_z)$$

b) Le son réfléchi à la même fréquence que celui entendu par la personne. Aucun changement dans la fréquence sur réflexion lorsque la source s'approche d'observateur, $f_{02} \uparrow$ .

c) $f_s = f_{01} = 863\ (H_z)$ , $v_s = 0$ , $v_0 = 25\ \left(\frac{m}{s}\right)$ , $v = 345\ \left(\frac{m}{s}\right), f_0 = ?$

$$f_{02} = f_{01} \frac{v + v_0}{v - v_s} = 863 \frac{345 + 25}{345} = 926\ (H_z)$$

Le conducteur du train entend la superposition du son d'origine à 800 ($H_z$) et du son réfléchi à 926 ($H_z$). Donc,

$$f_{0cond} = f_{02} - f_{01} = 926 - 800 = 126\ (\ H_z)$$

**Exemple 16.** Une ambulance se déplace sur une section droite de l'autoroute à la vitesse de100 $\left(\frac{km}{h}\right)$. Sa sirène émet un son à fréquence de 400 ($H_z$). Quelle fréquence est entendue par un passager dans une voiture voyageant sur la même autoroute à une vitesse de 100 $\left(\frac{km}{h}\right)$. Considérer toutes les possibilités.
La vitesse du son dans l'air est de 345 $\left(\frac{m}{s}\right)$.

**Solution :**

$f_s = 400\ H_z$ , $v_s = 100 \left(\frac{km}{h}\right) = 27.78 \left(\frac{m}{s}\right)$ , $v_0 = 80 \left(\frac{km}{h}\right) = 22.22 \left(\frac{m}{s}\right)$ ,
$v = 345 \left(\frac{m}{s}\right), f_0 = ?$

Cas I)
Source approchant de l'observateur $f_0 \uparrow$,
Observateur s'éloigne de la source $f_0 \downarrow$

$$f_0 = f_s \frac{v \pm v_0}{v \mp v_s}$$

$$f_0 = f_s \frac{v - v_0}{v - v_s} = 400 \frac{345 - 22.22}{345 - 27.78} = 407 \ (H_z)$$

Cas II)
Observateur approchant de la source $f_0 \uparrow$,
Source s'éloigne de l'observateur $f_0 \downarrow$

$$f_0 = f_s \frac{v + v_0}{v + v_s} = 400 \frac{345 + 22.22}{345 + 27.78} = 394 \ (H_z)$$

Cas III)
Observateur approchant de la source $f_0 \uparrow$,
Source approchant de l'observateur $f_0 \uparrow$

$$f_0 = f_s \frac{v + v_0}{v - v_s} = 400 \frac{345 + 22.22}{345 - 27.78} = 463 \ (H_z)$$

Cas IV)
 Observateur s'éloigne de la source $f_0 \downarrow$
Source s'éloigne de l'observateur $f_0 \downarrow$

$$f_0 = f_s \frac{v - v_0}{v + v_s} = 400 \frac{345 - 22.22}{345 + 27.78} = 346 \ (H_z)$$

**Exemple 17.** Une onde plane se transmettant dans la direction positive de l'axe « $z$ » formée de l'addition de deux ondes de polarisation linéaire. Cette onde a un champ électrique orienté selon l'axe « $x$ » et l'autre champ électrique orienté selon l'axe « $y$ ». Si,

$$E_x = 2 \left( \frac{V}{m} \right), \qquad \text{valeur crête}$$

$$E_y = 4 \left( \frac{V}{m} \right) , \qquad \text{valeur crête}$$

et que $E_y$ est en retard de phase sur $E_x$ dans le temps d'un angle $\varphi = \frac{\pi}{2}$.

a) Quelle est l'intensité de la puissance transporte par cette onde?
b) Faire un diagramme dans un plan perpendiculaire à la direction de propagation pour indiquer bien la polarisation de cette onde.
c) Quel est le sens de cette rotation du champ électrique résultant? (Pour un observateur qui voit l'onde s'éloigner de lui).

**Solution :**

[a) $< P > = 0.026 \left( \frac{W}{m^2} \right)$; b) La polarisation de cette onde est elliptique; c) Le vecteur champ électrique résultant tourne dans le sens horaire. Donc, c'est une polarisation elliptique à droite]

**Exemple 18.** Une tour de contrôle dans un aéroport est placée à une distance de 1 (*km*) d'un émetteur radar d'une puissance de 1 (*MW*) pendant les cycles d'émission. La fréquence de l'onde émise est de 6 (*GHz*) et sa polarisation est circulaire.

a) Quelle est la densité de puissance (puissance par unité de surface) incidente sur le mur de la tour sachant que, dû à la forme de l'antenne, cette dernière transmet de préférence vers l'avant produisant une densité de puissance dans cette direction de $10^4$ fois plus grande que si toute l'énergie était rayonnée uniformément dans toutes les directions?
b) Quelle est l'amplitude du champ électrique incident mesurée dans l'air près du mur de la tour?
c) Est-ce qu'un mur en béton d'une épaisseur de 2 (*cm*) est suffisant pour bien isoler un ordinateur placé dans la tour? (Le champ électrique incident sur l'ordinateur doit être inférieur à 1 (*V/m*). Les propriétés du béton sont :
$\sigma = 4 \left( \frac{S}{m} \right), \quad \varepsilon_r = 6, \ \mu_r = 1$ .
d) Comment pourrait – on protéger l'ordinateur contre les interférences ?
e) Quelle est la puissance dissipée dans le mur pour une aire de 1 ($m^2$) du mur?

**Solution:** [a] $P = 796 \left( \frac{W}{m^2} \right)$ ; b) $E = 547.8 \left( \frac{V}{m} \right)$ valeur crête ; c) Pas suffisant, l'isolation n'est pas adéquate; d) Il n'est pas nécessaire de construire une tour ayant des murs plus épais (couts plus élevés, etc., …). Aussi la tour est déjà construite. On peut faire un blindage en posant sur les murs une enveloppe protectrice qui a une conductivité élevée et qui va protéger l'ordinateur dans la salle;

e) $< P_t > = 0.0255 \left( \frac{W}{m^2} \right) \rightarrow$ Puissance moyenne transmise à l'interieur ,

$< P_d > = 567.62 \left( \frac{W}{m^2} \right) \rightarrow$ Puissance moyenne dissipée dans le mur ,

$< P_i > = 795.8 \left( \frac{W}{m^2} \right) \rightarrow$ Puissance moyenne icidente sur le mur ), $P = 567.6$ W )

**Exemple 19.** Une onde électromagnétique de polarisation linéaire selon « *y* » dont la fréquence est de 100 *MHz* se propage dans la direction « *+ x* » et frape perpendiculairement un plan parfaitement conducteur à *x* = 0. Supposons l'amplitude du champ électrique de 6 (*V/m*). Donner :

a) Les phaseurs et les valeurs instantanées pour les champs $\bar{E}_i$ et $\bar{H}_i$ de l'onde incidente.
b) Les phaseurs et les valeurs instantanées pour les champs $\bar{E}_r$ et $\bar{H}_r$ de l'onde réfléchie.
c) Les champs $\bar{E}_{Total}$ et $\bar{H}_{Total}$ dans l'air (phaseurs et valeurs instantanées)
d) La position la plus proche du plan conducteur dans l'air, où le champ $\bar{E}_{Total}$ est zéro.
e) A partir de ce résultat trouver la longueur d'onde dans l'air. Approuver ce résultat par une autre formule.

**Solution :**

[a) $\bar{E}_i = 6x10^{-3}\hat{y} \; e^{-j\frac{2\pi}{3}x} \; \left( \frac{V}{m} \right)$, $\bar{H}_i = \frac{1}{2\pi}10^{-4}\hat{z} \; e^{-j\frac{2\pi}{3}x} \; \left( \frac{A}{m} \right)$,

$\bar{E}_i(x,y,z) = 6x10^{-3}\hat{y} \; \cos\left( \omega t - \frac{2\pi}{3} \right) \left( \frac{V}{m} \right)$,

$\bar{H}_i(x,y,z) = \frac{10^{-4}}{2\pi}\hat{z} \; \cos\left( \omega t - \frac{2\pi}{3} \right) \left( \frac{A}{m} \right)$;

b) $\rho = -1$ , $\hat{n} = -\hat{x}$;

c) $\bar{E}_r = -6x10^{-3}\hat{y}\ e^{-j\frac{2\pi}{3}x}\ \left(\frac{V}{m}\right),\ \ \bar{H}_i = \frac{1}{2\pi}10^{-4}\hat{z}\ e^{-j\frac{2\pi}{3}x}\ \left(\frac{A}{m}\right),$

$\bar{E}_i(x,y,z) = 6x10^{-3}\hat{y}\ \cos\left(\omega t - \frac{2\pi}{3} - \mu\right)\left(\frac{V}{m}\right),$

$\bar{H}_i(x,y,z) = \frac{10^{-4}}{2\pi}\hat{z}\ \cos\left(\omega t - \frac{2\pi}{3}\right)\left(\frac{A}{m}\right);$

d) $\bar{E}_{Total} = \bar{E}_i + \bar{E}_r = -j12x10^{-3}\hat{y}\ \sin\left(\frac{2\pi}{3}\right)\left(\frac{V}{m}\right)$

$\bar{H}_{Total} = \bar{H}_i + \bar{H}_r = \frac{1}{\pi}x10^{-4}\hat{z}\ \cos\left(\frac{2\pi}{3}\right)\left(\frac{A}{m}\right)$

$\bar{E}_{Total}(x,y,z) = Re\left[\bar{E}_{Total}e^{j\omega t}\right] = 12x10^{-3}\hat{y}\ \sin\left(\frac{2\pi}{3}\right)\sin\left(\omega t\right)\left(\frac{V}{m}\right)$

$\bar{H}_{Total}(x,y,z) = Re\left[\bar{H}_{Total}e^{j\omega t}\right] = \frac{1}{\pi}x10^{-4}\hat{z}\ \sin\left(\frac{2\pi}{3}\right)\cos\left(\omega t\right)\left(\frac{V}{m}\right);$

e) $x = \frac{3}{2}\ n$, e) $n = 0, \pm1, \pm2, \pm\cdots.$, $n = 0 \rightarrow x_1 = 0, n = -1 \rightarrow\ x_2 = -\frac{3}{2}$,

$\frac{\lambda}{2} = x_1 - x_2 = \frac{3}{2} \rightarrow \lambda = 3\ (m).$ D'autre part $\lambda = \frac{2\pi}{\beta_0} = 3\frac{2\pi}{2\pi} = 3\ (m)]$

**Exemple 20.** Une onde plane uniforme à polarisation linéaire se propage dans un milieu sans perte ayant permittivité relative de 5. La fréquence est de 100 (*MHz*). L'onde frape a incidence oblique un second milieu. L'angle d'incidence est de $45^0$ et la polarisation est parallèle au plan d'incidence. À quelle profondeur le champ électrique de l'onde transmise dans le second milieu sera-t-il atténuée de 300 (*dB*) par rapport au niveau de champ électrique de l'onde incidente si :

   a) Ce milieu est de cuivre ayant une conductivité de $5.8x10^7$ (*S/m*) et $\varepsilon' = \varepsilon_0$,
   b) Ce milieu est de l'air.

Solution :

[a) $\bar{E}_{II}^t = \bar{E}_{II}^i T_{II}e^{-j(\beta_{tx}x+\beta_{tz}z)}$. On cherche « $z$ » tel que $20log_{10}\left(\left|\frac{\bar{E}_{II}^i}{\bar{E}_{II}^t(z)}\right|\right) = 300$ (*dB*), $20log_{10}\left[\left|T_{II}e^{-j\beta_{tz}z}\right|\right] = -300$ (*dB*),

$T_{II} = 2\frac{Z_2\cos\theta_i}{Z_2\cos\theta_t+Z_1\cos\theta_i}$, $\beta_{tx}^2 + \beta_{tz}^2 = \omega^2\mu_2\varepsilon_2 = \beta_t^2$, $\frac{\sigma}{\omega\varepsilon} \gg 1$, $\beta_{ix} = \beta_{tx}$

$\beta_{ix} = \beta_1 \sin(45^0) = \omega\sqrt{\mu_0\varepsilon_0}\ \sqrt{5}\ \sin(45^0) = \frac{\pi}{3}\sqrt{10} = \beta_{tx}$,

$\beta_{tz}^2 = \beta_t^2 - \beta_{tx}^2 = \omega^2\mu_0\left(\varepsilon_0 - j\frac{\sigma}{\omega}\right) - \frac{10}{9}\pi^2 \rightarrow \beta_{tz} \approx (1-j)\sqrt{\frac{\sigma\omega\mu_0}{2}}$

$$\beta_{tz} = \beta_t cos\theta_2 \rightarrow (1-j)\sqrt{\frac{\sigma\omega\mu_0}{2}} = \sqrt{\omega^2\mu_0\left(\varepsilon_0 - j\frac{\sigma}{\omega}\right)}\ cos\theta_2$$

$$\approx \sqrt{-j\sigma\omega\mu_0}\ cos\theta_2 \rightarrow\ cos\theta_2,\ Z_2 = 2.6x(1+j)10^{-3}\ (\Omega), Z_2 = 168.6\ (\Omega)$$

$$cos\theta_i = cos(45^0) \rightarrow\ T_{II} \approx 2\frac{Z_2}{Z_1} = 3.1x10^{-5}(1+j),\ |T_{II}| = 4.36x10^{-5}$$

$$|e^{-j\beta_{tz}z}| = \left|e^{-j(1-j)\sqrt{\frac{\sigma\omega\mu_0}{2}}z}\right| = e^{-1.5x10^5z} \rightarrow 10^{-\frac{300}{20}} = 3.1x10^{-5}xe^{-1.5x10^5z}$$

$$z = \ln\left(\frac{10^{-5}}{3.1x10^{-5}}\right) = 1.64x10^{-4}\ (m); \text{b) } z = 13.5\ (m)\ ]$$

**Exemple 21.** Une onde plane de polarisation circulaire se transmet dans un parfait diélectrique ($\varepsilon_r = 4, \mu_r = 1,\ \sigma = 0$). L'intensité de la puissance transporte par cette onde est de,

$$< P > = 24\pi\ (\frac{mW}{cm^2})$$

Le phaseur champ électrique est donné par :

$$\bar{E} = (A_1\hat{x} + B_1\hat{y} + C_1\hat{z})e^{jK(x-y)}(\frac{V}{m})$$

où $K = \frac{\pi\sqrt{2}}{30}\left(\frac{rad}{m}\right)$. Certains termes $A_1, B_1, C_1,$ pourraient être des quantités imaginaires ou complexes.

a) Calculer le facteur de phase de cette onde ainsi que $\beta_x, \beta_y\ et\ \beta_z$,
b) Quelle est la direction de propagation de l'onde?
c) Quelle est la fréquence de l'émetteur?
d) Calculer la longueur de cette onde ainsi que la vitesse de phase,
e) Calculer $\lambda_x, \lambda_y\ et\ \lambda_z$ ainsi que $v_{px}, v_{py}\ et$ ,
f) Vérifier les identités suivantes : $\frac{1}{\lambda^2} = \frac{1}{\lambda_x^2} + \frac{1}{\lambda_y^2} + \frac{1}{\lambda_z^2}$ ainsi que

$$\frac{1}{v_p^2} = \frac{1}{v_{px}^2} + \frac{1}{v_{py}^2} + \frac{1}{v_{pz}^2}$$

g) Calculer les valeurs $A_1, B_1$ et $C_1$.

**Solution :**
a) $$-\beta\left(n_x\hat{x} + n_y\hat{y} + n_z\hat{z}\right) = K(x-y)$$

ce qui donne,

$$\beta n_x = -K \ , \ \ \beta n_z = K, \ \ \beta n_z = 0$$

Étant donné que,

$$n_x^2 + n_y^2 + n_z^2 = 1$$

on obtient,

$$\beta^2(n_x^2 + n_y^2 + n_z^2) = 2K^2$$

ou

$$\beta = \sqrt{2}\,K = \frac{\pi}{15}\,\left(\frac{rad}{m}\right)$$

$$K(x-y) = -\frac{\pi}{15}(-x+y) = -\left(\hat{\beta}\cdot\hat{n}\right) = -(\frac{\pi}{15}\sqrt{2}\,\hat{x} + \frac{\pi}{15}\sqrt{2}\,\hat{y})(-\frac{x\hat{x}}{\sqrt{2}} + \frac{y\hat{y}}{\sqrt{2}})$$

d'où

$$\hat{\beta} = \frac{\pi}{15}\sqrt{2}\,\hat{x} + \frac{\pi}{15}\sqrt{2}\,\hat{y} \ \rightarrow \beta_x = \frac{\pi}{15}\sqrt{2} = \beta_y, \ \ \beta_z = 0$$

b)
$$\hat{n} = -\frac{\hat{x}}{\sqrt{2}} + \frac{\hat{y}}{\sqrt{2}} = \cos(A)\,\hat{x} + \cos(B)\,\hat{y} + \cos(C)\,\hat{z}$$

$$\cos(A) = -\frac{1}{\sqrt{2}} \ , \qquad \cos(B) = \frac{1}{\sqrt{2}}, \ \ \cos(C) = 0$$

c)
$$\beta = \omega\sqrt{\varepsilon\mu} = \frac{\omega}{c}\sqrt{\varepsilon_r} \ \rightarrow \omega = \frac{\beta c}{\sqrt{\varepsilon_r}} = \pi x 10^7 \left(\frac{rad}{s}\right)$$

$$f = \frac{\omega}{2\pi} = 5 \ (MH_z)$$

d)
$$\lambda = \frac{2\pi}{\beta} = \frac{2\pi}{\frac{\pi}{15}} = 30 \ (m)$$

$$v_p = \frac{\omega}{\beta} = 1.5 x 10^8 \left(\frac{m}{s}\right)$$

e)

$$\lambda_x = \frac{\lambda}{\cos(A)} = -30\sqrt{2}\,(m), \qquad \lambda_y = \frac{\lambda}{\cos(B)} = 30\sqrt{2}\,(m) \text{ et } \lambda_z = \infty$$

$$v_{px} = \frac{v_p}{\cos(A)} = -1.5x\sqrt{2}x10^8 \left(\frac{m}{s}\right), v_{py} = \frac{v_p}{\cos(B)} = 1.5x\sqrt{2}x10^8 \left(\frac{m}{s}\right)$$

et $v_{py} = \infty$

f)

$$\frac{1}{\lambda^2} = \frac{1}{\lambda_x^2} + \frac{1}{\lambda_y^2} + \frac{1}{\lambda_z^2}$$

$$\frac{1}{30^2} = \frac{1}{2x30^2} + \frac{1}{2x30^2} = \frac{1}{30^2}$$

$$\frac{1}{v_p^2} = \frac{1}{v_{px}^2} + \frac{1}{v_{py}^2} + \frac{1}{v_{pz}^2}$$

$$\frac{10^{-8}}{1.5^2} = \frac{10^{-8}}{2x1.5^2} + \frac{10^{-8}}{2x1.5^2} = \frac{10^{-8}}{1.5^2}$$

g)  On a une onde plane,

$$\hat{n} \cdot \bar{E} = 0 \qquad\qquad (I)$$

Aussi, l'intensité de la puissance transporte par une onde de polarisation circulaire est :

$$<P> = \frac{1}{2Z}\,|\bar{E}_1^2| + \frac{1}{2Z}\,|\bar{E}_2^2| \qquad (II)$$

$$<P> = 24\pi \left(\frac{mW}{cm^2}\right) = 240\pi \left(\frac{W}{m^2}\right)$$

L'onde se transmet suivant une direction (vecteur unitaire $\hat{n}$) qui se trouve dans le plan $x-y$.

Donc,

$C_1$ peut être associé à une onde et,

$\sqrt{A_1^2 + B_1^2}$ peut être associé à une autre onde.

Toutes deux sont déphasées de $\frac{\pi}{2}$. $\hspace{3cm}(III)$

La condition $(I)$ donne,

$$\hat{n} \cdot \bar{E} = \left[-\frac{\hat{x}}{\sqrt{2}} + \frac{\hat{y}}{\sqrt{2}}\right] \cdot [A_1\hat{x} + B_1\hat{y} + C_1\hat{z}] = 0$$

ou $\quad (-A_1 + B_1)\frac{1}{\sqrt{2}} = 0 \quad \rightarrow \quad A_1 = B_1$

La condition $(II)$ donne,

$$< P > = \frac{1}{2Z}\,|\bar{E}_1^2| + \frac{1}{2Z}\,|\bar{E}_2^2| = \frac{1}{2Z}[A_1^2 + B_1^2 + C_1^2] = 240\pi$$

$$Z = \sqrt{\frac{\mu}{\varepsilon}} = Z_0\,\frac{1}{\sqrt{\varepsilon_r}} = 60\pi\ (\Omega)$$

La condition $(III)$ donne,

$$A_1^2 + B_1^2 = C_1^2$$

De ces trois équations on a,

$$A_1 = B_1 = 60\sqrt{2}\pi, \qquad C_1 = 120\,\pi$$

**Exemple 22.** Une onde plane de polarisation linéaire et dont le champ électrique est perpendiculaire au plan d'incidence sur une surface de séparation du milieu 1 vers le milieu 2. L'intensité de la puissance transmise dans l'onde incidente est $< P > = 10\,(\frac{W}{m^2})$ et l'angle d'incidence est $\theta_i = 30^0$. Le milieu 1 est de l'air et le milieu 2 est un diélectrique de propriétés ($\varepsilon_r = 9,\ \mu_r = 1,\ \sigma = 0$).

   a) Quelle est l'amplitude du champ électrique réfléchi ?
   b) Quelle est la direction de propagation de l'onde réfléchie?
   c) Calculer l'intensité de de la puissance réfléchie et de la puissance transmise,

d) Vérifier si $< P_i > = < P_r > + < P_t >$. Est-ce que cette expression est correcte pour ce problème-ci? Expliquer.

Solution :

[a)$|E_r| = 47.41 \left(\frac{V}{m}\right)$; b) $\hat{n} = -0.5\hat{x} + 0.866\hat{y}$; c) $< P_r > = 2.98 \left(\frac{W}{m^2}\right)$,

$< P_t > = 6.154 \left(\frac{W}{m^2}\right)$; d) $< P_i > \neq < P_r > + < P_t >$, Chacune de ces expressions est une intensité de puissance mesurée dans un plan perpendiculaire à la direction de propagation. Puisque la direction de propagation de l'onde transmise différente de l'onde incidente, la projection de 1 ($m^2$) sur la surface de séparation est différente pour l'onde incidente et l'one transmise]

**Exemple 23.** Une onde plane se propageant dans l'air (milieu 1) avec une polarisation perpendiculaire est incidente avec un angle de $30^0$ sur un milieu (milieu 2) avec une permittivité relative $\varepsilon_r = 5$.

a) Trouver le coefficient de réflexion,
b) Si l'onde se propage dans le milieu 2 et touche le milieu 1, trouver l'angle critique.

Solution :
[a) $\rho = -0.431$; b) $\theta_c = 26.59^0$]

**Exemple 24.** Les deux prismes de la Figure II.34 sont an verre avec $\varepsilon_r = 2.25$. Quelle fraction de la densité de puissance portée par le rayon incident sur le prisme supérieur émerge du prisme inferieur? Négliger les réflexions multiples.

Solution :

En utilisant $Z = \frac{1}{n} Z_0$ aux interfaces 1 et 4, on a $Z = \frac{1}{\sqrt{\varepsilon_r}} Z_0$, $\rho_i = \frac{Z - Z_0}{Z + Z_0} = -0.2$

Aux interfaces 3 et 6, $\rho_t = -\rho_i = 0.2$. Aux interfaces 2 et 5, on a l'angle critique,

$$\theta_c = arcsin\left[\frac{1}{\sqrt{\varepsilon_r}}\right] = 41.81^0$$

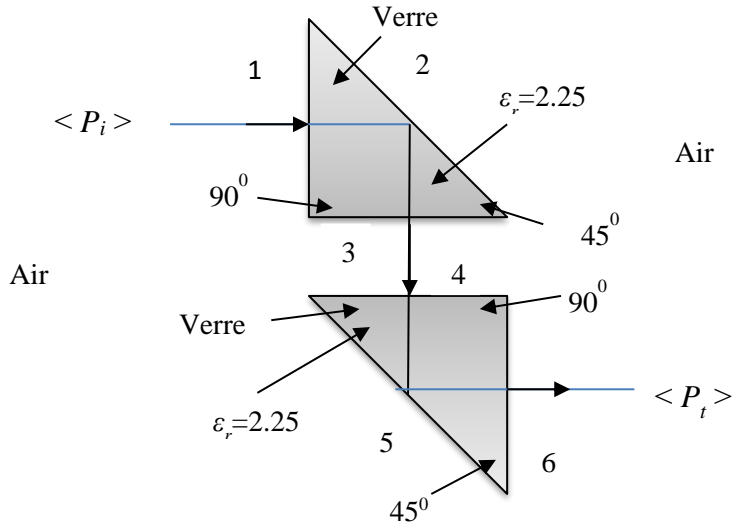

<div align="center">Figure II.34.</div>

Par conséquent, une réflexion interne totale a lieu à ces interfaces. Aux interfaces 1, 3, 4 et 6, le rapport de la densité puissance transmise à cet incident est $[\,1 - \rho^2]$.

Donc,

$$\frac{<P_t>}{<P_i>} = [\,1 - \rho^2]^4 = 0.85 = 85\ \%$$

**Exemple 25.** Une antenne de 0.5 ($MH_z$) portée par un avion survolant la surface de l'océan génère une onde qui s'approche de la surface de l'eau sous la forme d'une onde plane incidente avec une amplitude de champ électrique de 3000 ($V/m$). L'eau de mer a des caractéristiques suivantes $\varepsilon_r = 72$, $\mu_r = 1$, $\sigma = 4 \left(\frac{S}{m}\right)$. L'avion essaie de communiquer un message à un sous-marin immergé à une profondeur ($d$) sous la surface de l'eau. Si le récepteur du sous-marin une amplitude de signal minimal de 0.01 ($\mu V/m$) quelle est la profondeur maximale ($d$) à laquelle une communication réussie est encore possible?

**Solution :**

$[\frac{\sigma}{\omega\varepsilon} = 1000 \gg 1$, $\alpha_2 = 2.81 \left(\frac{Np}{m}\right)$, $\beta_2 = 2.81 \left(\frac{rad}{m}\right)$, $Z_2 = (1+j)\frac{\alpha_2}{\sigma}$

$Z_2 = 0.7(1 + j)\,(\Omega),\ \ \rho = (-0.9963 + j3.7x10^{-3}),$

$T = \ 1 + \rho = 5.24x10^{-3}e^{j44.98^0},\ |E_t| = |\,TE_0^i\,e^{-2\alpha_2 d}|,\ \text{Pour } E_t = 0.01\mu V/m\ \rightarrow$

$E_t = 10^{-8}V/m, \qquad 10^{-8} = \ 5.24x10^{-3}x3x10^3\,e^{-2x2.81d}\ \rightarrow\ \ d = 7.54\ (m)]$

**Exemple 26.** Une onde électromagnétique plane, linéairement polarisée passe d'un diélectrique a un autre. Le plan de séparation des milieux est le plan *xOz* des coordonnées cartésiennes. Le point de l'impact est le point *O* (l'origine des coordonnées). L'axe *Oy* constitue la normale au plan de séparation et le milieu «1 » se trouve du côté des « *y* » positifs. Le phaseur du vecteur électrique transmis est :

$$\bar{E}_t = 10\big(-\sqrt{0.7}\,\hat{x} + \sqrt{0.3}\,\hat{y}\big)e^{j\pi\left(x\sqrt{0.3}+y\sqrt{0.7}\right)}\,(\frac{V}{m})$$

La constante diélectrique relative du second milieu est 4.

    a) Identifier la valeur de la constante de phase $\beta_2$ dans l'expression de $\bar{E}_t$ et calculer par la suite la longueur dans le milieu «2 » ainsi que la fréquence du signal,

    b) Si l'angle d'incidence est supposé d'être l'angle de Brewster, calculer $\varepsilon_{r1}$ et $\theta_{iB}$ à partir de $\theta_i$ dans l'expression de $\bar{E}_t$,

    c) Quelle est le vecteur unitaire qui indique la direction de propagation dans le deuxième milieu?

    d) Trouver le vecteur phaseur du champ électrique incident dans le premier milieu ainsi que sa valeur instantanée,

    e) Calculer la puissance incidente, réfléchie et transmise par cette onde.

**Solution :**

[a] À partir du champ électrique transmis Il s'agit du cas parallèle (||) au plan d'incidence qui est *xOy*.

$\theta_t = \ 33.211^0,\ \ \beta_2 = \ \pi\left(\frac{rad}{m}\right),\ \ \lambda_2 = 2\ (m),\ \ f = (75\ MH_z); \text{b) } \varepsilon_{r1} = 1.714$

$\theta_{iB} = 56.789^0\ ;\ \ c)\ \theta_i = \ \theta_{iB} = 56.789^0\,,\beta_1 = \ 2.056\left(\frac{rad}{m}\right),T_{||} = 0.665;$

d) $\bar{E}_i = \frac{|\bar{E}_t|}{T_\parallel} \left(-cos\theta_i\,\hat{x} + sin\theta_i\,\hat{y}\right)e^{j\beta_1(sin\theta_i x + cos\theta_i y)}$

$\bar{E}_i = 15.28\left(-\sqrt{0.3}\,\hat{x} + \sqrt{0.7}\,\hat{y}\right)e^{j2.056(\sqrt{0.7}x + y\sqrt{0.3})}\left(\frac{V}{m}\right)$

$\bar{E}_i(x,y,z,t) = 15.28\left(-\sqrt{0.3}\,\hat{x} + \sqrt{0.7}\,\hat{y}\right)cos(150\pi x 10^6 t + 1.72x + 1.13y)\left(\frac{V}{m}\right)$

e) $<P_i> = \frac{1}{2Z_1}|\bar{E}_i|^2 = 0.405\left(\frac{W}{m^2}\right), <P_t> = \frac{1}{2Z_2}|\bar{E}_i|^2 = 0.265\left(\frac{W}{m^2}\right),$

$\rho_{\parallel} = -000043, <P_r> = \frac{1}{2Z_1}|\bar{E}_r|^2 = \frac{1}{2Z_1}\rho^2_{\parallel}|\bar{E}_i|^2 = 7.54x10^{-10}\left(\frac{W}{m^2}\right),$

La densité de puissance réfléchie est négligeable (angle de Brewster).

Remarque:    $<P_t> x \frac{cos\theta_t}{cos\theta_i} = <P_i> \rightarrow 0.222\left(\frac{W}{m^2}\right) = 0.222\left(\frac{W}{m^2}\right),$ !!! C'est la question de la surface qui convertie à l'interface].

**Exemple 27.** La lumière naturelle est polarisée de façon aléatoire, ce qui signifie qu'en moyenne, la moitié de l'énergie lumineuse est polarisée dans une direction donnée (dans le plan orthogonal à la direction de propagation) et l'autre moitié de l'énergie est polarisée le long de la direction orthogonale à la première direction de polarisation. Par conséquent, lors du traitement naturel incident de lumière sur une frontière plane, on peut considérer que la moitié de son énergie est la forme d'ondes polarisées en parallèle et l'autre moitié comme l'onde polarisée perpendiculaire. Déterminer la fraction de la puissance incidente réfléchie par la surface d'un morceau de verre avec $\varepsilon_{r2} = 2.25$ lorsque il est éclairé par la lumière naturelle à $70^0$.

**Solution :**
[Supposons que la puissance incidente soit de 1 ($W$). Donc, la puissance incidente avec la polarisation perpendiculaire est de 0.5 ($W$) et la puissance incidente avec polarisation parallèle est de 0.5 ($W$). Le rapport $\frac{\varepsilon_2}{\varepsilon_1} = 2.25$. $Z_1 = Z_0 = 120\pi$ $(\Omega)$,

$Z_2 = \frac{1}{\sqrt{\varepsilon_{r2}}} Z_0 = 80\pi$ $(\Omega)$, $\rho_\perp = -0.55$, $\rho_{\parallel} = 0.21$,

La puissance réfléchie avec polarisation perpendiculaire est $0.5(\rho_\perp)^2 = 151.3$ $(mW)$ et la puissance réfléchie avec polarisation parallèle est $0.5(\rho_{\parallel})^2 = 22$ $(mW)$. La puissance totale réfléchie est $(22 + 151.3)(mW) = 173.3(mW)$   $ou$ $17.33\%$]

**Exemple 28.** Ce problème combine les notions de l'incidence normale et l'incidence oblique. Une onde électromagnétique plane de fréquence 200 ($MH_z$) se propage dans deux diélectriques, puis frape une surface métallique a une incidence normale (Figure II.35). Le point de l'impact entre deux diélectriques est le point $O$, l'origine des coordonnés cartésiens. La surface de séparation entre les deux diélectriques est le plan ($xOy$). Les milieux « 1 » et « 2 » sont des diélectriques sans pertes ($\varepsilon_1 = \varepsilon_0 (12 - j4) \left(\frac{F}{m}\right)$, $\varepsilon_1 = \varepsilon_0 (3 - j) \left(\frac{F}{m}\right)$ ). Pour tous les milieux on considère $\mu = \mu_0$. Le champ électrique incident dans le milieu « 1 » se trouve dans un plan d'incidence ($xOy$) et fait un angle d'incidence avec l'axe tel que $sin\theta_i = 0.25$. Le troisième milieu qui est un parfait conducteur est placé de façon perpendiculaire au rayon transmis dans le deuxième milieu.

    a) Calculer les facteurs de propagation dans les deux milieux diélectriques,
    b) Calculer les longueurs d'onde dans ces milieux,
    c) Trouver l'angle de transmission $\theta_t$,
    d) Si la distance $OB$ est égale à $3\lambda$ ($\lambda$ : longueur d'onde dans le milieu « 2 », calculer en $dB$ la perte en puissance de l'onde polarisée linéairement pour un aller-retour sur le chemin $OB$.
    e) Si l'on garde l'angle d'incidence $\theta_i$ tel que $sin\theta_i = 0.25$ , comment faut-il changer $\varepsilon_1$ pour qu'il n'y a pas de réflexion dans le milieu « 1 » ?

**Solution :**

[a) $\gamma_1 = 14.876 e^{80.775^0} \left(\frac{1}{m}\right), \gamma_2 = 7.448 e^{80.775^0} \left(\frac{1}{m}\right)$; b) $\beta_1 = 14.703 \left(\frac{rad}{m}\right)$,

$\lambda_1 = 0.427\ (m)$, b) $\beta_2 = 7.351 \left(\frac{rad}{m}\right), \lambda_2 = 0.855\ (m)$; c) $\theta_t = 30^0$;

d) $OB = 3\lambda_2 = 2.565\ (m)$,    $\lambda_2 = Re(\gamma_2) = 1.194\ (\frac{Np}{m})$

À l'incidence normale sur un conducteur parfait il n'y a pas de puissance à la réflexion. Il y a seulement un changement de phase de $180^0$. Dans un aller-retour, les champs vont subir un affaiblissement de

$e^{-\lambda_2 x(2OB)} = Aff \rightarrow\ Aff = 0.00219\ (Np) \rightarrow Att(dB) = 20 log_2(0.00219) \rightarrow$

$Att(dB) = -53.201\ (dB)$;  e) $tan\theta_{iB} = 0.258 = \sqrt{\dfrac{\varepsilon_2}{\varepsilon_1}}$ ,

$\varepsilon_1 = 3.984 x 10^{-10} - j\, 2.656 x 10^{-11}]$

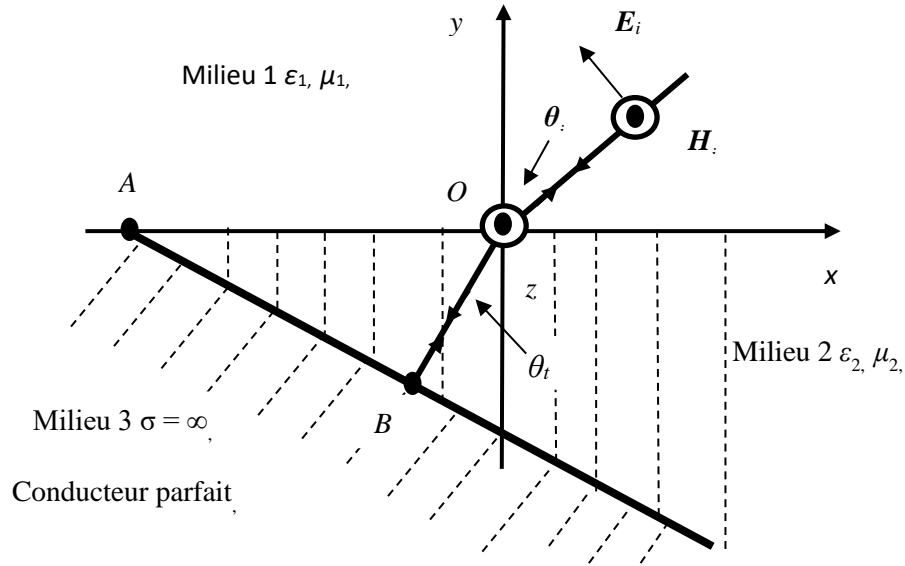

Figure II.35.

**Exemple 29.** Une onde plane de polarisation parallèle est incidente de l'air à un angle $\theta_i = 30^0$ sur une paire de couches diélectriques comme indique sur la Figure II.36.

- a) Déterminer les angles de transmission $\theta_2, \theta_3$ et $\theta_4$.
- b) Déterminer la distance latérale ($d$).

**Solution :**

[a) $\theta_2 = 11.54^0, \theta_3 = 19.48^0$ et $\theta_4 = 30^0$; b) $d = 2.97\ (cm)$]

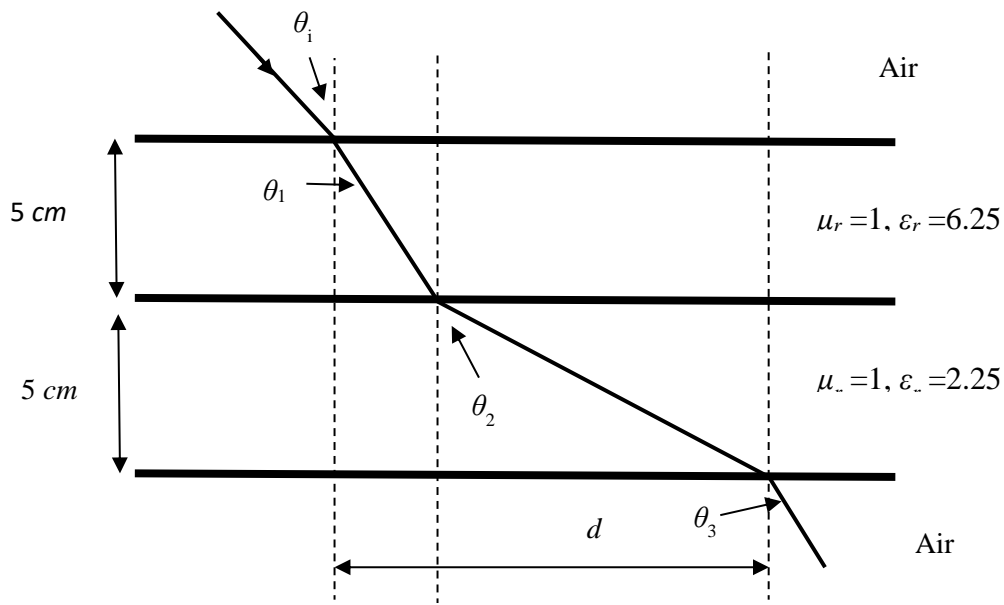

<div align="center">Figure II.36.</div>

**Exemple 30.** Une onde plane dans l'air avec,

$$\bar{E}_i = (2\hat{x} - 4\hat{y} - 6\hat{z})e^{-j(2x+3z)} \ (\frac{V}{m})$$

est incident sur la surface plane d'un matériau diélectrique, avec $\varepsilon_{r2} = 2.25$ , occupant le demi-espace z ≥ 0 (Figure II.37). Déterminer :

a)  L'angle d'incidence $\theta_i$ et l'angle $\theta_t$,

b)  La fréquence de l'onde,

c)  Le champ $\bar{E}_r$ de l'onde réfléchie,

d)  Le champ $\bar{E}_t$ de l'onde transmise dans le milieu diélectrique,

e)  La densité de puissance moyenne portée par l'onde dans le milieu diélectrique.

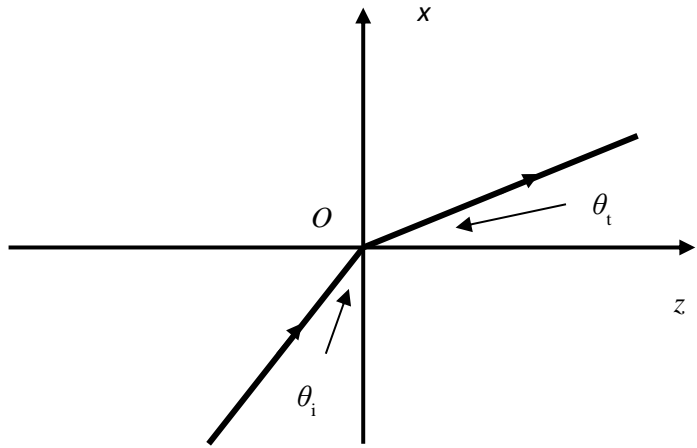

Figure II.37.

Solution :

[a) $\beta_1 = 3.6 \left(\frac{rad}{m}\right)$, $\theta_i = \theta_r = 33.7^0$, $\theta_t = 21.7^0$ ; b) $f = 173 \ (MH_z)$;  c) Du champ incident $\bar{E}_i$ on voit que c'est un mélange de parallèle et de perpendiculaire composantes de polarisation. $\bar{E}_i = \bar{E}_{i\perp} + \bar{E}_{i||}$ où,

$$\bar{E}_{i\perp} = -4\hat{y}e^{-j(2x+3z)} \ (\frac{V}{m})$$
$$\bar{E}_{i||} = (2\hat{x} - 6\hat{z})e^{-j(2x+3z)} \ (\frac{V}{m})$$

$\bar{E}_{i\perp 0} = -4 \left(\frac{V}{m}\right)$, $\bar{E}_{i||0} = \sqrt{2^2 + 6^2} = 6.32 \left(\frac{V}{m}\right)$, $\rho_\perp = 0.25$, $T_\perp = 0.75$

$\rho_{||} = -0.15$, $T_{||} = 0.76$, $\bar{E}_{r\perp} = \hat{y}\bar{E}_{r\perp 0}e^{-j\beta_1(sin\theta_r x - cos\theta_r z)} \left(\frac{V}{m}\right)$

$\bar{E}_{t\perp} = \hat{y}\bar{E}_{t\perp 0}e^{-j\beta_2(sin\theta_t x + cos\theta_t z)} \left(\frac{V}{m}\right)$,

$\bar{E}_{r||} = (cos\theta_r \ \hat{x} + sin\theta_r \hat{z})\bar{E}_{r||0}e^{-j\beta_1(sin\theta_r x - cos\theta_r z)} \left(\frac{V}{m}\right)$,

$$\bar{E}_{t\|} = (cos\theta_r\ \hat{x} - sin\theta_r\hat{z})\bar{E}_{t\|0}e^{-j\beta_2(sin\theta_r x + cos\theta_r z)}\ \left(\frac{V}{m}\right),$$

$$\beta_2 = 5.4\ \left(\frac{rad}{m}\right),\ \bar{E}_{r\perp0} = \rho_\perp\bar{E}_{i\perp0} = 1\left(\frac{V}{m}\right), \bar{E}_{t\perp0} = T_\perp\bar{E}_{i\perp0} = -3\left(\frac{V}{m}\right),$$

$$\bar{E}_{r\|0} = \rho_{|\ \ |}\bar{E}_{i\|0} = -0.95\ \left(\frac{V}{m}\right), \bar{E}_{t\|0} = T_{|\ \ |}\bar{E}_{i\|0} = 4.8\ \left(\frac{V}{m}\right),$$

$$\bar{E}_r = \bar{E}_{r\perp} + \bar{E}_{r\|} = (-0.79\hat{x} + \hat{y} - 0.53\hat{z})\hat{y}e^{-j(2x-3z)}\ \left(\frac{V}{m}\right);$$

d)

$$\bar{E}_t = \bar{E}_{t\perp} + \bar{E}_{t\|} = (4.46\hat{x} - 3\hat{y} - 1.78\hat{z})\hat{y}e^{-j(2x+5z)}\ \left(\frac{V}{m}\right);$$

e)

$$<P_t> = \frac{|\bar{E}_{t0}|^2}{2Z_2} = \frac{1}{2x251.33}[4.46^2 + 3^2 + 1.78^2] = 63.8x10^{-3}(\frac{W}{m^2})]$$

**Exemple 31.** Une onde plane de 50 ($MH_z$) avec une amplitude de champ électrique de 50 ($V/m$) est normalement incident dans l'air sur un milieu diélectrique parfait semi-infini avec $\varepsilon_r = 36$. Déterminer :
  a) Le coefficient de réflexion,
  b) Les densités de puissance moyennes de l'onde incidente et réfléchies,
  c) La distance dans le milieu aérien de la frontière jusqu'au minimum de la plus proche intensité du champ électrique, $|\bar{E}|$,
  d) Quelle est l'amplitude maximale du champ électrique total dans l'air?

**Solution :**
[a) $\rho = -0.71$; b) $<P_{moy}^i> = 3.32\ (W/m^2)$, $<P_{moy}^r> = 1.67\ (W/m^2)$;
c) $\lambda_0 = 6\ (m)$, $l_{max} = \frac{\theta_r\lambda_0}{4\pi}$, $l_{min} = l_{max} - \frac{\lambda_0}{4} = 0\ (m)$. Donc, sur la frontière;
d) $|\bar{E}_t| = (1 + |\rho|)|\bar{E}| = 85.5\ \left(\frac{V}{m}\right)$ à $l_{max} = 1.5\ (m)]$

**Exemple 32.** Démontrer que pour un milieu non-magnétique le coefficient de réflexion $\rho_\perp$ eut être écrit sous la forme $\rho_\perp = \frac{\sin(\theta_t - \theta_i)}{\sin(\theta_t + \theta_i)}$.

**Exemple 33.** Démontrer que pour un milieu non-magnétique le coefficient de réflexion $\rho_{\parallel}$ peut être écrit sous la forme $\rho_{\parallel} = \frac{\tan\,(\theta_i - \theta_t)}{\tan\,(\theta_t + \theta_i)}$.

**Exemple 34.** Démontrer que pour un milieu non-magnétique le coefficient de transmission $T_{\perp}$ peut être écrit sous la forme $T_{\perp} = 2\frac{\cos\theta_i \sin\,\theta_t}{\sin\,(\theta_t + \theta_i)}$.

**Exemple 35.** Démontrer que pour un milieu non-magnétique le coefficient de transmission $T_{\parallel}$ peut être écrit sous la forme $\rho_{\parallel} = 2\frac{\cos\theta_i \sin\,\theta_i}{\sin(\theta_t + \theta_i)\cos\,(\theta_i - \theta_t)}$.

**Exemple 36.** Une onde plane uniforme se propage à la fréquence de $10^5$ ($H_z$) dans un bon conducteur. On note que les champs s'atténuent par un facteur de $e^{-2\pi}$ sur une distance de 2.5 ($m$). Trouver :

a) La distance sur laquelle les champs subissent un changement de phase de $2\pi$ ($rad$) à un temps fixé.

b) La distance parcourue en 1($\mu s$) aux fréquences de $10^5$ ($H_z$) et $10^4$ ($H_z$) en supposant que les paramètres électriques n'ont pas changé.

Solution :
[a) 5 ($m$); b) à $10^5$ ($H_z$) $d$ = 5 ($m$); à $10^4$ ($H_z$) $d$ = 0.1581 ($m$)]

**Exemple 37.** La transmission de la lumière à traverse des fibres filamenteuses (guides d'onde diélectrique) est basée sur le phénomène de réflexion interne totale, qui se produit lorsque la lumière est incidente obliquement sur une interface entre deux milieux différents à un angle supérieur à l'angle critique. Considère la diélectrique fibre (Figure II.38) dont la permittivité diélectrique relative est inconnue. Il est nécessaire de déterminer la plage de valeurs de la permittivité relative de telle sorte que la réflexion interne se produise pour toute valeur de l'angle incident.

Solution : [ $\varepsilon_r \geq 1 + [sin\theta_i]^2$. Si cette condition était remplie pour l'incidence du pâturage ($\theta_i = \frac{\pi}{2}$) la lumière incidente serait passée par le tuyau ce qui nécessite $\varepsilon_r \geq 2$]

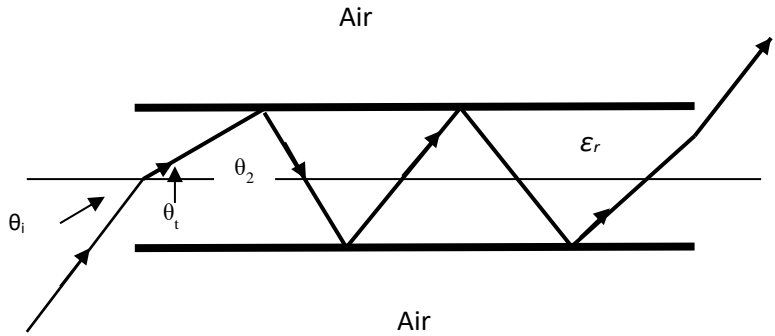

Air

θ$_i$     θ$_t$   θ$_2$        ε$_r$

Air

<div align="center">Figure II.38.</div>

**Exemple 38.** Déterminer la polarisation des champs suivants,

a)   $\bar{E}(z) = 5(\hat{x} - j\hat{y})e^{-j\beta z} \left(\frac{V}{m}\right)$

b)   $\bar{E}(z) = 5(\hat{x} + j\hat{y})e^{-j\beta z} \left(\frac{V}{m}\right)$

c)   $\bar{E}(z) = 5(\hat{x} - j\hat{y})e^{j\beta z} \left(\frac{V}{m}\right)$

d)   $\bar{E}(z) = 5(\hat{x} + j\hat{y})e^{j\beta z} \left(\frac{V}{m}\right)$

e)   $\bar{E}(z) = (\hat{x} - j2\hat{y})e^{-j\beta z} \left(\frac{V}{m}\right)$

f)   $\bar{E}(z) = (3\hat{x} + j\hat{y})e^{-j\beta z} \left(\frac{V}{m}\right)$

g)   $\bar{E}(z) = (\hat{x} - j2\hat{y})e^{j\beta z} \left(\frac{V}{m}\right)$

h)   $\bar{E}(z) = (3\hat{x} + j\hat{y})e^{j\beta z} \left(\frac{V}{m}\right)$

i)   $\bar{E}(z) = -j5\hat{x}e^{-j\beta z} \left(\frac{V}{m}\right)$

j)   $\bar{E}(z) = (2\hat{x} + 3\hat{y})e^{j\beta z} \left(\frac{V}{m}\right)$

k)   $\bar{E}(z) = (-5\hat{x} + 4\hat{y})e^{-j\beta z} \left(\frac{V}{m}\right)$

**Solution :** [a) Circulaire, polarisé à droite, vers l'avant; b) Circulaire, polarisé à gauche, vers l'avant; c) Circulaire, polarisé à gauche, reculé; d) Circulaire, polarisé à droite, reculé; e) Elliptique, polarisé à droite, vers l'avant; f) Elliptique, polarisé à gauche, vers l'avant; g) Elliptique, polarisé à gauche, reculé; h) Elliptique, polarisée à droite, reculé; i) Linéaire, avant; j) Linéaire, arrière; k) Linaire, avant]

## Exemple 39.

Il est souhaitable de réchauffer la purée de pommes de terre surgelée et les carottes cuites surgelées dans un four à micro-ondes fonctionnant à 2.45 GHz. Déterminer la profondeur de pénétration et évaluer son efficacité méthode de chauffage. De plus, déterminer l'atténuation du champ électrique (en dB et en absolu unités) à une profondeur de 1 cm de la surface de l'aliment. Les constantes diélectriques de la purée de pommes de terre et les carottes sont a) $\varepsilon_{r(p-t)} = (65 - j25)\varepsilon_0, \varepsilon_{r(car)} = (75 - j25)\varepsilon_0,$  b)  $\varepsilon_0 = \frac{1}{36\pi} \, x 10^{-9} \left(\frac{F}{m}\right).$

**Solution :**

[ a) $\beta_0 = 51.32 \left(\frac{rad}{m}\right),$      $\beta_{(p-t)} = 421 \left(\frac{rad}{m}\right),$ $\alpha_{(p-t)} = 78.19 \left(\frac{Np}{m}\right) =$

$8.686 x 78.19 \left(\frac{dB}{m}\right) = 679.16 \left(\frac{dB}{m}\right),$     $\delta_{(p-t)} = \frac{1}{\alpha_{(p-t)}} = 0.013 \, (m)$

$= 1.3 (cm),$   $Att_{(p-t)(z=1cm)}(dB) = 8.686 x \frac{z}{\delta_{(p-t)}} = 6.682 \, (dB),$

$10^{-\frac{Att_{(p-t)(z=1cm)}}{20}} = 0.4634 = 46.34\%;$

b) $\beta_{(car)} = 421 \left(\frac{rad}{m}\right), \beta_{(car)} = 450.27 \left(\frac{rad}{m}\right), \alpha_{(car)} = 72.99 \left(\frac{Np}{m}\right) =$

$8.686 x 72.99 \left(\frac{dB}{m}\right) = 633.99 \left(\frac{dB}{m}\right),$

$\delta_{(car)} = \frac{1}{\alpha_{(car)}} = 0.0137 \, (m) = 1.37 (cm),$ $Att_{(car)(z=1cm)}(dB) = 8.686 x \frac{z}{\delta_{(p-t)}}$

$= 6.34 \, (dB),$  $10^{-\frac{Att_{(car)(z=1cm)}}{20}} = 0.4819 = 48.19\%$ ]

# Chapitre III

# 1. Guides d'ondes

## 1.1 HISTOIRE

Les travaux de M. Chu dans les années 1939, où une de recherche était très riche en découvertes à cause des mises au point des premiers radars, ont montré que l'on pouvait transmettre à longue distance les ondes électromagnétiques à l'intérieur d'un tube métallique. C'est ainsi que fut créer les premiers guides d'onde à section rectangulaire, circulaire et elliptique.

## 1.2 AVANTAGE D'UTILISER LES GUIDES D'ONDE

Les principaux avantages d'une telle transmission résident dans les faits suivants:

1) Pertes relativement faibles par rapport à d'autres systèmes de transmission pour les plus hautes fréquences (ondes millimétriques), les guides diélectriques NRD (non radiating dielectrics) jouent le même rôle. Pour les fréquences optiques on est obligatoirement lié au choix d'une fibre optique qui est un diélectrique.

2) Permet de transmettre de grandes puissances. L'arrivée des supraconducteurs pratiques peut encore améliorer la situation actuelle.

3) Permet une bonne isolation avec le milieu environnant. Dans l'état actuel des choses, on a tendance à oublier cette vertu des guides d'onde. "Personnal Wireless Communication" thème à la mode souffre des interférences vu le trafic chaotique des communications rendues à la disposition de l'homme de la rue (téléphone cellulaire). En plus, un tel trafic en espace libre crée une pollution électromagnétique dont les effets biologiques à long terme sont en encore inconnus à la science.

4) Permet le design et la construction de toutes sortes de dispositifs fonctionnant à hautes fréquences.

Le but visé de cette étude peut se résumer en deux points importants:

a) Obtenir des expressions mathématiques concernant la distribution spatiale des ondes électromagnétiques à l'intérieur d'un guide.

b) Obtenir différents paramètres caractéristiques de la transmission propres aux guides d'onde.

La méthode consiste à reprendre le développement théorique qui a été utilisé pour les ondes planes avec certaines restrictions dues aux conditions frontières qui sont imposées par les parois du guide et avec un peu moins de simplification à cause de la géométrie présente.

# 2.  Guides d'onde rectangulaires

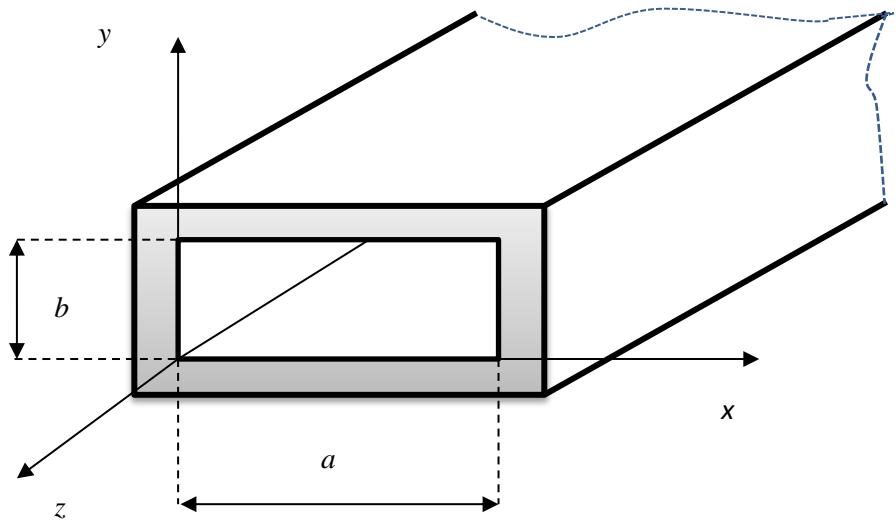

Figure III.1.

L'intérieur du guide (Figure III.1) est un milieu homogène, isotrope, sans charges d'espace et sans pertes (en général c'est de l'air)

Les équations d'Helmholtz donnent:

$$\nabla^2 \bar{E} + \omega^2 \mu\varepsilon \bar{E} = 0$$

$$\nabla^2 \bar{H} + \omega^2 \mu\varepsilon \bar{H} = 0$$

Dans ces équations, on a toujours trois composantes du champ électrique et trois composantes du champ magnétique ayant la même forme de solution.

On peut en choisir une (soit $H_z$) pour en obtenir une typique.

$$\frac{\partial^2 H_z}{\partial x^2} + \frac{\partial^2 H_z}{\partial y^2} + \frac{\partial^2 H_z}{\partial z^2} + \omega^2 \mu\varepsilon H_z = 0$$

En utilisant la séparation des variables, on pose

$$H_z(x, y, z) = X(x)Y(y)Z(z)$$

dans l'équation différentielle ce qui donne,

$$YZ \frac{d^2 X(x)}{dx^2} + XZ \frac{d^2 Y(y)}{dy^2} + XY \frac{d^2 Z(z)}{dz^2} = -\omega^2 \mu\varepsilon XYZ$$

En divisant partout par $XYZ$ on aura,

$$\frac{1}{X} \frac{d^2 X}{dx^2} + \frac{1}{Y} \frac{d^{2Y}}{dy^2} + \frac{1}{Z} \frac{d^2 Z}{dz^2} = -\omega^2 \mu\varepsilon$$

Pour pouvoir satisfaire cette égalité, sachant que chaque fonction est indépendante à cause des variables affectées il faut que chaque terme de la partie de gauche soit une constante.

$$\frac{1}{X} \frac{d^2 X}{dx^2} = -A^2 \quad \rightarrow \quad \frac{d^2 X}{dx^2} + A^2 X = 0$$

$$\frac{1}{Y} \frac{d^2 Y}{dy^2} = -B^2 \quad \rightarrow \quad \frac{d^2 Y}{dy^2} + B^2 Y = 0$$

$$\frac{1}{Z} \frac{d^2 Z}{dz^2} = +\gamma^2 \quad \rightarrow \quad \frac{d^2 Z}{dz^2} - \gamma^2 Z = 0$$

La direction « *z* » a été privilégiée pour la transmission. D'autre part il ne faut pas perdre de vue la condition importante

$$-A^2 - B^2 + \gamma^2 = -\omega^2\mu\varepsilon$$

Il est évident que le choix des signes a été fait en accord avec l'orientation du guide choisie précédemment. La direction axiale du guide est orientée selon l'axe « *z* ».

La solution générale pour ces équations différentes peut donc s'écrire:

$$H_z(x, y, z) = X(x)Y(y)Z(z)$$
$$= (C_1 sinAx + C_2 cosAx)(C_3 sinBy + C_4 cosBy)(C_5 e^{\gamma z} + C_6 e^{-\gamma z})$$

La même forme de solution générale est valable pour les 5 autres composantes électromagnétiques.

## 2.1 RELATIONS ENTRE LES COMPOSANTES DU CHAMP **EM**

$$\nabla \times \bar{E} = -j\omega\mu\bar{H} \qquad\qquad \nabla \times \bar{H} = j\omega\varepsilon\bar{E}$$

$$\nabla \times \bar{E} = \begin{bmatrix} \hat{x} & \hat{y} & \hat{z} \\ \dfrac{\partial}{\partial x} & \dfrac{\partial}{\partial y} & \dfrac{\partial}{\partial z} \\ E_x & E_y & E_z \end{bmatrix} \qquad \nabla \times \bar{H} = \begin{bmatrix} \hat{x} & \hat{y} & \hat{z} \\ \dfrac{\partial}{\partial x} & \dfrac{\partial}{\partial y} & \dfrac{\partial}{\partial z} \\ H_x & H_y & H_z \end{bmatrix}$$

$$\nabla \times \bar{E} = \hat{x}\left(\frac{\partial E_z}{\partial y} - \frac{\partial E_y}{\partial z}\right) - \hat{y}\left(\frac{\partial E_z}{\partial x} - \frac{\partial E_x}{\partial z}\right) + \hat{z}\left(\frac{\partial E_y}{\partial x} - \frac{\partial E_x}{\partial y}\right)$$

$$\nabla \times \bar{H} = \hat{x}\left(\frac{\partial H_z}{\partial y} - \frac{\partial H_y}{\partial z}\right) - \hat{y}\left(\frac{\partial H_z}{\partial x} - \frac{\partial H_x}{\partial z}\right) + \hat{z}\left(\frac{\partial H_y}{\partial x} - \frac{\partial H_x}{\partial y}\right)$$

$$\frac{\partial E_z}{\partial y} - \frac{\partial E_y}{\partial z} = -j\omega\mu H_x \qquad\qquad \frac{\partial H_z}{\partial y} - \frac{\partial H_y}{\partial z} = j\omega\varepsilon E_x$$

$$\frac{\partial E_z}{\partial x} - \frac{\partial E_x}{\partial z} = j\omega\mu H_y \qquad\qquad \frac{\partial H_z}{\partial x} - \frac{\partial H_x}{\partial z} = -j\omega\varepsilon E_y$$

$$\frac{\partial E_y}{\partial x} - \frac{\partial E_x}{\partial y} = -j\omega\mu H_z \qquad\qquad \frac{\partial H_y}{\partial x} - \frac{\partial H_x}{\partial y} = j\omega\varepsilon E_z$$

Puisque l'axe « z » a été choisi comme la direction de propagation, on portera plus d'attention sur l'onde qui se transmet dans la direction positive de « z ». Puisque chaque composante du champ EM varie, selon ce choix, en termes de $e^{\gamma z}$, toutes les dérivées par rapport à « z » ne représentent aucune difficulté. Le tableau ci-dessus se transforme aisément comme suit:

$$\frac{\partial E_z}{\partial y} + \gamma E_y = -j\omega\mu H_x \qquad\qquad \frac{\partial H_z}{\partial y} + \gamma H_y = j\omega\varepsilon E_x$$

$$\frac{\partial E_z}{\partial x} + \gamma E_x = j\omega\mu H_y \qquad\qquad \frac{\partial H_z}{\partial x} + \gamma H_x = -j\omega\varepsilon E_y$$

$$\frac{\partial E_y}{\partial x} - \frac{\partial E_x}{\partial y} = -j\omega\mu H_z \qquad\qquad \frac{\partial H_y}{\partial x} - \frac{\partial H_x}{\partial y} = j\omega\varepsilon E_z$$

On y distingue deux catégories de composantes:

- composantes dans la direction de propagation : $E_z$ , $H_y$

- composantes transversales à la propagation : $E_x$ , $E_y$ , $H_x$ , $H_y$

Dans la démarche étudiée on se propose d'exprimer les 4 composantes transversales en fonction de 2 composantes longitudinales, autrement dit:

$$E_x = f(E_z , H_z)$$

$$E_y = f(E_z , H_z)$$

$$H_x = f(E_z , H_z)$$

$$H_y = f(E_z , H_z)$$

On précise,

$$E_x = \frac{1}{j\omega\varepsilon}\left(\frac{\partial H_z}{\partial y} + \gamma H_y\right) = \frac{1}{j\omega\varepsilon}\left\{\frac{\partial H_z}{\partial y} + \frac{\gamma}{j\omega\mu}\left(\frac{\partial E_z}{\partial x} + \gamma E_x\right)\right\}$$

d'où

$$E_x\left(1 + \frac{\gamma^2}{\omega^2\mu\varepsilon}\right) = \frac{1}{j\omega\varepsilon}\frac{\partial H_z}{\partial y} - \frac{\gamma}{\omega^2\mu\varepsilon}\frac{\partial E_z}{\partial x}$$

$$\rightarrow E_x = \frac{\omega^2\mu\varepsilon}{j\omega\varepsilon(\omega^2\mu\varepsilon + \gamma^2)}\frac{\partial H_z}{\partial y} - \frac{\gamma\omega^2\mu\varepsilon}{\omega^2\mu\varepsilon(\omega^2\mu\varepsilon + \gamma^2)}\frac{\partial E_z}{\partial x}$$

En posant,

$$h^2 = \omega^2\mu\varepsilon + \gamma^2$$

on aura,

$$E_x = -\frac{j\omega\mu}{h^2}\frac{\partial H_z}{\partial y} - \frac{\gamma}{h^2}\frac{\partial E_z}{\partial x}$$

$$E_y = +\frac{j\omega\mu}{h^2}\frac{\partial H_z}{\partial x} - \frac{\gamma}{h^2}\frac{\partial E_z}{\partial y}$$

$$H_x = -\frac{\gamma}{h^2}\frac{\partial H_z}{\partial x} + \frac{j\omega\varepsilon}{h^2}\frac{\partial E_z}{\partial y}$$

$$E_x = -\frac{\gamma}{h^2}\frac{\partial H_z}{\partial y} - \frac{j\omega\varepsilon}{h^2}\frac{\partial E_z}{\partial x}$$

Il faut se rappeler que

$$\omega^2\mu\varepsilon + \gamma^2 = A^2 + B^2$$

$$h^2 = A^2 + B^2$$

À ce niveau on constate la possibilité de scinder le problème en deux parties :

a) On considère le cas où $E_z = 0$ . Il reste pour le champ électrique $E_x$ et $E_y$ toutes les deux perpendiculaires à la direction de propagation.

<p style="text-align:center;">MODE TE          (Transverse Electric)</p>

b) ou bien $H_z = 0$. Il reste alors pour le champ magnétique $H_x$ et $H_y$ $\perp$ à la direction de propagation.

<p style="text-align:center;">MODE TM          (Transverse Magnetic)</p>

c) Si l'on imagine simultanément $E_z = H_z = 0$

Le mode impossible (MODE TEM) pour un guide d'onde formé d'une seule pièce.

La solution la plus générale pour les ondes qui peuvent se transmettre dans un guide sera en fait la somme de tous les modes TE plus tous les modes TM. Ceci n'est pas en général le cas dans la pratique courante de l'utilisation des guides d'onde.

On est souvent intéressé à ne transmettre que quelques-uns de ces modes à la fois.

## 2.2 Mode TM

On s'intéresse aux champs dans le guide (Figure III.2).

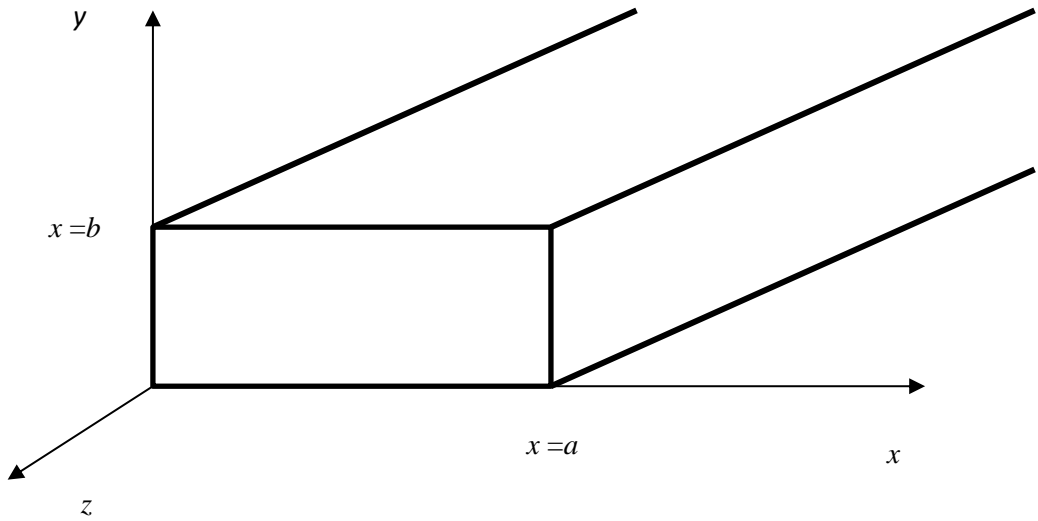

Figure III.2.

Une forme de solution générale pour $E_z$ sera comme:

$$E_z(x,y,z) = (C_1 \sin Ax + C_2 \cos Ax)(C_3 \sin By + C_4 \cos By\,)e^{\gamma z}$$

Conditions aux frontières (parois conducteurs) :

$E_{tan} = 0$ sur les surfaces parfaitement conductrices

$E_x$ tangentiel au mur du haut et du bas

$E_y$ tangentiel au mur de gauche et de droite

$E_z$ tangentiel à tous les murs

$E_z = 0$ $pour$ $x = 0$ $et$ $y = 0$

$E_z = 0$ $pour$ $x = a$ $et$ $y = b$

Lorsque $x = 0$ et $y = 0$ $\rightarrow$ $C_2 = 0$ , $C_4 = 0$

$$E_z(x,y,z) = C_1 C_3 \sin Ax \, \sin By \, e^{-\gamma z}$$

Lorsque,

$$y = b \rightarrow \sin Bb = 0$$

$$x = a \;\rightarrow\; \sin Aa = 0$$

Donc,

$$A = \frac{m\pi}{a} \qquad m = 1,2,3,\dots\infty$$

$$B = \frac{n\pi}{b} \qquad n = 1,2,3\dots\infty$$

Les valeurs de $m$ = 0 ou $n$ = 0 sont impossibles puisque toutes les composantes du champ *EM* deviennent nulles à l'intérieur du guide rectangulaire.

### 2.3 Mode TM$_{mn}$

$$E_z(x,y,z) = C \sin\frac{m\pi x}{a} \sin\frac{n\pi y}{b}\; e^{-\gamma x}$$

$$-\frac{\gamma}{h^2}\frac{\partial E_z}{\partial x} = E_x(x,y,z) = -\frac{\gamma}{h^2}\, C\left(\frac{m\pi}{a}\right)\cos\frac{m\pi x}{a}\sin\frac{n\pi y}{b}e^{-\gamma z}$$

$$-\frac{\gamma}{h^2}\frac{\partial E_z}{\partial y} = E_y(x,y,z) = -\frac{\gamma}{h^2}\, C\left(\frac{n\pi}{b}\right)\sin\frac{m\pi x}{a}\cos\frac{n\pi y}{b}e^{-\gamma z}$$

$$-\frac{j\omega\varepsilon}{h^2}\frac{\partial E_z}{\partial y} = H_x(x,y,z) = -\frac{j\omega\varepsilon}{h^2}C\left(\frac{n\pi}{b}\right)sin\frac{m\pi x}{a}cos\frac{n\pi y}{b}e^{-\gamma z}$$

$$-\frac{j\omega\varepsilon}{h^2}\frac{\partial E_z}{\partial x} = H_y(x,y,z) = -\frac{j\omega\varepsilon}{h^2}C\left(\frac{m\pi}{a}\right)cos\frac{m\pi x}{a}sin\frac{n\pi y}{b}e^{-\gamma z}$$

$$H_z(x,y,z) = 0$$

La solution générale implique la somme de tous les modes. Ce n'est pas un cas pratique.

### 2.4 Mode TE$_{mn}$

$$E_z = 0 \quad\quad H_z \neq 0$$

$$H_z(x,y,z) = (C_1 \sin Ax + C_2 \cos Ax)(C_3 \sin By + C_4 \cos By)e^{-\gamma z}$$

$$E_x = -\frac{j\omega\mu}{h^2}\frac{\partial H_z}{\partial y} \quad;\quad E_y = \frac{j\omega\mu}{h^2}\frac{\partial H_z}{\partial x}$$

où $E_x$ et $E_y$ sont les seules composantes du champ électrique qui peuvent être tangentielles aux murs parfaitement conducteurs du guide d'onde.

Les conditions frontières portent sur ces composantes.

Donc,

$$E_x = -\frac{j\omega\mu}{h^2}(C_1 \sin Ax + C_2 \cos Ax)(C_3 \cos By - C_4 \sin By)e^{-\gamma z}$$

$$E_y = \frac{j\omega\mu}{h^2}(C_1 A \cos Ax - C_2 A \sin Ax)(C_3 \sin By + C_4 \cos By)e^{-\gamma z}$$

Sur le mur $y = 0$, $E_x = 0 \rightarrow$ $C_3 = 0$

Sur le mur $x = 0$, $E_y = 0 \rightarrow$ $C_1 = 0$

Sur le mur $y = b$, $E_x = 0 \rightarrow$ $C_4 \sin Bb = 0$

$$B = \frac{n\pi}{b}; \qquad n = 0,1,2,3,\dots\infty$$

Sur le mur $x = a$, $E_y = 0 \rightarrow$ $C_2 \sin Aa = 0$

$$A = \frac{m\pi}{a}; \qquad m = 0,1,2,3,\dots\infty$$

$$H_z(x,y,z) = C \cos\frac{m\pi x}{a} \cos\frac{n\pi y}{b}\, e^{-\gamma z}$$

$$-\frac{\gamma}{h^2}\frac{\partial H_z}{\partial x} = H_x(x,y,z) = \frac{\gamma}{h^2}\, C\left(\frac{m\pi}{a}\right)\sin\frac{m\pi x}{a}\cos\frac{n\pi y}{b}e^{-\gamma z}$$

$$-\frac{\gamma}{h^2}\frac{\partial H_z}{\partial y} = H_y(x,y,z) = \frac{\gamma}{h^2}\, C\left(\frac{n\pi}{b}\right)\cos\frac{m\pi x}{a}\sin\frac{n\pi y}{b}e^{-\gamma z}$$

$$-\frac{j\omega\mu}{h^2}\frac{\partial H_z}{\partial y} = E_x(x,y,z) = \frac{j\omega\mu}{h^2}\, C\left(\frac{n\pi}{b}\right)\cos\frac{m\pi x}{a}\sin\frac{n\pi y}{b}e^{-\gamma z}$$

$$\frac{j\omega\mu}{h^2}\frac{\partial H_z}{\partial y} = E_y(x,y,z) = -\frac{j\omega\mu}{h^2}\, C\left(\frac{m\pi}{a}\right)\sin\frac{m\pi x}{a}\cos\frac{n\pi y}{b}e^{-\gamma z}$$

$$E_z(x,y,z) = 0$$

où l'on constate que $m = 0$ ou $n = 0$ est possible, mais pas m = 0 et $n = 0$ en même temps.

## 2.5   CARACTÉRISTIQUES DES MODES **TE$_{mn}$** ET **TM$_{mn}$** DANS LES GUIDES RECTANGULAIRES

$$-A^2 - B^2 + \gamma^2 = -\omega^2\mu\varepsilon$$

$$\gamma^2 = A^2 + B^2 - \omega^2\mu\varepsilon$$

$$\gamma^2 = \left(\frac{m\pi}{a}\right)^2 + \left(\frac{n\pi}{b}\right)^2 - \omega^2\mu\varepsilon$$

$$\gamma = \text{facteur de propagation}$$

$\gamma = \alpha + j\beta$

Si $\gamma$ est réel, l'onde est atténuée.
Si $\gamma$ est imaginaire, propagation sans perte.

Donc,

$Si\ \omega^2\mu\varepsilon > \left(\frac{m\pi}{a}\right)^2 + \left(\frac{n\pi}{b}\right)^2 \rightarrow \gamma : \text{imaginaire}$

$Si\ \omega^2\mu\varepsilon < \left(\frac{m\pi}{a}\right)^2 + \left(\frac{n\pi}{b}\right)^2 \rightarrow \gamma : \text{réel}$

La fréquence critique qui détermine le seuil lorsque,

$$\omega_c^2\mu\varepsilon = \left(\frac{m\pi}{a}\right)^2 + \left(\frac{n\pi}{b}\right)^2$$

$\omega_c$ = valeur critique de $\omega$, au-dessus de laquelle l'onde peut se transmettre ($\omega \geq \omega_c$) et en dessous de cette valeur ($\omega < \omega_c$) l'onde s'atténue énormément.

### 2.6 La fréquence de coupure

C'est cette valeur de $\omega$ qui définit la fréquence de coupure,

$$\omega_c^2\mu\varepsilon = \left(\frac{m\pi}{a}\right)^2 + \left(\frac{n\pi}{b}\right)^2$$

$$\omega_c = \frac{1}{\sqrt{\mu\varepsilon}}\sqrt{\left(\frac{m\pi}{a}\right)^2 + \left(\frac{n\pi}{b}\right)^2} = 2\pi f_c$$

$$f_c = \frac{1}{2\sqrt{\mu\varepsilon}}\sqrt{\left(\frac{m}{a}\right)^2 + \left(\frac{n}{b}\right)^2}$$

$\frac{1}{\sqrt{\mu\varepsilon}} = $ vitesse de phase de l'onde dans un milieu non guidé

$$\frac{1}{\sqrt{\mu_o \varepsilon_o}} = c = 3\times 10^8 (\frac{m}{s}) \quad (a \text{ et } b \text{ exprimés en mètre})$$

## 2.7 Facteur de phase

Lorsque $f > f_c$ le facteur de propagation devient imaginaire

$$\gamma = j\sqrt{\omega^2\mu\varepsilon - \left(\frac{m\pi}{a}\right)^2 - \left(\frac{n\pi}{b}\right)^2}$$

$$\beta = \sqrt{\omega^2\mu\varepsilon - \left(\frac{m\pi}{a}\right)^2 - \left(\frac{n\pi}{b}\right)^2}$$

$$\beta = \omega\sqrt{\mu\varepsilon}\sqrt{1 - \frac{\omega_c{}^2\mu\varepsilon}{\omega^2\mu\varepsilon}}$$

$$\beta = \omega\sqrt{\mu\varepsilon}\sqrt{1 - \left(\frac{f_c}{f}\right)^2} \ \ rad/m$$

Comparer cette expression à celle de l'onde plane en espace libre.

## 2.8 Vitesse de phase $(f > f_c)$

$$v_p = \frac{\omega}{\beta} = \frac{\omega}{\omega\sqrt{\mu\varepsilon}} \frac{1}{\sqrt{1 - \left(\frac{f_c}{f}\right)^2}}$$

$$v_p = \frac{1}{\sqrt{\mu\varepsilon}} \frac{1}{\sqrt{1 - \left(\frac{f_c}{f}\right)^2}}$$

Remarque: on constate une vitesse de phase en apparence plus grande que la vitesse de la lumière.

## 2.9 Longueur d'onde dans le guide

$$\lambda g = \frac{2\pi}{\beta} = \frac{2\pi}{\omega\sqrt{\mu\varepsilon}} = \frac{1}{\sqrt{1 - \left(\frac{f_c}{f}\right)^2}} = \frac{v_o}{f} \frac{1}{\sqrt{1 - \left(\frac{f_c}{f}\right)^2}}$$

$$\boxed{\lambda g = \frac{\lambda_o}{\sqrt{1 - \left(\frac{f_c}{f}\right)^2}}}$$

$v_o = $ vitesse de phase de l'onde $EM$ dans un milieu non guidé

$\lambda_o = $ longueur d'onde de l'onde $EM$ dans un milieu non guidé

## 2.10 Impédance des ondes TE et TM dans un guide rectangulaire

$$Z_{TM} = \frac{E_x}{H_y} = -\frac{E_x}{H_x} = \frac{\gamma}{j\omega\varepsilon} = \frac{j\beta}{j\omega\varepsilon} = \frac{\beta}{\omega\varepsilon} \quad si \ f > f_c$$

$$Z_{TE} = \frac{E_x}{H_y} = -\frac{E_y}{H_x} = \frac{j\omega\mu}{\gamma} = \frac{j\omega\mu}{j\beta} = \frac{\omega\mu}{\beta} \quad si \ f > f_c$$

En développant un peu plus ces expressions on aura,

$$Z_{TE} = \frac{\omega\mu}{\beta} = \frac{\omega\mu}{\omega\sqrt{\mu\varepsilon}} \frac{1}{\sqrt{1 - \left(\frac{f_c}{f}\right)^2}}$$

$$Z_{TE} = \sqrt{\frac{\mu}{\varepsilon}} \frac{1}{\sqrt{1 - \left(\frac{f_c}{f}\right)^2}} = \boxed{\frac{Z_0}{\sqrt{1 - \left(\frac{f_c}{f}\right)^2}} = Z_{TE}}$$

où $Z_0 = \sqrt{\frac{\mu}{\varepsilon}}$ correspond à la valeur obtenue pour une onde non guidée.

D'autre part,

$$Z_{TM} = \frac{\beta}{\omega\varepsilon} = \frac{\omega\sqrt{\mu\varepsilon}}{\omega\varepsilon}\sqrt{1-\left(\frac{f_c}{f}\right)^2}$$

$$Z_{TM} = \sqrt{\frac{\mu}{\varepsilon}}\sqrt{1-\left(\frac{f_c}{f}\right)^2} = \boxed{Z_0\sqrt{1-\left(\frac{f_c}{f}\right)^2} = Z_{TM}}$$

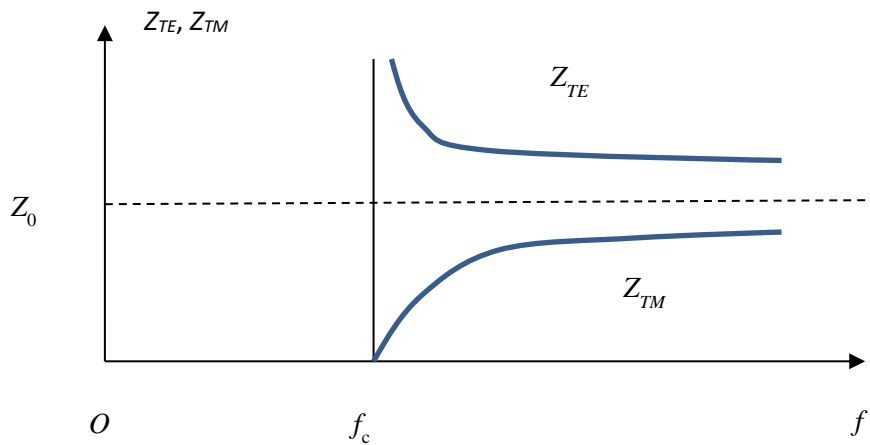

Figure III.3.

Remarque: $Z_{TE}$ et $Z_{TM}$ varient toutes les deux en fonction de la fréquence du générateur (Figure III.3). Pas très utile comme caractéristique d'un guide rectangulaire.

## 2.11 Diagramme $\omega - \beta$ (Figure III.4.)

$$\gamma^2 = \left(\frac{m\pi}{a}\right)^2 + \left(\frac{n\pi}{b}\right)^2 - \omega^2\mu\varepsilon$$

$$\gamma^2 = \omega_c^2\mu\varepsilon - \omega^2\mu\varepsilon = \mu\varepsilon(\omega_c^2 - \omega^2)$$

Pour une onde plane non guidée $\beta = \omega\sqrt{\mu\varepsilon}$

Figure III.4.

Lorsque

$$\omega < \omega_c \quad \alpha^2 = \omega_c^2 \mu\varepsilon - \omega^2 \mu\varepsilon$$

$$\alpha = \sqrt{\mu\varepsilon}\,\sqrt{\omega_c^2 - \omega^2} \qquad \text{onde atténuée}$$

## 2.12 Vitesse de groupe

C'est cette vitesse qui définit le transport d'information pour un signal modulé.

$$\text{vitesse de phase} \quad v_p = \frac{\omega}{\beta}$$

Pour les guides

$$v_g = \frac{\Delta\omega}{\Delta\beta} \qquad \text{vitesse à laquelle l'énergie est transmise}$$

$$v_g = \frac{\mathrm{d}\omega}{\mathrm{d}\beta} = \frac{1}{\dfrac{\mathrm{d}\beta}{\mathrm{d}\omega}} = v_p\sqrt{1 - \left(\frac{f_c}{f}\right)^2} \quad \text{si} \quad v_p = c, \qquad v_g = c\sqrt{1 - \left(\frac{f_c}{f}\right)^2}$$

## 2.13 MODE DOMINANT $TE_{10}$

a) Un seul mode qui peut être transmis sans risque d'interférence avec d'autres modes.

b) Fréquence de coupure la plus basse de tous les modes

c) Atténuation la plus faible (seulement guides d'onde rectangulaires!)

Si $m = 1$ et $n = 0$, on aura:

$$H_z = C \cos \frac{\pi x}{a} e^{-j\beta z}$$

$$H_x = \frac{j\beta}{h^2} \left(\frac{\pi}{a}\right) C \sin \frac{\pi x}{a} e^{-j\beta z}$$

$$E_y = -\frac{j\omega\mu}{h^2} \left(\frac{\pi}{a}\right) C \sin \frac{\pi x}{a} e^{-j\beta z}$$

avec

$$h^2 = \omega^2 \mu\varepsilon + \gamma^2 = \left(\frac{m\pi}{a}\right)^2 + \left(\frac{n\pi}{b}\right)^2$$

Pour le mode $TE_{10}$, $h^2 = \left(\frac{\pi}{a}\right)^2$

Donc,

✓ $\qquad H_z = C \cos \frac{\pi x}{a} e^{-j\beta z}$

✓ $\qquad H_x = j\beta \left(\frac{a}{\pi}\right) C \sin \frac{\pi x}{a} e^{-j\beta z}$

✓ $\qquad E_y = -j\omega\mu \left(\frac{a}{\pi}\right) C \sin \frac{\pi x}{a} e^{-j\beta z}$

## 2.14    REPRÉSANTATION GRAPHIQUE DE LA DISTRIBUTION DU CHAMP EM À L'INTERIEUR DU GUIDE : MODE **TE$_{10}$**

Le champ électrique instantané:

$$E_y(x, y, z) = Re\left\{-j\omega\mu\left(\frac{a}{\pi}\right)C\sin\frac{\pi x}{a}e^{-j\beta z}e^{-j\omega t}\right\}$$

$$= \omega\mu\frac{a}{\pi}C\sin\frac{\pi x}{a}\sin(\omega t - \beta z)$$

On peut choisir une valeur de « $t$ » particulière et une position « $z$» privilégiée de façon à avoir $\sin(\omega t - \beta z) = 1$

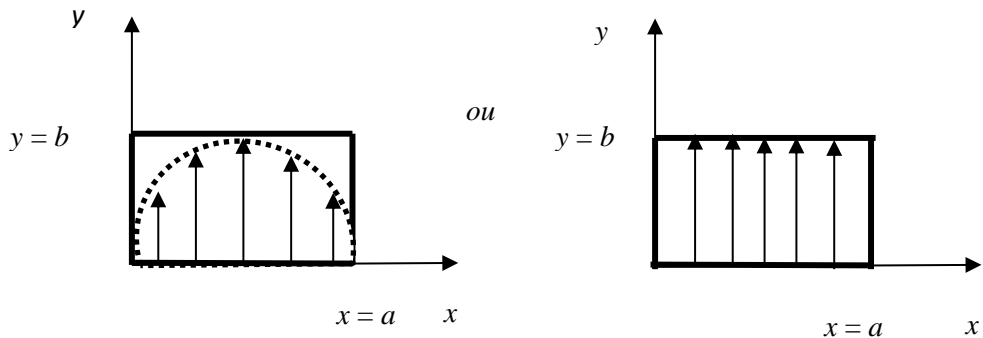

Figure III.5.

Si on pratique "une coupe" au centre du guide (Figure III.5), on peut observer la distribution du champ électrique selon « $z$ » pour une valeur de « $t$ » figée.

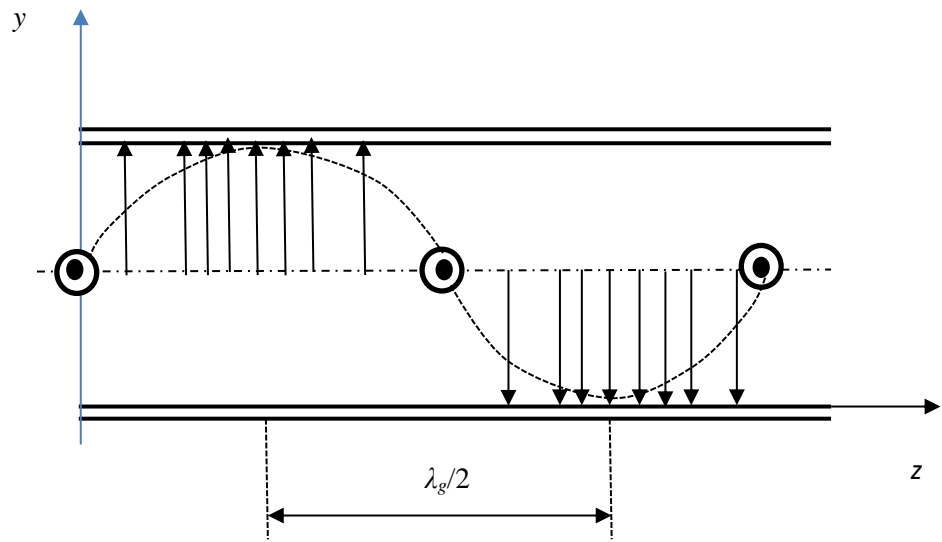

Figure III.6.

On peut également laisser varier « $t$ » et observer le déplacement de cette configuration (Figure III.6).

## 2.15 REPRÉSANTATION GRAPHIQUE CONCERNANT LE CHAMP MAGNETIQUE

$$H_z(x,y,z) = Re\left\{C\cos\frac{\pi x}{a}e^{-j\beta z}e^{-j\omega t}\right\} = C\cos\frac{\pi x}{a}\cos(\omega t - \beta z)$$

$$H_x(x,z,t) = Re\left\{j\beta\left(\frac{a}{\pi}\right)C\sin\frac{\pi x}{a}e^{-j\beta z}e^{j\omega t}\right\}$$

$$= -\beta\left(\frac{a}{\pi}\right)C\sin\frac{\pi x}{a}\sin(\omega t - \beta z)$$

Sur la Figure III.7 on donne graphiquement la distribution du champ magnétique.

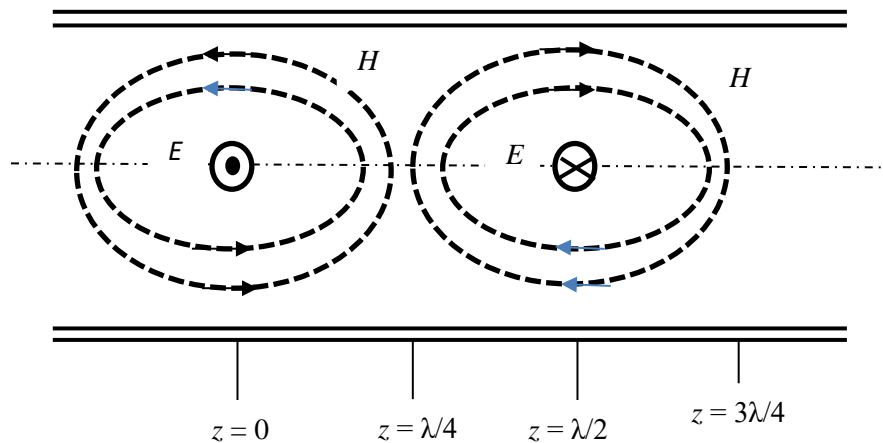

Figure III.7.

Si  $t = 0$  et  $z = 0$   $H_x = 0$   reste $H_z$

Si  $t = 0$  et  $z = \frac{\lambda}{4}$   $H_z = 0$   reste $H_x$

etc.

On additionne les deux composantes pour obtenir $H$ total résultant. On peut également faire varier « $t$ » si on veut.

$\rightarrow$ Alimentation du mode TE$_{10}$ (par boucle ou antenne "électrique") $\leftarrow$

## 2.16 PUISSANCE MOYENNE TRANSMISE DANS UN GUIDE RECTANGULAIRE

L'intensité de la puissance transmise dans la direction positive de « $z$ » est :

$$< P_z > = \frac{1}{2} Re\{E_x \times H_y^* - E_y \times H_x^*\}$$

Puisque

$$\frac{E_x}{H_y} = -\frac{E_y}{H_x} = \frac{\beta}{\omega \varepsilon} = Z_{TM} \text{ pour les modes } TM$$

$$= \frac{\omega \mu}{\beta} = Z_{TE} \text{ pour les modes } TE$$

Soit $Z^\Delta$ pour $Z_{TE}$ $ou$ $Z_{TM}$

$$< P_z > = \frac{1}{2} Re\left[Z^\Delta \left\{|H_y|^2 + |H_x|^2\right\}\right]$$

Si $H_x$ et $H_y$ sont uniformes sur toute la section du guide, il suffira de manipuler cette dernière expression par la surface de la section. Sinon, il faut procéder par des intégrales.

C'est ce qui est le cas :

$$< P_T \geq = \int_0^b \int_0^a \frac{1}{2} Z^\Delta \left\{|H_x|^2 + |H_y|^2\right\}$$

Cette expression est générale et valables pour tous les modes.

Pour le cas particulier du mode $TE_{10}$.

$$H_y = 0 \quad et \quad |H_x| = \beta \frac{a}{\pi} C \sin \frac{\pi x}{a}$$

Donc,

$$< P_T > = \frac{1}{2} Z^\Delta \left[ \beta \frac{a}{\pi} C \right]^2 \int_0^b \int_0^a \left| \sin \frac{\pi x}{a} \right|^2 dx dy$$

$$= \frac{1}{2} Z^\Delta \left[ \beta \frac{a}{\pi} C \right]^2 b \int_0^a \sin^2 \frac{\pi x}{a} dx$$

$$= \frac{1}{2} Z^\Delta \left[ \beta \frac{a}{\pi} C \right]^2 b \left\{ \frac{x}{2} \Big|_0^a + \frac{a}{4\pi} \sin \frac{2\pi x}{a} \Big|_0^a \right\}$$

$$= \frac{1}{2} Z^\Delta \left[ \beta \frac{a}{\pi} C \right]^2 b \left[ \frac{a}{2} + \frac{a}{4\pi} (0 - 0) \right]$$

$$= \frac{1}{2} Z^\Delta \left[ \beta \frac{a}{\pi} C \right]^2 \frac{ab}{2}$$

$$\boxed{< P_T > = \frac{1}{4} Z^\Delta \left[ \beta \frac{a}{\pi} C \right]^2 ab}$$

L'expression précédente est peu commune à celle qu'on trouve plus fréquemment dans les volumes. À l'aide des manipulations mathématiques, on peut obtenir une forme plus familière.

$$Z^\Delta = Z_{TE} = \sqrt{\frac{\mu}{\varepsilon}} \frac{1}{\sqrt{1 - \left( \frac{f_c}{f} \right)^2}} \quad ; \quad \beta = \omega \sqrt{\mu \varepsilon} \sqrt{1 - \left( \frac{f_c}{f} \right)^2}$$

Donc,

$$< P_t >_{(TE_{10})} = \frac{ab}{4} C^2 \left(\frac{a}{\pi}\right)^2 \left\{ \omega^2 \mu \varepsilon \left[ 1 - \left(\frac{f_c}{f}\right)^2 \right] \frac{\sqrt{\frac{\mu}{\varepsilon}}}{\sqrt{1 - \left(\frac{f_c}{f}\right)^2}} \right\}$$

$$= \frac{ab}{4} C^2 \sqrt{\frac{\mu}{\varepsilon}} \left(\frac{a}{\pi}\right)^2 \omega^2 \mu \varepsilon \sqrt{1 - \left(\frac{f_c}{f}\right)^2}$$

Si on utilise la longueur d'une onde de coupure $\lambda_c$

$$\lambda_c = \frac{2\pi}{\omega_c \sqrt{\mu \varepsilon}} = \frac{2\pi}{\left\{\left(\frac{m\pi}{a}\right)^2 + \left(\frac{n\pi}{b}\right)^2\right\}^{\frac{1}{2}}} = \frac{2\pi a}{\pi} = 2a$$

$(TE_{10})$

$$\lambda = \frac{2\pi}{\omega \sqrt{\mu \varepsilon}} \quad d'où$$

$$< P_T >_{TE_{10}} = \frac{abC^2}{4} \sqrt{\frac{\mu}{\varepsilon}} \left(\frac{\lambda_c}{\lambda}\right)^2 \sqrt{1 - \left(\frac{f_c}{f}\right)^2}$$

Cette expression montre la première puissance maximale possible de transmettre dans un guide d'onde rectangulaire fonctionnant au mode $TE_{10}$, mais il faut tenir compte des pertes possibles qui sont dues essentiellement à :

$$toujours\ f > f_c$$

1) Pertes dans le diélectrique non parfait placé dans le guide (en général on préfère l'air)

2) Pertes dans les murs du guide à cause de la conductivité finie de ces derniers.

## 2.17 Pertes dans le diélectrique

Si on envisage le facteur de propagation:

$$\gamma = \sqrt{\left(\frac{m\pi}{a}\right)^2 + \left(\frac{n\pi}{b}\right)^2 - \omega^2\mu\varepsilon^*}$$

avec une permittivité complexe,

$$\varepsilon^* = \varepsilon' - j\varepsilon'' = \varepsilon' - j\frac{\sigma_d}{\omega}$$

On peut inclure dans le terme $\varepsilon''$ l'effet de "$\sigma_d$" -la conductivité du matériau même s'il est très petit.

$$\gamma^2 = (\alpha_d + j\beta_d)^2 = \left(\frac{m\pi}{a}\right)^2 + \left(\frac{n\pi}{b}\right)^2 - \omega^2\mu\varepsilon^*$$

$$\alpha_d^2 - \beta_d^2 + j2\alpha_d\beta_d = \left(\frac{m\pi}{a}\right)^2 + \left(\frac{n\pi}{b}\right)^2 - \omega^2\mu\varepsilon' + j\omega\mu\sigma_d$$

$$\alpha_d^2 - \beta_d^2 = \left(\frac{m\pi}{a}\right)^2 + \left(\frac{n\pi}{b}\right)^2 - \omega^2\mu\varepsilon'$$

$$2\alpha_d\beta_d = \omega\mu\sigma \ \rightarrow \ \alpha_d = \frac{\omega\mu\sigma_d}{2\beta_d}$$

En admettant que $\alpha_d^2 \ll \beta_d^2 \rightarrow \ \alpha_d^2 - \beta_d^2 \approx -\beta_d^2$ ce qui donne,

$$\beta_d = \omega\sqrt{\mu\varepsilon'}\sqrt{1 - \left(\frac{f_c}{f}\right)^2} \ rad/m$$

et

$$\alpha_d = \frac{\omega\mu\sigma}{2\beta_d} = \frac{\mu\sigma_d}{2\sqrt{1-\left(\frac{f_c}{f}\right)^2}}\frac{1}{\sqrt{\mu\varepsilon'}} = \frac{\sqrt{\frac{\mu}{\varepsilon_r'}}}{2\sqrt{\varepsilon_r'}}\frac{\sigma_d}{\sqrt{1-\left(\frac{f_c}{f}\right)^2}} = \frac{1}{2\sqrt{\varepsilon_r'}}\frac{\sigma_d Z_0}{\sqrt{1-\left(\frac{f_c}{f}\right)^2}}$$

$$\alpha_d = \frac{1}{2\sqrt{\varepsilon_r'}}\frac{\sigma_d Z_0}{\sqrt{1-\left(\frac{f_c}{f}\right)^2}}(\frac{Np}{m})$$

Relation *Néper-dB* où *Np → Néper*

$$Np \times 20 \log e = 8.686 \times Np \ [dB]$$

## 2.18 Pertes dans le mur d'un guide rectangulaire

On a considéré jusqu'ici les murs du guide étant des parfaits conducteurs. En pratique, ils ne le sont pas tout à fait. Ceci a pour conséquence que les expressions qu'on a obtenus pour la distribution des ondes électromagnétiques à l'intérieur du guide ne sont pas tout à fait exactes. Néanmoins, elles représentent une bonne partie de la réalité et on peut se baser encore sur ces expressions pour calculer les pertes dans les murs. On peut travailler avec n'importe quel mode $TE_{mn}$ ou $TM_{mn}$. On choisit le mode le plus utilisé $TE_{10}$.

$$H_z = C \cos\frac{\pi x}{a} e^{-j\beta z}$$

$$H_x = j\beta \left(\frac{a}{\pi}\right) C \sin\frac{\pi x}{a} e^{-j\beta z}$$

$$E_y = -j\omega\mu \left(\frac{a}{\pi}\right) C \sin\frac{\pi x}{a} e^{-j\beta z}$$

Les pertes dans les murs sont associées aux courants de surface induits par les ondes qui se transmettent dans le guide (Figure III.8).

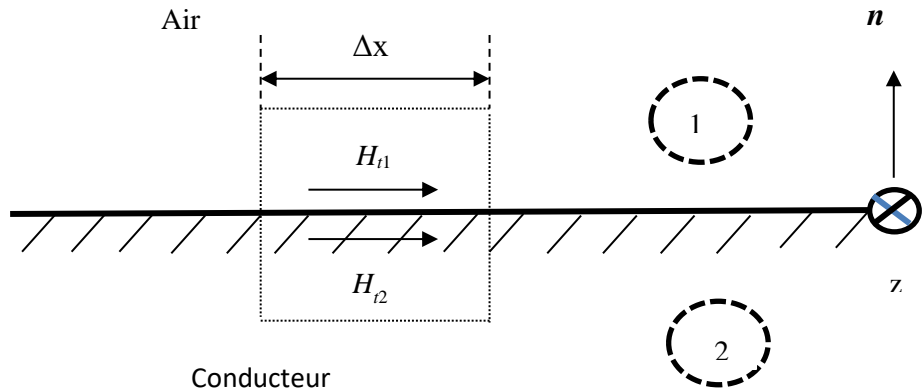

Figure III.8.

$$\oint \overline{H} d\overline{l} = I \qquad\qquad H_{tan1} = H_{t1}$$
$$H_{tan2} = H_{t2}$$

$$\overline{H}_{t1}\Delta_x - \overline{H}_{t2}\Delta_x = \overline{I} \qquad ; H_{t2} = 0$$

$$\overline{J}_s = \text{courant de surface } (A/m)$$

$$\boxed{\hat{n} \times \overline{H}_{tan1} = \overline{J}_s}$$

Cette dernière expression permet d'évaluer la distribution du courant de surface partout à l'intérieur du guide.

Calculons la puissance en moyenne dissipée dans le mur $x$=0 pour un guide d'onde de longueur de $1(m)$ (Figure III.9).

$$< P_d > |_{x=0} = \frac{1}{2} \int_0^1 \int_0^b |\overline{J}_s| R_s dy dz$$

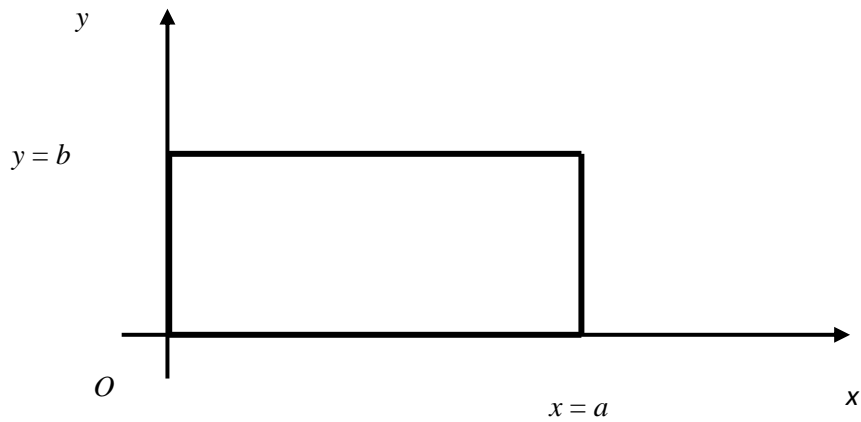

Figure III.9.

$R_s$ est la résistance de surface dans les parois. On l'a déjà calculé dans le cas d'une impédance dans le conducteur.

Rappel: $\sigma \gg \omega\varepsilon$

$$Z = \sqrt{\frac{j\omega\mu}{\sigma}} = \sqrt{\frac{\omega\mu}{\sigma}}\, e^{\frac{j\pi}{4}}$$

$$Z = R + jX = \sqrt{\frac{\omega\mu}{2\sigma}}\,(1+j)$$

$$R_s = R = \sqrt{\frac{\omega\mu}{2\sigma}}$$

D'autre part,
$$\hat{n} \times \bar{H}_{tan} = \bar{J}_s = \hat{x} \times \bar{H}_z$$

$$|\,\bar{H}_z|_{x=0} = C$$

Donc,

$$<P_d>|_{x=0} = \frac{1}{2}\int_0^1\int_0^b C^2\,R_s\,dy\,dz = \frac{C^2}{2}\,R_s\,b$$

$$avec \; \sigma \gg \omega\varepsilon$$

$$\text{Par symétrie} < P_d > |_{x=a} = \frac{C^2}{2} R_s b$$

D'autre part:

$$< P_d > |_{y=0} = \frac{1}{2} \int_0^1 \int_0^a \{|H_x^2| + |H_z^2|\} R_s \, dxdz$$

$$= \frac{R_s}{2} \int_0^a \left\{ \left| \frac{\beta a}{\pi} C \sin \frac{\pi x}{a} \right|^2 + \left| C \cos \frac{\pi x}{a} \right|^2 \right\} dx$$

$$< P_d > |_{y=0} = \frac{R_s}{2} \left\{ \left( \frac{\beta a}{\pi} C \right)^2 \frac{a}{2} + C^2 \frac{a}{2} \right\}$$

Aussi à cause de la symétrie,

$$< P_d > |_{y=0} = < P_d > |_{y=b}$$

d'où la valeur moyenne de la puissance perdue dans 4 murs est:

$$< P_L > = 2 < P_d > |_{x=0} + 2 < P_d > |_{y=0}$$

$$< P_L > = C^2 R_s b + R_s \left( \beta \frac{a}{\pi} C \right)^2 + R_s C^2 \frac{a}{2}$$

$$\boxed{< P_L >_{TE_{10}} = R_s C^2 \left\{ b + \frac{a}{2} \left[ 1 + \left( \frac{\beta a}{\pi} \right)^2 \right] \right\} \; W/m}$$

Dans le cas pratique on s'intéresse plutôt à obtenir le facteur d'atténuation "$\alpha$" à la place de la puissance perdue par unité de longueur.

Lorsqu'il y a des pertes, l'atténuation des amplitudes de chaque composante *EM* est gérée par le terme $e^{-\alpha z}$ à mesure qu'on avance dans le guide selon la direction de la propagation « *z* ». La puissance étant le produit du champ électrique par le champ magnétique, son évolution selon l'axe des « z » va être gérée par le terme $e^{-2\alpha z}$. La puissance perdue entre deux points de référence selon l'axe « z » avec une séparation de $1(m)$ peut être obtenue par la dérivée de la puissance transmise.

$$<P_L> = -\frac{\partial <P_t>}{\partial z} = -(2\alpha) <P_t>$$

$$= 2\,\alpha <P_t>$$

Donc,

$$\alpha = \frac{<P_L>}{2 <P_t>}$$

Pour le mode $TE_{10}$,

$$\alpha_m = \frac{R_s C^2 \left[b + \frac{a}{2}\left(1 + \frac{\beta^2 a^2}{\pi^2}\right)\right]}{\frac{2ab}{4}\, C^2 \sqrt{\frac{\mu}{\varepsilon}}\left(\frac{\lambda_c}{\lambda}\right)^2 \sqrt{1 - \left(\frac{f_c}{f}\right)^2}}\,\left(\frac{Np}{m}\right)$$

Si on remplace $\beta$ par,

$$\beta = \sqrt{\omega^2 \mu\varepsilon - \left(\frac{\pi}{a}\right)^2}$$

$$\alpha_m = \frac{R_s \left\{b + \frac{a}{2} + \left[\omega^2 \mu\varepsilon - \left(\frac{\pi}{a}\right)^2\right]\frac{a^3}{2\pi^2}\right\}}{\frac{ab}{2}\sqrt{\frac{\mu}{\varepsilon}}\,(2a)^2 \left(\frac{\omega\sqrt{\mu\varepsilon}}{2\pi}\right)^2 \sqrt{1 - \left(\frac{f_c}{f}\right)^2}}$$

On a,

$$\lambda_c = 2a$$

$$\lambda = \frac{2\pi}{\omega\sqrt{\mu\varepsilon}}$$

$$\frac{\lambda}{\lambda_c} = \frac{\pi^2}{a^2 \omega^2 \mu \varepsilon}$$

$$\alpha_m = \frac{R_s \left\{ b + \frac{\omega^2 \mu \varepsilon a^3}{2\pi^2} \right\}}{a^3 b \sqrt{\frac{\mu}{\varepsilon}} \frac{\omega^2 \mu \varepsilon}{2\pi^2} \sqrt{1 - \left(\frac{f_c}{f}\right)^2}}$$

$$\alpha_m = \frac{R_s \left\{ \frac{2\pi^2 b}{\omega^2 \mu \varepsilon a^3} + 1 \right\}}{b \sqrt{\frac{\mu}{\varepsilon}} \sqrt{1 - \left(\frac{f_c}{f}\right)^2}} = \frac{R_s \left\{ \frac{2b}{a} \left(\frac{\lambda}{\lambda_c}\right)^2 + 1 \right\}}{b \sqrt{\frac{\mu}{\varepsilon}} \sqrt{1 - \left(\frac{f_c}{f}\right)^2}}$$

$$\boxed{\alpha_m = \frac{R_s \left\{ \frac{2b}{a} \left(\frac{f_c}{f}\right)^2 + 1 \right\}}{b \sqrt{\frac{\mu}{\varepsilon}} \sqrt{1 - \left(\frac{f_c}{f}\right)^2}} = \alpha_m}$$

Si on a en même temps des pertes dans le diélectrique et dans les murs:

$$\alpha_T = \alpha_{d(di\acute{e}lectrique)} + \alpha_{m(murs)}$$

**Exemple III.1.**

Un guide d'onde rectangulaire creux doit être utilisé pour transmettre des signaux à une fréquence de 6 GHz. Choisir ses dimensions pour que la fréquence de coupure du mode TE dominant soit inférieure à la porteuse de 25% et celle du mode suivant soit au moins 25 % supérieure à la porteuse.

Solution :

Pour $m$=1 et $n$=0 (mode $TE_{10}$), $v_{p0}$ = c = 3x10$^8$ $(m/s)$.

$$f_{10} = \frac{c}{2a} \quad et \ f_{10} = 0.75 \ f_0 = 4.5 \ (GH_z)$$

ce qui donne,

$$a = \frac{c}{2f_{10}} = 3.33 \ (cm)$$

Si $b$ est choisi tel que,

$$a > b > a/2$$

Le deuxième mode sera $TE_{01}$, suivi de $TE_{20}$ à $f_{20} = 9 \ GH_z$.

Pour $TE_{01}$, $f_{01} = \frac{c}{2b}$.

Avec $f_{01} = 1.25 \ f_0 = 7.5 (GH_z)$, on obtient,

$$b = \frac{c}{2f_{01}} = 2 \ (cm)$$

## Exemple III.2.

Soit un guide d'onde rectangulaire, rempli d'air, de dimensions transversales $a$ et $b$. Une onde électromagnétique de mode $TE_{11}$ se propage dans ce guide. L'axe « $z$ » d'un système de coordonnées cartésiens est orienté selon la direction de propagation des ondes. La composante selon « $z$ » du champ magnétique de l'onde est donnée par :

$$H_z = C \cos \frac{m\pi}{a} \ x \cos \frac{n\pi}{b} \ y \ e^{-jBz}$$

a) Donner l'expression d'autres composants phaseurs des champs électromagnétiques dans ce guide au mode $TE_{11}$,

b) Si le maximum de l'amplitude du champ Hz est 1A/m dans le guide avec les dimensions $a = 2\pi$ (cm), $b = \pi$ (cm), trouver les endroits (donner les valeurs de *x* et de *y*) où ce maximum se produit dans une section transversale du guide au mode TE$_{11}$,

c) Calculer l'expression de la puissance transportée dans ce guide par le mode *TE*$_{11}$ en fonction de ($\lambda_c / \lambda$) et (*f*$_c$ / *f*) où $\lambda_c$ et *f*$_c$ représentent respectivement la longueur d'onde de coupure et la fréquence de coupure. *f* est la fréquence d'opération et $\lambda$ est la longueur d'onde en espace libre à cette fréquence.

**Solution :**

a)

$$H_z(x,y,z) = C \cos \frac{\pi x}{a} \cos \frac{\pi y}{b} \; e^{-\gamma z}$$

$$H_x(x,y,z) = \frac{\gamma}{h^2} \, C \left(\frac{\pi}{a}\right) \sin \frac{\pi x}{a} \cos \frac{\pi y}{b} e^{-\gamma z}$$

$$H_y(x,y,z) = \frac{\gamma}{h^2} \, C \left(\frac{\pi}{b}\right) cos \frac{\pi x}{a} sin \frac{\pi y}{b} e^{-\gamma z}$$

$$E_x(x,y,z) = \frac{j\omega\mu}{h^2} \, C \left(\frac{\pi}{b}\right) cos \frac{\pi x}{a} sin \frac{\pi y}{b} e^{-\gamma z}$$

$$E_y(x,y,z) = -\frac{j\omega\mu}{h^2} \, C \left(\frac{\pi}{a}\right) sin \frac{\pi x}{a} cos \frac{\pi y}{b} e^{-\gamma z}$$

$$E_z(x,y,z) = \; 0$$

$$h^2 = \left(\frac{\pi}{a}\right)^2 + \left(\frac{\pi}{b}\right)^2, \qquad h^2 = \omega^2 \mu\varepsilon + \gamma^2 = \left(\frac{m\pi}{a}\right)^2 + \left(\frac{n\pi}{b}\right)^2$$

$$\gamma = \; j\beta$$

$$\beta = \omega\sqrt{\mu\varepsilon} \; \sqrt{1 - \left(\frac{f_c}{f}\right)^2} \; \; (\frac{rad}{m})$$

b)

$$H_z(x, y, z) = C \cos \frac{x}{0.01} \cos \frac{y}{0.01} \, e^{-\gamma z}$$

$x$ et $y$ en mètres.

$$|H_{zmax}| = 1(\frac{A}{m}) \rightarrow \ C = 1(\frac{A}{m})$$

Les points,

$$\frac{x}{0.02} = 1 \ \rightarrow \ x = 0.02p\pi, \ \ p = 0,1$$

$$\frac{y}{0.01} = 1 \ \rightarrow \ x = 0.01k\pi, \ \ k = 0,1$$

$x_1 = 0, \ x_2 = 0.02\pi = 2\pi \ (cm)$

$y_1 = 0, \ y_2 = 0.01\pi = \pi \ (cm)$

Les points possibles sont : $(x_1, y_1)$, $(x_1, y_2)$, $(x_2, y_1)$ et $(x_2, y_2)$.

c)

$$< P_{T_{TE_{11}}} \ge = \int_0^b \int_0^a \frac{1}{2} Z_{TE} \left\{ |H_x|^2 + |H_y|^2 \right\}$$

$$Z_{TE} = \frac{\omega \mu}{\beta}$$

En remplaçant les expressions pour $H_x$ et $H_y$ et $Z_{TE}$ on obtient :

$$< P_{T_{TE_{11}}} \ge = \frac{1}{2} Z_{TE} \frac{ab}{4} \frac{\beta^2 C^2}{h^4} [(\frac{\pi}{a})^2 + (\frac{\pi}{b})^2] = \frac{ab}{8} Z_{TE} \frac{\beta^2 C^2}{(\frac{\pi}{a})^2 + (\frac{\pi}{b})^2}$$

$$< P_{T_{TE_{11}}} \ge = \frac{abC^2}{8} \frac{\omega \mu \omega \sqrt{\varepsilon \mu}}{(\frac{\pi}{a})^2 + (\frac{\pi}{b})^2} \sqrt{1 - \frac{f_c^2}{f^2}}$$

$$\lambda = \frac{1}{f}\frac{1}{\sqrt{\varepsilon\mu}} = \frac{2\pi}{\omega}\frac{1}{\sqrt{\varepsilon\mu}}$$

$$\lambda_c = \frac{1}{f_c}\frac{1}{\sqrt{\varepsilon\mu}} = \frac{2\pi}{\omega_c}\frac{1}{\sqrt{\varepsilon\mu}}$$

$$\frac{\lambda_c^2}{\lambda^2} = \frac{4\pi^2}{\left(\frac{\pi}{a}\right)^2 + \left(\frac{\pi}{b}\right)^2}\frac{\omega^2\varepsilon\mu}{4\pi^2} = \frac{\omega^2\varepsilon\mu}{\left(\frac{\pi}{a}\right)^2 + \left(\frac{\pi}{b}\right)^2}$$

$$< P_{T_{TE_{11}}} \ge = \frac{abC^2}{8}\frac{\lambda_c^2}{\lambda^2}\sqrt{\frac{\mu}{\varepsilon}}\sqrt{1 - \frac{f_c^2}{f^2}} \ (W)$$

$$\sqrt{\frac{\mu}{\varepsilon}} = 120\pi \ (\Omega) - \text{Air}$$

$$< P_{T_{TE_{11}}} \ge = 30\pi^3 \frac{\lambda_c^2}{\lambda^2}\sqrt{\frac{\mu}{\varepsilon}}\sqrt{1 - \frac{f_c^2}{f^2}} \ (W)$$

## Exemple III.3.

Calculer la fréquence de coupure pour les huit premiers modes d'un guide d'onde rectangulaire avec $a$ = 2.286 ($cm$) et $b$ = 1.524 ($cm$). Le guide est rempli d'air.

## Solution :

Pour un guide rempli d'air la fréquence de coupure est donnée par,

$$f_{c_{mn}} = \frac{c}{2}\sqrt{\frac{m^2}{a^2} + \frac{n^2}{b^2}}$$

Avec l'inspection on obtient les combinaisons suivantes pour $m$ = 0,1,2,3 et $n$ = 0,1,2,3 :

Tableau 1.

| MODE | $f_{Cmn}$ ($GH_z$) |
|------|--------------------|
| $TE_{10}$ | 6.56 |
| $TE_{01}$ | 9.84 |
| $TE_{11}$ | 11.83 |
| $TM_{11}$ | 11.83 |
| $TE_{20}$ | 13.12 |
| $TE_{21}$ | 16.40 |
| $TM_{30}$ | 19.69 |
| $TE_{02}$ | 19.69 |

### Exemple III.4.

Un guide rectangulaire en cuivre a les dimensions internes $a$ = 2.5 ($cm$) et $b$ = 1 ($cm$). Il est rempli du Téflon$^{TM}$ dont la constante diélectrique relative est $\varepsilon_r$ = 2.11 et les pertes diélectriques et la conductivité sont supposées être nulles. Pour le mode $TE_{10}$ :

a) Calculer la fréquence de coupure $f_c$ et l'atténuation (en $Np/m$ puis en $dB/m$) pour la condition $f$ = 1.453 $f_c$,

b) Comparer cette atténuation celle a un guide d'onde rectangulaire rempli d'air, ayant la même fréquence de coupure $f_c$, le même rapport des dimensions $a/b$, fonctionnant à la même fréquence,

c) Le même guide d'onde ($a$ = 2.5 ($cm$) et $b$ = 1 ($cm$)) en cuivre rempli d'air transmet une puissance moyenne de 0.3 $W$ a une fréquence de 9 $GH_z$. Trouver la valeur maximal (ou la valeur crête) du champ électrique dans le guide. (Note : la conductivité du cuivre : $\sigma = 5.8x10^7(\frac{S}{m})$, perméabilité du cuivre : $\mu = 4\pi x10^7(\frac{H}{m})$).

Solution :

a) Pour le mode TE$_{10}$    $f_c = \frac{1}{2\sqrt{\varepsilon_0\mu_0}}\frac{1}{\sqrt{\varepsilon_r}}\frac{1}{a} = 4.1306$ ($GH_z$)

$$f = 1.453 f_c = 6 \ (GH_z)$$

Il y a l'atténuation seulement à cause des pertes dans les murs :

$$\alpha_m = \frac{R_s \left\{ \dfrac{2b}{a} \left( \dfrac{f_c}{f} \right)^2 + 1 \right\}}{b \sqrt{\dfrac{\mu}{\varepsilon}} \sqrt{1 - \left( \dfrac{f_c}{f} \right)^2}} = \alpha_m$$

$$R_s = R = \sqrt{\frac{\omega \mu}{2\sigma}} = 0.020209 \ \left( \frac{\Omega}{m} \right)$$

Pour le cuivre $\mu = \mu_0 = 4\pi x 10^7 (H/m)$

ce qui donne, $\alpha = \alpha_m = 0.015 \ (Np/m)$

$$\alpha(dB) = \alpha_m x 8.686 = 0.129 \ (dB)$$

   b)

$$f_c = 4.1306 x 10^9 = \frac{1.5 x 10^8}{a'}$$

d'où

$$a' = 3.63 \ (cm)$$

$$b' = a' \frac{b}{a} = 1.45 \ (cm)$$

$$f = 6 \ (GH_z)$$

$$\alpha'^{(air)} = \frac{R_s\left\{\frac{2b}{a}\left(\frac{f_c}{f}\right)^2 + 1\right\}}{b\sqrt{\frac{\mu_0}{\varepsilon_0}}\sqrt{1 - \left(\frac{f_c}{f}\right)^2}} = 0.061\left(\frac{dB}{m}\right)$$

Le guide d'onde rectangulaire rempli d'air pour la même fréquence $f_c$ a des dimensions plus larges et l'atténuation moins forte.

Pour le mode $TE_{10}$ les champs $EM$ sont :

$$H_z = C \cos\frac{\pi x}{a}\, e^{-j\beta z}$$

$$H_x = \frac{j\beta}{h^2}\left(\frac{\pi}{a}\right) C \sin\frac{\pi x}{a}\, e^{-j\beta z}$$

$$E_y = -\frac{j\omega\mu}{h^2}\left(\frac{\pi}{a}\right) C \sin\frac{\pi x}{a}\, e^{-j\beta z}$$

$$h^2 = \left(\frac{\pi}{a}\right)^2$$

$$|E_{max}| = |E_{ymax}| = C\omega\mu\frac{a}{\pi}$$

lorsque $\sin\frac{\pi x}{a} = 1$.  Pour calculer la constante $C$ on utilise,

$$< P_T >_{TE_{10}} = \frac{abC^2}{4}\sqrt{\frac{\mu_0}{\varepsilon_0}}\left(\frac{\lambda_c}{\lambda}\right)^2\sqrt{1 - \left(\frac{f_c}{f}\right)^2} = 0.3\ (W)$$

$$f = 9\ (GH_z)$$

$$f_c = \frac{1}{2\sqrt{\mu_0\varepsilon_0}}\frac{1}{a} = \frac{1.5x10^8}{0.025} = 6(GH_z)$$

$$\frac{f_c}{f} = \frac{2}{3}\ ,\qquad \lambda = \frac{c}{f},\qquad \lambda_c = \frac{c}{f_c}\quad\rightarrow\quad \frac{\lambda_c}{\lambda} = \frac{3}{2}$$

En utilisant la formule pour la puissance moyenne on obtient $C = 2.739$ et

$$|E_{max}| = |E_{ymax}| = 2.739 \left(\frac{V}{m}\right)$$

## Exemple III.5.

Une onde *TE* se propageant dans un guide rectangulaire rempli d'un diélectrique de permittivité inconnue a des dimensions $a = 5$ (*cm*) et $b = 3$ (*cm*). Si la composante *x* du champ électrique est :

$$E_x = -36 \cos(40\pi x) \sin(100\pi y) \sin(2.4 \cdot 10^{10} - 52.9\pi z) \left(\frac{V}{m}\right)$$

a) Déterminer le mode d'onde *TE*,
b) La permittivité relative $\varepsilon_r$ du matériel dans le guide,
c) La fréquence de coupure,
d) L'expression pour le champ $H_y$.

Solution :

a) $\frac{m\pi}{a} = 40\pi \rightarrow m = 2$    $\frac{n\pi}{b} = 100\pi \rightarrow n = 3$

Le mode est *TE*$_{23}$.

b) De l'expression du champ électrique $E_x$ on a :

$$\omega = 2.4\pi 10^{10} (rad), f = 12 \, GH_z, \ \beta = 52.9\pi \left(\frac{rad}{m}\right)$$

Étant donné que $\gamma = j\beta$,

$$\omega^2 \mu \varepsilon - \beta^2 = \left(\frac{m\pi}{a}\right)^2 + \left(\frac{n\pi}{b}\right)^2$$

ou

$$\varepsilon = \frac{1}{\omega^2 \mu}\left[\beta^2 + \left(\frac{m\pi}{a}\right)^2 + \left(\frac{n\pi}{b}\right)^2\right] = \frac{c^2}{\omega^2}\left[\beta^2 + \left(\frac{m\pi}{a}\right)^2 + \left(\frac{n\pi}{b}\right)^2\right] = 2.25$$

c)

$$v_p = \frac{c}{\sqrt{\varepsilon_r}} = 2x10^8 (\frac{m}{s})$$

$$f_{c_{23}} = \frac{v_p}{2} \sqrt{\frac{2^2}{a^2} + \frac{3^2}{b^2}} = 10.77 \; (GH_z)$$

d)

$$Z_{TE} = \sqrt{\frac{1}{\varepsilon_r}} \frac{Z_0}{\sqrt{1 - \left(\frac{f_c}{f}\right)^2}} = \frac{377}{\sqrt{2.25}} \frac{1}{\sqrt{1 - \left(\frac{10.77}{12}\right)^2}} = 569.9 \; (\Omega)$$

$$H_y = \frac{E_x}{Z_{TE}} = -0.063 \cos(40\pi x) \sin(100\pi y) \sin(2.4\pi 10^{10} - 52.9\pi z)(\frac{A}{m})$$

### Exemple III.6.

Le guide d'onde rectangulaire, avec les dimensions $a = 1$ $(cm)$ et $b = 0.7$ $(cm)$, est utilisé au mode dominant. Trouver l'impédance d'onde pour ce mode dominant quand,

a) Le guide est vide,
b) Le guide est rempli de polyéthylène ($\varepsilon_r = 2.25$).

Solution :

a) Pour le mode dominant $TE_{10}$ la fréquence de coupure est,

$$f_{10} = \frac{1}{2\sqrt{\varepsilon_r \mu_0 \varepsilon_0}} \frac{1}{a} = \frac{c}{2a\sqrt{\varepsilon_r}}, \quad c = \frac{1}{\sqrt{\mu_0 \varepsilon_0}} = 3x10^8 (\frac{m}{s})$$

Pour le vide,

$$f_{10} = \frac{1}{2\sqrt{\mu_0 \varepsilon_0}} \frac{1}{a} = \frac{c}{2a} = 15 \; (GH_z)$$

$$Z_{TE} = \frac{Z_0}{\sqrt{1 - \left(\frac{f_c}{f}\right)^2}} = \frac{377}{\sqrt{1 - \left(\frac{15}{20}\right)^2}} = 570 \ (\Omega)$$

b) Pour le polyéthylène,

$$f_{10} = \frac{1}{2\sqrt{\varepsilon_r \mu_0 \varepsilon_0}} \frac{1}{a} = \frac{c}{2a\sqrt{\varepsilon_r}} = 10 \ (GH_z)$$

$$Z_{TE} = \frac{1}{\sqrt{\varepsilon_r}} \frac{Z_0}{\sqrt{1 - \left(\frac{f_c}{f}\right)^2}} = \frac{1}{\sqrt{2.25}} \frac{377}{\sqrt{1 - \left(\frac{10}{20}\right)^2}} = 290 (\Omega)$$

## Exemple III.7.

Un guide d'onde rectangulaire rempli d'air ayant une largeur $a$ = 5 ($cm$) et une hauteur $b$ = 3 ($cm$) est utilisé pour transmettre de l'énergie selon le mode $TE_{10}$. La fréquence de la source est telle que si l'énergie était rayonnée en espace libre, dans l'air à l'extérieur du guide, la longueur d'onde serait 4 ($cm$).

a) Quelle est la fréquence de la source?
b) Quelle est la fréquence de coupure du mode dominant dans le guide?
c) Calculer la longueur d'onde dans ce cas,
d) Faire la liste de tous les modes qui pourraient se transmettre dans ce guide si l'on décidait de remplir d'un parfait diélectrique (avec $\varepsilon_r = 2$) et que l'on conserve la même fréquence de source.

## Solution :

a) Le mode $TE_{10}$ et $\lambda_0 = 4 \ cm$.

$$f = \frac{c}{\lambda_0} = 7.5 \ (GH_z)$$

b)

$$f_{cTE_{10}} = f_c = \frac{c}{2a} = 3 \ (GH_z)$$

c)

$$\lambda_g = \frac{\lambda_0}{\sqrt{1 - \left(\frac{f_c}{f}\right)^2}} = 4.346 \ (cm)$$

d)

$$f_{cTE_{10}} = \frac{1}{2\sqrt{\varepsilon_r \mu_0 \varepsilon_0}} \frac{1}{a} = \ 2.121 \ (GH_z) \ \text{(mode possible)}$$

$$f_{cTE_{11}} = \frac{1}{2\sqrt{\varepsilon_r \mu_0 \varepsilon_0}} \sqrt{\left(\frac{\pi}{a}\right)^2 + \left(\frac{\pi}{b}\right)^2} = \ 4.12 \ (GH_z) \ \text{(mode possible)}$$

$$f_{cTE_{01}} = \frac{1}{2\sqrt{\varepsilon_r \mu_0 \varepsilon_0}} \frac{1}{b} = \ 3.535 \ (GH_z) \ \text{(mode possible)}$$

$$f_{cTE_{11}} = \ f_{cTM_{11}} \ \text{(mode possible)}$$

$$f_{cTE_{20}} = \frac{1}{2\sqrt{\varepsilon_r \mu_0 \varepsilon_0}} \frac{2}{a} = \ 4.424 \ (GH_z) \ \text{(mode possible)}$$

$$f_{cTE_{21}} = \frac{1}{2\sqrt{\varepsilon_r \mu_0 \varepsilon_0}} \sqrt{\left(\frac{2\pi}{a}\right)^2 + \left(\frac{\pi}{b}\right)^2} = \ 5.52 \ (GH_z) \ \text{(mode possible)}$$

$$f_{cTE_{21}} = \ f_{cTM_{21}} \ \text{(mode possible)}$$

$$f_{cTM_{12}} = \ f_{cTE_{12}} = 7.37 \ (GH_z) \ \text{(mode possible)}$$

$$f_{cTE_{30}} = \frac{1}{2\sqrt{\varepsilon_r \mu_0 \varepsilon_0}} \frac{3}{a} = \ 6.36 \ (GH_z) \ \text{(mode possible)}$$

$$f_{cTM_{22}} = \ f_{cTE_2} = 8.24 \ (GH_z) \ \text{(mode impossible)}$$

Les modes possibles : $TE_{10}$, $TE_{01}$, $TE_{11}$, $TM_{11}$, $TE_{20}$, $TE_{02}$, $TE_{21}$, $TM_{21}$, $TE_{12}$, $TM_{12}$, $TE_{30}$ et $TE_{03}$.

### Exemple III.8.

La vitesse de groupe dans un guide d'onde rempli d'air est inférieure à la vitesse de la lumière tandis que la vitesse de phase est plus grande. Pour un guide d'onde rectangulaire avec les dimensions $a$ = 7 ($cm$) et $b$ = 4($cm$), en supportant le mode $TE_{10}$ calculer $v$, $v_p$ et $v_g$ avec $f = 1.3\, f_c$ dans les cas suivants :

a) Le guide rempli d'air.
b) Le guide rempli de verre ($\varepsilon = 4\varepsilon_0$).
c) Répéter ces deux calculs pour $f = 1.1\, f_c$.

### Solution :

a) Air :

$$v = c = 3x10^8 (\frac{m}{s})$$

$$v_p = \frac{1}{\sqrt{\varepsilon\mu}} \frac{1}{\sqrt{1 - \left(\frac{f_c}{f}\right)^2}} = \frac{c}{\sqrt{1 - \left(\frac{f_c}{f}\right)^2}} \ , \ \frac{f_c}{f} = \frac{1}{1.3} = \frac{10}{13}$$

$$v_p = 4.67 10^8 (\frac{m}{s})$$

$$v_g = \frac{1}{\sqrt{\varepsilon\mu}} \sqrt{1 - \left(\frac{f_c}{f}\right)^2} = c \sqrt{1 - \left(\frac{f_c}{f}\right)^2}$$

$$v_g = 1.92x10^8 (\frac{m}{s})$$

b) Verre :

$$v = \frac{c}{\sqrt{\varepsilon_r}} = 1.5 x 10^8 (\frac{m}{s})$$

$$v_p = \frac{1}{\sqrt{\varepsilon_r}} \frac{c}{\sqrt{1 - \left(\frac{f_c}{f}\right)^2}} \, , \, \frac{f_c}{f} = \frac{1}{1.3} = \frac{10}{13}$$

$$v_p = 2.335 x 10^8 (\frac{m}{s})$$

$$v_g = \frac{c}{\sqrt{\varepsilon_r}} \sqrt{1 - \left(\frac{f_c}{f}\right)^2}$$

$$v_g = 0.96 x 10^8 (\frac{m}{s})$$

c) Air :

$$v = c = 3 x 10^8 (\frac{m}{s})$$

$$v_p = \frac{1}{\sqrt{\varepsilon \mu}} \frac{1}{\sqrt{1 - \left(\frac{f_c}{f}\right)^2}} = \frac{c}{\sqrt{1 - \left(\frac{f_c}{f}\right)^2}} \, , \, \frac{f_c}{f} = \frac{1}{1.1} = \frac{10}{11}$$

$$v_p = 7.2 x 10^8 (\frac{m}{s})$$

$$v_g = \frac{1}{\sqrt{\varepsilon \mu}} \sqrt{1 - \left(\frac{f_c}{f}\right)^2} = c \sqrt{1 - \left(\frac{f_c}{f}\right)^2}$$

$$v_g = 1.25 x 10^8 (\frac{m}{s})$$

Verre :

$$v = \frac{c}{\sqrt{\varepsilon_r}} = 1.5 x 10^8 (\frac{m}{s})$$

$$v_p = \frac{1}{\sqrt{\varepsilon_r}} \frac{c}{\sqrt{1 - \left(\frac{f_c}{f}\right)^2}} \; , \; \frac{f_c}{f} = \frac{1}{1.1} = \frac{10}{11}$$

$$v_p = 3.6 \times 10^8 \left(\frac{m}{s}\right)$$

$$v_g = \frac{c}{\sqrt{\varepsilon_r}} \sqrt{1 - \left(\frac{f_c}{f}\right)^2}$$

$$v_g = 0.625 \times 10^8 \left(\frac{m}{s}\right)$$

# 3. Guides d'onde circulaires[*]

Considérons le problème de décrire la propagation des ondes dans un guide d'ondes cylindrique de rayon $r_0$ (Figure III.10). Ce problème est plus facile à manipuler avec un système de coordonnées cylindriques. La procédure est similaire à celle utilisée dans la section précédente pour le guide d'onde rectangulaire. Nous supposons une variation temps-harmonique, aux parois parfaitement conductrices, et un milieu intérieur sans perte ($\sigma$ = 0) ne contenant aucune charge ($\rho$ = 0). Deux équations de Maxwell donnent six équations scalaires, et les équations de Maxwell de divergence donnent deux équations scalaires. En coordonnées cylindriques ceux-ci sont les suivantes:

$$\frac{1}{r}\frac{\partial E_z}{\partial \Phi} + \gamma E_\Phi - ZH_r = 0$$

$$-\gamma E_r - \frac{\partial E_z}{\partial z} - ZH_\Phi = 0$$

$$\frac{\partial E_\Phi}{\partial r} + \frac{1}{r}E_\Phi - \frac{1}{r}\frac{\partial E_r}{\partial \Phi} - ZH_z = 0$$

$$\frac{1}{r}\frac{\partial H_z}{\partial \Phi} + \gamma H_\Phi - YE_r = 0$$

$$-\gamma H_r - \frac{\partial H_z}{\partial z} - YE_\Phi = 0$$

$$\frac{\partial H_\Phi}{\partial r} + \frac{1}{r}H_\Phi - \frac{1}{r}\frac{\partial H_r}{\partial \Phi} - YE_z = 0$$

$$\frac{\partial E_r}{\partial r} + \frac{1}{r}E_r + \frac{1}{r}\frac{\partial E_\Phi}{\partial \Phi} - \gamma E_z = 0$$

$$\frac{\partial H_r}{\partial r} + \frac{1}{r}H_r + \frac{1}{r}\frac{\partial H_\Phi}{\partial \Phi} - \gamma H_z = 0$$

[*]Traduit du livre: « Electromagnetics » John D. Kraus third edition McGraw-Hill Book Company.

où

$$Z \text{ (impédence en série)} = -j\mu\omega \frac{\Omega}{m}$$

$$Y \text{ (admittance shunt)} = -j\varepsilon\omega \frac{S}{m}$$

$$E_z = E_1 e^{-\gamma z}$$

$$\gamma = \alpha + j\beta \; (constante \; de \; propagation)$$

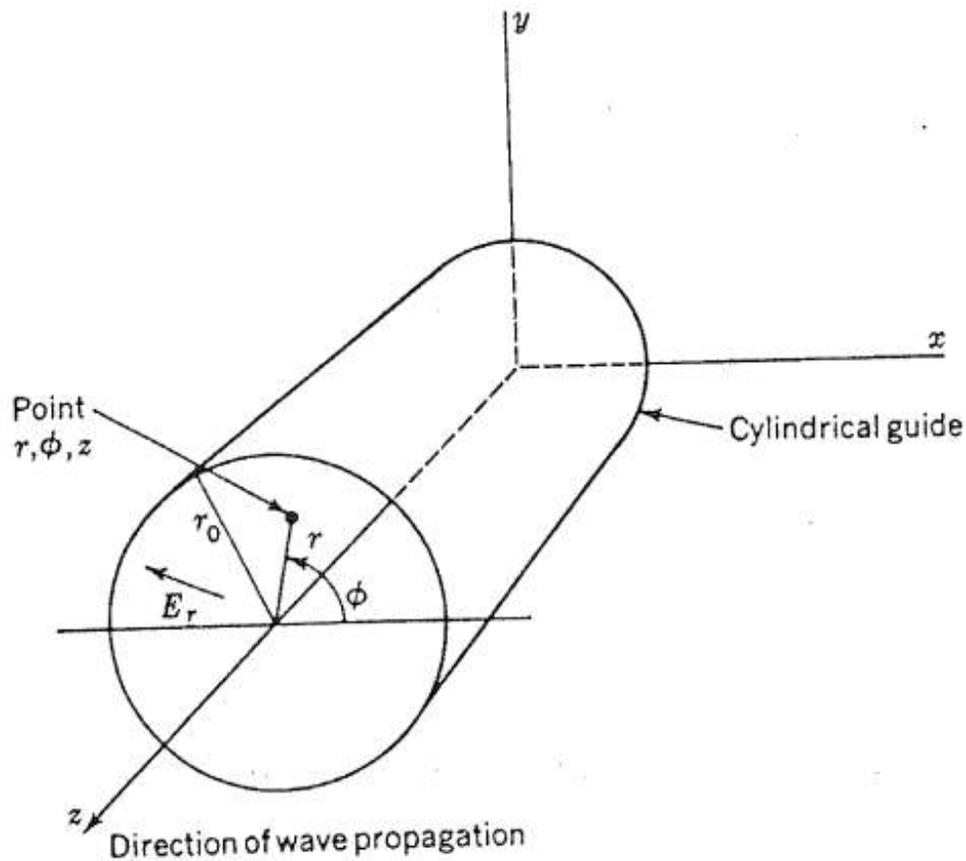

Figure III.10.

Ces huit relations sont les équations générales pour le domaine du régime permanent pour une onde se propageant dans la direction « $z$ », exprimée en coordonnées cylindriques. À ce stade, nous pouvons limiter le problème soit à une onde *TE* ou à une onde *TM*. Sélectionnant l'onde *TE* et procédant de la même manière que pour un guide d'onde rectangulaire, nous trouvons la solution pour la composante $H_z$.

$H_z = c_1 c_3 \sin(n\Phi) J_n(kr) + c_2 c_3 \cos(n\Phi) J_n(kr) + c_1 c_4 \sin(n\Phi) N_n(kr) + c_2 c_4 \cos(n\Phi) N_n(kr)$   où

$$k = \sqrt{\gamma^2 + YZ} = \sqrt{\gamma^2 + \omega^2 \varepsilon \mu}$$

$$n \rightarrow entier$$

L'équation précédente est une solution ou combinaison de termes qui peuvent former une solution. La solution appropriée doit satisfaire les conditions aux limites, qui sont que $E_\Phi = 0$ à $r = r_0$ et que les champs dans le guide sont finis. La fonction de Neumann devient infinie à $r = 0$ et n'est donc pas appropriée comme solution. Ainsi, l'équation précédente peut être réduite à,

$$H_z = H_0 [\cos(n\Phi) + j\sin(n\Phi)] J_n(kr)$$

où

$$H_0 = c_2 c_3 = \frac{c_1 c_3}{j}$$

Lorsque le terme sinus tombe, l'équation précédente se simplifie en

$$H_z = H_0 \cos(n\Phi) J_n(kr)$$

Ce choix signifie que $H_z$ est à son maximum où $\phi = 0$ et $\phi = \pi$ et est nul là où $\phi = \pi/2$ et $\phi = 3\pi2$ à condition que $n \neq 0$. Si nous avions laissé tomber le terme en cosinus au lieu du terme sinus. Les conditions pour $H$, serait tourné de $90^0$ dans le guide. En admettant que l'orientation du champ est arbitraire, chacune des solutions aurait suffi (mais chacune serait requise pour une onde à polarisation circulaire dans le guide).

\* Les conditions aux frontières $E_\Phi = 0$ à $r = r_0$ peuvent aussi être exprimées comme étant,

$$\boldsymbol{n} \times \boldsymbol{E} = 0 \text{ à } r = r_0$$

Les six composantes sont :

$$E_r = H_0 \frac{n\gamma Z_{r\Phi}}{k^2 r} \sin(n\Phi) J_n(kr) e^{-\gamma z} = Z_{r\Phi} H_\Phi$$

$$E_\Phi = H_0 \frac{\gamma Z_{r\Phi}}{k^2} \cos(n\Phi) \frac{dJ_n(kr)}{dr} e^{-\gamma z} = -Z_{r\Phi} H_r$$

$$E_z = 0 \quad \textit{requis pour mode TE}$$

$$H_r = -H_0 \frac{\gamma}{k^2} \cos(n\Phi) \frac{dJ_n(kr)}{dr} e^{-\gamma z}$$

$$H_\Phi = H_0 \frac{n\gamma}{k^2 r} \sin(n\Phi) J_n(kr) e^{-\gamma z}$$

$$H_z = H_0 \cos(n\Phi) J_n(kr) e^{-\gamma z}$$

Les expressions alternatives pour $E_\phi$ et $H_r$ peuvent être écrites en utilisant les relations de récurrence,

$$\frac{dJ_n(kr)}{dr} = k[\frac{n}{kr} J_n(kr) - J_{n+1}(kr)]$$

Les équations ci-dessus sont des expressions générales pour une onde mode *TE* dans un guide cylindrique. Toutefois, la condition limite $E_\Phi = 0$ à $r = r_0$ n'a pas été imposée. L'application de cette condition exige que,

$$\frac{dJ_n(kr)}{dr} = 0 \quad \text{à} \quad r = r_0$$

et donc que,

$$k = \frac{k'_{nr}}{r_0}$$

où $k'_{nr}$ est la nième racine de la dérivée de la fonction Bessel d'ordre *n*. Les trois premières racines de la dérivée de la fonction de Bessel du premier ordre se produisent à $kr_0 = k'_{nr} = 1.84$; 5.33 *et* 8.54 (Figures III.11-a et III.11-b). Ainsi, les racines sont $k_{11}'$=1.84, $k_{12}'$=5.33 $et$ $k_{13}'$=8.54 correspondantes à des modes d'onde $TE_{11}$, $TE_{12}$ $et$ $TE_{13}$ dans le guide. En général, un mode $TE$ dans un guide cylindrique est désigné $par$ $TE_{nr}$, où l'indice $n$ indique l'ordre de la fonction de Bessel et « $r$ » indique le rang de la racine. Pour les modes $TE_{01}$ et $TE_{02}$, $n$ = 0 et les racines de l'équation précédente pour ce cas sont $k_{nr}$=3.832 $et$ 7.016. Ce sont les mêmes que les zéros de $J_1(kr)$.

Note: $Z_{r\Phi} = \dfrac{E_r}{H_\Phi} = -\dfrac{E_\Phi}{H_r} = -\dfrac{Y}{Z} = j\dfrac{\mu\omega}{\gamma}$

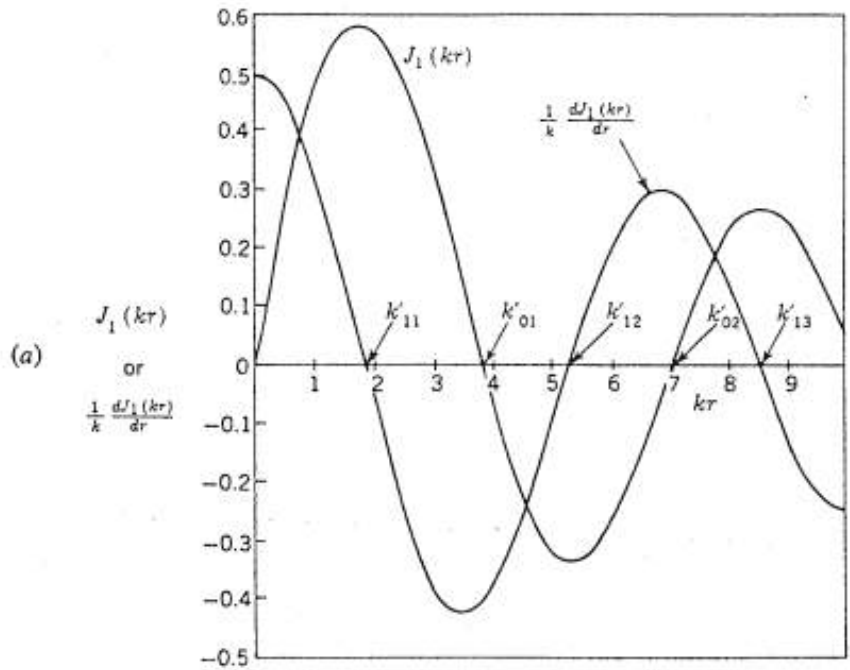

Figure III.11-a.

La Figure III.11-a présente la fonction de Bessel du premier ordre et de sa dérivée en fonction de $kr$. La Figure III.11-b présente les relations de la fonction de Bessel impliqués dans la détermination de $E_r$, et de la variation de $E_\Phi$ pour le mode $TE_{11}$.

Se référant aux équations précédentes, la variation de $E_\Phi$, avec $r$ à $\phi$=0 pour le mode $TE_{11}$ est tel que représenté dans la Figure III.11-$b$.

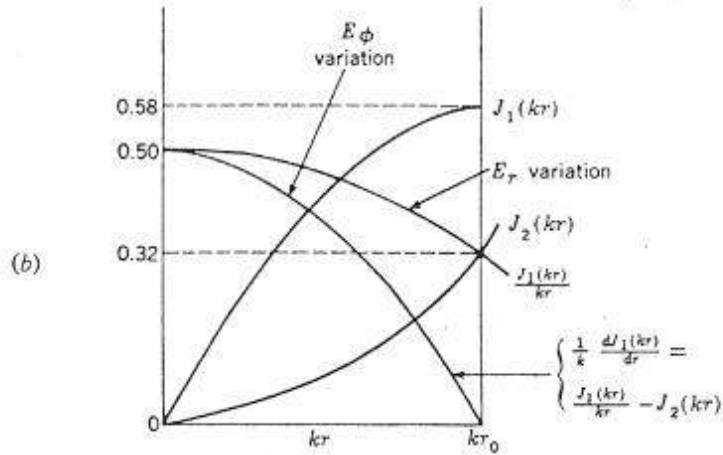

Figure III.11-b.

Des équations précédentes, la variation de $E_r$ avec $r$ à $\phi$=90° est donnée par la courbe de $J_1(kr)/kr$. Les amplitudes des deux composantes du champ, $E_r$ et $E_\Phi$ peuvent ainsi être représentées par des courbes, comme dans la Figure III.11-a, ou par des flèches, comme dans la Figure III.11- b.  Il faut noter que $E_\Phi$ est un maximum à $\phi$=0° $et$ 180° et de zéro à $\phi$=90° et 270°, tandis que $E_r$, est un maximum à $\phi$=90° $et$ 270° et de zéro à $\phi$=0° $et$ 180 °. On note également qu'au centre du guide ($r = 0$) $E_r$ et $E_\Phi$ tous les deux ont la même direction et l'amplitude. Lorsque les $E_r$ et $E_\Phi$ composants sont ajoutées, le champ électrique total de mode $TE_{11}$ dans le guide cylindrique est tel que suggéré par les lourdes flèches dans la Figure III.11-b avec la direction du champ électrique totale, comme indiqué par les lignes courbes (Figure III.12). La Figure III.12-a en haut présente une variation du champ électrique des composantes $E_r$ $et$ $E_\Phi$ avec une distance radial « $r$» pour le mode $TE_{11}$ dans un guide d'onde cylindrique. La Figure III.12-b présente des vecteurs de champ électrique pour le mode $TE_{11}$ dans un guide d'onde cylindrique avec la composante de champs $E_r$ $et$ $E_\Phi$. D'une manière similaire, la configuration champ magnétique transversal pour le mode $TE_{11}$ peut être déduite de la variation de $E_r$ $et$ $E_\Phi$ avec « $r$ » dont le résultat est montré dans la Figure III.12-a. La solution que l'on avait obtenu pour le mode $TE_{11}$ mais est l'un d'un nombre infini de modes d'ondes possibles dans un guide cylindrique.  Le ou les modes particuliers qui sont effectivement présents dans tous les cas, dépendent des dimensions de guide, la méthode d'exciter le guide, et des discontinuités dans le guide. Le champ résultant dans le guide est égal à la somme des champs de tous les modes actuels. La configuration du champ pour le mode $TE_{11}$ est illustrée à la Figure III.12-b.

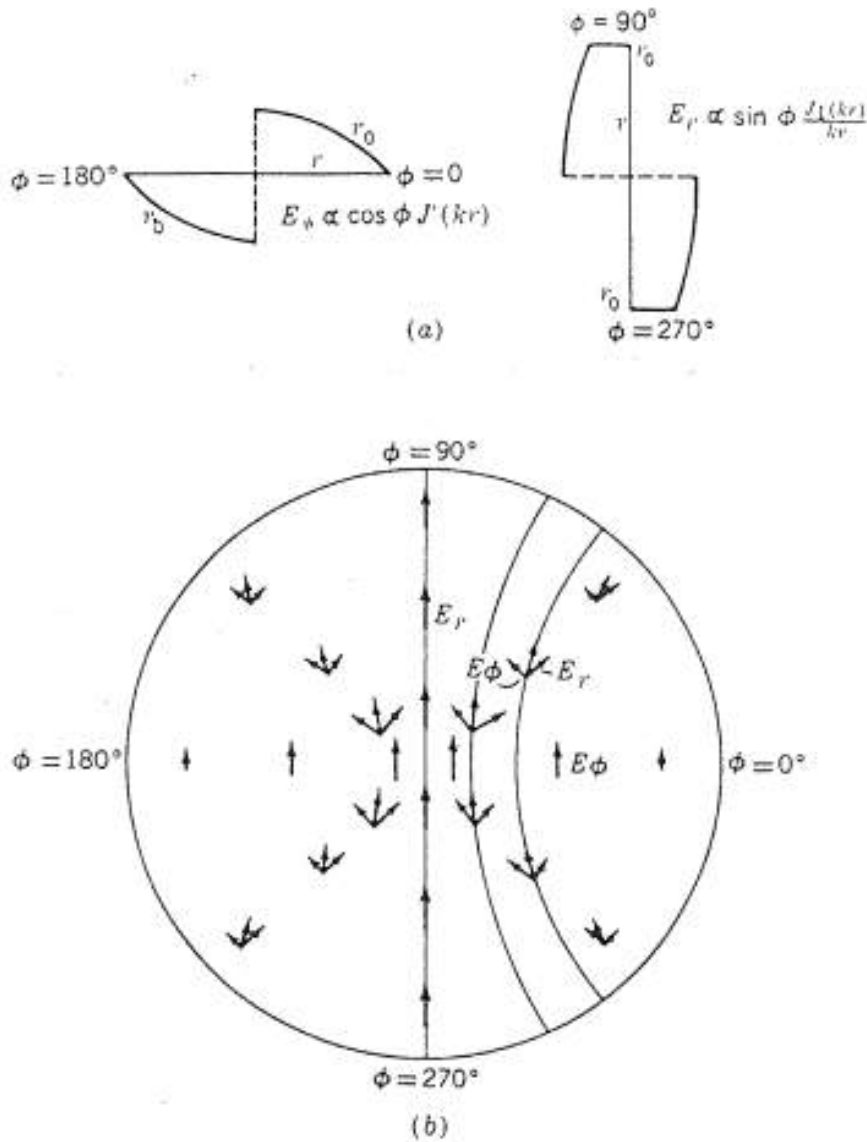

Figures III.12-a et III.12-b.

Jusqu'à présent, seuls les modes *TE* dans un guide cylindrique ont été envisagés. Pour trouver les relations sur le terrain pour les ondes TM-mode nous laissons $H_z$ = 0 et exprimer les composantes du champ restants en termes de $E_z$ de champ longitudinal. Une équation d'onde dans $E_z$ est développée et résolu de satisfaire la condition de limite.

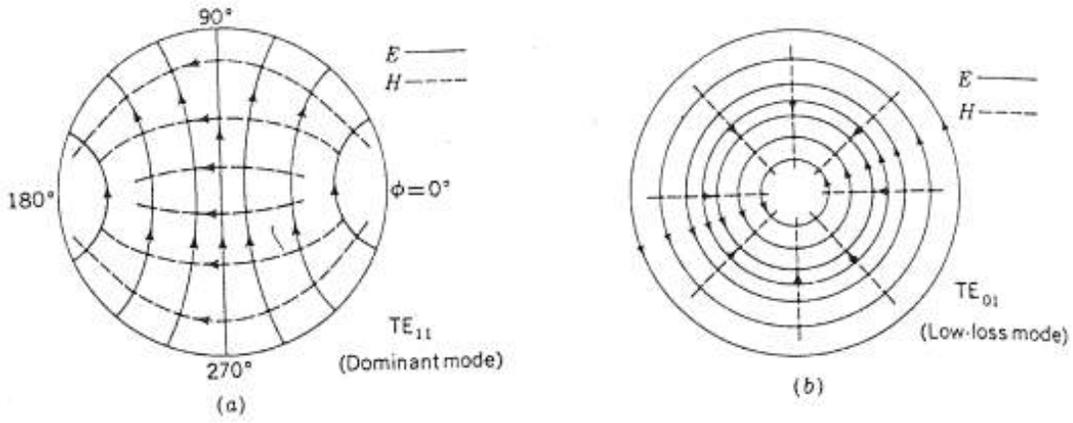

Figures III.13-a et III.13-b.

Les figures montrent les lignes de champ électrique (solide) et le champ magnétique alignés (en pointillés) du guide d'onde cylindrique (Figure III.13-a) pour le mode $TE_{11}$ (dominant) et (Figure III.13-b) pour le mode $TE_{01}$ (faible perte).

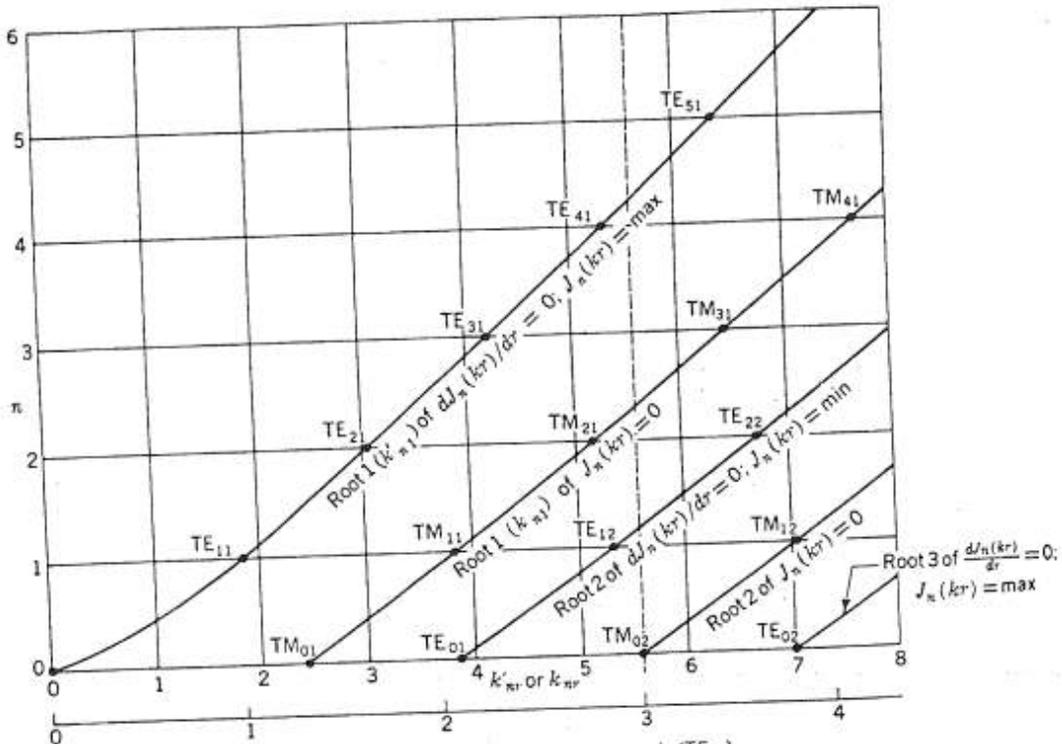

Figure III.14.

La Figure III.14 présente les modes possibles *TE* et *TM* d'un guide d'onde cylindrique en fonction de la fréquence. À trois fois la fréquence de coupure pour le mode $TE_{11}$, il y a neuf modes qui passent et un mode ($TM_{02}$) au seuil de la coupure.

Jusqu'à présent on a considéré seulement les modes *TE*. Pour les champs dans les modes *TM* on doit maintenant considérer $H_z = 0$ et exprimer le restant des composantes en fonction de $H_z$.
Une équation d'onde est développée pour $E_z$ et solutionnée pour satisfaire les conditions aux limites.

Ceci qui impose $J_n(kr) = 0$ *pour* $r=r_0$. Les trois premières racines $k_{nr}$ (= $kr_0$) pour le cas $n = 1$ sont $k_{11}$= 3.832, $k_{12}$=7.016 *et* $k_{13}$=10.173.
Les racines (ou valeurs propres) pour les modes *TM*, $k_{nr}$ écrites (non apprêté) correspondent à des valeurs nulles de la fonction de Bessel $J_n(kr)$, alors que les racines pour les modes *TE* écrites $k_{nr}'$ (prime) correspondent aux racines de la dérivée (par rapport à « $r$ ») de la fonction de Bessel. Les relations pour les modes *TE* et *TM* sont affichées dans la Figure III.15 et leurs valeurs numériques sont indiquées dans le tableau.

Comme on le voit dans la Figure ci-dessus, le mode $TE_{01}$ devrait logiquement être désigné par le mode $TE_{02}$ car il se réfère à la seconde racine de la dérivée de la fonction de Bessel. De même, le mode $TE_{02}$ devrait bien être désigné par $TE_{03}$. La première racine est $k_{nr}'$=0, correspondant à un mode trivial (inexistant). Cette incohérence dans la nomenclature des guides cylindriques a été soulignée par C-T Tai, mais les désignations communes et incorrectes persistent et le Tableau 2 et la Figure III.14 se conforment à cet usage commun.

Examinons maintenant les conditions nécessaires pour la propagation dans un guide d'onde cylindrique. En substituant l'équation en $k = \sqrt{\gamma^2 + \omega^2 \varepsilon \mu}$ et la résolution pour la constante de propagation, on obtient

$$\gamma = \pm \sqrt{\frac{k_{nr}'^2}{r_0^2} - \omega^2 \varepsilon \mu} = \alpha + j\beta$$

Comme dans le guide rectangulaire, il y a trois conditions:

TABLEAU 2. LES MODES DANS LES GUIDES D'ONDE CIRCULAIRE

| LE MODE * | Valeurs propres | | Longueur d'onde de coupure |
|---|---|---|---|
| | $k'_{nr}$ | $k_{nr}$ | $\lambda_{oc}$ |
| TM$_{01}$ | | 2.405 | 2.61r$_0$ |
| TE$_{01}$ (low loss) | 3.832 | | 1.64 r$_0$ |
| TM$_{02}$ | | 5.520 | 1.14r$_0$ |
| TE$_{02}$ | 7.016 | | 0.89r$_0$ |
| TE$_{11}$ (dominant) | 1.840 | | 3.41r$_0$ |
| TM$_{11}$ | | 3.832 | 1.64r$_0$ |
| TE$_{12}$ | 5.330 | | 1.18r$_0$ |
| TM$_{12}$ | | 7.016 | 0.89r$_0$ |
| TE$_{21}$ | 3.054 | | 2.06r$_0$ |
| TM$_{21}$ | | 5.135 | 1.22r$_0$ |
| TE$_{22}$ | 6.706 | | 0.94r$_0$ |
| TE$_{31}$ | 4.201 | | 1.49r$_0$ |
| TM$_{31}$ | | 6.379 | 0.98r$_0$ |
| TE$_{41}$ | 5.318 | | 1.18r$_0$ |
| TM$_{41}$ | | 7.588 | 0.83r$_0$ |
| TE$_{51}$ | 6.416 | | 0.98r$_0$ |

\* Les lettres $n_r$ comme dans $TE_{nr}$ ou $k_{nr}$ ont la signification suivante :

$n$ = fonction Bessel d'ordre $n$
$r$ = l'ordre de la racine de la fonction Bessel de nième ordre

1. Aux basses fréquences, $\omega$ faible, $\gamma$ réel, le guide opaque (ondes ne se propagent pas).
2. À une fréquence intermédiaire, $\omega$ intermédiaire, $\gamma = 0$, l'état de transition (coupure).
3. Aux hautes fréquences, avec un grand $\omega$, $\gamma$ imaginaire, un guide transparent (propage les ondes).

En mettant $\gamma=0$ dans l'équation ci-dessus, nous trouvons pour la fréquence de coupure et la longueur d'onde de coupure,

$$f_c = \frac{1}{2\pi\sqrt{\varepsilon\mu}}\frac{k'_{nr}}{r_0} \ (H_z)$$

$$\lambda_{0c} = \frac{2\pi r_0}{k'_{nr}} \ (m)$$

Pour le mode $TE_{11}$ $k_{nr}'=k_{11}'=1.84$, de sorte que $\lambda_{0c} = \frac{2\pi r_0}{k'_{nr}} = \frac{2\pi r_0}{1.84} = 3.41 r_0$. Ainsi, la longueur d'onde de coupure pour le mode $TE_{11}$ correspond à une longueur d'onde 3.41 fois le rayon du guide. La longueur d'onde de coupure pour différents modes dans un guide cylindrique est énumérée dans le Tableau 2.

Aux fréquences supérieures, au seuil de la coupure on a :

$$\beta = \sqrt{\omega^2\varepsilon\mu - \frac{k_{nr}'^2}{r_0^2}} \ (rad/m)$$

À partir des deux équations précédentes, nous avons pour une longueur d'onde dans le guide (selon la direction z).

$$\lambda_g = \frac{\lambda_0}{\sqrt{1 - \left(\frac{\lambda_0}{\lambda_{0c}}\right)^2}}$$

où

$\lambda_0$ = longueur d'onde dans un milieu sans bornes de même type qui occupe le guide.

$\lambda_{0c}$ = longueur d'onde de coupure.

Pour la vitesse de phase dans le guide,

$$v_p = \frac{\omega}{\beta} = \frac{v_0}{\sqrt{1 - \left(\frac{\lambda_0}{\lambda_{0c}}\right)^2}} \quad \left(\frac{m}{s}\right)$$

où

$$v_0 = \frac{1}{\sqrt{\varepsilon\mu}}$$

Ces équations sont identiques à ceux obtenus précédemment pour le guide d'onde rectangulaire. Elles s'appliquent également aux ondes dans les guides creux de section quelconque.

On notera que les racines $k_{nr}'$ (aussi appelée valeurs propres) ne sont pas espacés régulièrement, contrairement au cas du guide rectangulaire. La Figure III.14, donne les racines et les longueurs d'onde de coupure pour certains modes *TE* et *TM* dans un guide cylindrique. Ces modes sont également illustrés dans le Tableau 2. Le mode $TE_{11}$ se propage à une fréquence inférieure à n'importe quel autre mode (y compris les modes *TM*) et est donc appelé mode dominant un guide cylindrique. Le mode $TE_{01}$ est intéressant en raison de ses faibles caractéristiques d'atténuation dans les guides pratiques ayant une conductivité finie. Pour ce mode, l'atténuation diminue de façon monotone avec une fréquence croissante.

**Exemple III.9.** Un guide d'onde circulaire fonctionnant au mode $TE_{11}$ est suivi d'un guide d'onde rectangulaire dont la largeur *a* correspond au diamètre interne du guide circulaire (voir Figure III.15 pour la transition). Le diélectrique dans les guides est de l'air.

a) Si un signal de 1.5 ($GH_z$) passe du guide circulaire dans le guide rectangulaire dont les dimensions latérales sont *a* = 12 (*cm*) et *b* = 6 ($cm$), quel mode sera excité dans le guide rectangulaire pour une transmission avec peu de pertes ? Pourquoi ?

b) Calculer les longueurs d'onde dans chaque portion du guide en cas de transmission du signal de 1.5 ($GH_z$),

c) Pour le même signal et les mêmes modes de propagation dans les guides, on remplirait la partie cylindrique avec un diélectrique sans perte de façon à avoir la même longueur d'onde dans les deux sections de cette ligne de transmission. Calculer la valeur de la constante diélectrique pour cette opération.

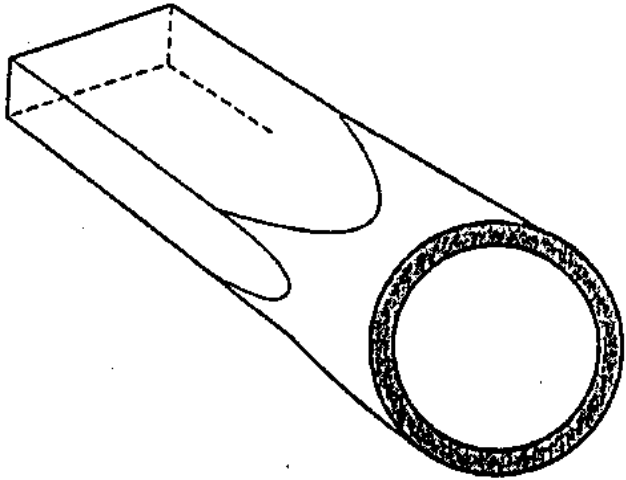

<p style="text-align:center;color:#29ABE2;">Figure III.15</p>

<span style="color:#29ABE2;">Solution :</span>

a) Pour le guide circulaire la fréquence de coupure du mode $TE_{11}$ est :

$$f_{cTE_{11}} = \frac{1}{2\pi\sqrt{\varepsilon\mu}} \frac{k'_{nr}}{r_0} = \frac{c}{2\pi} \frac{k'_{nr}}{r_0}$$

où  $k'_{nr} = 1.84$ (Voir le tableau 2), $r_0 = a/2 = 3$ ($cm$),

$$f_{cTE_{11}} = 1.464 \ GH_z$$

Pour le mode dominant $TE_{10}$ dans un guide d'onde rectangulaire la fréquence de coupure est :

$$f_{cTE_{10}} = \frac{1}{2\sqrt{\varepsilon\mu}}\frac{1}{a} = \frac{c}{2a} = 1.25 \ (GH_z)$$

et pour le mode $TE_{01}$ :

$$f_{cTE_{01}} = \frac{1}{2\sqrt{\varepsilon\mu}}\frac{1}{b} = \frac{c}{2b} = 2.5 \ (GH_z)$$

Ensuite,

$$f_{cTE_{11}} = f_{cTM_{11}} = \frac{c}{2}\sqrt{\left(\frac{1}{a}\right)^2 + \left(\frac{1}{b}\right)^2} = 2.795 \ (GH_z)$$

En inspectant d'autres modes on trouve que le seul mode possible qui se propage dans le guide rectangulaire est le mode dominant $TE_{10}$ avec $f_{cTE_{10}} = 1.5 \ (GH_z)$.

b)  Pour la partie circulaire au mode $TE_{11}$,

$$\lambda_{gTE_{11}} = \frac{\lambda_0}{\sqrt{1 - \left(\frac{\lambda_0}{\lambda_{0c}}\right)^2}} \ , \qquad \lambda_0 = \frac{c}{f} = 0.2 \ (m)$$

$$\lambda_{0c} = \frac{2\pi r_0}{k'_{nr}} = 0.20487 \ (m) \ \rightarrow \ \lambda_{gTE_{11}} = 0.92125 \ (m)$$

Dans le guide rectangulaire

$$\lambda_{gTE_{10}} = \frac{\lambda_0}{\sqrt{1 - \left(\frac{f_c}{f}\right)^2}} \ \text{avec} \ \frac{f_c}{f} = \frac{1.25}{1.5} = 0.833$$

ce qui donne

$$\lambda_{gTE_{10}} = 0.36181 \ (m)$$

c)

Il y aura le changement de $\lambda_g$ dans la partie cylindrique. Il faut vérifier d'abord que devient la fréquence de coupure avec cette nouvelle donnée.

$$f'_{cTE_{11}} = \frac{1}{2\pi\sqrt{\varepsilon\mu}}\frac{k'_{nr}}{r_0} = \frac{c}{2\pi\sqrt{\varepsilon_r}}\frac{k'_{nr}}{r_0}$$

Pour l'instant on ne connaît pas la permittivité relative $\varepsilon_r$. On calcule,

$$\lambda_{gTE_{11}} = 0.36181 = \frac{\lambda'_0}{\sqrt{1 - \left(\frac{\lambda'_0}{\lambda'_{0c}}\right)^2}}$$

$$\lambda'_{0c} = \frac{v}{f'_c} = \frac{v}{f'_{cTE_{11}}} \ , \quad v = \frac{1}{\sqrt{\varepsilon\mu}} = \frac{c}{\sqrt{\varepsilon_r}}$$

$$\lambda'_{0c} = \frac{c}{\sqrt{\varepsilon_r}}2\pi\frac{\sqrt{\varepsilon_r}}{c}\frac{r_0}{k'_{nr}} = 2\pi\frac{r_0}{k'_{nr}}$$

Donc, il n'y a pas de changement pour $\lambda'_{0c}$.

$$\lambda'_{0c} = \lambda_{0c}$$

$$\lambda'_0 = \frac{c}{f\sqrt{\varepsilon_r}}$$

$$\frac{\lambda'_0}{0.36181} = \sqrt{1 - \left(\frac{\lambda'_0}{\lambda_{0c}}\right)^2}$$

ce qui donne,

$$\lambda'^2_0 = \frac{0.36181^2}{1 - 0.36181^2} \ \rightarrow \lambda'_0 = \frac{c}{f\sqrt{\varepsilon_r}} = \frac{0.36181}{\sqrt{1 - 0.36181^2}}$$

$$\varepsilon_r = \frac{c^2}{f^2}\frac{1 - 0.36181^2}{0.36181^2} = 1.32781$$

**Exemple III.10.**

Un cylindre en aluminium de diamètre intérieur 1.952 ($cm$) sert de guide d'onde à 10 ($GH_z$).

a) Si on veut utiliser le mode dominant $TE_{11}$ ($k'_{nr}$ = 1.84) pour la transmission de ce signal, calculer la longueur d'onde le guide à la fréquence de 10 ($GH_z$),

b) On veut changer de mode de transmission et adopter le mode qui procure moindre perte. Est-ce que le mode $TE_{01}$ ($k'_{nr}$ = 3.832) convient? Sinon trouver la valeur de la constante $\varepsilon_r$ d'un diélectrique qu'il faut introduire pour avoir la même fréquence de coupure qu'en a) pour le mode $TE_{01}$.

Solution :

a) $2r_0$ = 1.952 ($cm$) ce qui donne $r_0$ = 0.976 ($cm$). Pour le mode $TE_{11}$ la fréquence de coupure est dans le vide,

$$f_{cTE_{11}} = \frac{c}{2\pi}\frac{k'_{nr}}{r_0} = 9.0014 \ (GH_z)$$

$$\lambda_{cTE_{11}} = \frac{2\pi r_0}{k'_{nr}} = 0.0333 \ (m), \qquad \lambda_0 = \frac{c}{f} = 0.3 \ (m),$$

$$\lambda_{gTE_{11}} = \frac{\lambda_0}{\sqrt{1 - \left(\frac{\lambda_0}{\lambda_{0c}}\right)^2}} = 6.883 \ (cm);$$

*b*)

$$f'_{cTE_{01}} = \frac{c}{2\pi}\frac{k'_{nr}}{r_0} = \frac{3x10^8}{2\pi}\frac{3.832}{0.00976} = 18.7464 \ (GH_z)$$

Non. Il faut y mettre un diélectrique. Donc,

$$f''_{cTE_{01}} = f_{cTE_{11}} = 9.0014 = \frac{18.7464}{\sqrt{\varepsilon_r}} \rightarrow \varepsilon_r = 4.337$$

# 4. Ondes qui se propagent entre deux plaques métalliques parallèles

## 4.1 Guides plaques parallèles (Figure III.16):

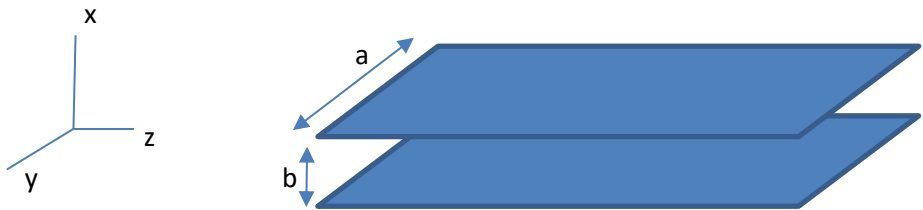

$b \gg a$  Il n'y a  pas de variation des champs selon y

Figure III.16.

Conditions aux frontières :

$$\vec{E}_{tang} = 0 \qquad \vec{H}_{norm} = 0$$

sur les surfaces conductrices.

Milieu entre les plaques : $\varepsilon \ et \ \mu - Cte$.

Simple ($\sigma = 0 \ et \ \varepsilon = réel$)

Équation de Maxwell,

$$\nabla \times \bar{H} = j\omega E$$

$$\nabla \cdot \bar{E} = 0$$

$$\nabla \times \bar{E} = -j\omega\mu H$$

$$\nabla \cdot \bar{H} = 0$$

Équation d'onde,

$$\nabla^2 \bar{E} + \omega^2 \mu \varepsilon \bar{E} = 0 \quad ou \quad \nabla^2 \bar{E} = -\beta^2 \bar{E}$$

$$\nabla^2 \bar{H} + \omega^2 \mu \varepsilon \bar{H} = 0 \quad ou \quad \nabla^2 \bar{H} = -\beta^2 \bar{H}$$

avec,

$$\beta = \omega \sqrt{\mu \varepsilon}$$

$$\frac{\partial H_z}{\partial y} - \frac{\partial H_y}{\partial z} = j\omega \varepsilon E_x \; ; \frac{\partial E_z}{\partial y} - \frac{\partial E_y}{\partial z} = -j\omega \mu H_x$$

$$\frac{\partial H_x}{\partial z} - \frac{\partial H_z}{\partial x} = j\omega \varepsilon E_y \; ; \frac{\partial E_x}{\partial z} - \frac{\partial E_z}{\partial x} = -j\omega \mu H_y$$

$$\underbrace{\frac{\partial H_y}{\partial x} - \frac{\partial H_x}{\partial y} = j\omega \varepsilon E_z}_{\nabla \times \bar{H} = j\omega \varepsilon E} \; ; \underbrace{\frac{\partial E_y}{\partial x} - \frac{\partial E_x}{\partial y} = -j\omega \mu H_z}_{\nabla \times \bar{E} = -j\omega \mu H}$$

On peut supposer que les plaques conductrices peuvent se prolonger jusqu'à l'infini sans changement dans les ondes : selon « $y$ » → $\partial(\dots)/\partial y = 0$.

Propagation selon « $z$ » avec une constante de propagation $\bar{\gamma} = \bar{\alpha} + j\bar{\beta}$

Avec $\bar{\alpha}$ et $\bar{\beta}$ réel (pour différencier de $\gamma$ qui est dans l'espace libre)

Deux cas spéciaux à distinguer,

1) $\bar{\alpha} = 0$
2) $\bar{\beta} = 0$

Donc,

$$Re\left\{ e^{-\bar{\gamma}z} e^{j\omega t} \right\} = \begin{cases} \cos(\omega t - \beta z) & \text{onde propageante} \\ \bar{e}^{\bar{\alpha}z} \cos(\omega t) & \text{onde évanescente} \end{cases}$$

$$\nabla^2 \bar{E} = -\omega^2 \mu\varepsilon\bar{E} \;\rightarrow\; \frac{\partial^2 \bar{E}}{\partial x^2} + \frac{\partial^2 \bar{E}}{\partial y^2} + \frac{\partial^2 \bar{E}}{\partial z^2} = -\omega^2 \mu\varepsilon\bar{E}$$

$$\nabla^2 \bar{H} = -\omega^2 \mu\varepsilon\bar{H} \;\rightarrow\; \frac{\partial^2 \bar{H}}{\partial x^2} + \frac{\partial^2 \bar{H}}{\partial y^2} + \frac{\partial^2 \bar{H}}{\partial z^2} = -\omega^2 \mu\varepsilon\bar{H}$$

avec,

$$\frac{\partial^2 (\dots)}{\partial y^2} = 0$$

Conditions frontières pour $x = 0$ et $x = a$

**Exemple** : $H_y(x, y)$

$$\text{On propose } H_y^0(x) \cdot e^{-\bar{\gamma}z}$$

Alors,

$$\frac{\partial}{\partial z} H_y(x, y) = -\bar{\gamma} H_y^0(x) e^{-\bar{\gamma}z} = -\bar{\gamma} H_y(x, z)$$

Donc,

$$\frac{\partial}{\partial z} \rightarrow -\bar{\gamma}$$

ce qui donne deux équations d'onde :

$$\frac{\partial^2 \bar{E}}{\partial x^2} + \bar{\gamma}^2 \bar{E} = -\omega^2 \mu\varepsilon\bar{E}$$

$$\frac{\partial^2 \bar{H}}{\partial x^2} + \bar{\gamma}^2 \bar{H} = -\omega^2 \mu\varepsilon\bar{H}$$

et les équations de Maxwell,

$$\bar{\gamma} E_y = -j\omega\mu H_x \quad ; \quad \bar{\gamma} H_y = j\omega\varepsilon E_x$$

$$-\bar{\gamma} - \frac{\partial E_z}{\partial x} = -j\omega\mu H_y \quad ; \quad -\bar{\gamma} H_x - \frac{\partial H_z}{\partial x} = j\omega\varepsilon E_y$$

$$\frac{\partial E_y}{\partial x} = -j\omega\mu H_z \qquad ; \qquad \frac{\partial H_y}{\partial x} = j\omega\varepsilon E_z$$

On peut exprimer les quatre composantes transverses des champs en fonction de deux composantes axiales (i.e. $E_z$ *et* $H_z$)  Définissons $h^2 = \bar{\gamma}^2 + \omega^2\mu\varepsilon$.

On trouve,

$$H_x = -\frac{\bar{\gamma}}{h^2}\frac{\partial H_z}{\partial x}$$

$$H_y = -\frac{j\omega\varepsilon}{h^2}\frac{\partial E_z}{\partial x}$$

$$E_x = -\frac{\bar{\gamma}}{h^2}\frac{\partial E_z}{\partial x}$$

$$E_y = \frac{j\omega\mu}{h^2}\frac{\partial H_z}{\partial x}$$

Catégories :

$$\text{Ondes type "transverse electric" } (TE) \begin{cases} E_z = 0 \\ H_z \neq 0 \end{cases}$$

$$\text{Ondes type "transverse magnetic" } (TM) \begin{cases} E_z \neq 0 \\ H_z = 0 \end{cases}$$

$$\text{Ondes type "transverse electromagnetic" } (TEM) \begin{cases} E_z = 0 \\ H_z = 0 \end{cases}$$

## 4.2 Ondes *TE*

$$E_z = 0 \;\; ; \;\; H_z \neq 0$$

$$\frac{\partial^2 E_y}{\partial x^2} + \bar{\gamma}^2 E_y = -\omega^2\mu\varepsilon E_y \rightarrow \frac{\partial^2 E_y}{\partial x^2} = -h^2 E_y$$

On suppose encore la fonction $E_y(x, y)$ comme multiplication de 2 fonctions indépendantes : $E_y(x, y) = E_y^0(x) \cdot e^{-\bar{\gamma}z}$

Alors on aura,

$$\frac{d^2 E_y^0}{dx^2} = -h^2 E_y^0(x)$$

ce qui donne comme solution :

$$E_y^0(x) = C_1 \sin(hx) + C_2 \cos(hx)$$

et finalement,

$$E_y(x, y) = [C_1 \sin(hx) + C_2 \cos(hx)]e^{-\bar{\gamma}z}$$

Avec les conditions aux frontières :

$$\forall \, y \; et \; \forall \, z \begin{cases} x = 0 & E_y = 0 \\ x = a & E_y = 0 \end{cases}$$

ce qui implique $C_2 = 0$

et

$$h = \frac{m\pi}{a} \qquad m = 1,2,3, \dots$$

$$E_y(x, y) = C_1 \sin(hx) \, e^{-\bar{\gamma}z}$$

Toutes les autres composantes peuvent être obtenues à partir de $E_z = 0$ et de $E_y$

Donc :
$$E_x = H_y = 0 \quad \text{et}$$

$$E_y = C_1 \sin\left(\frac{m\pi}{a}x\right) e^{-\bar{\gamma}z}$$

$$H_z = -\frac{1}{j\omega\mu}\frac{\partial E_y}{\partial x} = -\frac{m\pi}{j\omega\mu a} C_1 \cos\left(\frac{m\pi}{a}x\right) e^{-\bar{\gamma}z}$$

$$H_x = -\frac{\bar{\gamma}}{j\omega\mu} E_y = --\frac{\bar{\gamma}}{j\omega\mu} C_1 \sin\left(\frac{m\pi}{a}x\right) e^{-\bar{\gamma}z}$$

On a alors des modes plaques parallèles *TE*

$$m = \pm 1, \pm 2, \ldots$$

## 4.3 Ondes *TM*

On a,

$$H_z = 0 \text{ et } E_z \neq 0$$

On procède d'une façon similaire en utilisant l'équation d'onde pour $\overline{H}$ , ce qui donne pour $H_y$

$$H_y = [C_3 \sin(hx) + C_4 \cos(hx)]e^{-\overline{\gamma}z}$$

Pour appliquer les conditions tangentielles de $\overline{E} = E_z$ en fonction de $H_z$

$$E_z = \frac{1}{j\omega\varepsilon}\frac{\partial H_y}{\partial x} = \frac{h}{j\omega\varepsilon}[C_3 \cos(hx) - C_4 \sin(hx)]e^{-\overline{\gamma}z}$$

Conditions aux frontières :

$$E_z = 0 \text{ } pour \text{ } x = 0 \rightarrow C_3 = 0$$

$$E_z = 0 \text{ } pour \text{ } x = a \rightarrow h = \frac{m\pi}{a} \text{ } ; m = 0,1,2,3,\ldots$$

Parmi les composantes restantes seulement $E_x$ est différent de zéro. Donc on aura :

$$H_y = C_4 \cos\left(\frac{m\pi}{a}\right)e^{-\overline{\gamma}z}$$

$$E_x = \frac{\overline{\gamma}}{j\omega\varepsilon}\,H_y = \frac{\overline{\gamma}}{j\omega\varepsilon}C_4 \cos\left(\frac{m\pi}{a}\right)\,e^{-\overline{\gamma}z}$$

## 4.4 Ondes *TEM*

On remarque que contrairement aux modes *TE* les composantes des modes *TM* ne s'annulent pas pour *m*=0. Le mode *TM*₀ est un mode *TEM*.

$$H_y = C_4 e^{-\overline{\gamma}z}$$

$$E_x = \frac{\bar{\gamma}}{j\omega\varepsilon}\, C_4 e^{-\bar{\gamma}z}$$

$$E_z = 0$$

**Plaque parallèle mode *TEM*.**

## 4.5 Fréquence de coupure, vitesse de phase et longueur d'onde

Les ondes qui se propagent entre des plaques parallèles aux modes *TE* et *TM* présentent pour $\bar{E}$ et $\bar{H}$ une distribution sinusoïdale stationnaire selon « *x* » et n'importe quel plan « *xy* » est un plan où la phase demeure constante pour un « *z* » donné. Ce qui veut dire que les ondes forment une propagation d'une onde plane avec la vitesse de phase :

$$\overline{v_p} = \frac{\omega}{\bar{\beta}} \;\; avec \;\; \bar{\gamma} = j\bar{\beta}$$

On prend le cas de $E_y(x, z, t)$ $\;\; avec\; C_1$ réel

$$E_y(x, z, t) = C_1 \sin\left(\frac{m\pi}{a}x\right) \cos\left(\omega t - \bar{\beta}z\right)$$

Comme,

$$h^2 = \bar{\gamma}^2 + \omega^2\mu\varepsilon \rightarrow \bar{\gamma}, = \sqrt{h^2 - \omega^2\mu\varepsilon} = \sqrt{\left(\frac{m\pi}{a}\right)^2 - \omega^2\mu h}$$

avec « *h* » valable aussi bien pour *TE* et *TM*.

On peut en déduire les caractéristiques suivantes en considérant,

$$\bar{\gamma} = 0$$

on obtient,

$$f_{em} = \frac{mv_p}{2a} = \frac{m}{2a\sqrt{\mu\varepsilon}} \;\;\; avec\; v_p = \frac{1}{\sqrt{\mu\varepsilon}} + \overline{v_P}$$

et

$$\lambda_{cm} = \frac{v_p}{f_{cm}} = \frac{2a}{m}$$

Pour $f > f_{cm}$

$$\bar{\gamma} = j\beta_m = j\sqrt{\omega^2\mu\varepsilon - \left(\frac{m\pi}{a}\right)^2} = j\beta\sqrt{1 - \left(\frac{f_{cm}}{f}\right)^2}$$

$\bar{\beta}_m \neq \beta$ et dépend de l'indice « $m$ ». Elle est la constante de phase pour la propagation guidée entre deux plaques parallèles.

Pour $f < f_{cm}$  $\bar{\gamma}$ est réel

$$\bar{\gamma} = \bar{\alpha}_m = \sqrt{\left(\frac{m\pi}{a}\right)^2 - \omega^2\mu\varepsilon} = \beta\sqrt{\left(\frac{f_{cm}}{f}\right)^2 - 1}$$

Avec,

$$\beta = \omega\sqrt{\mu\varepsilon}$$

Par exemple pour le mode *TE* avec $f < f_{cm}$ et $C_1$ réel on aura

$$E_y(x,z,t) = C_1 e^{-\bar{\alpha}z} \sin\left(\frac{m\pi}{a}x\right)\cos(\omega t)$$

C'est une onde évanescente.

Pour $f > f_c$, l'onde qui se propage dans la structure à une longueur d'onde $\bar{\lambda}_m$ et une vitesse de phase $\bar{v}_{pm}$

$$\bar{\lambda}_m = \frac{2\pi}{\bar{\beta}_m} = \frac{\lambda}{\sqrt{1 - \left(\frac{f_{cm}}{f}\right)^2}}$$

$$\bar{v}_{pm} = \frac{\omega}{\bar{\beta}_m} = \frac{v_p}{\sqrt{1 - \left(\frac{f_{cm}}{f}\right)^2}}$$

où $\lambda = \frac{2\pi}{\beta} = \frac{v_p}{f}$ est la longueur d'onde dans l'espace libre avec $\mu$ *et* $\varepsilon$ inclus.

Remarque : pour le mode *TM*$_0$ ou *TEM*

$$\bar{v}_p = v_p = \frac{1}{\sqrt{\mu\varepsilon}} \quad \text{indépendante de la fréquence}$$

Pour ce mode on a aussi $\bar{\beta}_m = \beta$ et $\bar{\lambda}_m = \lambda$. Il n'y a pas de fréquence coupure pour le mode *TEM*.

### 4.6 Tension, courant et impédance pour le guide plaques parallèles

Considérer le cas TEM seulement,

$$V_{TEM}(z) = -\int_{x=0}^{a} \bar{E}\, d\bar{l} = \int_{a}^{0} E_x(z)dx$$

$$= \frac{\beta}{\omega\varepsilon} C_4 e^{-j\beta z} \int_{a}^{0} dx = V^+ e^{-j\beta z}$$

$$I_{TEM}(z) = |\bar{J}_s|(1m) = \left|\hat{n} \times \hat{H}\right| = \left|\mp \hat{x} \times \hat{y}[H_x]\right|_{x=0}$$

$$= \left|\hat{x} \times \hat{y}\, C_4 e^{-j\beta z}\right| = C_4 e^{-j\beta z} = \frac{V^+}{Z_{TEM}} e^{-j\beta z}$$

Pour *TEM*

$$\bar{\beta} = \beta = \omega\sqrt{\mu\varepsilon} \text{ et } V^+ = -\beta C_4 \frac{a}{\omega\varepsilon} = -\eta\, a\, C_4$$

Donc,

$$Z_{TEM} = \frac{E_x}{H_y} = \frac{\left[\frac{\beta}{\omega\varepsilon}\right] C_4 e^{-j\beta z}}{C_4 e^{-j\beta z}} = \sqrt{\frac{\mu}{\varepsilon}} = \eta$$

En considérant $V(z) = V_{TEM}$ *et* $I(z) = I_{TEM}$ et $Z_0 = Z_{TEM}$ toute l'analyse applicable au mode de transmission *TEM* pour pour les plaques parallèles.

Pour les autres modes *TE* et *TM* la définition d'une tension ou courant équivalent devient plus ambiguë. Prenons le cas des modes *TEM*, le champ électrique est dans la direction de « *y* », alors :

$$-\int_0^a \dot{\bar{E}} \cdot d\bar{l} = -\int_0^a \hat{y} E_y \cdot \hat{x} dx = 0$$

Pour le courant, on pourrait penser qu'il traversera la plaque dans la direction « *z* » mais il se trouve confiné dans la direction « *y* »!

$$\bar{J}_s = \hat{n} \times \hat{z}[H_z]_{x=0} = \hat{x} \times \hat{z}\left(-\frac{m\pi}{j\omega\mu a}\right)e^{-j\bar{\beta}z}$$

## 4.7 Plaques en parallèle – Atténuation due aux pertes ohmiques et diélectriques

### 4.7.1 Atténuation : pertes ohmiques

Sur les surfaces conductrices la densité de courant due aux champs magnétiques s'exprime :

$$\hat{n} \times \bar{H} = \bar{J}_s$$

La puissance moyenne varie en fonction du terme $e^{-2\alpha_c z}$ (avec $\alpha_c$ due à la conduction).

Donc,

$$-\frac{\partial P_{av}}{\partial z} = +2\alpha_c P_{av} = P_{loss}$$

La puissance dissipée par unité de longueur ($P_{loss}$) donne $\alpha_c$ :

$$\alpha_c = \frac{P_{loss}}{2P_{av}}$$

Pour *TEM − PP* (plaques en parallèle)

$$H_y = c_4 e^{-j\bar{\beta}z} \quad ; \quad E_x = \frac{\beta}{\omega\varepsilon} C_4 e^{-j\bar{\beta}z}$$

$$\bar{J}_s = \hat{n} \times \hat{H} \rightarrow J_s = C_4 e^{-j\bar{\beta}z}$$

d'où

$$|J_s| = C_4 \qquad \text{(positif, réel)}$$

Impédance de surface pour le métal

$$\eta_c = Z_s = R_s + jX_s = (\sigma\delta)^{-1}(1 + \bar{J})$$

avec

$$\delta = \sqrt{\frac{2}{\omega\mu_0\sigma}}$$

d'où

$$R_s = \sqrt{\frac{\omega\mu_0}{2\sigma}}$$

D'autre part,

$$\bar{S}_{av} = \frac{1}{2} Re\{\bar{E} \times \bar{H}^*\}$$

$$|E_x| = \eta|H_y| \text{ avec } C_4 \text{ supposé réel}$$

$$|\bar{S}_{av}| = \frac{1}{2}\eta C_4^2$$

Avec les dimensions $a$ $et$ $b$ la surface de la section transversale est $ab$, d'où,

$$P_{av} = \frac{1}{2}\eta C_4^2 ab$$

Donc, PP $-$TEM

$$\alpha_{c_{TEM}} = \frac{C_4^2 R_s b}{2\left(\frac{1}{2}\eta C_4^2 ba\right)} = \frac{R_s}{\eta\alpha} = \frac{1}{\eta\alpha}\sqrt{\frac{\omega\mu_0}{2\sigma}}$$

Considérons maintenant les ondes *TE − PP*,

$$\overline{H} \text{ tangentiel à } x = 0 \text{ et } x = \alpha \quad [H_z]$$

$$|J_{sy}| = |H_z| = \frac{m\pi C_1}{\omega\mu\alpha}$$

$$P_{loss} = 2 \int_0^1 \int_a^b \left[\frac{1}{2}|J_{sy}|^2 R_s\right] dy\, dz \ = \frac{2bm^2\pi^2 C_1^2 \sqrt{\frac{\omega\mu_0}{2\sigma}}}{2\omega^2\mu^2\alpha^2}$$

et

$$|\bar{S}_{av}| = \frac{1}{2} Re[\bar{E} \times \bar{H}^*] \cdot \hat{z} = -\frac{1}{2} E_y H_x^* = \frac{\bar{\beta} C_1^2}{2\omega\mu} \sin^2\left(\frac{m\pi}{a} x\right)$$

La puissance moyenne qui passe à travers d'une section large de 1 (*m*) et de hauteur:

$$P_{av} = \int_b^a \frac{\bar{\beta} C_1^2}{2\omega\mu} \sin\left(\frac{m\pi}{2} x\right) dx = \frac{\bar{\beta} C_1^2}{4\omega\mu}$$

Alors, la constante d'atténuation conductrice pour les modes *TE − PP*

$$\alpha_{c_{TEM}} = \frac{2m^2\pi^2 \sqrt{\frac{\omega\mu_0}{2\sigma}}}{\bar{\beta}\omega\mu\alpha^3} = \frac{2R_s \left(\frac{f_{cm}}{f}\right)^2}{\eta\alpha \sqrt{1 - \left(\frac{f_{cm}}{f}\right)^2}}$$

où

$$\bar{\beta} = \beta \sqrt{1 - \left(\frac{f_{cm}}{f}\right)^2}$$

### 4.7.2  Atténuation due aux pertes diélectriques

Si le milieu entre les a des pertes diélectriques on aura,

$$\varepsilon^* = \varepsilon' - j\varepsilon''$$

Pour le mode *TEM*, considéré toujours uniforme la substitution de $\sigma \ avec \ \omega\varepsilon''$ laisse prédire les champs électromagnétiques.

Pour les *TE* et *TM* on avait présumé,

$$\bar{\gamma}^2 \text{ étant réel } \left(\bar{\gamma}^2 = \bar{\alpha}^2 \ ou \ -\bar{\beta}^2\right)$$

Dans la plupart des cas les pertes sont minimes et les configurations des champs restent parallèles et les fréquences de coupure se calculent comme s'il n'y avait pas de pertes.

On considère,

$$\bar{\gamma} = j[\mu\varepsilon(\omega^2 - \omega_{cm}^2)]^{\frac{1}{2}}$$

En substituant $\varepsilon$ par $\varepsilon^* = e' - je''$ et quelques manipulations mathématiques on obtient,

$$\bar{\gamma} = \bar{a}_{dm} + j\bar{\beta}_{dm} \cong \frac{\omega\sqrt{\mu_0\varepsilon'}\dfrac{\varepsilon''}{\varepsilon}}{2\sqrt{1 - \left(\dfrac{\omega_{cm}}{\omega}\right)^2}} + j\sqrt{(\omega^2 - \omega_{cm})\mu\varepsilon'}$$

d'où pour $\varepsilon'' \ll \varepsilon'$

$$\bar{\beta}_{dm} = \sqrt{(\omega^2 - \omega_{cm}^2)\mu\varepsilon'}$$

et

$$\bar{a}_{dm} = \frac{\omega\sqrt{\mu_0\varepsilon'}\dfrac{\varepsilon''}{\varepsilon}}{2\sqrt{1 - \left(\dfrac{\omega_{cm}}{\omega}\right)^2}}$$

Pour *TEM* mode $\omega_{cm} = 0$, alors :

$$\alpha_d = \frac{\omega\varepsilon''}{2}\sqrt{\frac{\mu}{\varepsilon'}}$$

avec en général $\mu = \mu_0$

## 4.8  Lignes coaxiales – mode TEM

### 4.8.1 Les champs EM

C'est la ligne la plus utilisé en bas de 5 ($GH_z$). Des constructions spéciales peuvent permettre jusqu'à 20 ($GH_z$). Le diélectrique qui supporte le condensateur central est le polyéthylène ou téflon. Pour le mode *TEM* on considère seulement les surfaces conductrices du conducteur interne et externe. On utilise les coordonnées cylindriques qui conviennent à la géométrie. Donc on doit avoir,

$$E_\phi = 0 \ et \ H_r = 0 \ \ \text{à} \ r = a, b$$

On pourrait imaginer que des composants $E_\phi$ et $H_r$ puissent exister entre les conducteurs mais à ce moment-là il faut qu'ils puissent varier avec « r ». Cela est impossible car, par exemple : si $E_\phi$ varie avec « $r$ » on aurait une composante dans la direction « $z$ » selon

$$[\nabla \times \bar{E}]_z = \frac{1}{r}\left[\frac{\partial}{\partial r}\left(rE_\phi\right) - \frac{\partial E_r}{\partial \phi}\right]$$

ce qui est incompatible avec le mode *TEM*. Donc on a seulement

$$\bar{E} = \hat{r} \, E_r(r, z)$$

$$\bar{H} = \hat{\phi} \, H_\phi(r, z) \ \ \text{à } cause \ de \ la \ symétrie$$

Selon $\phi$ on suppose une propagation selon « $z$ ». Ceci donne pour les composantes transversales :

$$-\frac{\partial H_\phi}{\partial z} = j\omega\varepsilon E_r \rightarrow j\beta H_\phi^\circ(r) = j\omega E_r^\circ(r)$$

$$\frac{1}{r} \, H_\phi + \frac{\partial H_\phi}{\partial r} = 0 \rightarrow \frac{1}{r} \, H_\phi^\circ(r) + \frac{\partial H_\phi^\circ(r)}{\partial r} = 0$$

On a considéré les expressions de $\overline{H}$ et de $\overline{E}$ comme multiplication de deux fonctions indépendantes :

$$\overline{H} = \hat{\phi} \, H_\phi(r,z) = \hat{\phi} H_\phi^\circ(r) \cdot e^{-j\beta z}$$

$$\overline{E} = \hat{r} \, E_r^\circ(r) \cdot e^{-j\beta z} \quad \text{avec } \beta = \omega\sqrt{\mu\varepsilon}$$

Ces équations donnent tout de suite :

$$H_\phi^\circ(r) = \frac{H_0}{r} \to \overline{H} = \hat{\phi} \frac{H_0}{r} e^{-j\beta z}$$

$$\overline{E} = \hat{r} \, E_r(r,z) = \hat{r} \frac{H_0 z}{r} e^{-j\beta z}$$

## 4.8.2 Atténuation du mode TEM dans la ligne coaxiale

Les courants de surface prennent naissance en considérant les densités de courant générées par les champs magnétiques près des conducteurs.

$$\bar{J}_s(z) = \hat{n} \times \overline{H} = \begin{cases} \hat{z} \, H_\phi(x,z) \; cond.\,int.\,(\hat{n} = \hat{r}) \\ -\hat{z} \, H_\phi(b,z) \; cond.\,ext.\,(\hat{n} = -\hat{r}) \end{cases}$$

Ces courants vont causer des pertes dans les conducteurs.

Puissance transmise :

$$P_{av} = \frac{1}{2} Re \left[ \int_a^b \int_0^{2\pi} E_r H_\phi^* r d\phi dr \right]$$

$$= \pi\eta H_0^2 [lnr]|_a^b = \pi\eta H_0^2 \ln\left(\frac{b}{a}\right)$$

Puissance perdue $= P_{loss}(z)$ par unité de longueur,

$$P_{loss}\big|_{z=1}^{1m} = \frac{R_s}{2} \oint |J_s|^2 ds$$

$$= \frac{R_s}{2} \int_0^1 \left[ \int_0^{2\pi} \big|H_\phi(x,z)\big|^2 a d\phi \right.$$

$$\left. + \int_0^{2\pi} \big|H_\phi(b,z)\big|^2 b d\phi \right] dz = \pi R_S H_0^2 \left[ \frac{1}{a} + \frac{1}{b} \right]$$

d'où la constante d'atténuation :

$$\alpha_{c_{TEM}} = \frac{R_s}{2\eta la\left(\frac{b}{a}\right)} \left[ \frac{1}{a} + \frac{1}{b} \right] \; néper/m$$

$$\eta = \sqrt{\frac{\mu}{\varepsilon}}; \qquad R_s = \sqrt{\frac{\omega\mu}{2\sigma}}$$

En général $\mu = \mu_0$ pour les pertes diélectriques.

Pour ceci on peut utiliser l'expression générale obtenue pour les plaques parallèles et considérer toujours $f_c = 0$ (pas de coupure).

$$\bar{a}_d = \left[ \frac{\omega\sqrt{\mu_0\varepsilon'}\frac{\varepsilon''}{\varepsilon}}{2\sqrt{1 - \left(\frac{\omega_c}{\omega}\right)^2}} \right]_{\omega_c=0} = \frac{\omega\varepsilon''\sqrt{\mu_0}}{2\sqrt{\varepsilon'}} \approx \frac{\omega\varepsilon''\eta}{2}$$

où $\eta$ pour les matériaux avec peu de pertes,

$$\eta \approx \sqrt{\frac{\mu_0}{\varepsilon'}}$$

## 4.8.3 La tension, le courant et l'impédance caractéristique pour les lignes coaxiales

### Tension

$$V(z) = -\int_a^b \bar{E} \cdot d\bar{l} = -\int_a^b E_r \hat{r} \cdot \hat{r} dr = \int_a^b E_r(r,z) dr$$

$$= -\eta H_0 \ln\left(\frac{b}{a}\right) e^{-j\beta z} = V^+ e^{-j\beta z}$$

## Courant

$$I(z) = \oint \overline{H} \cdot d\overline{l} = \int_0^{2\pi} H_\phi(a,z) \, \hat{\phi} \cdot \hat{\phi} a d\phi$$

$$= 2\pi H_0 e^{-j\beta z} = \frac{2\pi V^+ e^{-j\beta z}}{\eta \ln\frac{b}{a}} = \frac{V^+}{Z_0} e^{-j\beta z}$$

## Impédance caractéristique $Z_0$

$$V^+ = \eta H_0 \ln\left(\frac{b}{a}\right)$$

$$Z_0 = \frac{\eta \ln\left(\frac{b}{a}\right)}{2\pi} \approx 60 \ln\frac{b}{a}$$

Puisque $\eta \approx 120\pi$, et si l'on remplace $V^+ = \eta H_0 \ln\left(\frac{b}{a}\right)$ dans les expressions de $P_{av}(z)$ et $P_{loss}(z)$ on aura :

$$P_{av}(z) = \frac{1}{2}\frac{|V^+|^2}{Z_0} \quad et$$

$$P_{loss}(z) = \frac{R_s |V^+|^2}{4\pi Z_0^2}\left(\frac{1}{a} + \frac{1}{b}\right)$$

# 5. Problèmes

**Exemple 1.** Dans un guide d'onde rectangulaire rempli d'air des dimensions $a$ = 5 ($cm$) et $b$ = 2 ($cm$) la fréquence d'opération est de 15 ($GH_z$). Si la composante du champ électrique,

$$E_z = 20 \sin(40\pi x)\sin(50\pi y)e^{-j\beta z} \left(\frac{V}{m}\right). \text{ Trouver:}$$

a) Le mode de propagation,

b) La fréquence de coupure,

c) La constante de phase,

d) La longueur d'onde,

e) L'impédance d'onde pour ce mode,

f) La vitesse de phase,

g) La vitesse de groupe.

Solution:

[a) $TM_{21}$; b) 9.6 ($GH_z$); c) 241.4 ($rad/m$); d) 0.026 (m); e) 289.67 ($\Omega$);

f) 3.9($m/s$); g) 2.31($m/s$)].

**Exemple 2.** Dans un guide d'onde carré avec $a$ = 1.2 (cm) le champ électrique est,

$$E_x = -10 \sin\left(\frac{2\omega y}{a}\right) \sin\left(\omega t - 150z\right)\left(\frac{V}{m}\right). \text{ Trouver:}$$

a) Le mode de propagation,

b) La fréquence de coupure,

c) La fréquence d'opération,

d) Les composantes de champ $H$ et $E$.

Solution:

[a) $TE_{02}$; b) $f_{c02}$ = 25 $GH_z$; c) $f$ = 26 $GH_z$; d) $E_y = E_z = H_x$ = 0,

$$H_y = -7.286 \sin\left(\frac{\pi y}{a}\right) \sin(\omega t - 150z) \; x10^{-3} \left(\frac{A}{m}\right)),$$

$$H_z = 25.43 \cos\left(\frac{2\pi y}{a}\right) \cos(\omega t - 150z) \; x10^{-3} \left(\frac{A}{m}\right)]$$

**Exemple 3.** Dans un guide d'onde rectangulaire rempli d'air, la fréquence de coupure d'un mode $TE_{10}$ est de 5($GH_z$) tandis que celle du mode $TE_{01}$ est de 12 ($GH_z$). Trouver :

a) Les dimensions du guide $a$ et $b$,

b) Les fréquences de coupure des trois prochaines modes $TE$ supérieurs.

Solution: [a) $a = 3(cm)$, $b = 1.25$ ($cm$); b) $TE_{20}$ -10 ($GH_z$), $TE_{11}$ - 13 ($GH_z$), $TE_{30}$ - 15 ($GH_z$)]

**Exemple 4.** Un guide rectangulaire ayant des dimensions internes $a$ = 15 ($cm$) et $b$ = 8($cm$) est utilisé dans le mode $TE_{32}$. On choisit une fréquence d'opération comme étant 5 ($GH_z$).

a) Si le guide est rempli d'air, quelles sont la fréquence de coupure et la longueur d'onde dans le guide à ce mode?

b) Si la fréquence de 5 ($GH_z$) à la fréquence de l'onde porteuse transmise dans le guide et qu'on module cette onde porteuse en fréquence, quelle largeur de bande utile peut-on espérer obtenir tout en conservant le mode $TE_{32}$?

c) Si l'on introduit dans le guide un diélectrique sans perte, de constante diélectrique relative $\varepsilon_r = 1.44138$ , que deviendra la bande en question en b).

d) Dans la condition en c) quelle est la plus petite valeur qu'on peut envisager pour la dimension $b$ du guide pour continuer à fonctionner au mode $TE_{32}$ avec une fréquence différente de 5 ($GH_z$) ?

Solution:

[a) $f_c$ = 4.8023 ($GH_z$), $\lambda_g$ =21.553 ($cm$); b) $f \pm (f - fc)$, 4.8023 ($GH_z$) $< f$

$< 5.198 \ (GH_z); \quad L.d.B = 2x1.98 \ (MH_z),$ attantion: il faut verifier

s'il y a un mode avec $f_c \ < \ 5.198 \ (GH_z);$ c) $f_c' = \ 4 \ (GH_z),$

$N.B. \rightarrow 4 \ (GH_z) < f \ < 4.567 \ (GH_z), \ L.d.B = 567(MH_z); \ d) \ b' = 11.3 \ (cm)].$

**Exemple 5.** Un guide d'onde rectangulaire rempli d'air de dimensions $a$ = 4 ($cm$) et $b$ = 2 ($cm$) transporte l'énergie en mode dominant à un taux de 2 ($mW$). Si la fréquence d'opération est de 10 ($GH_z$), trouver :

>    a) Le mode de propagation,
>    b) La fréquence de coupure,
>    c) L'impédance du mode de propagation,
>    d) La valeur crête du champ électrique dans le guide.

**Solution:**

[a) $TE_{10}$; b) $f_c$ = 3.75 ($GH_z$); c) $Z_{TE}$ = 406.7 ($\Omega$); $d$) $P_m = E_0^2 ab/(4Z_{TE}) \rightarrow E_0$ =63.77 ($V/m$)].

**Exemple 6.** Un guide d'onde rectangulaire rempli d'air de dimensions $a$ = 6 cm et $b$ = 3 cm. Étant donné que le champ électrique est,

$$E_z = \ 5 \sin\left(\frac{2\pi x}{a}\right) \sin\left(\frac{3\pi y}{b}\right) \sin\left(10^{12}t - \beta z\right)\left(\frac{V}{m}\right). \quad \text{Trouver :}$$

>    a) Le mode de propagation,
>    b) La fréquence d'opération,
>    c) La fréquence de coupure,
>    d) La constante de phase,
>    e) La vitesse de groupe,
>    f) La longueur d'onde,
>    g) L'impédance du mode de propagation,
>    h) La puissance moyenne transmise dans le guide.

**Solution:**

[a)$TM_{23}$; b) $f$ =15.92 ($GH_z$); c) $f_c$ = 15.81 ($GH_z$); d) $\beta = 391.2 \left(\frac{rad}{m}\right)$;

e) $v_p = 25.56x10^8 \left(\frac{m}{s}\right)$; f) $\lambda_g = $ ; g) $Z_{TM} = 44.24(\,\Omega)$; h) $P_m = 0.25(\frac{mW}{m^2})$]

**Exemple 7.** Un guide d'onde rectangulaire en cuivre ($\sigma_{cu} = 5.8x10^7 \ (S/m)$) fonctionne à une fréquence de 4.8 ($GH_z$). Si le guide est rempli de polystyrène ($\sigma_d = 10^{-17} \ (S/m)$, $\varepsilon = 2.55\varepsilon_0$) et ses dimensions $a$ = 4.2 ($cm$) et $b$ = 2.6 ($cm$), calculer :

    a) L'atténuation causée par les pertes diélectrique,

    b) L'atténuation causée par les pertes ohmiques,

    c) L'atténuation totale.

Le mode d'opération est $TE_{10}$.

Solution:

[a] $f_c$ = 2.234 $GH_z$; $Z_{TE}$ = 236.1($\Omega$), $\frac{\sigma_d}{\omega\varepsilon} = 1.47x10^{-17} \ll 1$

(milieu dielectrique sans pertes), $\alpha_d = 1.334x10^{-15} \left(\frac{Np}{m}\right)$; b) $R_s = 1.808x10^{-2}(\Omega)$ , $\alpha_c = 4.218x10^{-5} \left(\frac{Np}{m}\right)$; c) Étant donné que $\alpha_d \ll \alpha_c$, on conclue que les pertes due à la conductivité finie des parois de guidage sont plus importantes que les pertes due au milieu diélectrique,

$$\alpha_t = \alpha_c = 4.218x10^{-5} \ (Np/m) \ ]$$

**Exemple 8.** Dans des conditions normales, le champ électrique maximal avant le claquage est autour de 29 ($kV/cm$) dans l'air. Un guide d'onde rectangulaire de dimensions $a$ = 0.010668 ($m$), $b$ = 0.004318 ($m$) opère entre 16 ($GH_z$) et 25 ($GH_z$) dans le mode dominant. Le guide est rempli d'air. Estimer :

    a) La fréquence de coupure,

    b) Les impédances du mode dominant pour les fréquences 16 ($GH_z$) et 25 ($GH_z$),

    c) La puissance maximale transportée.

Solution:

[a] $f_c$ = 14.06 ($GH_z$) ; b) $f_1$ = 16 ($GH_z$), $Z^{(1)}_{TE}$ = 789.9 ($\Omega$); $f_2$ = 26.5 ($GH_z$), $Z^{(1)}_{TE}$ = 444.76 ($\Omega$); c) 122.6 ($kW$) $< P_t <$ 217.76 ($kW$)]

**Exemple 9.** Un guide d'onde circulaire rempli d'air a le diamètre interne $d_0 = 0.06096$ ($m$). Déterminer la bande passant pour fonctionner en monomode $TE$.

Solution:

[Le mode $TE_{11}$ possède la fréquence de coupure la plus basse. Il constitue le mode fondamental $TE$ ou $TM$. Le mode suivant est le mode $TM_{10}$ qui possède la seconde fréquence de coupure la plus basse. C'est le mode fondamental en $TM$ seulement. Le mode $TE_{21}$ est le mode suivant en $TE$. Les fréquences de coupure pour les modes mentionnés sont: $f_{cTE_{11}} = 2.884$ ($GH_z$) ,$f_{cTM_{10}} = 3.767$ ($GH_z$), $f_{cTE_{21}} = 4.784$ ($GH_z$). Le guide fonctionne en monomode $TE$ :$2.884$ ($GH_z$) $< f < 4.784$ ($GH_z$)]

**Exemple 10.** Un signal de 10 ($GH_z$) est transmis dans un guide d'onde circulaire rempli d'air.

    a) Sachant que la fréquence de coupure la plus basse du guide est inférieure de 20% à la fréquence du signal transmis, trouver le diamètre du guide,

    b) Si le guide fonctionne à 15 ($GH_z$), quels modes sont autorisés à se propager?

Solution:

[a)$d_0 = 2$ ($cm$) ; b) $TM_{01}$, $TE_{11}$, $TE_{21}$].

**Exemple 11.** Concevoir un guide d'onde circulaire rempli d'air de telle sorte que le mode dominant se propage sur une bande passant de 10 ($GH_z$).

Solution:

[La fréquence de coupure du mode $TE_{11}$ est le plus bas limite de la bande passante, $f_{cTE_{11}} = \frac{ck'_{nr}}{2\pi a} = \frac{ck'_{13}}{2\pi a} = \frac{cx1.841}{2\pi a}$. Le mode suivant est le mode avec la fréquence de coupure $f_{cTM_{01}} = \frac{ck_{nr}}{2\pi a} = \frac{ck_{1r}}{2\pi a} = \frac{cx2.405}{2\pi a}$ , $BW = f_{cTM_{01}} - f_{EcTM_{11}} = \frac{c}{2\pi a}(2.405 - 1.841) = 10$ ($GH_z$) $\rightarrow a = 2.69$ ($cm$), Donc, $f_{cTE_{11}} = 32.7$ ($GH_z$) et $f_{cTM_{01}} = 42.76$ ($GH_z$)].

**Exemple 12.** Pour un guide d'onde circulaire le mode fondamental est le mode $TE_{11}$. Le second mode est le mode $TM_{01}$. Si le rayon du guide circulaire est $a$ = 3 ($cm$),

   a)  Quelle serait la bande passante?
   b)  Quelle serait la bande passante fractionnelle?

Solution:

[a) $BW = f_{cTM_{01}} - f_{cTM_{11}} = \frac{c}{2\pi a}(2.405 - 1.841) = 0.897\ (GH_z)$,

b) $BW_{fr} = 2\frac{BW}{f_{cTM_{11}}+f_{cTM_{01}}} = 2\frac{f_{cTM_{01}}-f_{cTM_{11}}}{f_{cTM_{11}}+f_{cTM_{01}}}\ = 0.266 = 26.6\%$].

**Exemple 13.** Considérer un guide d'onde circulaire rempli d'air avec un rayon de 1 ($cm$) et fonctionnant à 19 ($GH_z$).

   a)  Quels modes sont autorisés à se propager?
   b)  Obtenir l'expression du vecteur de Poynting du mode $TM_{01}$.
   c)  Déterminer la gamme de fréquences maximale d'un signal qui se propage uniquement dans ce mode.

Solution:

[a) $TM_{01}$, $TE_{01}$, $TE_{21}$, $TM_{11}$ et $TE_{11}$ ;   b) $f_{cTE_{11}} = \frac{cx1.841}{2\pi a} = 8.79\ (GH_z)$, $f_{cTM_{01}} = \frac{cx2.405}{2\pi a} = 11.48\ (GH_z)$ La plage de fréquences de 8.79 ($GH_z$) à 11.48 ($GH_z$)].

**Exemple 14.** Une impulsion rectangulaire étroite superposée à un support avec une fréquence de 9.5 ($GH_z$) a été utilisé pour exciter tous les modes possibles dans un guide d'onde creux avec $a$ = 3 ($cm$) et $b$ = 2 ($cm$). Si le guide mesure 100 ($m$) de long, combien de temps cela prendre-t-il des modes excités pour arriver à la réception?

Solution:

[Avec les valeurs données on obtient $f_{10} = 5\ (GH_z), f_{01} = 7.5\ (GH_z), f_{10} = 9.01\ (GH_z), f_{10} = 10\ (GH_z)$. L'impusion avec une porteuse de 9.5 ($GHz$) peut exciter les trois premiers modes. Leurs vitesse de groupe sont:

$v_{g_{TE10}} = 2.55x10^8 \left(\frac{m}{s}\right)$, $v_{g_{TE11}} = 1.84x10^8 \left(\frac{m}{s}\right)$, $v_{g_{TE01}} = v_{g_{TM11}} = 0.95x10^8 \left(\frac{m}{s}\right)$. $T = \frac{l}{v_{gi}}$, $i = 1,2,3.$].

**Exemple 15.** Un guide d'onde à plaques parallèles séparées de 1 ($cm$) est rempli d'un diélectrique à permittivité relative à 12 ($GH_z$). Le champ électrique à l'intérieur du guide est caractérisé par,

$$E_z = 40 \sin(100\pi y)\, e^{-j\beta z} \left(\frac{V}{m}\right)$$

Déterminer :
a) Le mode qui se propage et sa fréquence de coupure ?
b) La constante de phase,
c) La vitesse de phase,
d) $\bar{E}$ et $\bar{H}$,
e) $< \bar{P}_{moy} >$.

Solution:

[a] $TM_1$, $f_c$ = 7.5 ($GH_z$); b) $\beta = 392.4 \left(\frac{rad}{m}\right)$; c) $v_p = 1.92x10^8 \left(\frac{m}{s}\right)$;

d) $\bar{E} = 40[-j1.25 \cos(100\pi y)\,\hat{y} + \sin(100\pi y)\,\hat{z}]e^{-j392.4z} \left(\frac{V}{m}\right)$,

$\bar{H} = j0.34 \cos(100\pi y)\, e^{-j392.4z}\hat{x} \left(\frac{A}{m}\right)$;

e) $< \bar{P}_{moy} >= 8.5[\cos(100\pi y)]^2 \hat{z} \left(\frac{W}{m^2}\right)]$.

**Exemple 16.** Un guide d'onde circulaire de rayon $a$ =3 ($cm$) rempli de polystyrène ($\varepsilon_r = 2.56$) est utilisé à une fréquence de 2($GH_z$). Pour le mode dominant $TE_{mn}^z$ déterminer :
a) La fréquence de coupure,
b) La longueur d'onde dans le guide.
c) Le rapport de la longueur d'onde dans le guide et celle dans un milieu infinie rempli de polystyrène,
d) La constante de phase,
e) L'impédance d'onde,
f) La bande passante en mode monomode, en supposant uniquement les modes $TE^z$.

**Solution:**

[a)]Mode dominant $TE_{11}$ $f_c$=1.832($GH_z$); b) $\lambda_{guide} = 0.2334\ (m)$,

c) $\dfrac{\lambda_{guide}}{\lambda_{inf}} = 2.49$, d) $\beta = 269.2\ \left(\dfrac{rad}{m}\right)$; e) $Z_{TE_{11}} = 586.56(\Omega)$,

f) Le mode suivant d'ordre supérieur est $TE_{21}$ la bande passante du fonctionnement en mode $TE_{11}$ unique est $BW = \dfrac{3.0542}{1.8412} = \dfrac{1.6588}{1}$ ].

# 6. Littérature

1) J.D.Jackson, "Classical Electrodynamics," John Wiley and Sons Inc., 3rd Edition, New York, 1998.

2) M. F. Iskander: Electromagnetic Fields and Waves, PRENTIC HALL, Englewood Cliffs, New Jersey.

3) J. D. Kraus: Electromagnetics, Third edition McGraw-Hill Book Company.

4) D. M. Pozar: Microwave Engineering, Second Edition, JOHN WILEY & SONS, INC.

5) D. K. Cheng: Field and Wave Electromagnetics, Second Edition, Pearson new International Edition.

6) C. A. Balanis: Advanced Engineering Electromagnetics, Second Edition, Arizona State University, WILEY, John Wiley & Sons, Inc.

7) T. Ditchi: Éléments d'Électromagnétisme et Antennes, UPMC, Sorbonne Universités.

8) J. Mevel: Les Ondes Électromagnétiques, Presse Universitaires de France.

9) Maria Inês Barbosa de Carvahlho: Problems on Plane Wave Incidence, Waveguides and Transmission Lines, FEUP, Faculdado de Engenharia Universidado du Porto, November 2008.

10) B. D. Braaten: Applied Electromagnetics - ECE 351, North Dakota State University, Fargo, ND, USA, 2017.

11) D. Grenier: Électromagnétisme et transmission des ondes, Université Laval, Québec, 2019.

12) S. J. Orfanidis: Electromagnetic Waves and Antennas, Rutgers University Copyright © 1999–2016.

13) K.Kagonyera: Chapter 8 : Reflection, Transmission and Waveguies, https://files.transtutors.com/cdn/uploadassignments/738670_4_45793-ulabyismch08.pdf.

14) J. E. Schutt-Aine: Advanced Microwave Measurements. Circular and Coaxial Waveguides, University of Illinois.

15) S. Babic et C. Akyel : Logiciel Adaptation (Matlab), 2019.

16) Abramowitz, M. and I. A. Stegun: Handbook of Mathematical Functions, National Burea of Standards Applied Mathematics, Series 5, Washington DC, Dec. 1972.

# 7. Constantes, unités et grandeurs

Dans ce manuel on donne des constantes physiques, unités électriques que l'on utilise en SI.

## Constantes physiques

| Grandeur | Symbol | Valeur | Unité |
|---|---|---|---|
| Vitesse de la lumière dans le vide | $c$ | 299 792 458 | (m/s) |
| Permittivité du vide | $\varepsilon_0$ | 8.854 187 817 $\times 10^{-12}$ | (F/m) |
| Perméabilité du vide | $\mu_0$ | $4\pi x 10^{-7}$ | (H/m) |
| Impédance caractéristique | $Z_0$ | 376.730 313 461 | ($\Omega$) |
| | | | |

## Unités électriques

| Grandeur | Symbol | Unité | Symbol |
|---|---|---|---|
| Tension | $U$ | Volt | (V) |
| Courent électrique | $I$ | Ampère | (A) |
| Puissance | $P$ | Watt | (W) |
| Résistance (Impédance) | $R, Z$ | Ohm | ($\Omega$) |
| Capacité | $C$ | Farad | (F) |
| Inductance | $L$ | Henri | (H) |
| Période | $T$ | Seconde | (s) |
| Fréquence | $f$ | Hertz | ($H_z$) |
| Énergie | $E$ | Joule | (J) |
| Champ électrique | $\bar{E}$ | Volt par mètre | (V/m) |
| Champ magnétique | $\bar{P}$ | Ampère par mètre | (A/m) |

## Grandeurs physiques et unités de mesure associées

| Longueur ($l$) | Unité de longueur (m, cm, mm, $\mu$m, $n$m, $p$m) |
|---|---|
| Temps ($t$) | Unité de temps (seconds) |

# Spectre RF

| | Désignations de bande | Fréquence | Longueur d'onde |
|---|---|---|---|
| ELF | Fréquence extrêmement basse | 30–300 Hz | 1–10 Mm |
| VF | Fréquence vocale | 300–3000 Hz | 100–1000 km |
| VLF | Très basse fréquence | 3–30 kHz | 10–100 km |
| LF | Basse fréquence | 30–300 kHz | 1–10 km |
| MF | Fréquence moyenne | 300–3000 kHz | 100–1000 m |
| HF | Haute fréquence | 3–30 MHz | 10–100 m |
| VHF | Très haute fréquence | 30–300 MHz | 1–10 m |
| UHF | Ultra haute fréquence | 300–3000 MHz | 10–100 cm |
| SHF | Super haute fréquence | 3–30 GHz | 1–10 cm |
| EHF | Extrêmement haute fréquence | 30–300 GHz | 1–10 mm |
| | Submillimètre | 300-3000 GHz | 100–1000 µm |

# Bandes micro-ondes

| Bande | Fréquence |
|---|---|
| L | 1–2 GHz |
| S | 2–4 GHz |
| C | 4–8 GHz |
| X | 8–12 GHz |
| Ku | 12–18 GHz |
| K | 18–27 GHz |
| Ka | 27–40 GHz |
| V | 40–75 GHz |
| W | 80–100 GHz |

# Spectre visible

| Couleur | Longueur d'onde | Fréquence |
|---|---|---|
| Rouge | 780–620 nm | 385–484 THz |
| Orange | 620–600 nm | 484–500 THz |
| Jaune | 600–580 nm | 500–517 THz |
| Verte | 580–490 nm | 517–612 THz |
| Blue | 490–450 nm | 612–667 THz |
| Violette | 450–380 nm | 667–789 THz |